\pdfoutput=1

\documentclass[12pt]{article}
\usepackage{amssymb,latexsym,amsmath,color}
\usepackage{epsfig}

\def\modu{|\hbox{\bf u}|}

\def\previewstrip#1#2{
\begin{figure}
\begin{center}
   \leavevmode{\hbox{\epsfig{
      figure=#1.pdf, scale=0.22
   }}}
   \caption{\label{#1}#2}
\end{center}
\end{figure}
}

\newcommand{\rem}[1]{}

\def\figfour#1#2#3#4#5{
\begin{figure}
\begin{center}
   \leavevmode{\hbox{\epsfig{
       figure=#1.pdf, scale=0.30
   }}}
   \caption{
      \label{#1}
      Evolution of the #2 ``#3'' initial velocity profile with $\alpha=#4$.#5
   }
\end{center}
\end{figure}
}

\def\figfourthree#1#2#3#4#5{
\begin{figure}
\begin{center}
   \leavevmode{\hbox{\epsfig{
       figure=#1.pdf, scale=0.36
   }}}
   \caption{
      \label{#1}
      Evolution of the #2 ``#3'' initial velocity profile with $\alpha=#4$.#5
   }
\end{center}
\end{figure}
}

\def\figfourreverse#1#2#3#4{
\begin{figure}
\begin{center}
   \leavevmode{\hbox{\epsfig{
       figure=#1.pdf, scale=0.4
   }}}
   \caption{
      \label{#1}
      Reconstituted initial velocity profile after time reversal
      for the #2 ``#3'' case with
      $\alpha=\sigma$   (upper left),
      $\alpha=\sigma/2$ (upper right),
      $\alpha=\sigma/4$ (lower left),
      $\alpha=\sigma/8$ (lower right).#4
   }
\end{center}
\end{figure}
}

\newcommand{\bfi}{\bfseries\itshape}

\begin{document}

\title{
\vbox to 0pt {\vskip -1cm \rlap{\hbox to \textwidth {\rm{\small
{}
\hfil {}
 $\quad$ June 4, 2004}}}}
Interaction Dynamics of Singular Wave Fronts
}
\author{Darryl D. Holm${}^{1,2}$ and Martin F. Staley${}^1$
\\
${}^1$Theoretical Division
\\Los Alamos National Laboratory, MS B284
\\ Los Alamos, NM 87545
\\ {\footnotesize email: dholm@lanl.gov}
\\ {\footnotesize email: staley@lanl.gov}
\and
${}^2$Mathematics Department
\\Imperial College London
\\ SW7 2AZ, UK
\\ {\footnotesize email: d.holm@imperial.ac.uk}
}
\date{}
\maketitle

\begin{abstract}\noindent
Some of the most impressive singular wave fronts seen in Nature are the
transbasin oceanic internal waves, which may be observed from the Space
Shuttle as they propagate and interact with each other, for example, in the
South China Sea. The characteristic feature of these strongly nonlinear
wavefronts is that they reconnect when two of them collide transversely.
We derive the EPDiff equation, and use it to model this phenomenon as
elastic collisions between singular wave fronts (solitons) whose momentum is
distributed along curves moving in the plane. Numerical methods for EPDiff
based on compatible differencing algorithms (CDAs) are used for simulating
these collisions among curves. The numerical results show the same nonlinear
behavior of wavefront reconnections as that observed for internal waves in
the South China Sea. We generalize the singular solutions of EPDiff for
other applications, in computational anatomy and in imaging science, where
the singular wavefronts are evolving image outlines, whose momentum may be
distributed on surfaces moving though space in {\it three dimensions}. The
key idea is always momentum exchange during collisions of the wavefronts.
A suite of 2d and 3d numerical simulations provide collision rules for the
wavefront reconnection phenomenon in a variety of scenarios.

\end{abstract}

\tableofcontents

\section{Synopsis}
\label{synop-sec}

Space Shuttle observations show great lines in the sea, which show up in
Synthetic Aperture Radar (SAR) images as in Figure
\ref{SCS-pan-fronts}. These are internal waves, whose fronts appear in SAR
images as unbroken curves extending for hundreds of kilometers. For
example, tidal flows and undulations of the Japanese Current in the region
of the Luzon Strait between Taiwan and the Philippines generate internal
wavefronts which are well over one hundred kilometers in length and which
propagate westward for hundreds of kilometers all the way across the South
China Sea. These wave fronts show strong coherence and possess the
characteristic feature of reconnecting with each other whenever any two of
them intersect transversely. This reconnection is the hallmark of a
nonlinear process. Weaker waves may intersect, but not reconnect. We seek
to encode the motion of these internal wave fronts and their collision
rules mathematically in a minimal PDE (partial differential equation)
model, and to investigate these wave front reconnections in numerical
simulations of the model. Establishing a simplified PDE model whose
solutions encode the motion of these internal wave fronts and their
collision rules will enable understanding the effects of the
nonhydrostatic processes that govern them, without requiring the full
numerical simulation and analysis of the 3d fluid motion equations.
\bigskip

We shall also derive more detailed PDE models for the interactions of
these nonlinear wavefronts when topography and boundaries are also
included. However, we shall defer developing the numerical methods needed
to deal with these additional effects until elsewhere.
\bigskip

Our derivation of the 2d multilayer Euler-Poincar\'e (EP) equations
begins by vertically integrating the variational principle for Euler's
equations in 3d. These 2d equations must be nonhydrostatic, because the
wave fronts they model possess strong horizontal gradients of vertical
acceleration. These nonhydrostatic terms are included in a hierarchy of EP
equations which is derived by making a series of simplifying
approximations in the exact variational principle. The resulting
multilayer 2d equations preserve the EP properties of energy balance,
circulation laws and potential vorticity (PV) conservation. PV analysis in
a multilayer fluid system is essential for assessing its baroclinic
instability \cite{Pedlosky[1987]}. However, PV for wave fronts is a new
concept, whose implications are hardly understood yet. We will show that
the new hierarchy of EP equations recovers a sequence of previously known
models of internal wave dynamics when specialized in various ways.
\bigskip

Next, from the full hierarchy we extract a simple, minimal PDE description,
called EPDiff, which models internal wave fronts as delta functions of
momentum distributed on moving curves in the plane. This corresponds to
modeling internal wave fronts as contact discontinuities in velocity.
EPDiff thus explains the creation and stability of wave fronts as the
development of singular momentum solutions from continuous velocity
distributions in a PDE initial value problem (IVP). This description links
the shape of the wavefront velocity profiles to the Green's function
associated with the nonlocal relation in the PDE between its singular
momentum and its continuous velocity. Hence, EPDiff explains the uniform
width of the wave front velocity profiles that is observed in internal wave
trains. Namely, the momentum of the wavefront is concentrated on delta
functions supported on the curved evolving wavefronts. The corresponding
velocity profile is then obtained via a Green's function relation between
wave momentum and fluid velocity. This minimal description of internal wave
fronts as singular momentum solutions of the evolutionary equation EPDiff
encompasses their reconnections and provides the geometric mechanism
underlying their propagation and collision interactions. In 1d, these
collisions are understood as elastic solition collisions, whose solutions
are obtained by the inverse scattering transform (IST) for an associated
isospectral linear eigenvalue problem. As far as we know, the machinery
of IST is not available in higher spatial dimensions for EPDiff. However,
the elastic scattering interactions seen in its wavefront solutions in 2d
(and in 3d) are explained by another geometric mechanism. As we shall
explain, this geometric mechanism for propagation and collisions turns out
to be Hamiltonian geodesic flow on the time-dependent smooth maps
(diffeomorphisms), defined with respect to the kinetic energy metric that
also determines the wave profile.
\bigskip

To solve the EPDiff equation, we develop a numerical method called the
compatible differencing algorithm (CDA) that is able to capture the
collisions among these weak solutions of EPDiff, and we characterize the
wave front interactions we observed numerically in a variety of scenarios.
\bigskip

In summary, we use EP variational theory to derive the EPDiff equation,
whose solution shows singular wave fronts, and we use new
CDAs for its numerical simulations. EPDiff is the
first mathematical explanation of the observed 2d internal wave front
reconnections.
\bigskip

\paragraph{Geometric setting.}
We have found that modeling the reconnection process in internal wave
front dynamics requires a class of PDEs that are both nonlinear and
nonlocal. EPDiff is a not exactly a hyperbolic equation. It has a
characteristic velocity, but the relation its EP variational principle
defines between the fluid's velocity and momentum in Newton's Law is
nonlocal. That is, the fluid's velocity is determined from its momentum by
solving an elliptic equation. Physically, the elliptic equation arises from
nonhydrostatic processes that cause linear, or nonlinear, dispersion or
focusing. This nonlocal relationship between velocity and momentum is
reminiscent of the Biot-Savart relationship between velocity and vorticity.
However, like solitons, internal wave fronts carry momentum and inertia,
while fluid vortices do not.
\bigskip

Our work has some close similarities with soliton theory and some
important differences from it. The Korteweg-de Vries (KdV) equation at
linear order in the asymptotic expansion for shallow water waves and the
Camassa-Holm (CH) equation at quadratic order are both soliton equations,
and they are both associated with geodesic
flow \cite{KhMi2003,Ko1999,Mi1998}. That is, they each make optimal use of
their kinetic energy, which provides a norm for their velocity. The EPDiff
equation has this same characteristic. Moreover, the singular solution
ansatz for the geodesic flow of momentum associated with the EPDiff
equation which we discuss below turns out to be a momentum map, as
discovered in Holm and Marsden \cite{HoMa2004}. The momentum map property of
these singular solutions means they comprise an invariant manifold,
preserved by the flow of EPDiff. This property allows us to reduce the
dimension of the wave front interactions to an invariant manifold of
singular momentum solutions of EPDiff. These solutions describe the
observed propagation and reconnection phenomena of the wave fronts, but
they have no internal degrees of freedom, and thus they have no mechanism
for wave breaking to occur.
\bigskip

In this paper, we model internal wave fronts as contact discontinuities in
velocity, whose motion is governed by applying time-dependent smooth maps
which act on the curves in the plane that outline the wave fronts and
specify their momentum vectors at points along these curves. The motion of
the internal wave fronts is governed by the EPDiff equation. This equation
is the condition for the smooth maps that act on these curves to evolve by
geodesic flow, with respect to a metric determined by their kinetic energy,
which also determines the functional form of their wave profile. The wave
front motion may thus be determined numerically as an initial value problem
by solving the EPDiff equation.
\bigskip

In such a strongly geometric setting, we expect that combining finite
element methods and discrete exterior calculus may provide future
improvements in both modeling and simulations, by providing a useful
setting for integration of the analysis with the numerics in the
description of internal wave fronts as singular weak solutions, or
limiting solutions (contacts) for nonlinear, nonlocal PDEs. However, we
defer the full use of discrete exterior calculus methods for numerically
solving EPDiff until these methods have been further developed. See
\cite{Hirani2003,Leok2004} for introductions to these methods.

\paragraph{Outline of the paper.}
Mathematical modeling problems for internal wave fronts, and our approaches
to solving them, are summarized in Section \ref{prob-state-sec}.
A preview of our numerical results appears in Section \ref{num-prev-sec}.
Previous work in multilayer descriptions of internal waves is reviewed in
Section \ref{model-sec}.
The columnar motion ansatz for multilayer fluids is explained in Section
\ref{multi-setup-sec}, and is used there to derive the nonhydrostatic 2d
MultiLayer Columnar Motion (MLCM) equations and their natural boundary
conditions.
A series of weakly nonlinear approximations is introduced into the EP
variational principle for the MLCM equations in Section \ref{weak-nonlin-sec}.
The resulting weakly nonlinear 2d multilayer equations are also compared to the
1d multilayer equations of Choi and Camassa \cite{ChCa1999} and others.
In Section \ref{ke-limit-sec}, we derive the EPDiff geodesic equation by
neglecting dispersion due to potential energy. Singular solutions of EPDiff
and their canonical dynamics in the Lagrangian fluid representation are also
introduced in Section \ref{ke-limit-sec}.

The main numerical results of this paper are given in Sections
\ref{num-appr-sec}, \ref{num-2d-sec} and \ref{num-3d-sec}.
Section \ref{num-appr-sec} describes our numerical approach using
compatible differencing algorithms.
Section \ref{num-2d-sec} provides a large suite of numerical simulations
investigating the collision rules for the interactions of wave-front solutions
of EPDiff in 2d.
Section \ref{num-3d-sec} extends our numerical solutions of EPDiff to 3d,
and shows that its stable codimension-one singular solution behavior persists
in higher dimensions.
Finally, we discuss conclusions, future directions and remaining outstanding
problems for internal wavefront interactions in Section \ref{future-direx-sec}.

\rem{
The conservation properties for energy, circulation and potential vorticity
of the MLCM equations are also derived in Section \ref{multi-setup-sec} as
consequences of the EP variational principle. The Lie-Poisson Hamiltonian
formulation of the MLCM equations and its connections to stability analysis of
their equilibrium solutions are also discussed in Section \ref{multi-setup-sec}.
}

\section{Problem statement, approach, and main results}
\label{prob-state-sec}

\begin{figure}
\begin{center}
   \leavevmode{\hbox{\epsfig{
      figure=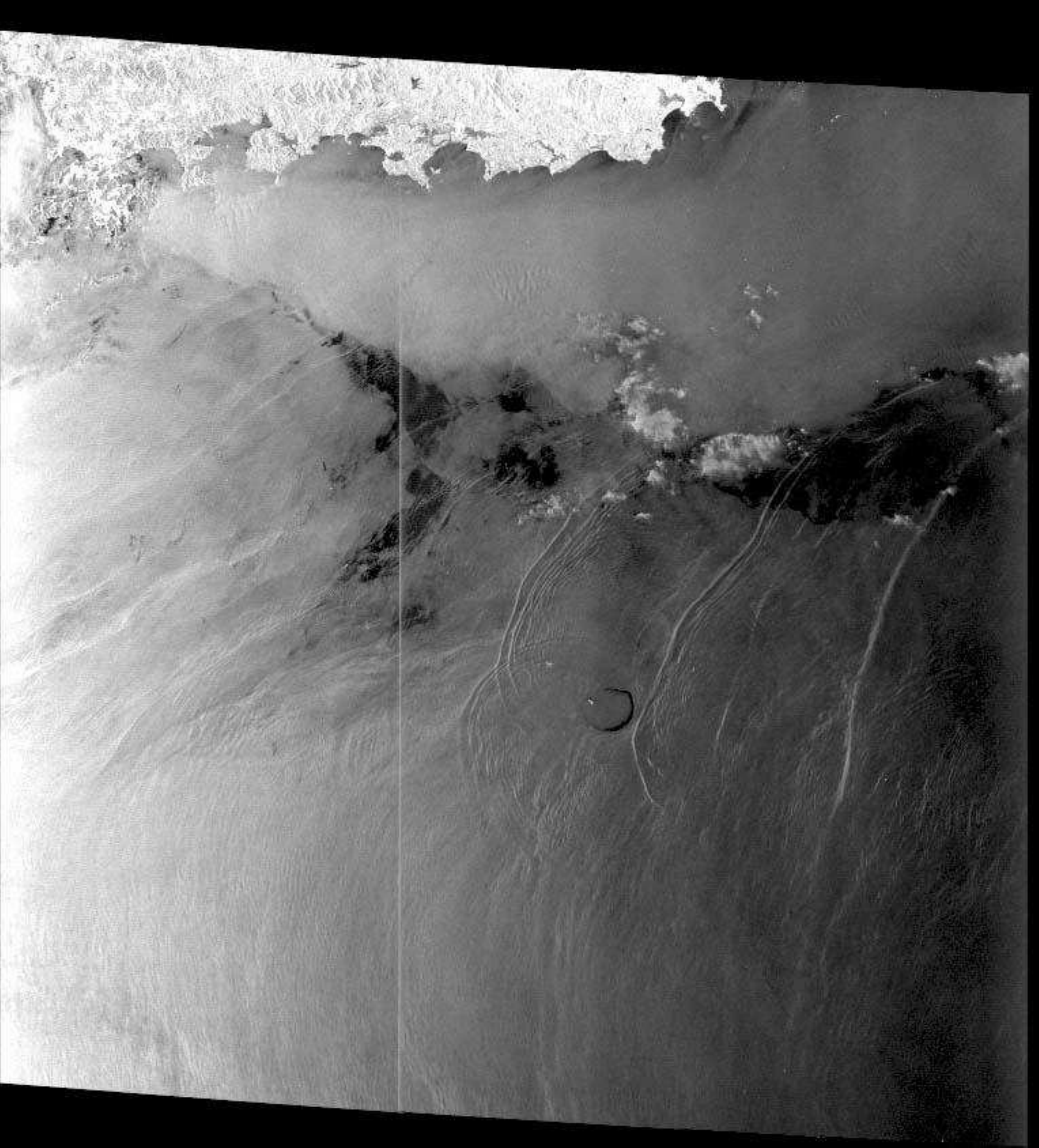, scale=0.5
   }}}
   \caption{\label{SCS-pan-fronts}
      Synthetic Aperture Radar (SAR) images of the South China Sea
      surface taken from the Space Shuttle show a sequence of large
      amplitude internal waves at basin scale. These wave fronts are moving
      westward after being created by tides and currents running through
      the Luzon Strait on the eastern side of the basin. These wave fronts
      encounter the Dongsha Atoll in the center of the basin and re-emerge
      after colliding with it.
   }
\end{center}
\end{figure}

\begin{figure}
\begin{center}
   \leavevmode{\hbox{\epsfig{
       figure=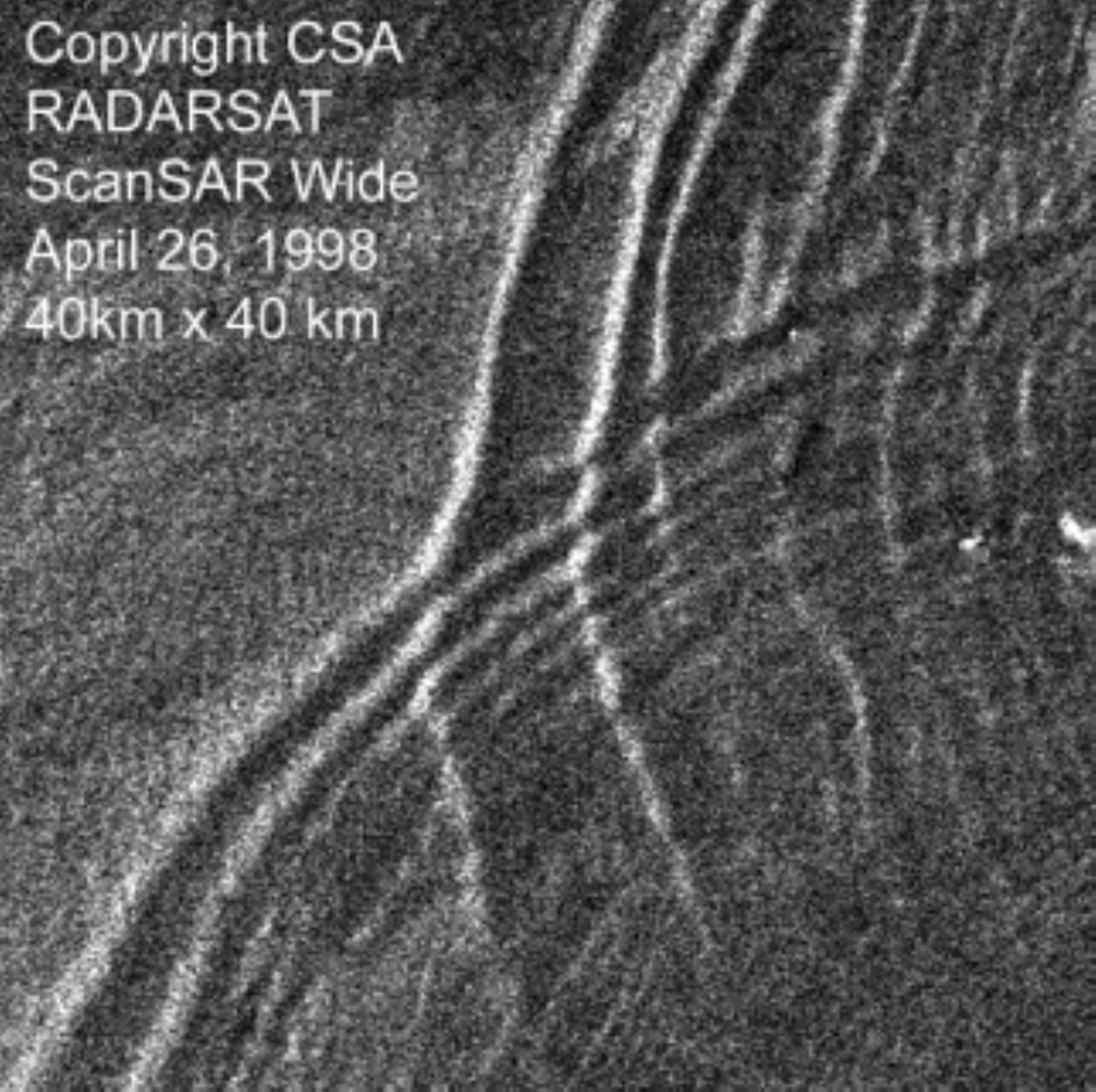, scale=0.7
   }}}
   \caption{\label{SCS-zoom-reconnex}
      The transverse interactions of the wave fronts re-emerging from
      their encounter with the Dongsha Atoll are so strong that the wave
      fronts reconnect, rather than pass through each other. This indicates
      transfer of momentum.
   }
\end{center}
\end{figure}

Space Shuttle observations of the South China Sea surface show a sequence
of large amplitude internal waves at basin scale moving westward after being
created by tides and currents running through the Luzon Strait on the
eastern side of the basin. These internal waves appear in Synthetic
Aperture Radar (SAR) images as long, slightly curved, internal wave fronts
100-200 {\it km} in length and separated by 50-80 {\it km}. When these
isolated internal wave fronts sweep across Dongsha Atoll in the middle of
the South China Sea, they are perturbed and subsequently re-radiated as
two sets of wave front trains propagating westward, each with greater
curvature than the incoming wave front. The SAR images of this process are
shown in Figure \ref{SCS-pan-fronts}.
The transverse interactions of the re-radiated wave fronts are so
strong that they reconnect, rather than pass through each
other. This reconnection is shown in Figure \ref{SCS-zoom-reconnex}. The
reconnection phenomenon for nonlinear internal wave fronts in the ocean
indicates transfer of momentum and
is one of the primary motivations of the present study.
Internal wave interactions have been well studied in one dimension, often
by using the weakly nonlinear Boussinesq approximation, usually resulting
in a variant of the Korteweg-de Vries (KdV) equation and its soliton
solutions \cite{Whitham1967}. However, the complex wave front interactions
shown in Figures \ref{SCS-pan-fronts} and \ref{SCS-zoom-reconnex} are
plainly two dimensional. Moreover, their reconnection is not captured by
the 2d extension of KdV for weakly nonlinear waves with slow transverse
variations, known as the Kadomsev-Petviashvili (KP) equation. (The KP
equation assumes weak gradients in the direction transverse to
propagation. However, this assumption does not hold during the
wavefront reconnections we seek to model.) Thus, we begin our study of
internal wave front reconnection by developing a new set of equations that
extends, to multilayer fluids in two dimensions, the Su-Gardner
\cite{SuGa1969} or Green-Naghdi \cite{GN1976} equations for fully nonlinear
waves on the free surface of a single-layer fluid in one dimension.
Related earlier derivations of one dimensional equations for
multilayer fluids appear in \cite{ChCa1999,LiMaWe1995,LiWe1997}.

Our approach to developing these equations takes advantage of the
Euler-Poincar\'e (EP) theory of Hamilton's variational principle for ideal
fluids, in the Eulerian representation \cite{HoMaRa1998}, expressed as
$\delta{S}=0$ with $S=\int \ell(\mathbf{u})\,dt$ for Eulerian
horizontal fluid velocity $\mathbf{u}$. The advantage of the EP approach
for our present purpose is that it provides a hierarchy of equations at
various levels of approximation that preserves the EP mathematical
structure of the highest level theory.  This approach enables us to strike
the appropriate balance between accuracy and computational tractability, by
choosing the appropriate level in the hierarchy of approximations. The
symmetries of the EP variational principle at each level then endow its
resulting evolutionary equations with conservation laws via the
Kelvin-Noether theorem
\cite{HoMaRa1998}. For example, Kelvin's circulation theorem for the
multilayer fluid theory in 2d leads to local conservation of potential
vorticity (PV) in each fluid layer, even though these layers are strongly
coupled.

The starting point in the derivation of these equations is the assumption
of columnar motion; so that the horizontal velocity in each layer is
independent of the vertical coordinate. We explain how substituting this
columnarity into the EP Hamilton's principle for multilayer fluids reduces
their description from 3d to 2d, and results in a hierarchy of equations
which arises when a sequence of further approximations are introduced into
Hamilton's principle in the EP framework.

The common features of the equations in this hierarchy are (1) they
express Newton's Law for the evolution of momentum in the Eulerian fluid
representation; (2) their nonlinearity involves both momentum and fluid
velocity; and (3) their momentum and fluid velocity satisfy a nonlocal
relationship, which must be solved by inversion of an elliptic operator at
each time step. While keeping these features, we simplify the description
by making a series of approximations in Hamilton's principle, until we
finally arrive at a minimal description which still captures the wave front
reconnection phenomenon. We then employ this minimal description to
investigate and classify the wave front interactions analytically and in
numerical simulations. Our application is for internal wave fronts, which
we compare with our numerical simulations in two dimensions. In a further
developmental step, we also consider numerical solutions of our description
in three dimensions. As mentioned earlier, this 3d extension yields
singular solutions of EPDiff whose momentum is defined on contact surfaces.

\paragraph{The EPDiff equation and its weak wavefront solutions.}
The EP equation at which we eventually arrive by this sequence of
variational approximations is the following evolutionary integral-partial
differential equation, expressed in vector form as
\cite{HoSt2003,HoMa2004},
\begin{equation}\label{H1-EP-eqn-Intro}
\frac{\partial}{\partial t}
\mathbf{m}
+\,
\mathbf{u}\cdot\nabla
\mathbf{m}
+\,
\nabla \mathbf{u}^T\cdot
\mathbf{m}
+
\mathbf{m}
({\rm div\,}{\mathbf{u}})
=
0
\,,\quad\hbox{with}\quad
\mathbf{m}
=
\mathbf{u}
-
\alpha^2\Delta{\mathbf{u}}
\,,
\end{equation}
for which we assume periodic boundary conditions. This is the EPDiff
equation, for ``Euler-Poincar\'e equation on the diffeomorphisms,'' where
$\hbox{\bf m}$ and $\hbox{\bf u}$ are vectors, $\partial_t\hbox{\bf m}$
denotes the time derivative of $\hbox{\bf m}$, and $\alpha$ is a constant
parameter with dimensions of length. EPDiff is the $n-$dimensional
version of the 1d shallow water equation introduced in \cite{CaHo1993}.
EPDiff arises from a variational principle based on the fundamental
dynamical properties of fluids. Mathematically, EPDiff describes {\it
geodesic} motion on the diffeomorphism group with respect to
$\|\mathbf{u}\|_{H^1}^2$, the $H^1$ norm of the fluid velocity, which is
the kinetic energy of this vertically averaged model fluid
\cite{HoMaRa1998}.

The EPDiff equation (\ref{H1-EP-eqn-Intro}) may be written equivalently
as {\it differential Riemann invariance}; namely, as invariance of the
momentum one-form density along the fluid velocity characteristics. That
is,
\begin{equation}\label{H1-EP-eqn-Riemann}
\frac{d}{dt}\big(
\mathbf{m}\cdot d\mathbf{x}\otimes dVol\big)
=0
\,,\quad\hbox{along}\quad
\frac{d\mathbf{x}}{dt}
=
\mathbf{u}
\equiv
G*\mathbf{m}
\,,\end{equation}
where $\mathbf{u}=G*\mathbf{m}$ denotes convolution of the momentum
$\mathbf{m}$ with the Green's function $G$ to produce the fluid velocity
$\mathbf{u}$. This convolution is the elliptic-solve step in the
determination of the fluid's velocity from its momentum for this class of
equations. This property is the nonlocal relationship mentioned above. In
the particular case of EPDiff, we use the elliptic Helmholtz operator in
$\mathbf{m}=\mathbf{u}-\alpha^2\Delta{\mathbf{u}}$, with length scale
$\alpha$. The Helmholtz operator relationship between velocity
$\mathbf{u}$ and momentum $\mathbf{m}$ is derived in Section
\ref{multi-setup-sec} for strongly nonlinear columnar motion of shallow
water.

\paragraph{The relation of EPDiff to 1d soliton equations.}
In 1d, EPDiff becomes
   \[\partial_tm+m\partial_xu+\partial_x(mu) = 0
   \,,\quad\hbox{with}\quad
   m = u-\alpha^2\partial_x^2u
   \,.\]
This is the dispersionless limit of the Camassa-Holm (CH) equation
   \[\partial_tm+m\partial_xu+\partial_x(mu)
   = -c_0\partial_xu-\Gamma\partial_x^3u
   \,,\quad\hbox{with}\quad
   m = u-\alpha^2\partial_x^2u
   \,,\]
which is a completely integrable soliton equation \cite{CaHo1993}.
The CH equation reduces to the familiar KdV equation when $\alpha^2\to0$.

EPDiff in 1d with viscosity $\nu$ is expressed as
   \[\partial_tm+m\partial_xu+\partial_x(mu) = \nu\partial_x^2m
   \,,\quad\hbox{with}\quad
   m = u-\alpha^2\partial_x^2u
   \,.\]
When $\alpha^2\to0$ this becomes the well known Burgers equation,
   \[\partial_tu+\partial_x(\frac{3}{2}u^2)=\nu\partial_x^2u\,.\]
The Burgers equation is hyperbolic and supports weak solutions (shocks) in
the limit $\nu\to0$.

EPDiff is a new departure for nonlinear hyperbolic equations. It is
conservative in the absence of viscosity and it may then be written in
Riemann invariant form (\ref{H1-EP-eqn-Riemann}). However, it is nonlocal,
because obtaining its characteristic velocity $\hbox{\bf u}$ from its
momentum $\hbox{\bf m}$ by inverting the Helmholtz operator via
$\mathbf{u}=G*\mathbf{m}$ requires an elliptic solve at each time step
when $\alpha\ne0$. Our analysis \cite{HoSt2003,HoMa2004} shows that EPDiff
has interesting and unusual solution properties which are only beginning
to be studied. The properties of its solutions will provide great
challenges for analysis and numerics.

\paragraph{Singular momentum solutions of EPDiff.}
For example, EPDiff has weak {\it singular momentum solutions} that are
expressed as \cite{HoSt2003,HoMa2004},
\begin{equation}\label{EP-sing-mom}
\mathbf{m}(\mathbf{x},t)
=
\sum_{a=1}^N\int_{S_a}
\mathbf{P}_a(t,S_a)\delta\big(\mathbf{x}-\mathbf{Q}_a(t,S_a)\big)\,dS_a
\,,
\end{equation}
where $S_a$ is a Lagrangian coordinate defined along a set of curves in the
plane by the equations $\mathbf{x}=\mathbf{Q}_a(t,S_a)$ supported on the delta
functions in the EPDiff solution (\ref{EP-sing-mom}). Thus, the singular
momentum solutions of EPDiff are vector valued curves representing evolving
wave fronts defined by the Lagrange-to-Euler map (\ref{EP-sing-mom}) for their
momentum.

The Green's function for the Helmholtz operator relates the fluid velocity to
the momentum in EPDiff (\ref{H1-EP-eqn-Intro}). Thus, substituting the defining
relation $\mathbf{u}\equiv G*\mathbf{m}$ into the singular momentum solution
(\ref{EP-sing-mom}) yields the velocity representation for the wave fronts as
another superposition of integrals,
\begin{equation}\label{EP-sing-vel}
\mathbf{u}(\mathbf{x},t)
=
\sum_{a=1}^N\int_{S_a}
\mathbf{P}_a(t,S_a)
G\big(\mathbf{x},\mathbf{Q}_a(t,S_a)\big)\,dS_a
\,.
\end{equation}
The Green's function $G$ for the second order Helmholtz operator in this
expression has a discontinuity in slope across each Lagrangian curve moving with
the velocity of the flow. Being discontinuities in the gradient of velocity that
move along with the flow, these singular solutions for the velocity at the wave
fronts are classified as {\it contact discontinuities} in the fluid.

Thus, the weak solutions of EPDiff are contacts, rather than shocks. This
behavior challenges numerics and leads to a wide range of potential physical
applications. In addition to describing internal wave fronts, the EPDiff
equation describes the interaction of contacts in a variety of
other situations ranging from solitons \cite{CaHo1993} to turbulence
\cite{Chen-etal1998,FoHoTi2001} to computational anatomy
\cite{HoRaTrYo2004,MiTrYo2002}. The EPDiff fluid dynamics equation in
(\ref{H1-EP-eqn-Intro}) describes the limiting (pressureless) cases of
{\it all} of these applications.

\paragraph{Relation between contact solutions of EPDiff and solitons.}
The EPDiff singular solutions (\ref{EP-sing-vel}) for the velocities of the
internal wave fronts represent the third of the three known types of fluid
singularities: shocks, vortices and contacts. The key feature of these contacts
is that they carry momentum; so the wave front interactions they represent are
{\it collisions}, in which momentum is exchanged. This is very reminiscent of
the soliton paradigm in one dimension. Indeed, in one dimension the singular
solutions (\ref{EP-sing-vel}) of EPDiff are true solitons that undergo elastic
collisions and are solvable by the inverse scattering transform for an
isospectral eigenvalue problem \cite{CaHo1993}.

\paragraph{Applications of EPDiff in turbulence.}
Lagrangian averaging is a
promising approach in turbulence closure modeling. This approach has the
advantage that Lagrangian averaging commutes with the advective time
derivative in fluid mechanics. Thus, Lagrangian averaging preserves the
vorticity stretching process in the resulting approximate equations.
For example, after Lagrangian averaging the Navier-Stokes equation and
using Taylor's hypothesis that the fluctuations move with the mean flow,
one finds the following turbulence closure model \cite{Chen-etal1998},
\begin{eqnarray}\label{LANS-alpha-eqn-Intro}
&&
\frac{\partial}{\partial t}
\mathbf{v}
+
\mathbf{u}\cdot\nabla
\mathbf{v}
+\,
\nabla \mathbf{u}^T\cdot
\mathbf{v}
+
\nabla p
=
\nu\Delta\mathbf{v} + \mathbf{F}
\,,
\\&&
\hbox{with}\quad
\mathbf{v}
=
\mathbf{u}
-
\alpha^2\Delta{\mathbf{u}}
\quad\hbox{and}\quad
\nabla \cdot\mathbf{u}=0
\,.
\nonumber
\end{eqnarray}
These are the equations of the Lagrangian averaged Navier-Stokes alpha
(LANS-alpha) model of turbulence, whose properties are reviewed, for
example, in \cite{FoHoTi2001}. EPDiff (\ref{H1-EP-eqn-Intro}) is recovered
from the LANS-alpha equations (\ref{LANS-alpha-eqn-Intro}) when the
constraint of incompressibility $(\nabla \cdot\mathbf{u}=0)$ is relaxed
(so that pressure gradient $\nabla p$ may be dropped), and the viscous and
forcing terms on the right hand side are absent. Thus, these
equations possess the following analogy: EPDiff is to LANS-alpha, as
Burgers is to Navier-Stokes. Namely, Burgers is a simplified model of
compressible Navier-Stokes which allows shocks as singular solutions
of its initial value problem, when $\mathbf{F}$ is absent and $\nu\to0$;
and EPDiff is a simplified model of compressible LANS-alpha which allows
contact discontinuities as singular solutions of its initial value
problem, when $\mathbf{F}$ is absent and $\nu\to0$. The reduction of
the singular solutions from Burgers shocks to EPDiff contacts is a result
of the Lagrangian averaging process, which tempers the nonlinearity in the
Navier-Stokes equations.

\paragraph{Applications of EPDiff in computational anatomy.}
Applications of EPDiff (\ref{H1-EP-eqn-Intro}) in computational image
science (CIS) focus on computational anatomy, based on pattern
matching algorithms and smooth morphing of planar figure outlines, called
templates. For a review of this approach, see \cite{MiTrYo2002}. An
interesting objective of this CIS application is to reconstruct a 3d map
of the brain in the spatial region between two parallel 2d PET Scans. In
this application, the figure outlines are contact curves, which form a
finite-dimensional invariant manifold of EPDiff, as in equation
(\ref{EP-sing-mom}). A 3d brain map is constructed by treating two 2d PET
scans in parallel planes as initial and final conditions, and then flowing
by EPDiff to interpolate in the 3d region between them as an optimization
problem. The 2d solution for the contact curves evolving by EPDiff
smoothly reconstructs the outlines of the PET scan images on any parallel
plane between them. The EPDiff contact interactions also allow
reconnection of planar outlines, corresponding to changes of topology in
planar sections of the 3d object being imaged
\cite{HoRaTrYo2004}. The representation (\ref{EP-sing-mom}) of
the singular momentum solutions of EPDiff encodes the image contours for
this application.

Importantly, the momentum representation of the image contours
(\ref{EP-sing-mom}) is complete and nonredundant (one-to-one). Another
advantage of the momentum representation of image contours as singular
solutions of EPDiff is that this representation is {\it linear} in nature,
being dual to the velocity vectors. Thus, linear combinations of either
velocity fields, or momenta are meaningful mathematically and physically,
provided they are applied to the same template \cite{HoRaTrYo2004}. This
means the average of a collection of momenta, of their principal
components, or time derivatives of momenta at a fixed template are all
well-defined quantities. The linearity the momentum representation also
allows for example the statistics of an ensemble of images to be analyzed,
or the results of adding noise to the image outlines to be computed using
EPDiff. All of these advantages of the singular momentum representation for
2d images also apply in the corresponding representation of 3d images.
Evolution of singular momentum solutions of EPDiff as surfaces in 3d
corresponding in medical imaging to growth, or changes in 3d shape with
time. The last part of this paper will deal with the computation of 3d
solutions of EPDiff.

\paragraph{Summary of the analytical approach.}
The rich array of possible applications of EPDiff motivates our
investigation of its initial value problems in 1d, 2d and 3d.
The paradigm raised in our analysis of internal wave front
interactions is the generalization of soliton collisions from 1d to 2d.
The class of shallow water equations we derive and study in 2d has a
sequence of simplifications that yield two dimensional generalizations of
all the weakly nonlinear shallow water equations that have historically
been used to study solitons and solitary waves in one dimension. We first
argue that by using the highest, most accurate level of the fully
nonlinear description in 2d, one should be able to reproduce the observed
phenomenon of wave front reconnection observed for internal waves in the
South China Sea. We then simplify the description by stages until we
finally arrive at EPDiff (\ref{H1-EP-eqn-Intro}) in 2d, while preserving
what we believe is the key feature responsible for this wave front
reconnection phenomenon -- its momentum transfer during collisions. Because
of its other potential applications, particularly its linear encoding of
information for imaging science as momentum, we also consider the initial
value problem for 3d singular solutions of EPDiff. Its imaging science
applications are envisioned as optimization problems. However, the first
step in understanding the role of momentum exchange for EPDiff in imaging
science is the solution of its initial value problem for the emergence of
its singular momentum solutions in 2d and 3d.

\paragraph{Brief sketch of the paper.}
The first few sections of this paper motivate using the much simpler equation
EPDiff (\ref{H1-EP-eqn-Intro}), whose properties we study analytically and
numerically in later sections, as a simplified but realistic model of wave front
interactions of nonlinear internal waves in the ocean. The last few sections
present numerical results for EPDiff in both 2d and 3d, which we now preview.

\section{Preview of numerical results}
\label{num-prev-sec}

\subsection{CFD methods for fluid singularities}

\noindent
Shocks, vortices and contacts are a compressible fluid's singular
nonlinear responses to strong applied forces. Capturing these singular
responses accurately has always been the grand challenge of computational
fluid dynamics (CFD). Shocks move {\it through\/} the flow and carry
inertia. Vortices move {\it with\/} the flow, but they have no inertia.
Contacts are discontinuities in the derivatives of velocity and density
that, like vortices, move with the fluid flow, and which {\it do\/} have
inertia. Consequently, it makes dynamical sense to speak of momentum
exchange in contact-contact collisions. Today, various CFD methods exist
that accurately capture shocks and vortices. However, considerably less is
known about designing numerical methods for capturing contacts and
characterizing their nonlinear interactions, especially in higher
dimensions, when vorticity is present.

One might consider Lagrangian numerical methods as an obvious approach,
because contact discontinuities move with the flow. However, for head-on
contact collisions, parallel studies showed that the Lagrangian approach
suffers in comparison to compatible differencing algorithms (CDAs) that we
apply here in the Eulerian framework. In particular, the head-on collision
of oppositely moving contacts produces an elastic bounce seen in the weak
solution of EPDiff as a mutual annihilation and recreation of generalized
functions, but only the annihilation was captured well by Lagrangian
methods in the parallel studies \cite{Li2003}. In the future, we may also
consider developing new methods for EPDiff based on discrete exterior
calculus (DEC) \cite{Hirani2003,Leok2004}.

\subsection{Numerical approach}

The second half of this paper numerically investigates and classifies the EPDiff
dynamics of contact interactions in various cases of its initial value problem.
This is accomplished by applying compatible differencing algorithms to
EPDiff in both 2d and 3d. One reason for choosing CDAs as the preferred
simulation method is that the EPDiff equation (\ref{H1-EP-eqn-Intro}) may be
rewritten as
\begin{equation}\label{H1-EPcurl-eqn-intro}
\frac{\partial}{\partial t}
\mathbf{m}
-\,
{\mathbf{u}}\times{\rm curl\,}{\mathbf{m}}
+
{\rm grad}({\mathbf{u}}\cdot{\mathbf{m}})
+
\mathbf{m}
({\rm div\,}{\mathbf{u}})
=
0
\,.
\end{equation}
This equation contains the divergence, gradient and curl operators, whose
identities must be treated properly for accurate computations. Preserving
these vector calculus identities is the basis of CDAs and is the first
fundamental property of discrete exterior calculus. Namely, these
identities form the discrete analog of the identity $d^2=0$ for the
exterior derivative. As mentioned earlier, EPDiff represents geodesic
motion, which naturally involves exterior calculus and variational
principles. A framework for designing DEC methods with promising potential
for simulations of EPDiff has been developed and advanced recently in
\cite{Hirani2003,Leok2004}.

\subsection{Interaction dynamics of contacts}

This paper takes advantage of recent developments in compatible
differencing algorithms by applying CDAs to new classes of problems
involving contact-contact interaction phenomena. For the report of a recent
conference on CDAs, see \cite{IMA-CDA-HotTopics2004}. We characterize the
emergence of contacts from smooth initial velocity distributions, and
describe their subsequent evolution, propagation and interaction dynamics.
We address dynamical issues for basic interactions among
contacts in one, two and three dimensions, as follows.

{\bf 1d.}
An integrable shallow water equation whose peaked soliton solutions emerge from
a smooth spatially confined initial conditions for velocity.

{\bf 2d.}
Interacting contact curve segments in the plane: trains of contact curves
emerging from an initially continuous fluid velocity distribution,
propagating, and interacting nonlinearly through fundamental collision
rules. The 2d collision rules for singular solutions of EPDiff are
elucidated by plotting 1d linear sections through the 2d solutions which
show their spatial profiles in various directions. The solution behavior
along these sections shows the same elastic collision properties and
momentum exchange as seen for the 1d peaked soliton solutions (peakons) in
1d for the dispersionless CH equation.

{\bf 3d.}
Interacting contact surfaces in space: sheaves of contact surfaces are shown
emerging from an initially continuous fluid velocity distribution. We
investigate their propagation and interaction by plotting level surfaces of
speed, as well as plotting 2d planar slices through these surfaces.
\bigskip

More specifically, we examine dynamical issues for the following basic
interactions among the contacts, which are weak solutions of EPDiff, in
one, two, and three dimensions.

\begin{figure}
\begin{center}
   \leavevmode{\hbox{\epsfig{
       figure=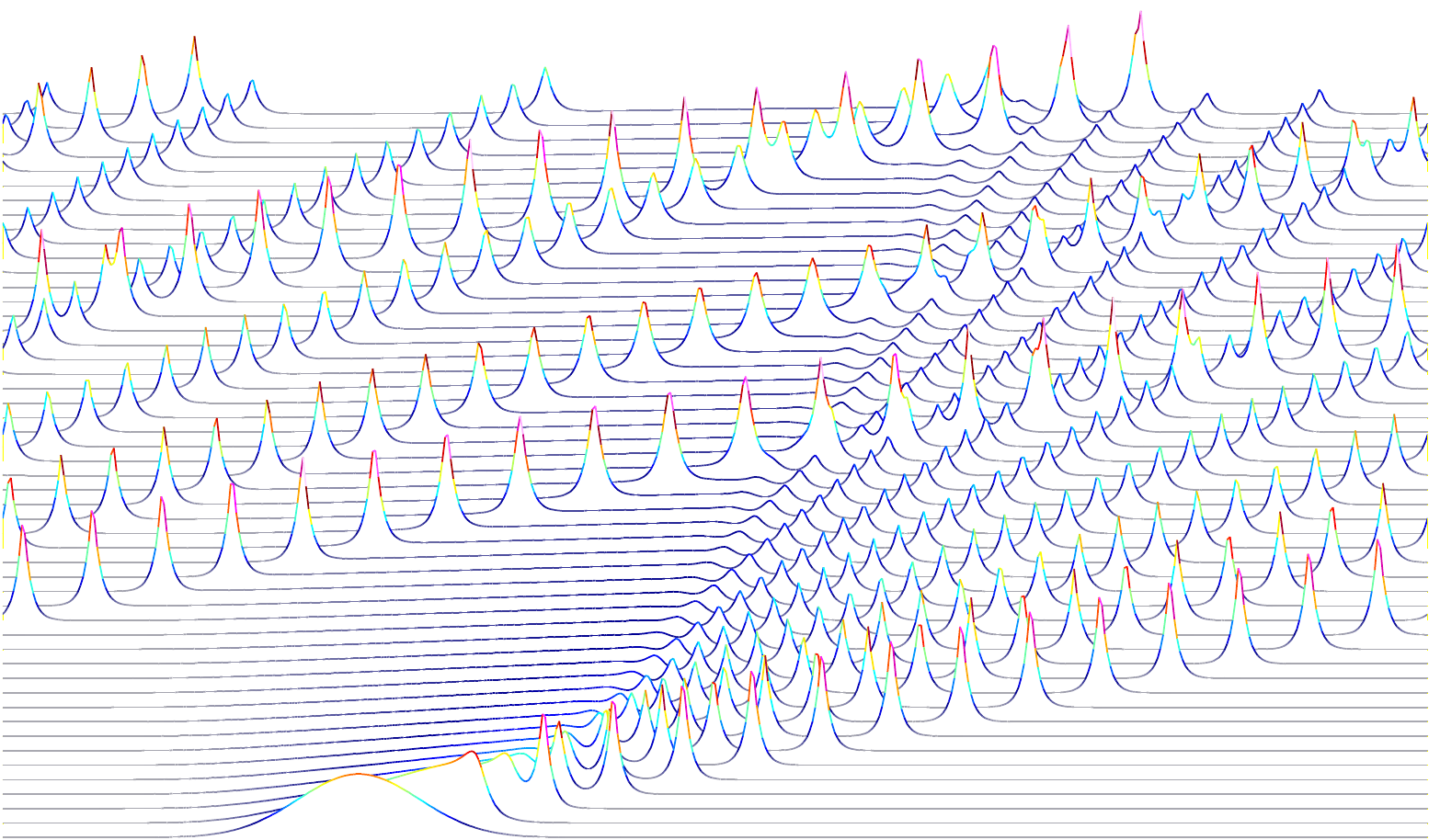, scale=0.7
   }}}
   \caption{\label{1d-peakons}
       A wave train consisting entirely of peakons emerges in 1d from an
initially Gaussian distribution of positive (rightward) velocity.
   }
\end{center}
\end{figure}

\paragraph{1d.}
%
In 1d, contacts move as points on a line, and EPDiff reduces to the well known
dispersionless Camassa-Holm (dCH) equation for shallow water waves, whose
contacts are solitons with sharp peaks, or peakons, as shown in the time
evolution in Figure \ref{1d-peakons}. The $N-$peakon problem is known to be
completely integrable \cite{CaHo1993}.

\paragraph{2d.}
%
In 2d, contacts move as segments of curves in the plane. These EPDiff
contacts correspond to oceanic internal waves in one application, or to
planar image outlines in another. We shall address the following specific
2d scenarios.

\paragraph{Evolution and interactions of EPDiff contact segments which are
initially straight line segments.}
Contacts move at the local fluid velocity; so a contact segment must terminate
with zero velocity. Hence, an initially straight-line contact segment does not
remain straight. Instead, it evolves under EPDiff into a contact curve segment
whose length increases, as shown in Figure \ref{preview-2d-plate}. The speed,
$\modu$, is displayed as colors in this figure, while the bottom panels
shows profiles in the east (black), north (dotted red), northeast (green), and
southeast (dotted blue) directions through the center of the box.
An arrow shows the initial direction of {\bf u}.
See Section \ref{num-2d-sec} for a fuller description of the contents of the
figures.

\previewstrip{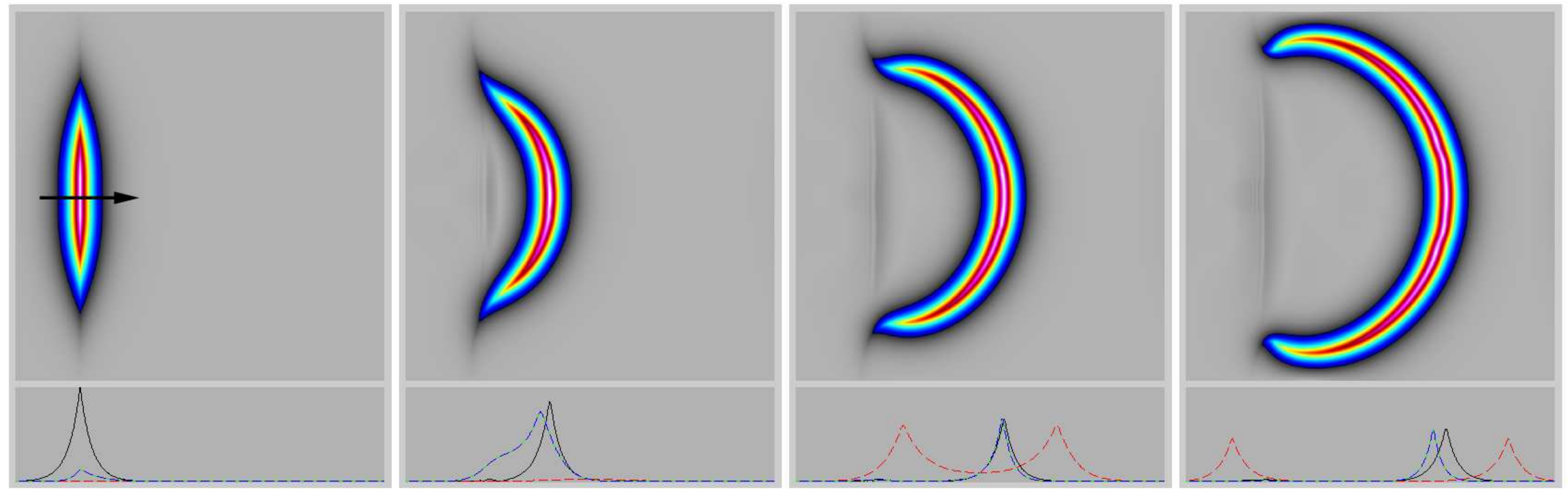}
{An initially straight, rightward moving peakon segment balloons outward and
stretches as it expands in 2d, but keeps its integrity as a single wave front.}

The transverse collision of two contact curve segments may result in
reconnection (merger, or melding) of the curves, which is admitted by the
rapid evolution of EPDiff along the directions tangential to the curves.
In Figure \ref{preview-2d-skew}, note the striking similarity between
the observed behavior, particularly in the second frame, and the behavior
seen in the earlier image of ocean internal waves in Figure
\ref{SCS-zoom-reconnex}. We shall investigate the ``collision rules'' under
which such reconnections of wave fronts occur.

\previewstrip{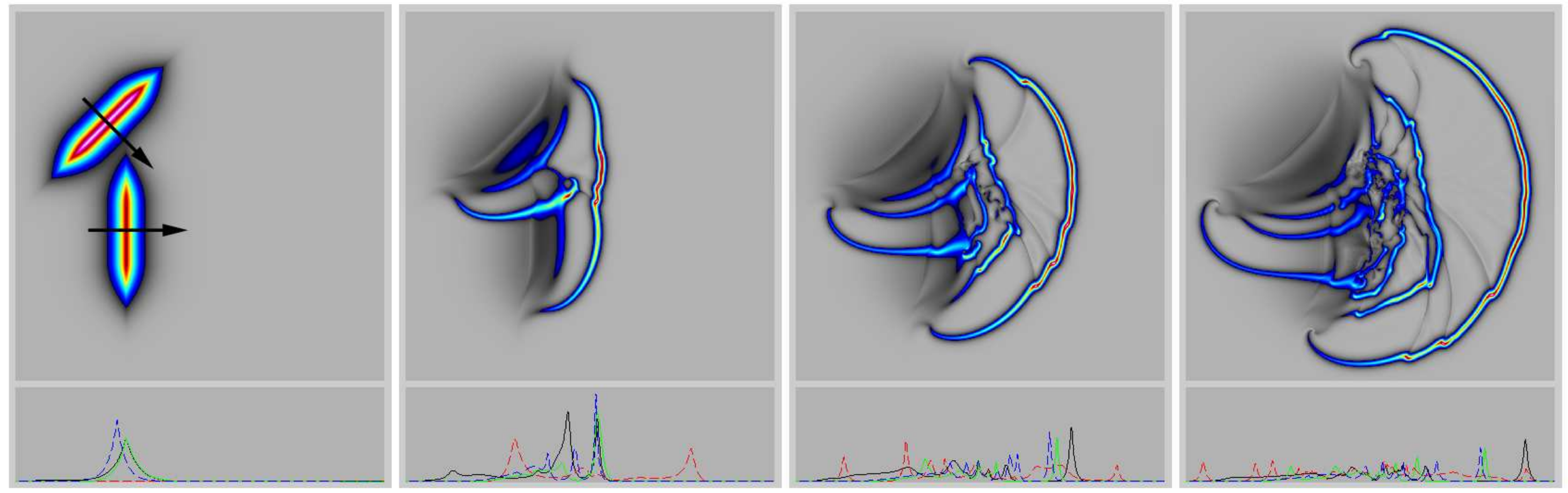}
{A skew overtaking collision of two peakon segments shows reconnection in which
the wavefronts merge repeatedly upon intersecting transversely.}

We will also consider offset collisions of initially parallel contact
segments, as shown in Figure \ref{preview-2d-headon}. In that figure,
notice the recreation (after annihilation) of the portions of the contact
curves in the regions where a head-on collision takes place.

\previewstrip{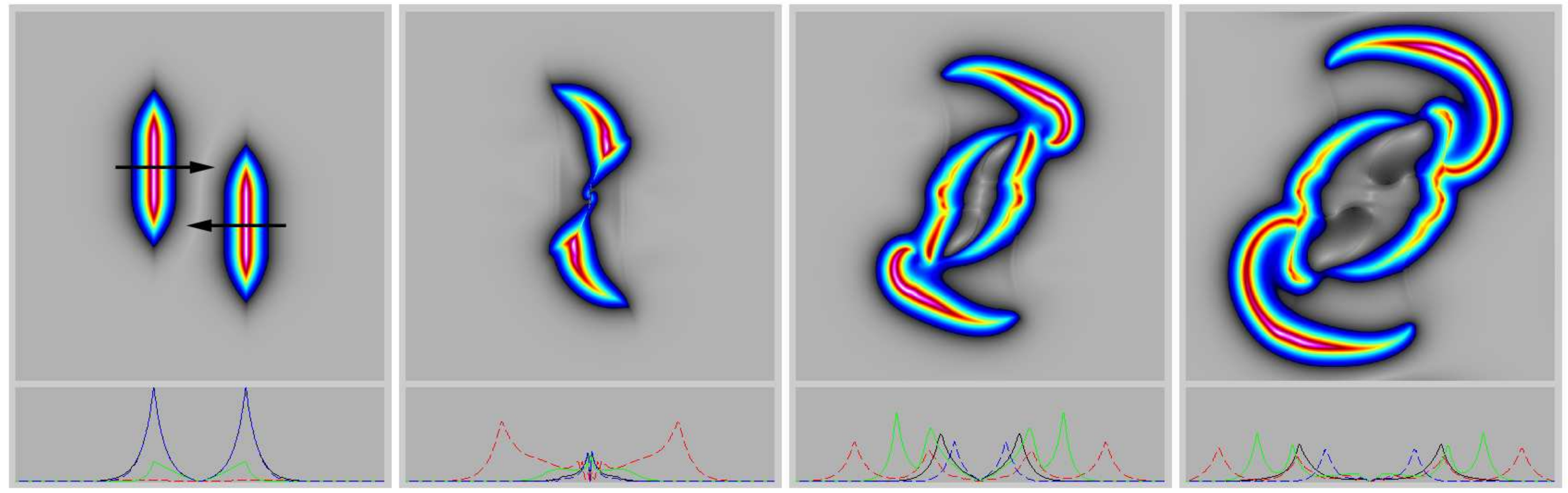}
{Head-on collisions of peakon segments first annihilate and then recreate their
wavefronts.}

\paragraph{Emergence of contact curve segments from representative
classes of smooth initial fluid velocity distributions and their
subsequent interactions.}
For example, consider an initial velocity distribution whose magnitude
(speed) is a circular Gaussian ring (annulus) and whose
direction is chosen in the following ways.

{\it Uniform translation (broken angular symmetry)\/}.
This simulation, shown in Figure \ref{preview-2d-right}, demonstrates
that the initial value problem for EPDiff tends to produce only contact
solutions. It also shows the differences in their propagation under
convergent (left half of annulus) and divergent (right half of annulus)
geometry, and illustrates the exchange of momentum in an overtaking
collision. The collision as the rightmost contact curve is overtaken from
the left transfers some of the overtaking curve's momentum forward and
causes the rightmost curve to bulge slightly, as seen in the last two
frames. (Thus, while momentum is transferred in this collision, the
overtaking contact curve does not ``pass through'' the rightmost contact
curve.)

\previewstrip{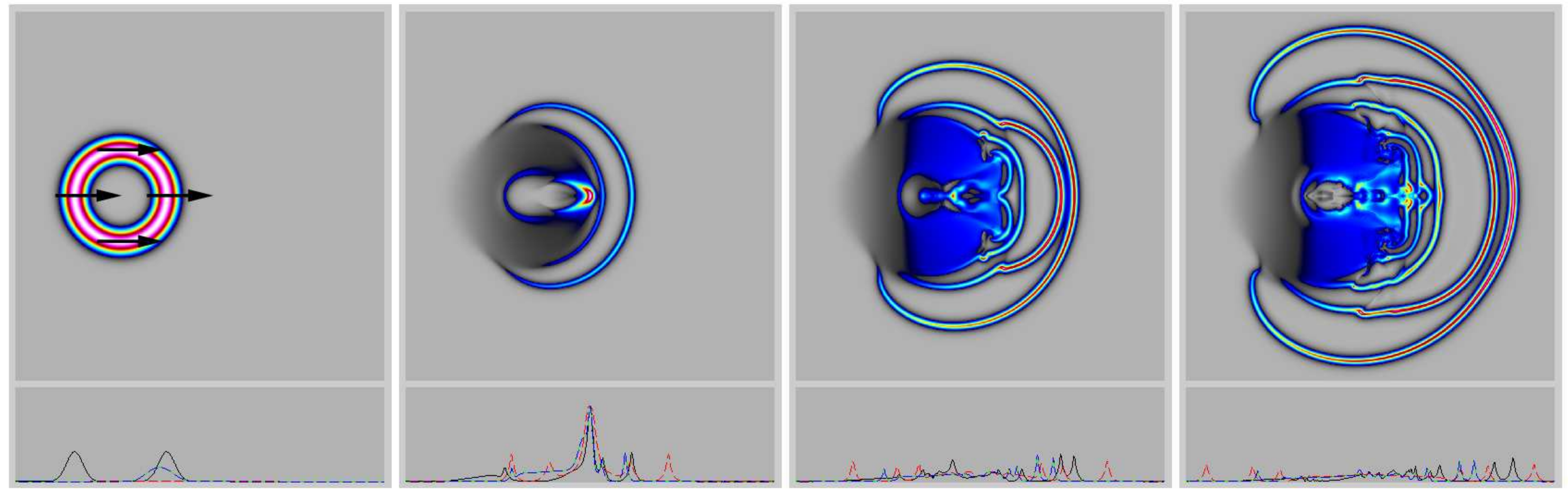}
{A rightward moving circular Gaussian ring of initial velocity breaks into
divergent and convergent contact wave fronts.}

{\it Uniform rotation (preserving angular symmetry)\/}.
Figure \ref{preview-2d-rotate} illustrates how the radial and
azimuthal components of EPDiff are coupled for radially symmetric
solutions. The figure also shows collapse to the origin and reflection
back outward, which presents an extreme test of the accuracy of our
Cartesian numerical algorithm. The radial collapse and bounce can be
computed semi-analytically for comparison, \cite{HoPuSt2004}.

\previewstrip{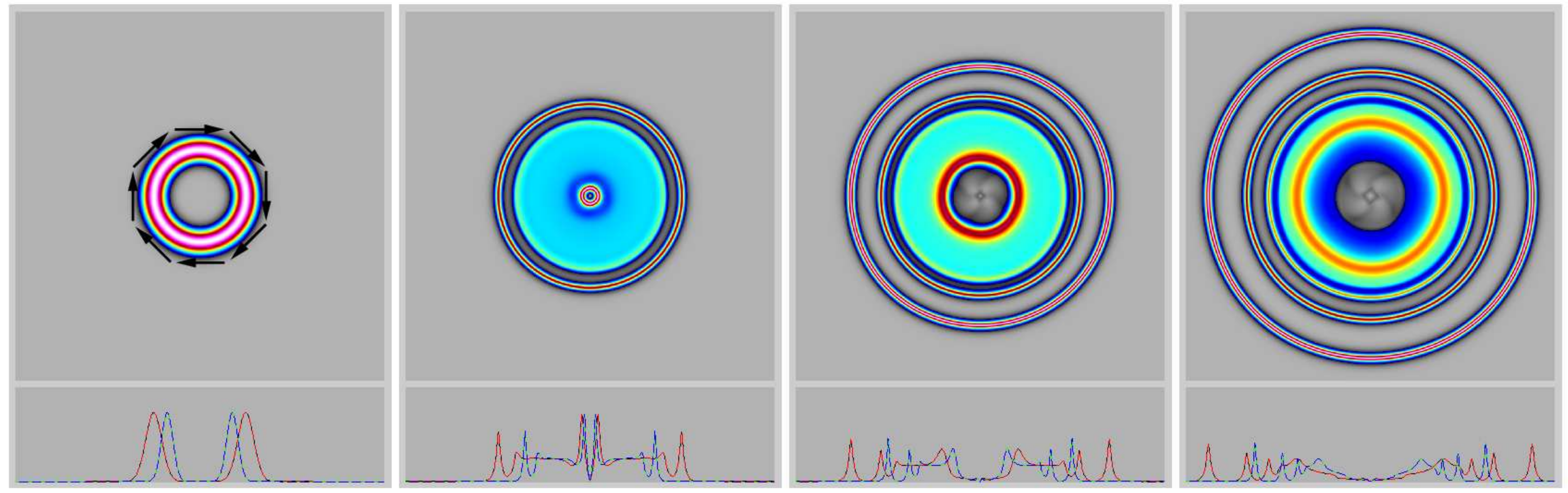}
{An initially rotating circular Gaussian ring of velocity couples its angular
motion to the radial motion of contact wavefronts which propagate both inward
and outward. The collapsing circular wavefronts reflect from the center of
symmetry.}

{\it M-fold discrete angular symmetry\/}.
Formation of contacts in the outward divergent part of the flow, and
the emergence, annihilation, and recreation of contacts collapsing to the
diagonal, are all illustrated in Figure \ref{preview-2d-inout}.
This process is typically followed by very complex patterns of mixing flow
involving peakon profiles along each 1d section.

\previewstrip{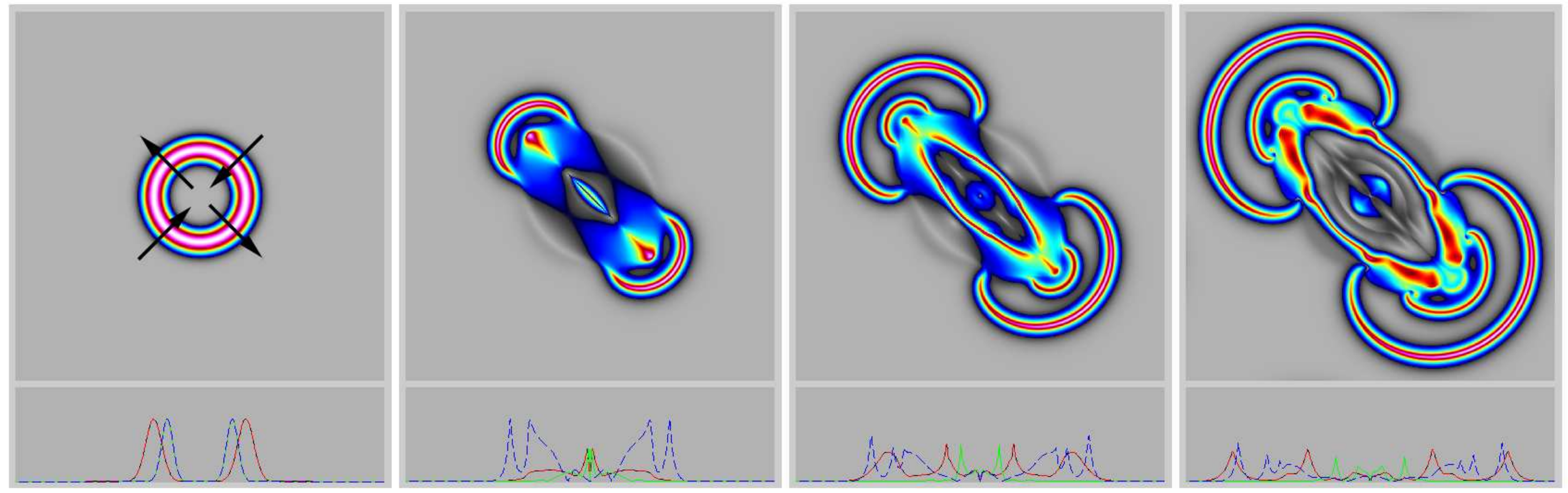}
{An initially circular Gaussian ring of velocity is diverging along one diagonal
and converging along the other. This initial condition breaks into interacting
contact wavefronts which propagate outward along one diagonal and converge
inward along the other to annihilate and re-emerge, leaving behind a complex
mixing flow in the center.}

\paragraph{3d.}
%
In 3d, the initial value problem for EPDiff produces contact surfaces
moving through space. The interactions (collisions) of these contact
surfaces produce very complex but coherent patterns. In the figures we
discuss here, the left, back, and bottom panels of each box show 2d planar
slices through the center $x$ (southeast), $y$ (northeast), and $z$ (north)
values of $\modu$. The corresponding planar slices through the
level surfaces of $\modu$ are shown on the side panels of each
box. 2d slices that form a plane of symmetry in 3d invariably
reflect the behavior of the corresponding 2d problem. We will address the
following scenarios in 3d.

\paragraph{Evolution and interactions of contact surfaces that are
initially disc-shaped velocity distributions.}
Initially flat contact surfaces evolve by ballooning into curved contact
surfaces, as shown in Figure \ref{preview-3d-plate}.

Because EPDiff is isotropic, it allows equally rapid evolutions in the
directions normal and tangential to the contact surfaces. In particular,
its tangential evolution admits significant stretching. It also allows
reconnection of contact surfaces which intersect transversely as they
collide. The reconnection is caused by transfer of momentum. This
phenomenon is illustrated in Figure \ref{preview-3d-skew}. We shall
investigate the ``collision rules'' under which such reconnections of
surfaces occur.

\previewstrip{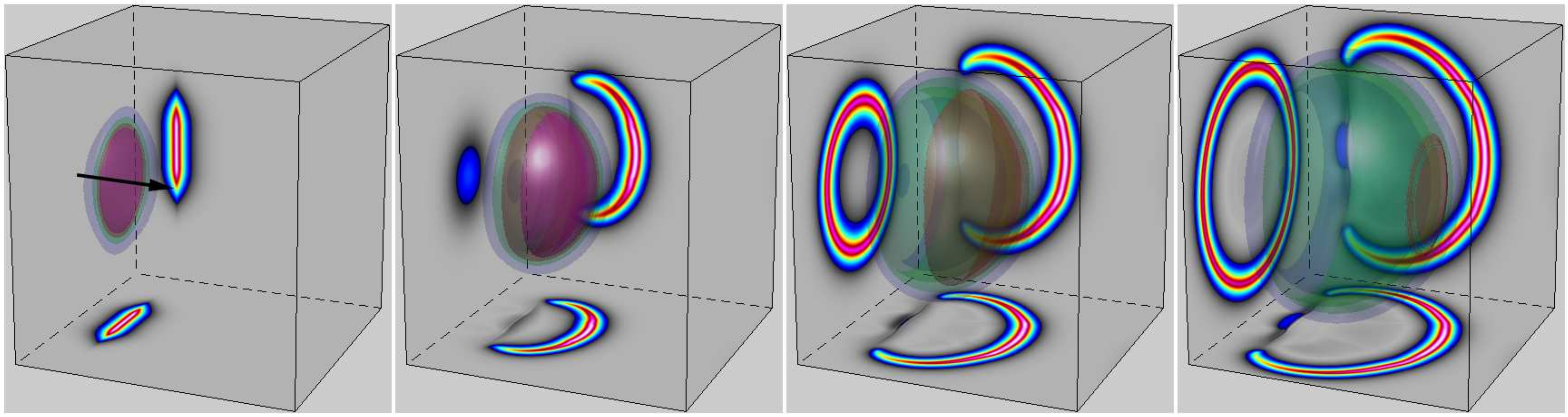}
{In 3d, an initially rightward moving disc with exponential transverse velocity
profile (of correct width alpha) balloons outward as a contact wavefront which
retains its integrity.}

\previewstrip{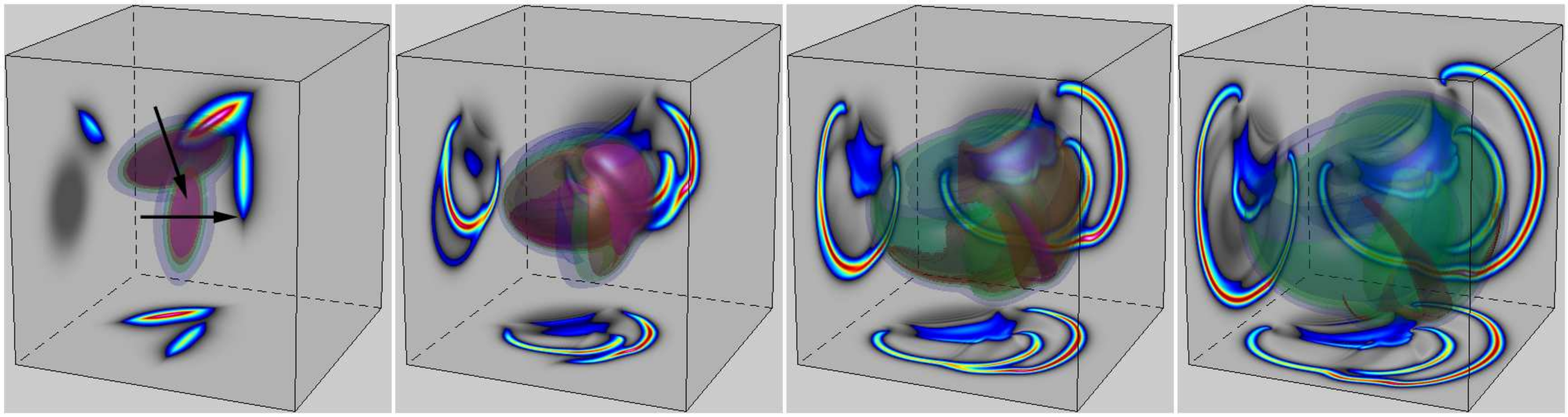}
{The skew collision of two initially disc-shaped contact wavefronts shows
reconnection in 3d.}

An offset collision of initially parallel contact surfaces is shown in
Figure \ref{preview-3d-headon}. As in 2d, we observe the recreation
(after annihilation) of a contact surface after a head-on collision takes
place. Notice also how the outer edges (away from the head-on collision)
of the contact surfaces merge to form a ring.

\previewstrip{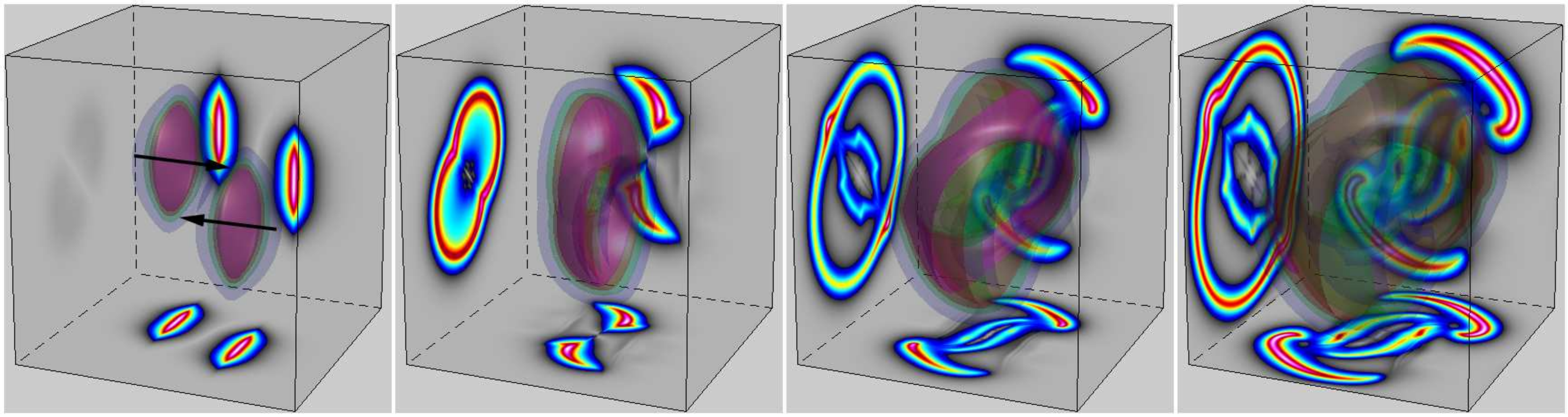}
{The head-on collision of two initially disc-shaped contact wavefronts shows
annihilation and reconnection in 3d.}

\paragraph{Emergence of contact curve surfaces from representative
classes of smooth initial fluid velocity distributions and their
subsequent interactions.}
For example, consider an initial velocity distribution whose magnitude is
a radially symmetric Gaussian shell (spherical annulus) and whose
direction is chosen in the following ways.

{\it Uniform translation (broken angular symmetry)\/}.
The simulation shown in Figure \ref{preview-3d-right}, as in the
corresponding figure for 2d, demonstrates that EPDiff tends to produce
only contact solutions, and illustrates the differences in their
propagation under convergent (left half of spherical annulus) and
divergent (right half of spherical annulus) geometry.

\previewstrip{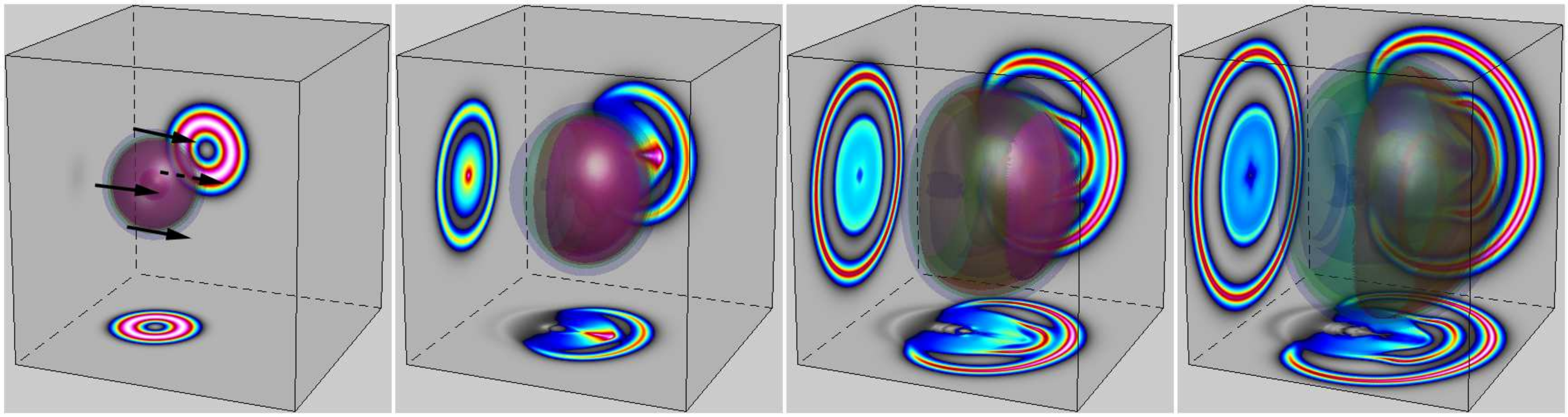}
{A rightward moving spherical Gaussian shell of initial velocity breaks into
divergent and convergent contact wave fronts.}

{\it Uniform rotation (preserving angular symmetry about the vertical
direction)\/}. The frames in Figure \ref{preview-3d-rotate} show
collapse of the spherical contact surfaces to the origin, and their
reflection back outward. This shows the coupling between angular and
radial motion.

\previewstrip{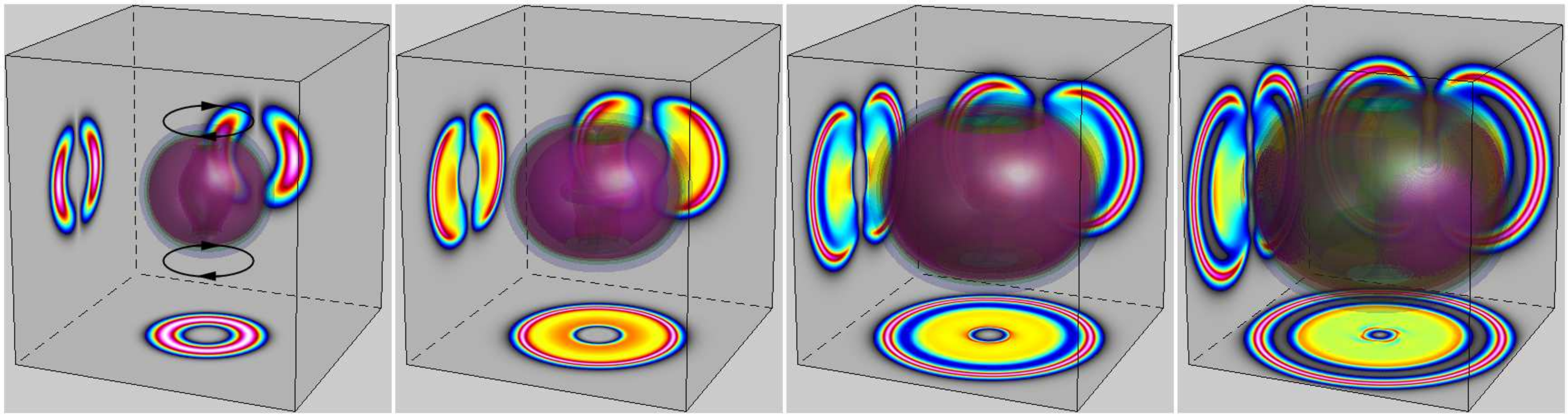}
{An initially spherical Gaussian shell of velocity rotating about the vertical
axis expands outward and collapses inward with cylindrical symmetry, as it
breaks into divergent and convergent contact wave fronts.}

{\it Two-fold discrete angular symmetry\/}.
Figure \ref{preview-3d-inout} illustrates the formation of contacts
for the outward divergent part of the flow and the emergence, annihilation
and recreation of contacts collapsing along lines of discrete symmetry.
The 2d slice through the horizontal midplane (which is projected on the
bottom panel) shows 2d behavior similar to that in Figure
\ref{preview-2d-inout}.

\previewstrip{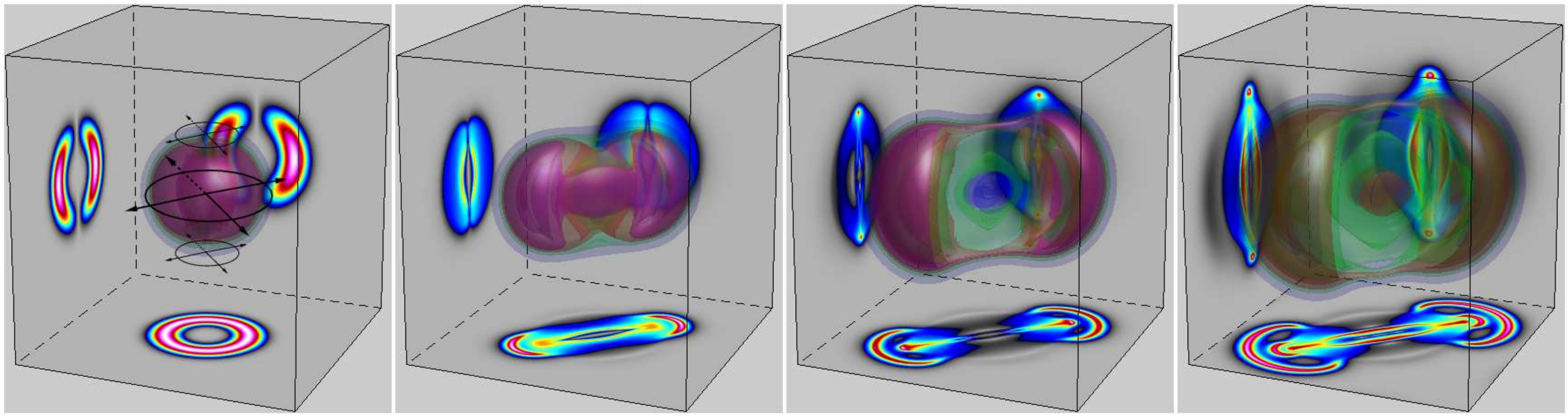}
{An initially spherical Gaussian shell of velocity diverging in one diagonal
vertical plane and converging in the other breaks into interacting contact
surfaces.}

\section{History of modeling internal wave fronts}
\label{model-sec}

Modeling observed internal waves propagating over real topography
requires a fully two-dimensional, multilayer description. Strong,
complex, two dimensional wave-wave and wave-topography interactions have
long been observed, for example, in internal waves propagating through the
Strait of Gibraltar \cite{Zieg1970}. Even more complex interactions were
recently observed from the Space Shuttle for internal wave trains
propagating in the vicinity of Dongsha Island in the South China Sea by
Liu et al. \cite{Liu-etal[1998]}; see Figures \ref{SCS-pan-fronts} and
\ref{SCS-zoom-reconnex}. These interactions produce remarkable nonlinear
phenomena. In particular, they produce wave front reconnection, as well as
the diffraction and refraction expected of large amplitude internal waves
interacting with bathymetry and boundaries such as straits, coasts,
shoals, islands and atolls.

Equations for strongly nonlinear dispersive waves on the free surface of
a {\it single} layer of incompressible fluid with topography are usually
attributed to Green and Naghdi \cite{GN1976}, although the same equations
were derived earlier by Su and Gardner \cite{SuGa1969}. These equations
for single layer columnar motion (1LCM) generalize the Boussinesq family
of shallow-water equations to allow for strong nonlinearity. For finite
wave amplitudes, the 1LCM equations capture strongly nonlinear effects,
such as the upstream emission of solitary waves and the downstream surface
depression and oscillations due to flow over a obstacle at Froude numbers
greater than unity \cite{MaNaSm1996,LiWe1997}.
References \cite{MaNaSm1996,LiWe1997} show that numerical simulations of
the 1LCM equations tend to be faithful representations of the
corresponding Euler solutions, provides that wave breaking does not
occur. The works of Choi and Camassa \cite{ChCa1996,ChCa1999} extended
the 1LCM description to a {\it two-layer} fluid with fixed horizontal upper
and lower boundaries, including the cases of two thin layers and of a thin
layer over an infinitely deep layer. The CC equations admit bi-directional
solutions and they may be derived from a variational principle by
following the averaged Lagrangian methods pioneered in Whitham
\cite{Whitham1967}. (See also Miles and Salmon \cite{MiSa1985}, who
derived the 1LCM equations using Hamilton's principle in the material
representation.) As a consequence of their variational principle, the CC
equations possess conservation laws for mass, momentum, and energy. The
same equations were derived earlier and analyzed for well-posedness in
Liska, Margolin and Wendroff \cite{LiMaWe1995,LiWe1997} by substituting
directly in the governing equations the columnar motion assumption that
the vertical velocity is a linear function of vertical coordinate. Choi
and Camassa \cite{ChCa1999} showed explicitly, by assuming first weak
nonlinearity and then unidirectional wave propagation, that these
equations recover all the known weakly nonlinear evolution equations for
interfacial waves. The well-posedness issue identified by Liska et al.
\cite{LiMaWe1995,LiWe1997} for these equations is similar to the
steepening lemma result proved for the CH equation in \cite{CaHo1993}.
Namely, an initial velocity distribution possessing an inflection point of
negative slope will develop a vertical slope in finite time. This loss of
well-posedness in finite time is part of the mechanism for the formation of
the singular solutions we seek to investigate here. These singular
solutions dominate the initial value problems we study and one should
expect to deviate from standard concepts of well-posedness in studying
their emergence from smooth initial conditions.

In the next few sections of this paper, we shall use the Euler-Poincar\'e
(EP) variational principle for continuum motion in the spatial, or
Eulerian, representation \cite{HoMaRa1998} to extend the CC equations by
deriving multilayer columnar motion (MLCM) equations. MLCM describes
strongly nonlinear internal waves propagating on the interfaces of
layer-stratified incompressible fluid driven by gravity, moving over
topography and possessing a free surface. The MLCM equations are derived
here by imposing columnar motion in the EP variational principle for the
Euler equations of a multilayer incompressible fluid with a free surface
and variable topography. In one dimension, when the free surface and
bathymetry are neglected, the MLCM equations recover the CC equations
\cite{ChCa1999}.

Strictly speaking, the CC equations are for vertically averaged horizontal
velocities, not simply for columnar motion. Thus, they need not have
emerged from an action principle for multilayer columnar motion. The
vertically averaged horizontal velocities do undergo columnar motion, but
this alone is insufficient to expect {\it a priori} that the CC equations
would arise by substituting columnarity into the action principle.
However, Choi and Camassa noticed {\it a posteriori} that their equations
do satisfy a variational principle \cite{ChCa1999}. We have used their
observation, combined with the suggestive results of Liska, Margolin and
Wendroff \cite{LiMaWe1995,LiWe1997} and of Miles and Salmon
\cite{MiSa1985} to extend the CC equations to the more general
MLCM situation required for modeling the large scale internal wave
interactions observed in the Gibraltar Strait and the South China Sea.

The new MLCM equations (\ref{EP-MLCM-eqns}) derived here generalize the CC
equations for the interfacial motion between two layers with fixed
horizontal upper and lower boundaries, to allow for waves at the top free
surface as observed by the Space Shuttle, to include an arbitrary number
of fluid layers, and to account for variable bottom topography. These
equations provide fundamental insight into how topography and the multiple
layers of a stably stratified incompressible fluid are coupled by
nonlinear, nonhydrostatic, dynamical effects. This dynamical coupling
arises in addition to the more familiar multilayer hydrostatic effects. We
derive these nonhydrostatic equations and their natural boundary
conditions from the Euler-Poincar\'e variational principle for continuum
motion of Holm, Marsden and Ratiu \cite{HoMaRa1998} by including in the
fluid Lagrangian the kinetic energy due to vertical oscillations in the
columnar motion approximation. Their traveling waves, stability properties,
well-posedness, weakly nonlinear aspects, relations to other
approximations and numerical simulations will be addressed elsewhere.

After deriving the highest level nonhydrostatic MLCM equations
for multilayer internal waves, we make a series of Boussinesq-like
approximations for weakly nonlinear waves that allows comparison with
previous work and which eventually results in a minimal description,
EPDiff (\ref{H1-EP-eqn-Intro}). We then discuss some of the geometrical
properties and singular solutions of EPDiff. Finally, we describe our
numerical approach and present a large suite of 2d and 3d simulation
results for EPDiff.

\section{Nonhydrostatic multilayer columnar motion (MLCM) equations}
\label{multi-setup-sec}

Consider a multilayer fluid consisting of $N$ immiscible layers moving
under the constant vertical acceleration of gravity. Regard the top layer
(whose rest position is $z=0$) as first and the bottom layer as last
($N$th), so that $i$ increases with the depth of the layer. The $i-$th
layer has constant density $\rho_i$, horizontal velocity
${\mathbf{u}}_i(x,y,z,t)$, vertical velocity $w_i(x,y,z,t)$, upper surface
at $z=h_i(x,y,t)$ and lower surface at $z=h_{i+1}(x,y,t)$, so the layer
thicknesses are
$D_i=h_i-h_{i+1}$, with $i=1,\dots,N$. The $N$th layer (on the bottom) has
density $\rho_N$, horizontal velocity ${\mathbf{u}}_N$, upper surface at
$z=h_N(x,y,t)$ and lower surface at $z=h_{N+1}=-b(x,y)$, the fixed bottom
topography. We shall assume the multilayer fluid is stably stratified, so
that $\rho_i<\rho_{{i+1}}$, for $i=1,\dots,N$ (density increases downward).

\subsection{Implications of the columnar motion ansatz}

The Lagrangian in Hamilton's principle for a multilayer fluid is the
difference of its kinetic and potential energies,
\begin{equation}\label{Lag1-multi}
\ell
=
\int dxdy\,\sum_{i=1}^N\rho_i\int_{h_{i+1}}^{h_i}
\Big[
\frac{1}{2}|{\mathbf{u}}_i|^2
+
\frac{1}{2}w_i^2
-
gz
\Big]\,dz
\,.
\end{equation}
One may also include the effects of Coriolis force due to rotation by
adding the term $\sum_{i=1}^ND_i{\mathbf{R}}(x,y)\cdot{\mathbf{u}}_i +
D_iS(x,y)w_i$ to the Lagrangian density, where $2\Omega = {\rm
curl\,}\big({\mathbf{R}}(x,y) + S(x,y)\,{\mathbf{\hat{z}}}\big)$ is twice
the rotation vector.

We note that incompressibility $\nabla\cdot{\mathbf{u}}_i=- \partial_zw_i$
relates the horizontal and vertical velocities in each layer. When one
also assumes the {\bfi columnar motion ansatz}, namely,
\begin{equation}\label{column-ansatz}
\frac{\partial{\mathbf{u}}_i}{\partial z}=0\,,
\end{equation}
then incompressibility also implies linear dependence of the vertical
velocity on the vertical coordinate in each layer, as
\begin{equation}\label{w-eqn}
w_i=-z\nabla\cdot{\mathbf{u}}_i
+
\partial_t h_{i+1} + \nabla\cdot (h_{i+1}{\mathbf{u}}_i)
\,,
\end{equation}
for $h_{i+1}\le z \le h_i$.
The thickness $D_i=h_i-h_{i+1}$ of the $i-$th layer obeys a
continuity equation,
\begin{equation}\label{continuity}
\partial_t D_i + \nabla\cdot (D_i{\mathbf{u}}_i)=0
\,.
\end{equation}
Hence, the volume of each constant density layer is conserved and the layer
thicknesses remain positive. We also note that
$
h_{i+1}=-\,b(x,y)+\sum_{j=i+1}^ND_j
\,,
$
as the sums of differences $D_j$ cancel in pairs and $h_{N+1}=-\,b(x,y)$. The
total depth is $\sum_{j=1}^ND_j=h_1(x,y,t)+b(x,y)$.

As a result of these relations, the vertical velocity in the $i-$th layer for
$h_{i+1}\le z \le h_i$ may be expressed in terms of the horizontal velocities,
the layer thicknesses, and the prescribed bathymetry, $b(x,y)$, as
\cite{footnote0}
\begin{eqnarray}\label{vert-col}
w_i
&=&
-z\,\nabla\cdot{\mathbf{u}}_i
-
\nabla\cdot b(x,y)\,{\mathbf{u}}_i
-
\sum_{j=i+1}^N \nabla\cdot D_j\big({\mathbf{u}}_j-{\mathbf{u}}_i\big)
\,.
\end{eqnarray}
We shall substitute this expression obtained from columnarity into the
Lagrangian (\ref{Lag1-multi}), then perform the vertical integrals and use the
{\bfi Euler-Poincar\'e} theory to obtain the motion equation in each layer, as
the EP equation \cite{HoMaRa1998},
\begin{equation}\label{EP-multilayer-eqn}
\frac{\partial}{\partial t}
\frac{\delta \ell}{\delta {\mathbf{u}}_i}
+\,
{\mathbf{u}}_i\cdot\nabla
\frac{\delta \ell}{\delta {\mathbf{u}}_i}
+\,
\nabla {\mathbf{u}}_i^T\cdot
\frac{\delta \ell}{\delta {\mathbf{u}}_i}
+
\frac{\delta \ell}{\delta {\mathbf{u}}_i}
\nabla \cdot{\mathbf{u}}_i
-
D_i\nabla
\frac{\delta \ell}{\delta D_i}
=
0
\,.\end{equation}
This procedure will produce the {\bfi multilayer columnar motion} (MLCM)
equations (\ref{EP-MLCM-eqns}), which are completed by the
corresponding continuity equation (\ref{continuity}) in each layer.

\subsection{Variational derivatives of the Lagrangian}

Columnarity (\ref{column-ansatz}) and its implied formula (\ref{vert-col}) for
$w_i$ allow the vertical integrals in the Lagrangian (\ref{Lag1-multi}) to be
performed as,
\begin{eqnarray}\label{Lag-hydro-multi}
\ell
&=&
\frac{1}{2}\int \sum_{i=1}^N\rho_i
\bigg[
D_i|{\mathbf{u}}_i|^2
-\
g\Big(h_i^2-h_{i+1}^2\Big)
\\
&+&
\frac{D_i}{6}\Big(
B_i^2
+
(D_iA_i+B_i)^2
+
(D_iA_i+2B_i)^2 \Big)
\bigg]dxdy\,.
\nonumber\end{eqnarray}
Perhaps not unexpectedly, the integrated kinetic energy is a
layer-thickness-weighted sum of squares of the horizontal velocities and their
divergences. The additional notation is defined as
$A_i\equiv\nabla\cdot{\mathbf{u}}_i\,,$
so that
\[D_iA_i
=
-\,\Big(\partial/\partial t
+ {\mathbf{u}}_i\cdot\nabla\Big)D_i
\equiv
-\,D_i\!{\boldsymbol{\dot{}}}
=\ {\boldsymbol{\dot{}}}\!\!\!h_{i+1}
-\ {\boldsymbol{\dot{}}}\!\!\!h_{i}\,,\] and
\begin{eqnarray}
B_i
&\equiv&
{\mathbf{u}}_i\cdot\nabla b
+
\sum_{j=i+1}^N\nabla\cdot (D_j{\mathbf{u}}_j)
\nonumber\\
&=&
-\,\Big(\frac{\partial}{\partial t}
+ {\mathbf{u}}_i\cdot\nabla\Big)h_{i+1}
=
w_i|_{z=h_{i+1}}
\equiv
-\,\ {\boldsymbol{\dot{}}}\!\!\!h_{i+1}
\,.\label{B-def}
\end{eqnarray}
Thus, $D_iA_i$ is the rate of expansion of the $i-$th layer, and $B_i$ is the
vertical velocity at its lower interface. Moreover, the vertical velocity at
its upper interface is
$
D_iA_i+B_i
=
-\,\ {\boldsymbol{\dot{}}}\!\!\!h_i
\,.$
The differences of squares in the potential energy of (\ref{Lag-hydro-multi})
may also be written in terms of the layer thicknesses upon substituting for
$h_{i+1}$, as
\begin{eqnarray}\label{diff-squares}
\frac{1}{2}\Big(h_i^2-h_{i+1}^2\Big)
&=&
\frac{1}{2}D_i^2 + D_ih_{i+1}
\\
&=&
\frac{1}{2}D_i^2 + D_i\,\Big(-\,b(x,y)+\sum_{j=i+1}^ND_j\Big)
\,.\nonumber
\end{eqnarray}
Thus, columnarity (\ref{column-ansatz}) allows the Lagrangian (\ref{Lag1-multi})
to be expressed solely in terms of the horizontal velocities $\{{\mathbf{u}}\}$,
their (weighted) divergences, and the layer thicknesses $\{D\}$. The potential
energy of each layer is coupled hydrostatically to all the layers beneath it by
the last term in (\ref{diff-squares}). We leave the top layer free. For a rigid
lid \cite{footnote1} one would add the constraint
$h_1=-b(x,y)+\sum_{i=1}^ND_i=0$, imposed by a Lagrange multiplier $p_s$ (the
surface pressure) determined by preservation of $h_1=0$.

We rearrange the sums by using the following identity, which holds for arbitrary
quantities $Q_i$ and $R_i$, with $i=1,2,\dots,N$,
\begin{equation}\label{sum-id}
\sum_{i=1}^N\left(Q_i\sum_{j=i+1}^NR_j\right)
=
\sum_{i=1}^N\left(R_i\sum_{j=1}^{i-1}Q_j\right)
\,.
\end{equation}
Consequently, we find that the variational derivatives of the Lagrangian
(\ref{Lag1-multi}) under columnarity (\ref{column-ansatz}) are given by
\cite{footnote2}
\begin{eqnarray}
\delta\ell
&=&
\int dxdy\,\sum_{i=1}^N\rho_i
\Bigg\{
\Big(
\frac{1}{2}|{\mathbf{u}}_i|^2
-
g h_i
-\frac{g}{\rho_i}
\sum_{j=1}^{i-1}\rho_jD_j\Big)\delta D_i
\nonumber\label{vert-KE}\\
&+&
\bigg[
\frac{1}{2}(D_iA_i+B_i)^2
+
\sum_{j=1}^{i-1}\nabla\cdot C_j{\mathbf{u}}_j
-
{\mathbf{u}}_i\cdot\nabla\sum_{j=1}^{i-1} C_j
\bigg]\delta D_i
\\
&+&
\bigg[D_i{\mathbf{u}}_i
-
\nabla (D_i^2F_i)
-
C_i\nabla h_{i+1}
-
D_i\nabla \sum_{j=1}^{i-1}C_j
\bigg]\cdot\delta{\mathbf{u}}_i
\Bigg\}
\nonumber
\end{eqnarray}
where we have introduced notation for two more linear combinations of the
interface velocities,
\begin{eqnarray}\label{F-def}
F_i\equiv
\frac{1}{3}D_iA_i+\frac{1}{2}B_i
=
-\,\frac{1}{6}(2\,\ {\boldsymbol{\dot{}}}\!\!\!h_i+\,\
{\boldsymbol{\dot{}}}\!\!\!h_{i+1})
\,,\end{eqnarray}
and $C_i \equiv D_iG_i\,,$ with average interface velocity,
\begin{eqnarray}\label{CG-def}
G_i
\equiv
\frac{D_iA_i}{2}
+
B_i
=
-\,\frac{1}{2}(\,\ {\boldsymbol{\dot{}}}\!\!\!h_i
+\,\ {\boldsymbol{\dot{}}}\!\!\!h_{i+1})
\,.
\end{eqnarray}

\subsection{Euler-Poincar\'e motion equation for MLCM}

Substituting the variational derivatives from the formula for $\delta\ell$ in
(\ref{vert-KE}) into the Euler-Poincar\'e equation (\ref{EP-multilayer-eqn})
yields the desired {\it multilayer columnar motion} (MLCM) system of
equations, which may be manipulated into the simple form,
\begin{eqnarray}\label{EP-MLCM-eqns}
&&\frac{d{\mathbf{u}}_{\,i}}{d t}
+
g\nabla H_i
=
\frac{1}{D_i}\nabla\Big(D_i^2\,\frac{dF_i}{d t}\Big)
-\,\frac{dG_i}{d t}\,\nabla \, b(x,y)
\\&&
+\
\nabla\
\sum_{j=1}^{i-1}
D_j\Big(
\frac{\partial G_j}{\partial t} + {\mathbf{u}}_j\cdot\nabla G_j
\Big)
+
\frac{dG_i}{d t}\,\nabla\sum_{j=i+1}^ND_j
\,.\nonumber
\end{eqnarray}
The top line of equation (\ref{EP-MLCM-eqns}) recovers the 1LCM equation
of Su-Gardner \cite{SuGa1969} or Green-Naghdi \cite{GN1976} in each layer.
The new terms involve sums on $j\ne i$, which couple the layers. The
nonhydrostatic contributions on the right side are {\it dynamical}, with
   $d/dt=(\partial/\partial t + {\mathbf{u}}_i\cdot\nabla)$.
The hydrostatic pressure in the $i-$th layer is given by
$p_i\equiv\rho_igH_i\,,$ with $H_i$ as in $\delta\ell$ (\ref{vert-KE}),
\begin{equation}\label{hydro-head}
H_i\equiv  h_i
+
\frac{1}{\rho_i}
\sum_{j=1}^{i-1}\rho_jD_j
 \quad\hbox{and}\quad
h_i=-\,b+\sum_{j=i}^ND_j
\,.
\end{equation}
The quantities $F_i\,$ and $G_i$ in (\ref{EP-MLCM-eqns}) are defined in terms of
the layer thicknesses and velocities by equations (\ref{F-def},\ref{CG-def}).
The resulting MLCM equations (\ref{EP-MLCM-eqns}) are closed by the layer
continuity relations (\ref{continuity}) for $D_i$.

The MLCM motion equations (\ref{EP-MLCM-eqns}) reduce to the standard 1LCM
equations \cite{SuGa1969,GN1976} in the single layer case, in which there
are no sums on $j\ne i$. In one dimension, the MLCM equations generalize
equations (3.19-3.22) of Choi and Camassa \cite{ChCa1999} for the
interfacial motion between two layers with fixed horizontal upper and
lower boundaries, to allow for waves at the top free surface, to include
an arbitrary number of fluid layers and to account for variable bottom
topography. One recovers the CC equations of
\cite{ChCa1999,LiMaWe1995,LiWe1997} by specializing to two layers with fixed
upper and lower boundaries in one dimension. This amounts to setting $N=2$ and
neglecting terms involving $B_i$, $G_i$ and $G_j$ in equations
(\ref{vert-KE}-\ref{EP-MLCM-eqns}).

\paragraph{Nonhydrostatic contributions to momentum and pressure.}
In the variational formula for $\delta\ell$ (\ref{vert-KE}), we see that the
horizontal momentum in the columnar motion equation for the $i-$th layer
involves horizontal gradients,
\begin{eqnarray}\label{mom-def}
{\mathbf{m}}_i
&\equiv&
\frac{\delta \ell}{\delta {\mathbf{u}}_i}
\equiv
\sum_{j=1}^N{\mathcal{L}}_{ij}(\{D\},b)\,
{\mathbf{u}}_j
\\
&=&
\rho_iD_i\Big({\mathbf{u}}_i
-
\frac{1}{D_i}\nabla(D_i^2F_i)
-\
G_i\nabla h_{i+1}
-
\nabla\sum_{j=1}^{i-1}
D_jG_j\Big)
\,.\nonumber
\end{eqnarray}
The last equality defines the symmetric, positive definite operator in
${\mathbf{m}}_i=\sum_{j=1}^{i-1}{\mathcal{L}}_{ij}(\{D\},b){\mathbf{u}}_j$,
which depends on the set of layer thicknesses $\{D\}$ and the bottom topography
$b$. In terms of this operator, the multilayer fluid Lagrangian
(\ref{Lag1-multi}) under columnarity (\ref{column-ansatz}) becomes
\begin{eqnarray}\label{Lag-multi}
\ell
&=&\frac{1}{2}
\!\!\int\sum_{i=1}^N
\Big({\mathbf{m}}_i\cdot
{\mathbf{u}}_i
-
g(h_i^2-h_{i+1}^2)\Big) dxdy
\,.
\end{eqnarray}
Because the multilayer Lagrangian (\ref{Lag-multi}) depends on the horizontal
spatial coordinate through the bathymetry $b(x,y)$, the spatial integral of the
total momentum (\ref{mom-def}) is not conserved, except in translation-invariant
directions of $b(x,y)$. The total pressure in the $i-$th layer,
$\delta\ell/\delta D_i$, also contains nonhydrostatic contributions arising from
the vertical columnar oscillations and velocity differences among the layers, in
addition to its hydrostatic component. All of these additional nonhydrostatic
contributions to the momentum and the pressure gradient arise from the kinetic
energy of vertical motion and are proportional to the velocities of the
interfaces \cite{footnote3}.

\paragraph{Kelvin circulation theorem and potential vorticity.}
Although the layer motions are strongly coupled, each layer has its own Kelvin
circulation theorem,
\begin{equation}\label{Kel-circ-ith}
\frac{d}{dt}\oint_{c({\mathbf{u}}_i)}
\frac{{\mathbf{m}}_i}{D_i}
\cdot d\mathbf{x}
=
0
\,,
\end{equation}
where the closed loop $c({\mathbf{u}}_i)$ moves with the horizontal velocity
${\mathbf{u}}_i$ in the $i-$th layer. In addition, each layer also locally
conserves its own potential vorticity (PV), i.e.,
\begin{equation}\label{PV-cons}
\frac{\partial q_i}{\partial t}
 +
{\mathbf{u}}_i\cdot\nabla q_i
=
0
\quad\hbox{where}\quad
q_i=\frac{1}{D_i}\, {\mathbf{\hat{z}}}\cdot\,{\rm
curl}\,\frac{{\mathbf{m}}_{\,i}}{D_i}
\,.\end{equation}
Consequently, one has an infinite set of conservation laws,
\begin{equation}\label{Casimirs}
C_\Phi
=
\int
D_i\,\Phi(q_i)\,dxdy
\,,\quad\hbox{for any function}\quad\Phi
\,.\end{equation}
The $C_\Phi$ are Casimirs of the Lie-Poisson Hamiltonian operator for the
MLCM system and they play a role in classifying the MLCM equilibrium solutions,
as in \cite{Holm1988,HoMaRaWe1985}.

Both the Kelvin circulation theorem and its associated local conservation of
potential vorticity follow from the invariance properties of the EP variational
principle. Namely, the EP variational principle is invariant under fluid parcel
relabeling that preserves Eulerian quantities. For full details of the
Kelvin-Noether theorem, see \cite{HoMaRa1998}. The implications of PV
conservation for multilayer internal waves remain to be investigated. PV
conservation is a new element of the MLCM equations, which differs fundamentally
from the standard Charney-Drazin non-acceleration theorem approach for purely
potential waves \cite{ChDr1961}, in allowing solutions representing waves, bores
and circulations to coexist and interact nonlinearly.

\paragraph{Lie-Poisson Hamiltonian structure of MLCM.}
The Euler-Poincar\'e variational formulation implies the Lie-Poisson Hamiltonian
structure of MLCM, upon Legendre transforming the multilayer Lagrangian $\ell$ in
(\ref{Lag-multi}), as shown in \cite{HoMaRa1998}. Hence, the conserved
Hamiltonian for the MLCM equations is
$
H_{MLCM}
=
\int \sum_{i=1}^N{\mathbf{m}}_i\cdot{\mathbf{u}}_i\, dxdy
\, - \,
\ell
$.
Thus, the MLCM equations of motion are expressible in Lie-Poisson Hamiltonian
form using the standard {\bfi Lie-Poisson bracket} in terms of the momenta
$\{{\mathbf{m}}\}$ and the layer depths $\{D\}$, as in
\cite{HoMaRa1998,Holm1988,HoMaRaWe1985}. As expected, the Casimirs for this
Lie-Poisson bracket are the potential vorticity functionals
$C_\Phi=\int D_i\Phi(q_i)\,dxdy$, which satisfy $\{H,C_\Phi\}=0$ for any
Hamiltonian $H$. The corresponding treatment of the Lie-Poisson Hamiltonian
structure for the single layer GN equations is given in
\cite{Holm1988,CaHoLe1996}.

\paragraph{Comparison with alternative 2d averaged shallow water equations.}
In addition to possessing conservation properties for energy,
circulation and potential vorticity, the EP motion equation
(\ref{EP-MLCM-eqns}) takes a simpler form than many other depth-integrated
motion equations discussed in the literature, such as in
\cite{LyLiu2002,GoKiWe2000,MaAg2003}. Whether MLCM will be as successful
in simulating internal wave interactions remains to be seen. As in all
columnar motion equations, a key feature of MLCM is the elliptic operator
in (\ref{mom-def}) relating velocity and momentum. Our numerical
simulations in section \ref{num-2d-sec} show that a weakly nonlinear
approximation of the elliptic inversion in the MLCM model does capture the
characteristic aspects of the internal wave-front collisions and
reconnections which were observed in the South China Sea by Liu et al.
\cite{Liu-etal[1998]}.

\section{Weakly nonlinear limit equations}
\label{weak-nonlin-sec}

\subsection{EP derivation of Boussinesq-like equations}
\label{Bouss-subsec}

One may rewrite the total Lagrangian (\ref{Lag-hydro-multi}) without
approximation as
\begin{eqnarray}\label{Lag-hydro-multi-1}
\ell
&=&
\frac{1}{2}\int \sum_{i=1}^N\rho_i
\bigg[
D_i|{\mathbf{u}}_i|^2
-\
g\Big(h_i^2-h_{i+1}^2\Big)
\nonumber\\
&&\hspace{15mm}+\
\frac{D_i}{3}\Big(
(D_iA_i)^2
+
3B_{i-1}B_i
\Big)
\bigg]dxdy
\,.\label{multi-D-Lag}
\end{eqnarray}
The two-dimensional CC equations arise as EP equations from this Lagrangian upon
neglecting its final term,
$\frac{1}{2}\int \sum_{i=1}^N\rho_iD_iB_{i-1}B_i\,dxdy$. We introduce an
alternative approximation of the last term, by approximating it in the
weakly nonlinear limit, as
\begin{eqnarray}
\ell
&=&
\frac{1}{2}\int \sum_{i=1}^N\rho_i
\bigg[
D_i|{\mathbf{u}}_i|^2
-\
g\Big(h_i^2-h_{i+1}^2\Big)
\nonumber\\
&&\hspace{15mm}+\
\frac{d_i}{3}\Big(
({\rm div\,}D_i{\mathbf{u}}_i)^2
+
3B_{i-1}B_i
\Big)
\bigg]dxdy
\,,
\label{Lag-Bouss-like}
\end{eqnarray}
where $\{d_i:\,i=1,2,\dots,N\}$ are a set of $N$ constants, differing only
slightly from the corresponding $D_i$. In this alternative approximation, the
Lagrangian (\ref{Lag-Bouss-like}) has the following variational derivatives
\begin{eqnarray}\label{Bouss-mom-def}
\mathbf{m}_i
\equiv
\frac{\delta \ell}{\delta \mathbf{u}_i}
&=&
\rho_iD_i\Big(\mathbf{u}_i
-
\frac{d_i}{3}\nabla({\rm div\,}D_i{\mathbf{u}}_i)
+\
\frac{1}{D_i}\frac{\delta{\mathcal{B}}_i}{\delta{\mathbf{u}}_i}
\Big)
\,\end{eqnarray}
for the momentum density, where
${\mathcal{B}}_i=({d_i}/{2})B_{i-1}B_i\,,$ and
\begin{eqnarray}\label{Bouss-pot-def}
\frac{1}{\rho_i}\frac{\delta \ell}{\delta D_i}
&=&
\frac{1}{2}|\mathbf{u}_i|^2
-
g H_i
-
\frac{d_i}{3}\nabla({\rm div\,}D_i{\mathbf{u}}_i)
+
\frac{\delta{\mathcal{B}}_i}{\delta{D_i}}
\,,\end{eqnarray}
for the Bernoulli potential.

From these variational derivatives, the multilayer EP equations
(\ref{EP-multilayer-eqn}) produce
\begin{eqnarray}\label{Bouss-EP-multilayer-eqn}
\frac{\partial}{\partial t}{\mathbf{u}_i}
\!\!&+&\!\!
{\mathbf{u}_i}\cdot\nabla{\mathbf{u}_i}
+\
g\nabla H_i\
+\
\frac{d_i}{3}\nabla\frac{\partial^2D_i}{\partial t^2}
\nonumber\\
\!\!&=&\!\!
{\mathbf{u}_i}\times{\rm curl\,}
\frac{1}{D_i}\frac{\delta{\mathcal{B}}_i}{\delta{\mathbf{u}}_i}
\
-\
\nabla\Big(
\frac{1}{D_i}{\mathbf{u}}_i\cdot
\frac{\delta{\mathcal{B}}_i}{\delta{\mathbf{u}}_i}
-
\frac{\delta{\mathcal{B}}_i}{\delta{D_i}}
\Big)
\,.\end{eqnarray}
We call these the ``Boussinesq-like'' multilayer equations, because the
approximation
$D_iA_i=D_i{\rm div\,}{\mathbf{u}}_i
\simeq{\rm div\,}D_i{\mathbf{u}}_i=-\partial D_i/\partial t$
in the Lagrangian (\ref{Lag-Bouss-like}) replaces the strongly nonlinear
dispersive term in the MLCM equations with Boussinesq-like linear
dispersion, as
\begin{eqnarray}
\frac{1}{3D_i}\nabla\Big(d_i^{\,2}\,\frac{d(D_iA_i)}{d t}\Big)
\to
-\,
\frac{d_i}{3}\frac{\partial }{\partial t}
\nabla({\rm div\,}D_i{\mathbf{u}}_i)
=
\frac{d_i}{3}\nabla\,\frac{\partial^2D_i}{\partial t^2}
\,.
\end{eqnarray}
This term is linear, by virtue of the continuity equation for each layer
thickness. The completion of this approximation depends on the treatment
of the ${\mathcal{B}}_i$ terms in equation (\ref{Bouss-EP-multilayer-eqn}),
which are neglected in the CC equations. In the approximation that one neglects
the ${\mathcal{B}}_i$ terms, the multilayer Boussinesq-like equations become
\begin{eqnarray}\label{Bouss-EP-multilayer-approx}
\frac{\partial}{\partial t}{\mathbf{u}_i}
+
{\mathbf{u}_i}\cdot\nabla{\mathbf{u}_i}
+\
g\nabla H_i\
+\
\frac{d_i}{3}\nabla\frac{\partial^2D_i}{\partial t^2}
=
0
\,.\end{eqnarray}
These equations are still coupled by the multilayer hydrostatic pressure
gradient, where hydrostatic pressure head $H_i$ is given in equation
(\ref{hydro-head}).

The contributions of the ${\mathcal{B}}_i=({d_i}/{2})B_{i-1}B_i\,,$ terms
to the EP equations (\ref{Bouss-mom-def}-\ref{Bouss-EP-multilayer-eqn})
may be obtained by computing the required variational derivatives, as
follows.
\begin{eqnarray}\label{Lag-Bouss-var}
&&\delta\int \sum_{i=1}^N\rho_i\frac{d_i}{2}
B_{i-1}B_i
dxdy
=
\int \sum_{i=1}^N\rho_i\frac{d_i}{2}
(B_{i-1}+B_{i+1})\delta{B_i}
\nonumber\\&&
=
\int \sum_{i=1}^N\rho_i\frac{d_i}{2}
(B_{i-1}+B_{i+1})\nabla{b}\cdot\delta{\mathbf{u}_i}
\\&&\qquad
-
\sum_{i=1}^N
({\mathbf{u}_i}\,\delta{D_i}+D_i\,\delta{\mathbf{u}_i})
\cdot\nabla
\sum_{j=1}^{i-1}
\rho_j\frac{d_j}{2}
(B_{j-1}+B_{j+1})
\,.
\nonumber
\end{eqnarray}
In the first line, we have used the definition of $B_i$ in equation
(\ref{B-def}). In the last line, we have dropped boundary terms when
integrating by parts and used the summation identity (\ref{sum-id}). These
formulas determine the contributions of the
${\mathcal{B}}_i$ terms to the EP equations
(\ref{Bouss-mom-def}-\ref{Bouss-EP-multilayer-eqn}).
In particular, they couple the horizontal motion of the $i-$th layer to
the vertical interface velocity of its {\it next} nearest layer below,
since
$B_{i-1}+B_{i+1}=
-\,\ {\boldsymbol{\dot{}}}\!\!\!h_i
-\,\ {\boldsymbol{\dot{}}}\!\!\!h_{i+2}$. Thus, ignoring
the contributions of the ${\mathcal{B}}_i$ terms corresponds to ignoring
interactions among next nearest layers in the weakly nonlinear limit.

\subsection{Other weakly nonlinear EP equations}
\label{weak-nonlin-subsec}

We invoke the weakly nonlinear limit to make a further approximation in the
multilayer Lagrangian (\ref{multi-D-Lag}), by representing its kinetic energy
due to vertical motion as
\begin{eqnarray}
\ell
=
\frac{1}{2}\int \sum_{i=1}^N\rho_i
\bigg[
D_i|{\mathbf{u}}_i|^2
-\
g\Big(h_i^2-h_{i+1}^2\Big)
+
\frac{d_i^{\,2}D_i}{3}\big(
{\rm div\,}{\mathbf{u}}_i
\big)^2
\bigg]dxdy
\,.
\label{weak-nlin-Lag}
\end{eqnarray}
Consequently, its variational derivatives become
\begin{eqnarray}\label{weak-nlin-mom-def}
\mathbf{m}_i
\equiv
\frac{\delta \ell}{\delta \mathbf{u}_i}
&=&
\rho_iD_i\Big(\mathbf{u}_i
-
\frac{d_i^{\,2}}{3D_i}\nabla(D_i{\rm div\,}{\mathbf{u}}_i)
\Big)
\,,\end{eqnarray}
for the momentum density, and
\begin{eqnarray}\label{weak-nlin-pot-def}
\frac{1}{\rho_i}\frac{\delta \ell}{\delta D_i}
&=&
\frac{1}{2}|\mathbf{u}_i|^2
-
g H_i
+
\frac{d_i^{\,2}}{6}\,({\rm div\,}{\mathbf{u}}_i)^2
\,,\end{eqnarray}
for the Bernoulli potential. The corresponding EP equations are,
\begin{eqnarray}\label{EP-weak-nlin-step}
\frac{\partial}{\partial t}{\mathbf{u}_i}
\!\!&+&\!\!
{\mathbf{u}_i}\cdot\nabla{\mathbf{u}_i}
+\
g\nabla H_i
\nonumber\\
&-&
\frac{\partial}{\partial t}\,
\frac{d_i^{\,2}}{3D_i}\,
\nabla(D_i{\rm div\,}{\mathbf{u}_i})
+
{\mathbf{u}_i}\times{\rm curl\,}
\frac{d_i^{\,2}}{3D_i}\,
\nabla(D_i{\rm div\,}{\mathbf{u}_i})
\nonumber\\
\!\!&=&\!\!
-\,\frac{d_i^{\,2}}{3}\,
\nabla\Big(
\frac{1}{2}\,({\rm div\,}{\mathbf{u}}_i)^2
+
\frac{1}{D_i}{\mathbf{u}}_i\cdot
\nabla(D_i{\rm div\,}{\mathbf{u}_i})
\Big)
\,.\end{eqnarray}
In the {\it weakly nonlinear limit}, one neglects terms in
$\nabla{D_i}$ to find
\begin{eqnarray}\label{EP-weak-nlin-approx}
\frac{\partial}{\partial t}\Big({\mathbf{u}_i}
\!\!&-&\!\!
\frac{d_i^{\,2}}{3}\,
\nabla({\rm div\,}{\mathbf{u}_i})\Big)
+
{\mathbf{u}_i}\cdot\nabla{\mathbf{u}_i}
+\
g\nabla H_i
\nonumber\\
\!\!&=&\!\!
-\,\frac{d_i^{\,2}}{3}\,
\nabla\Big(
\frac{1}{2}\,({\rm div\,}{\mathbf{u}}_i)^2
+
{\mathbf{u}}_i\cdot
\nabla({\rm div\,}{\mathbf{u}_i})
\Big)
+
O(\nabla{D_i})
\,.\end{eqnarray}
Upon neglecting terms of order $O(\nabla{D_i})$, choosing the center of volume
frame of reference with $\sum_{i=1}^N D_i{\mathbf{u}}_i=0$ for $N=2$, and
specializing to one dimension, one recovers the weakly nonlinear limit of the
CC equations in \cite{ChCa1999}. These weakly nonlinear limit equations were
shown in \cite{ChCa1999} to recover all of the various Boussinesq approximations
for one dimensional shallow water theory.

\paragraph{Remark on inversion of the second order elliptic operator.}
As with all the equations in the GN family, the evolution equation
(\ref{EP-weak-nlin-approx}) requires inversion of a second order operator
in solving for fluid velocity ${\mathbf{u}_i}$ from momentum density
${\mathbf{m}_i}$ at each time step. A simplification occurs in the weakly
nonlinear approximation, because the operator
$(1-\frac{d_i^{\,2}}{3}\,\nabla{\rm div\,})$
appearing in this approximation is independent of the layer thicknesses, $D_i$,
which do enter at the fully nonlinear level as in equations (\ref{mom-def}),
(\ref{Bouss-mom-def}) and (\ref{weak-nlin-mom-def}). The nonlocality and
smoothing associated with inversion of such elliptic operators is a
characteristic feature of the {\it entire family} of GN equations.

The remainder of this paper is devoted to using the EP theory in characterizing
the two-dimensional effects of this elliptic operator inversion on the solutions
of equations in the GN family of shallow water equations. To focus our attention
on this aspect of the investigation, we shall not require the effects of
potential energy.

\section{Kinetic energy Lagrangians and the EPDiff geodesic equation}
\label{ke-limit-sec}

\subsection{Kinetic energy Lagrangians}

Neglecting potential energy entirely in the weakly nonlinear limit multilayer
Lagrangian (\ref{weak-nlin-Lag}), by setting $D_i\to d_i$, gives
\begin{eqnarray}
\ell
=
\frac{1}{2}\int \sum_{i=1}^N\rho_id_i
\bigg[|{\mathbf{u}}_i|^2
+
\frac{d_i^{\,2}}{3}\big(
{\rm div\,}{\mathbf{u}}_i
\big)^2
\bigg]dxdy
\,.
\label{KE-lim-Lag}
\end{eqnarray}
The corresponding momentum density involves the second order elliptic operator,
\begin{eqnarray}\label{KE-lim-mom-def}
\mathbf{m}_i
\equiv
\frac{\delta \ell}{\delta \mathbf{u}_i}
&=&
\rho_id_i\Big(\mathbf{u}_i
-
\frac{d_i^{\,2}}{3}\nabla({\rm div\,}{\mathbf{u}}_i)
\Big)
\,.\end{eqnarray}
Without potential energy, the layers decouple and the EP equation
(\ref{EP-multilayer-eqn}) takes the same form in every layer, so we may drop the
layer index $i$ and write the EP equation for the kinetic energy Lagrangian
(\ref{KE-lim-Lag}) as
\begin{equation}\label{KE-EP-eqn}
\frac{\partial}{\partial t}
\mathbf{m}
+\,
\nabla \mathbf{u}^T\cdot
\mathbf{m}
+
\mathbf{m}
({\rm div\,}{\mathbf{u}})
=
0
\,,
\quad\hbox{with}\quad
\mathbf{m} = \mathbf{u}
-
\frac{d^{\,2}}{3}\nabla({\rm div\,}{\mathbf{u}})
\,.\end{equation}
By design, this equation has no contribution from potential energy. In addition,
its evolution conserves the kinetic energy,
\begin{eqnarray}
\ell
=
\frac{1}{2}\int
\bigg[|{\mathbf{u}}|^2
+
\frac{d^{\,2}}{3}\big(
{\rm div\,}{\mathbf{u}}
\big)^2
\bigg]dxdy
\,.
\label{KE-def}
\end{eqnarray}
Evolution by kinetic energy in Hamilton's principle results in geodesic motion,
with respect to the velocity norm provided by the kinetic energy Lagrangian.

\paragraph{Reduction to the Camassa-Holm (CH) equation in 1d.}
In one dimension, equation (\ref{KE-EP-eqn}) simplifies to
\begin{equation}\label{H1-EP1D-eqn}
\frac{\partial m}{\partial t}
+
\frac{\partial}{\partial x}(mu)
+
m
\frac{\partial u}{\partial x}
=
0
\,,\quad\hbox{with}\quad
m
=
u
-
\alpha^2\frac{\partial^2 u}{\partial x^2}
\,.
\end{equation}
This is the dispersionless limit of the Camassa-Holm (CH) equation
\cite{CaHo1993}, which is known to be the equation at quadratic order in the
shallow water asymptotic expansion, one full order beyond KdV, whereas KdV
appears at linear order in this expansion
\cite{DuGoHo2001,DuGoHo2003,DuGoHo2004}. (The dispersionless limit of the CH
equation appears because we are ignoring potential energy in this part of our
investigation.)

\paragraph{Strengthening the kinetic energy norm to $H_\alpha^1$.}
The kinetic energy (\ref{KE-def}) is only part of the $H_\alpha^1$ norm of the
velocity, defined as
\begin{eqnarray}
\|{\mathbf{u}}\|^2_{H_\alpha^1}
&=&
\int
\bigg[|{\mathbf{u}}|^2
+
\alpha^2\big(
{\rm div\,}{\mathbf{u}}
\big)^2
+
\alpha^2\big(
{\rm curl\,}{\mathbf{u}}
\big)^2
\bigg]dxdy
\nonumber\\
&=&
\int
\bigg[|{\mathbf{u}}|^2
+
\alpha^2|\nabla{\mathbf{u}}|^2
\bigg]dxdy
\,.
\label{H1-eqn}
\end{eqnarray}
Here we assume ${\mathbf{u}}$ is tangential on the boundaries upon integrating
by parts in Cartesian geometry. We have also simplified the notation slightly by
replacing $d^{\,2}/3$ with $\alpha^2$, where $\alpha$ is a length scale. In
anticipation that mathematical analysis will be facilitated by controlling the
entire $H_\alpha^1$ norm of the velocity, we shall choose our kinetic energy
Lagrangian to be
\begin{eqnarray}
\ell
=
\frac{1}{2}\|{\mathbf{u}}\|^2_{H_\alpha^1}
=
\frac{1}{2}\int
\bigg[|{\mathbf{u}}|^2
+
\alpha^2|\nabla{\mathbf{u}}|^2
\bigg]dxdy
\,.
\label{H1-norm}
\end{eqnarray}
The corresponding EP equation (\ref{EP-multilayer-eqn}) is EPDiff, which
involves the familiar {\it Helmholtz} operator in the relation between fluid
velocity and momentum density,
\begin{equation}\label{H1-EP-eqn}
\frac{\partial}{\partial t}
\mathbf{m}
+
\mathbf{u}\cdot\nabla\mathbf{m}
+
\nabla \mathbf{u}^T\cdot
\mathbf{m}
+
\mathbf{m}
({\rm div\,}{\mathbf{u}})
=
0
\,,\quad\hbox{with}\quad
\mathbf{m}
\equiv
\frac{\delta \ell}{\delta \mathbf{u}}
=
\mathbf{u}
-
\alpha^2\Delta{\mathbf{u}}
\,.\end{equation}
We shall assume {\it periodic boundary conditions} for the remainder of our
investigations in this paper. An alternative way of writing EPDiff
(\ref{H1-EP-eqn}) is
\begin{equation}\label{H1-EPcurl-eqn}
\frac{\partial}{\partial t}
\mathbf{m}
-\,
{\mathbf{u}}\times{\rm curl\,}{\mathbf{m}}
+
{\rm grad}({\mathbf{u}}\cdot{\mathbf{m}})
+
\mathbf{m}
({\rm div\,}{\mathbf{u}})
=
0
\,,\end{equation}
which involves the three differential operators curl, gradient and
divergence in two dimensions. This is the EPDiff equation whose solution
behavior for the initial value problem is studied in the remainder of the
paper.

\subsection{Momentum maps for singular solutions of EPDiff}
\label{mom-map-subsec}

Substituting the singular momentum solution formula (\ref{EP-sing-mom}) for
$s\in{\mathbb{R}^1}$ and its corresponding velocity (\ref{EP-sing-vel}) into
EPDiff, then integrating against a smooth test function, implies the following
Lagrangian wave front equations,
\begin{eqnarray}
\frac{\partial }{\partial t}\mathbf{{Q}}_a (s,t)
\!\!&=&\!\!
\!\!\sum_{b=1}^{N} \int\mathbf{P}_b(s^{\prime},t)\,
G(\mathbf{Q}_a(s,t),\mathbf{Q}_b(s^{\prime},t)\,\big)ds^{\prime}
\,,\label{IntDiffEqn-Q}\\
\frac{\partial }{\partial t}\mathbf{{P}}_a (s,t)
\!\!&=&\!\!
-\,\!\!\sum_{b=1}^{N} \int
\big(\mathbf{P}_a(s,t)\!\cdot\!\mathbf{P}_b(s^{\prime},t)\big)
\, \frac{\partial }{\partial \mathbf{Q}_a(s,t)}
G\big(\mathbf{Q}_a(s,t),\mathbf{Q}_b(s^{\prime},t)\big)\,ds^{\prime}
\,.
\nonumber
\end{eqnarray}
in which summation is explicit on $b\in1,2,\dots N,$ and there is no sum on $a$.
The dot product $\mathbf{P}_a\cdot\mathbf{P}_b$ denotes the inner, or scalar,
product of the two vectors $\mathbf{P}_a$ and $\mathbf{P}_b$ in
${\mathbb{R}}^2$. Thus, the momentum solution formula (\ref{EP-sing-mom}) yields
a closed set of integro-partial-differential equations (IPDEs) given by
(\ref{IntDiffEqn-Q}) for the vector parameters $\mathbf{Q}_a(s,t)$ and
$\mathbf{P}_a(s,t)$ with $i=1,2\dots N$.

\paragraph{Canonical Hamiltonian dynamics of wave fronts in ${\mathbb{R}^2}$.}
The singular momentum solution formula
(\ref{EP-sing-mom}) is shown to be a {\it momentum map} in
\cite{HoMa2004}. This fact guarantees the following result,

\paragraph{Theorem.}
{\it The Lagrangian wave front equations (\ref{IntDiffEqn-Q}) are canonical
Hamiltonian equations,}
\begin{equation} \label{IntDiffEqns-Ham}
\frac{\partial }{\partial t}\mathbf{{Q}}_a (s,t)
=
\frac{\delta H_N}{\delta \mathbf{P}_a}
\,,\qquad
\frac{\partial }{\partial t}\mathbf{{P}}_a (s,t)
=
-\,\frac{\delta H_N}{\delta \mathbf{Q}_a}
\,.
\end{equation}
{\it The corresponding Hamiltonian function $H_N$ is,}
\begin{equation} \label{H_N-def}
H_N = \frac{1}{2}\!\int\!\!\!\!\int\!\!\sum_{a\,,\,b=1}^{N}
\big(\mathbf{P}_a(s,t)\cdot\mathbf{P}_b(s^{\prime},t)\big)
\,G\big(\mathbf{Q}_a(s,t),\mathbf{Q}_{\,b}(s^{\prime},t)\big)
\,ds\,ds^{\prime}
\,.
\end{equation}
Because the solution formula (\ref{EP-sing-mom}) is a momentum map the
singular solution dynamics is {\it collective}. That is, the Hamiltonian
$H_N$ arises by substituting the singular momentum solution formula
(\ref{EP-sing-mom}) into the $H^1$ kinetic energy norm (\ref{H1-norm}) and
using the delta functions to perform the integrals. Thus, the evolutionary
IPDE system (\ref{IntDiffEqn-Q}) represents canonically Hamiltonian motion
on the space of curves in ${\mathbb{R}}^2$. Moreover, this Hamiltonian
motion is geodesic with respect to the co-metric given on these curves in
(\ref{H_N-def}) by the Green's function $G$. The Hamiltonian
$H_N=\frac{1}{2}\|\mathbf{P}\|^2$ in (\ref{H_N-def}) for this motion
defines the norm $\|\mathbf{P}\|$ in terms of this co-metric. This
momentum map result helps organize the theory and provides new avenues of
exploration, as suggested in \cite{HoMa2004}. The remainder of this paper,
however, will deal with numerical simulations which capture the
momentum exchange properties of these singular EPDiff solutions.

\section{Numerical approach}
\label{num-appr-sec}

\begin{figure}
\begin{center}
   \leavevmode{\hbox{\epsfig{
       figure=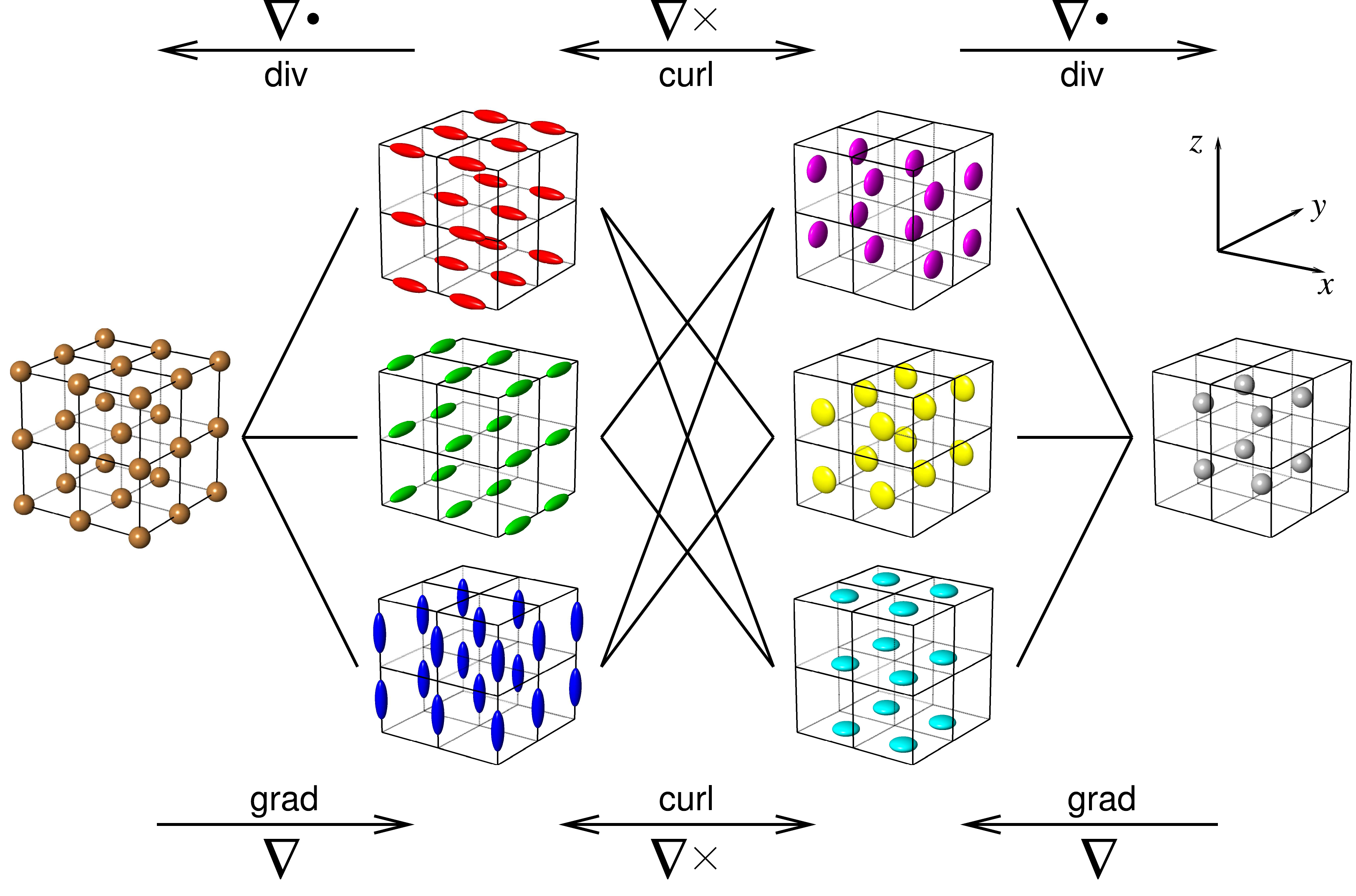, scale=0.40
   }}}
   \caption{\label{figure-mimetic}
       This schematic illustrates a particular compatible differencing algorithm
       for quantities defined on uniform, logically rectangular grids. Reading
       left to right, we have nodes, edges, faces, and cells. Nodes and cells support
       scalar-valued functions, while edges and faces support vector-valued
       functions. Divergence, gradient, and curl operators map between nodes,
       edges, faces, and cells.
   }
\end{center}
\end{figure}

Our numerical studies of EPDiff were performed on uniform, logically rectangular
Eulerian grids in 2d and 3d, using the compatible differencing algorithm (CDA)
described in \cite{HySh1997} and sketched in Figure \ref{figure-mimetic}. In
contrast to our experience with Lagrangian methods, our numerics using this CDA
have captured the elastic bounce expected in head-on collisions with only small
distortions observed in the recreated contact curves. Future investigations may
allow us to improve CDAs by developing related {\it variational\/} integrators
based on additional ideas from discrete exterior calculus (DEC)
\cite{Hirani2003,Leok2004}.

In this CDA, scalar and vector quantities
are defined at locations that are naturally appropriate for the domains and
ranges of the discrete divergence, gradient, and curl operators. Eight
{\it spaces\/}, or grid centerings, include a node space, at left, which
supports scalar functions. An edge space, consisting of $x$-direction edges,
$y$-direction edges, and $z$-direction edges, supports vector functions; $x$
components of vectors exists on $x$-direction edges, etc. A face space,
consisting of faces perpendicular to the $x$, $y$, and $z$ directions, also
supports vectors. Finally, a cell space supports scalars.

The set of spaces shown in the figure supports two versions each of divergence,
gradient, and curl, as described in \cite{HySh1997}.
Divergence maps vectors to scalars, gradient maps scalars to
vectors, and curl maps vectors to vectors. Lines between particular spaces
in Figure \ref{figure-mimetic} illustrate how the quantities defined on
different spaces contribute to one another through the various operators.

Node-space scalars map to edge-space vectors through the $x$, $y$, and $z$
components of the discrete gradient operator, while
cell-space scalars map to face-space vectors through the $x$, $y$, and $z$
components of another discrete gradient.
The $x$, $y$, and $z$ components of edge-space vectors each contribute a term
($\partial u/\partial x$, $\partial u/\partial y$, or $\partial u/\partial z$)
to a discrete divergence defined on the nodes, while the $x$, $y$, and $z$
components of face-space vectors each contribute a term to the cell-based
divergence.
Finally, the discrete curl operators map between edges and faces in the manner
one expects: $y$ and $z$ inputs contribute to $x$ outputs, $x$ and $z$ inputs
to $y$ outputs, and $x$ and $y$ inputs to $z$ outputs.

Different components of the same vectors have different
numbers of discrete points in this CDA. If the number of nodes is
$N \times M \times P$, then
the number of $x$-direction edges is $N-1 \times M \times P$,
the number of $y$-direction edges is $N \times M-1 \times P$, and
the number of $z$-direction edges is $N \times M \times P-1$.
The different numbers of discrete points and their slightly different locations
must be managed appropriately in any code that uses this scheme.

At present, our numerics are limited to uniform, logically rectangular Eulerian
grids with periodic boundary conditions. Our results are promising, but future
work will be needed to continue investigating numerical methods and identifying
the best candidates for capturing contact behavior on non-uniform or
unstructured grids and on non-rectangular domains with boundary conditions other
than periodic. For example, when studying the relation of EPDiff to internal
waves, one might examine the behavior of contact segments as they interact with
islands or atolls placed into the domain.

For our 2d and 3d numerical simulations, we advanced the momentum $\mathbf{m}$
in (\ref{H1-EPcurl-eqn}) with an explicit, variable time step Runge-Kutta type
predictor-corrector. We selected the time step for numerical stability by trial
and error, while our code selected the time step for numerical accuracy (not to
exceed the time step for numerical stability) according to the following
well-known formula,
\begin{equation}\label{vartimestep-control}
h_i = \gamma h_{i-1}\left(
\frac{\epsilon|h_{i-1}|}{||\bar u_i - \hat u_i||}
\right)^{1/p}
\,,
\end{equation}
which is used in the following way. At step $i$ of the calculation, we know the
predicted solution $\bar u_i$, the corrected solution $\hat u_i$, and the
previous time step $h_{i-1}$.
The predictor's order of accuracy is $p$, while the
corrector's order of accuracy is $p+1$. For our 2d simulations we used $p=4$,
while we used $p=3$ for our 3d simulations because this reduced the number of
large, 3d temporary arrays needed for the calculations, and thereby allowed us
to have a higher resolution while still using a reasonably accurate time
integration scheme.

A new time step $h_i$ is chosen from (\ref{vartimestep-control})
based on the old
time step $h_{i-1}$ and the norm of the difference between the current predicted
and corrected solutions. For both 2d and 3d, we used a relative error tolerance
per time step of $\epsilon=10^{-6}$, a safety factor $\gamma=0.9$, and the $L_2$
norm, $||\cdot||_2$.

Divergence, gradient, and curl were computed at second order accuracy
according to the
CDA outlined above, with the vectors $\mathbf{m}$ and $\mathbf{u}$ defined on
the edges (shown as the red, green, and blue spaces in Figure
\ref{figure-mimetic}). We observed that the quality of our numerics did not
improve markedly with fourth or sixth order operators, and as expected, the
higher-order operators were somewhat slower to compute. Note that our second
order operators, in addition to mimicking important properties of their
continuum analogs, also have greater accuracy than one might expect at second
order. This is because the staggered nature of the grid allows for a smaller
``$\Delta x$'' in the difference computations. For example,
$\partial u/\partial x$ mapping nodes to $x$-direction edges uses the stencil
$\partial u/\partial x(i+\frac{1}{2},j,k) = (u(i+1,j,k) - u(i,j,k))/\Delta x$,
whereas the analogous computation on a strictly nodal grid is
$\partial u/\partial x(i,j,k) = (u(i+1,j,k) - u(i-1,j,k))/(2\Delta x)$.

In two dimensions, one must regard the individual spaces in the schematic of
Figure \ref{figure-mimetic} as compressed vertically, so that they are flat.
This corresponds to having no $z$ component. In this case, $x$-direction edges
(red) are identical to $y$-direction faces (yellow), $y$-direction edges (green)
are identical to $x$-direction faces (purple), $z$-direction edges (dark blue)
are identical to nodes (brown), and $z$-direction faces (light blue) are
identical to cells (gray). (This is {\it not} generally true on nonuniform
grids.) Even so, the proper treatment of quantities defined on the different
spaces requires that we regard the different spaces as distinct in 2d.

In our formulation, the curl operator appearing in (\ref{H1-EPcurl-eqn}) is
still meaningful in 2d. In particular, ${\rm curl}(\mathbf{m})$ takes the 2d
function $\mathbf{m}$, defined on the $x$-direction edges (red) and
$y$-direction edges (green), and maps it to a face-space quantity with $x$
and $y$ components of 0 (which are not stored in a computer code, of course),
and a nonzero $z$ component (light blue in Figure \ref{figure-mimetic}), which
is regarded as a vector that is normal to the plane.

For our 2d simulations, we used a resolution of $1024^2$ zones, or $1025^2$
nodes. For our 3d simulations, we used $256^3$ zones ($257^3$ nodes). To invert
the Helmholtz operator in transforming between $\mathbf{m}$ and $\mathbf{u}$,
we convolved $m(x,t)$ with the Green function in Fourier space. No artificial
viscosity or other numerical tricks proved to be necessary for our simulations.

\section{Numerical results for EPDiff in 2d}
\label{num-2d-sec}

Upcoming sections describe 2d simulations of evolution under EPDiff for each
of nine initial velocity distributions. For each simulation we have figures for
$\alpha=\sigma$,
$\alpha=\sigma/2$,
$\alpha=\sigma/4$, and
$\alpha=\sigma/8$.
Each figure contains six frames, showing the initial magnitude (speed) of
velocity, $\modu$, in the upper left frame, followed by plots of $\modu$ at
future times, reading across and then down. For each
simulation, the domain is $[-1,1]\times[-1,1]$, $x$ is toward the right, and
$y$ is toward the top.

\begin{figure}
\begin{center}
   \leavevmode{\hbox{\epsfig{
       figure=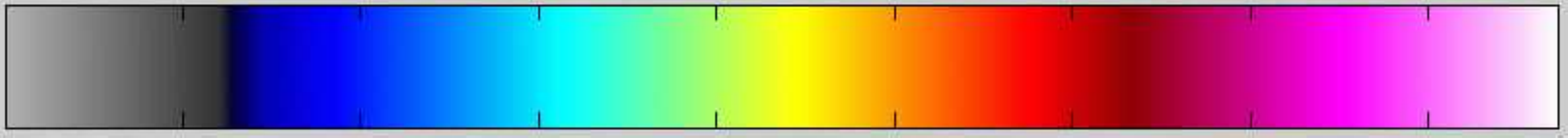, scale=0.45
   }}}
   \caption{\label{color-bar}
       Color scheme for the magnitude of velocity, $\modu$,
       in the upcoming figures.
   }
\end{center}
\end{figure}

Colors in each frame indicate the magnitude of the velocity, beginning with gray
for $\modu=0$ and ending with white for the maximum value of $\modu$, as shown
in the color bar in Figure \ref{color-bar}. Maximum values of $\modu$ are
determined for each frame individually, not over all frames in a figure, so that
the colors in frames with smaller maximum values of $\modu$ are not washed out.
Notice how the use of the color black for small $\modu$, just above gray for
$\modu=0$ in the color scheme, allows us to etch the outlines of the spatially
confined velocity distributions.

\begin{figure}
\begin{center}
   \leavevmode{\hbox{\epsfig{
       figure=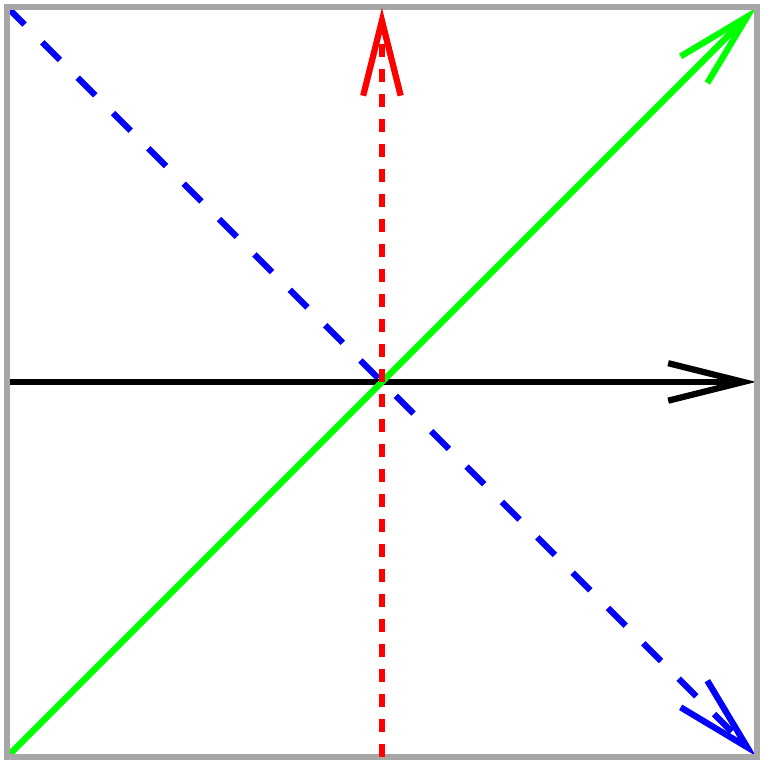, scale=0.75
   }}}
   \caption{\label{figure-scheme2d}
       Locations of 1d profiles of $\modu$ shown in the bottom panels
       in the upcoming 2d figures.
   }
\end{center}
\end{figure}

The transverse profile of the velocity distribution along the horizontal midline
of each frame is shown as the black (solid) graph in the lower panel of each
frame, while the red (dotted) graph shows the vertical midline. The profiles
along the northeast and southeast diagonals are plotted in green (solid) and
blue (dotted), as sketched in \ref{figure-scheme2d}. Unlike the colors in the
full 2d plots, which are scaled according to the maximum of $\modu$ in each
frame, the vertical axis of the profiles in the lower panels are set between 0
and the maximum over all frames (in each figure) of the black, red, green, and
blue profiles. So, for example, in the profiles of Figure
\ref{2d-plate-sigma-1},
$\modu$ as seen in the lower panels is seen to decrease as the evolution
proceeds, even though the colors in the 2d plots always vary between gray
(minimum) and white (maximum) in order to show maximum detail.

\subsection{Plate}\label{2d-plate}

Each figure we shall discuss compares evolution under EPDiff with various values
of alpha, starting from the {\it same} standard initial velocity distribution,
which for Figures \ref{2d-plate-sigma-1}-\ref{2d-plate-sigma-8} we call
``plate.'' The velocity in these four 2d plate flows is initially rightward. The
initial speed in each of these cases is distributed along a line segment in 2d,
which in 3d will be a disc, or plate. Hence, the term plate flow. This initial
speed is constant along most of the segment's length, then falls off
as a Gaussian at either end. It falls off exponentially in the transverse
directions. The width $\sigma$ of the initial exponential profile
$e^{-|x|/\sigma}$ is the {\it same} in each figure. We start the EPDiff
evolution in each figure with this profile, and we vary the parameter $\alpha$,
in the relation $\mathbf{m}=\mathbf{u}-\alpha^2\Delta{\mathbf{u}}$, from problem
to problem in fractions of the standard initial profile's exponential width
$\sigma$.

In Figures \ref{2d-plate-sigma-1}-\ref{2d-plate-sigma-1},
the values of $\alpha$ are,
respectively, $\sigma$, $\sigma/2$, $\sigma/4$, and $\sigma/8$. In Figure
\ref{2d-plate-sigma-1}, the initially straight velocity segment evolves to
```balloon'' outward, while maintaining its transverse peakon profile. In
Figures \ref{2d-plate-sigma-2}-\ref{2d-plate-sigma-4},
the initially straight velocity segment
in each case evolves into a series of curved peakon strips of width
$\alpha$ with transverse peakon profiles. The lower panels in each case
confirm that the velocity profiles in every transverse direction have the
characteristic exponential peakon shape, $e^{-|x|/\alpha}$. The first
strip to emerge travels fastest and each subsequent strip moves slower.
Consequently, the distances between the strips increases. These peakon
strips are curved because their endpoints are nearly fixed, while their
middle regions are still moving. They stretch during their evolution and
increase their lengths. Because of their peakon cross-sections in velocity
profile, each peakon strip corresponds to a singular momentum density
supported on a curve through the center of the strip. The speed varies
along each strip according to its height at the peak at a given point
along the curve. Consequently, the ``hotspots'' appearing in the velocity
color scheme as red inhomogeneities along the strip are moving faster than
the adjoining yellow regions. The extension of these hotspots along the
peakon strips also indicates order
$O(1)$ stretching.

One sees kinks near the endpoints of the curved peakon strips that first
arise at a distance of order $\sigma$, the matching length in the initial
velocity profile. Thus, the connection along the segment to the zero speed
background influences the stretching of the peakon segments over a finite
length scale. These kinks near the endpoints are more pronounced for the
larger values of $\alpha$ than for the smaller values, indicating that the
length scale $\alpha$ plays a role in the stretching process, as well as
in the shape of the transverse profile. These peakon segments also rotate
around their nearly fixed endpoints as their expansion and stretching
proceeds.

The black (horizontal) transverse profiles in the lower panels of the
figures show a steady rightward progression of peakon profiles. The
green and blue profiles show a similar progression of peakon profiles
along the diagonals. However, the red (vertical) profiles show a symmetric
progression, both upward and downward. This means the peakon strips are
stretching as they balloon out from the initial rightward motion. Being
roughly circular at late times, this stretching of the strip length is
comparable to the distance they propagate horizontally. Therefore, their
stretching is an order $O(1)$ effect.

\paragraph{Stability and an open problem.}
For $\alpha = \sigma$ in Figure
\ref{2d-plate-sigma-1}, the peakon curve segment is stable and it retains
its integrity. The other cases, for $\alpha < \sigma$, are unstable and they
break into narrower curved peakon segments each of width $\alpha$, as the
evolution tends toward the peakon profile $e^{-|x|/\alpha}$. In fact, any
smooth confined initial velocity distribution tends eventually to a set of
peakon strips. These are {\it contact curves\/},
along which the discontinuity in
slope moves with the fluid velocity. Hence, the momentum density tends to
a set of moving {\it curves\/}
on the plane, with each curve embedded in the flow.
The latter claim is proved by the isospectral
problem in 1d, but for now is only an observation in 2d and 3d. The proof
of this observed tendency remains an open problem from the analytical
viewpoint.

As $\alpha < \sigma$ decreases in Figures
\ref{2d-plate-sigma-2}-\ref{2d-plate-sigma-8},
the number of emerging contact curves increases. We emphasize that, at
sufficiently late times, {\it only} these contact curves are observed
emerging from a confined initial velocity distribution with width greater
than $\alpha$. However, the process occurs does require a certain amount
of time to reach completion, as the peakons are successively formed at
definite intervals. In the cases \ref{2d-plate-sigma-2} and
\ref{2d-plate-sigma-4} of $\alpha=\sigma/2$ and $\alpha=\sigma/4$ at the
times shown, some vestiges of the initial conditions still remain as
ramps. As time progresses further, these ramps will eventually decay
into a sequence of successively slower moving (lower amplitude) peakon
contact curves. No such vestiges remain in the case when
$\alpha=\sigma$. This behavior is in accord with the 1d steepening lemma
of \cite{CaHo1993}, which states that an initial velocity profile possessing
an inflection points of negative slope will develop a vertical slope in
finite time. The formation of the vertical slope is part of the nonlinear
steepening mechanism which creates the train of peakons from the ramp
velocity configuration.

\paragraph{Time reversal.}
Starting from the final velocity distribution for each value of $\alpha$,
we integrate back to the starting time in the EPDiff evolution numerics.
This procedure tests the reversibility of the numerical algorithm. (The
EPDiff equation itself is reversible.) Each case reverses accurately to its
initial condition, as is evident visually in Figure  \ref{2d-plate-back},
and as measured in the $L^1$, $L^2$, and $L^\infty$ norms shown in Table
\ref{norm-2d}.
For the simple plate evolutions, without
any of the head-on collisions of contact curves that will produce the more
complicated flows seen below, this reversibility affirms the numerical
scheme. It is not a severe test, however, in the sense that reversibility
from the endpoint back to the beginning
does not guarantee accuracy over the entire forward evolution.

\figfour{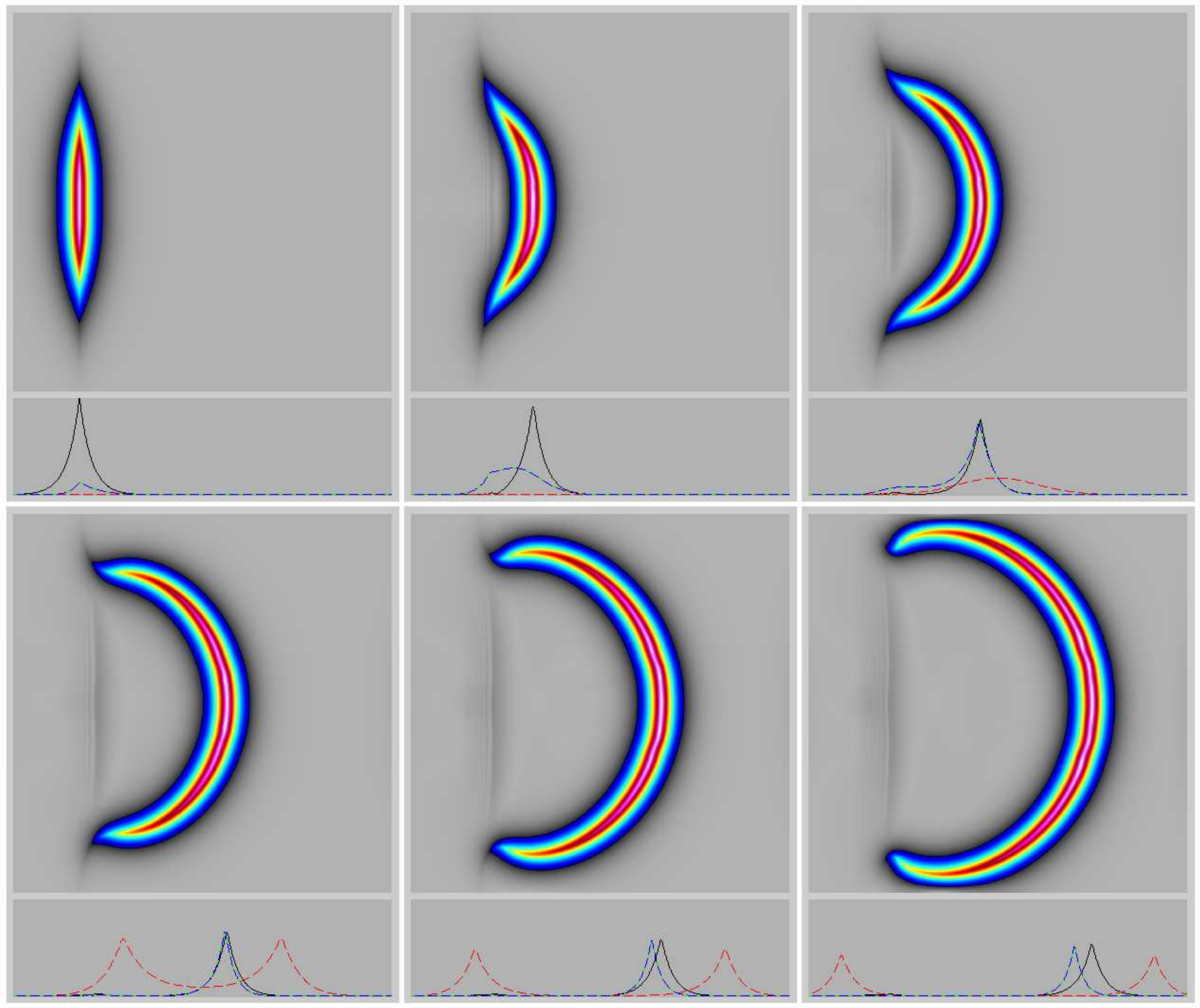}{2d}{plate}{\sigma  }{}
\figfour{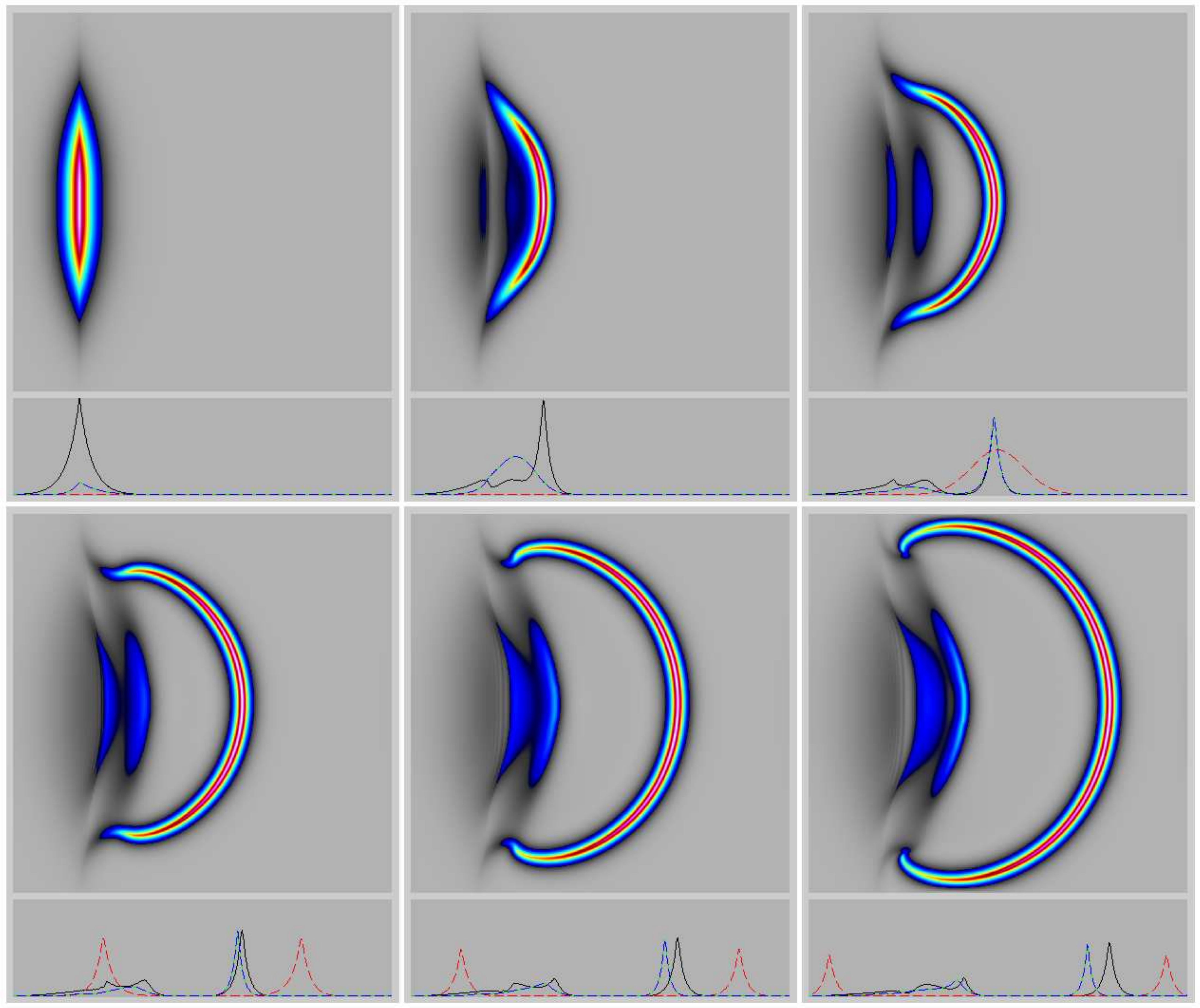}{2d}{plate}{\sigma/2}{}
\figfour{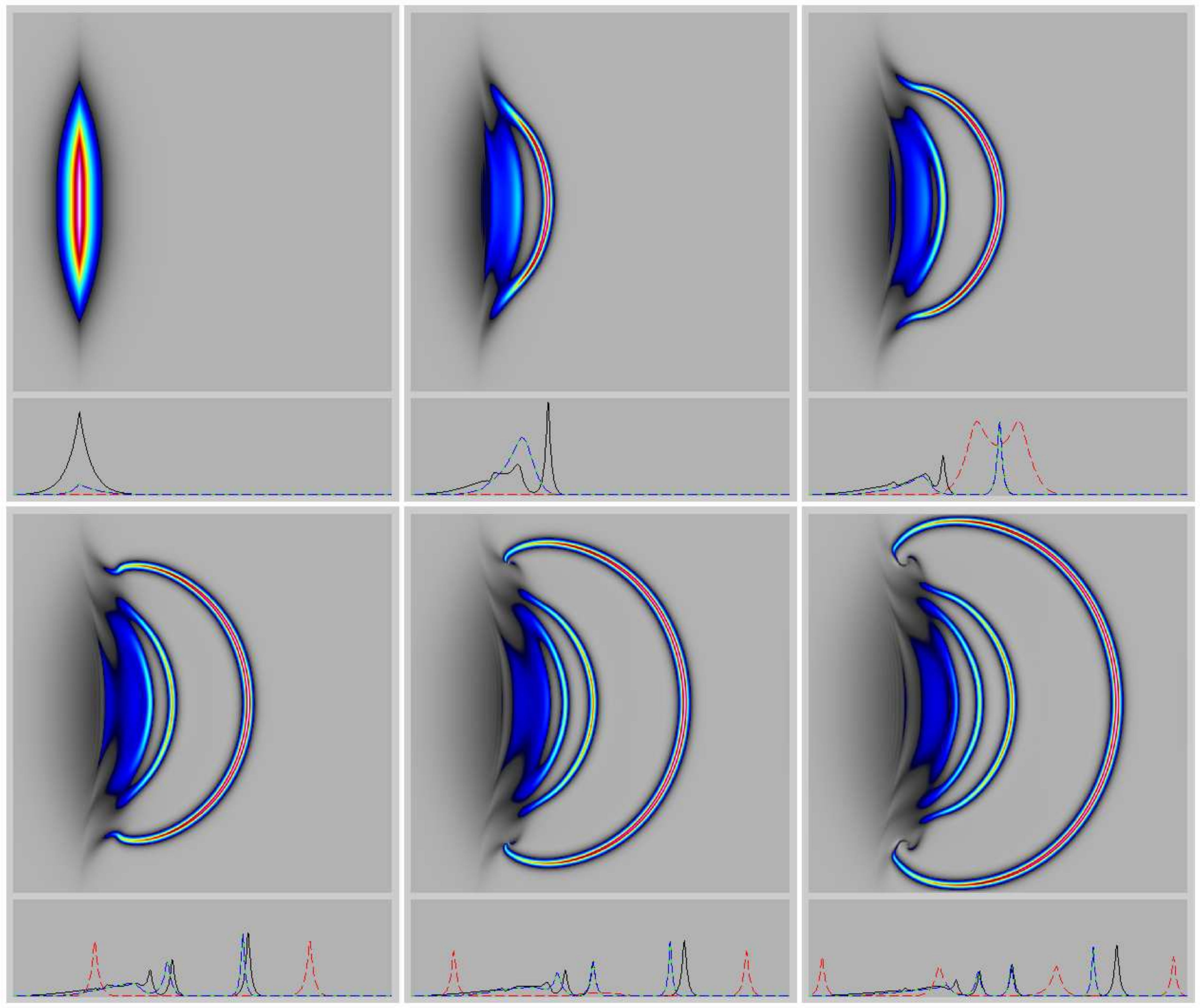}{2d}{plate}{\sigma/4}{}
\figfour{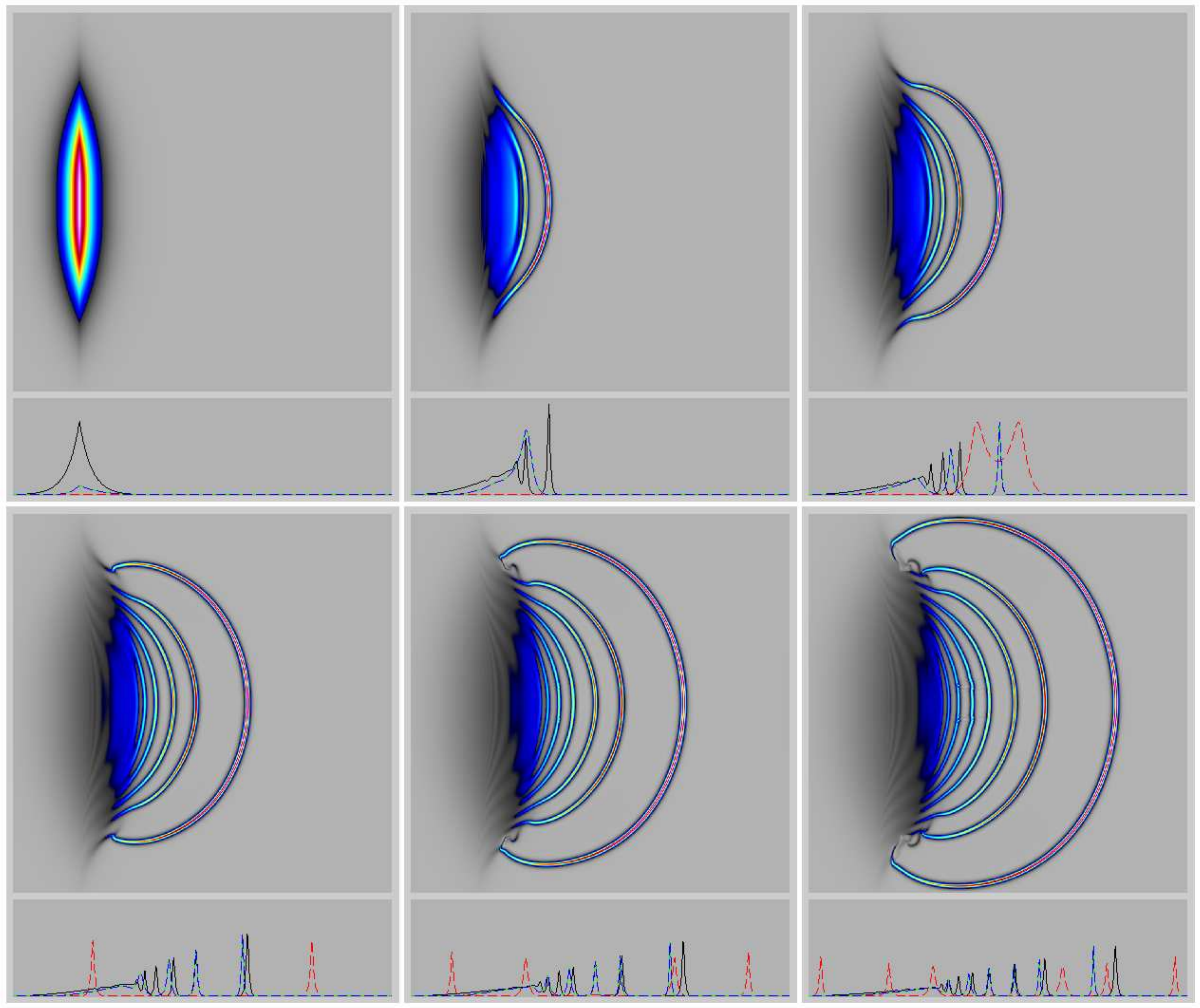}{2d}{plate}{\sigma/8}{}

\figfourreverse{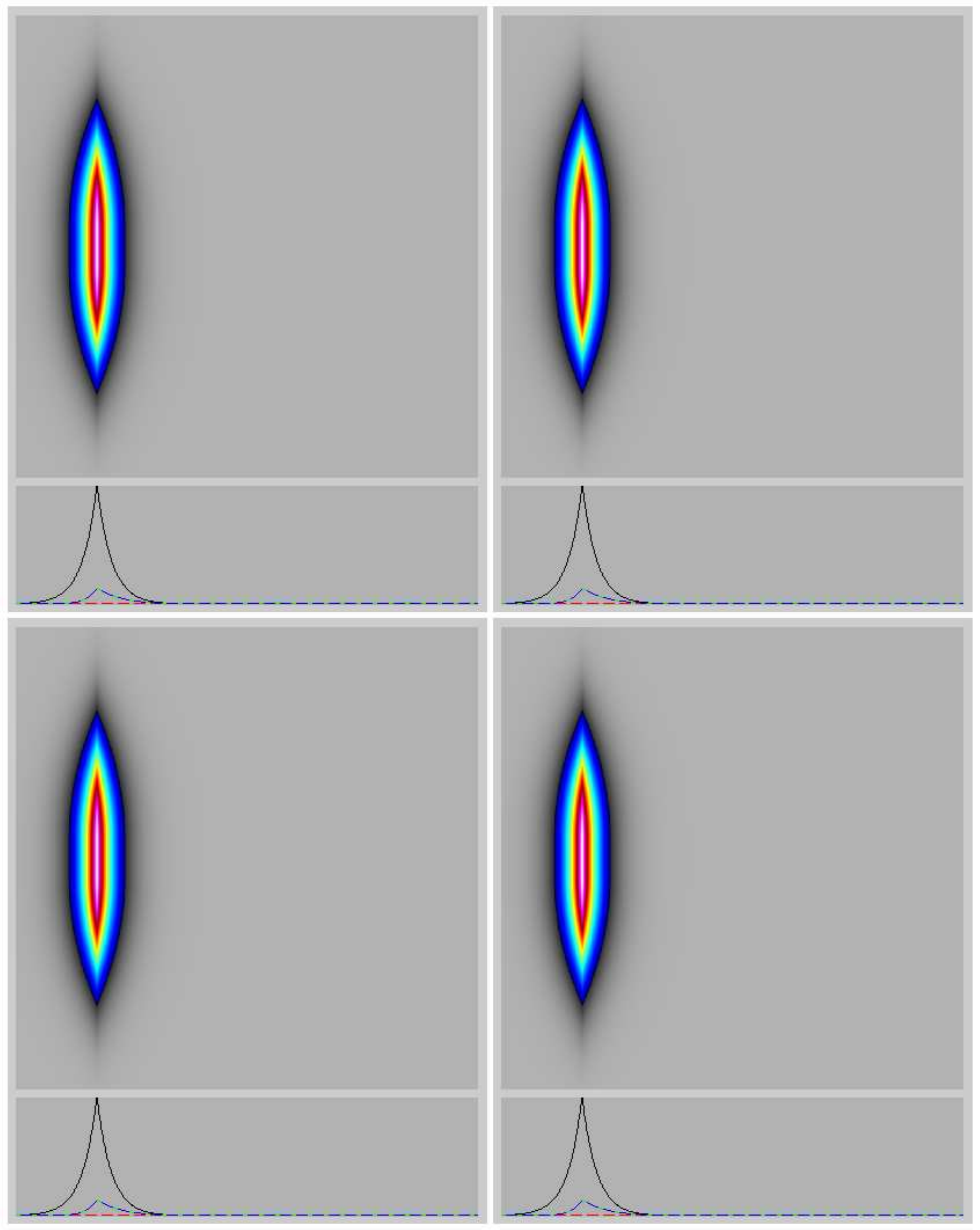}{2d}{plate}{This initial condition
reconstitutes well for all values of $\alpha$}

\subsection{Parallel}\label{2d-parallel}

In Figures \ref{2d-parallel-sigma-1}-\ref{2d-parallel-sigma-8}, two straight segments are
initialized moving rightward. The one behind has twice the speed of the
one ahead, and the two segments are offset in the vertical direction. The
segments each break into curved strips of width $\alpha$ and these undergo
overtaking collisions. For $\alpha=\sigma$, the segments retain their
integrity as they expand until the overtaking collision occurs. Upon
overtaking the slower segment, the faster segment transfers its momentum
to the slower one ahead and a remarkable ``reconnection'' or ``melding''
of the segments
occurs. This reconnection also shows rapid transverse stretching in which
``hot spots'' of velocity arise and then spread out along the wave front.
Also, remarkably, the
figures show a low amplitude ``peakon wisp'' connecting the endpoint of an
earlier reconnection to a point (usually to a hotspot) on the leading peakon
segment. This wisp apparently
provides the ``memory'' of the previous reconnection,
which is required for the evolution to remain reversible. Its reversibility affirms
the nondissipative nature of the reconnection process. Moreover, this
reconnection
must be reversible, because the evolution is Hamiltonian in the continuum
limit.

The lower panels of Figures
\ref{2d-parallel-sigma-1}-\ref{2d-parallel-sigma-8} for $\alpha=\sigma$ show that the profiles remain
exponential both before and after the overtaking collision. For decreasing
values of $\alpha<\sigma$, the evolution develops increasing complexity,
with numerous overtaking collisions and corresponding reconnections. In
each case, once sees the trailing memory wisps arising from these
reconnections. These trailing wisps often connect to kinks at hotspots
along the curve, indicating that their interaction is nontrivial, even
though they have small amplitude. (This may also indicate that the hotspot
is the {\it source} of the trailing wisp.)

In the time-reversed runs shown in Figure \ref{2d-parallel-back}, all of the
overtaking collisions reconstituted their initial conditions accurately in the
various norms shown in Table \ref{norm-2d}.
This indicates the accurate reversibility of the numerics for
overtaking collisions.

\figfour{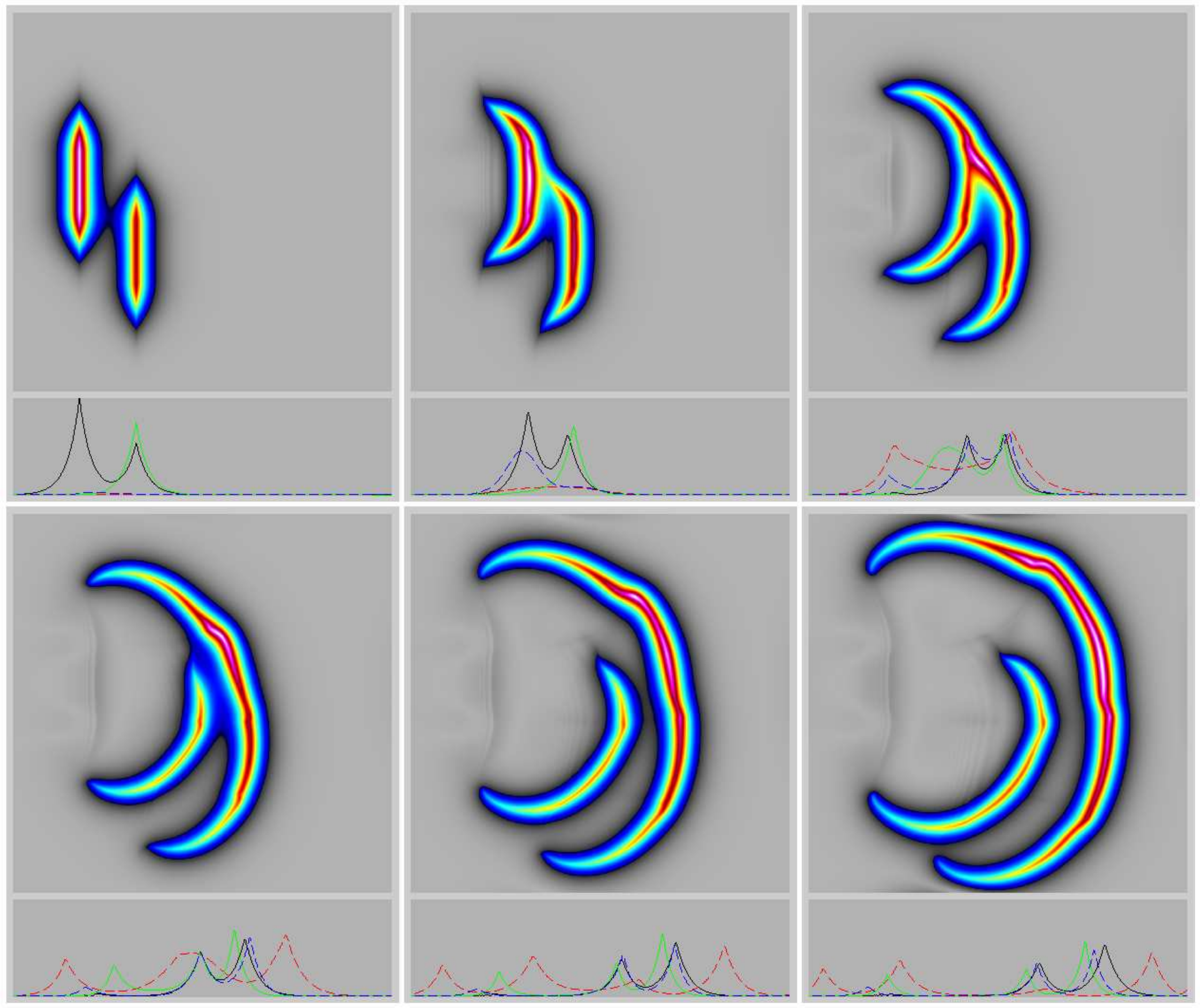}{2d}{parallel}{\sigma  }{}
\figfour{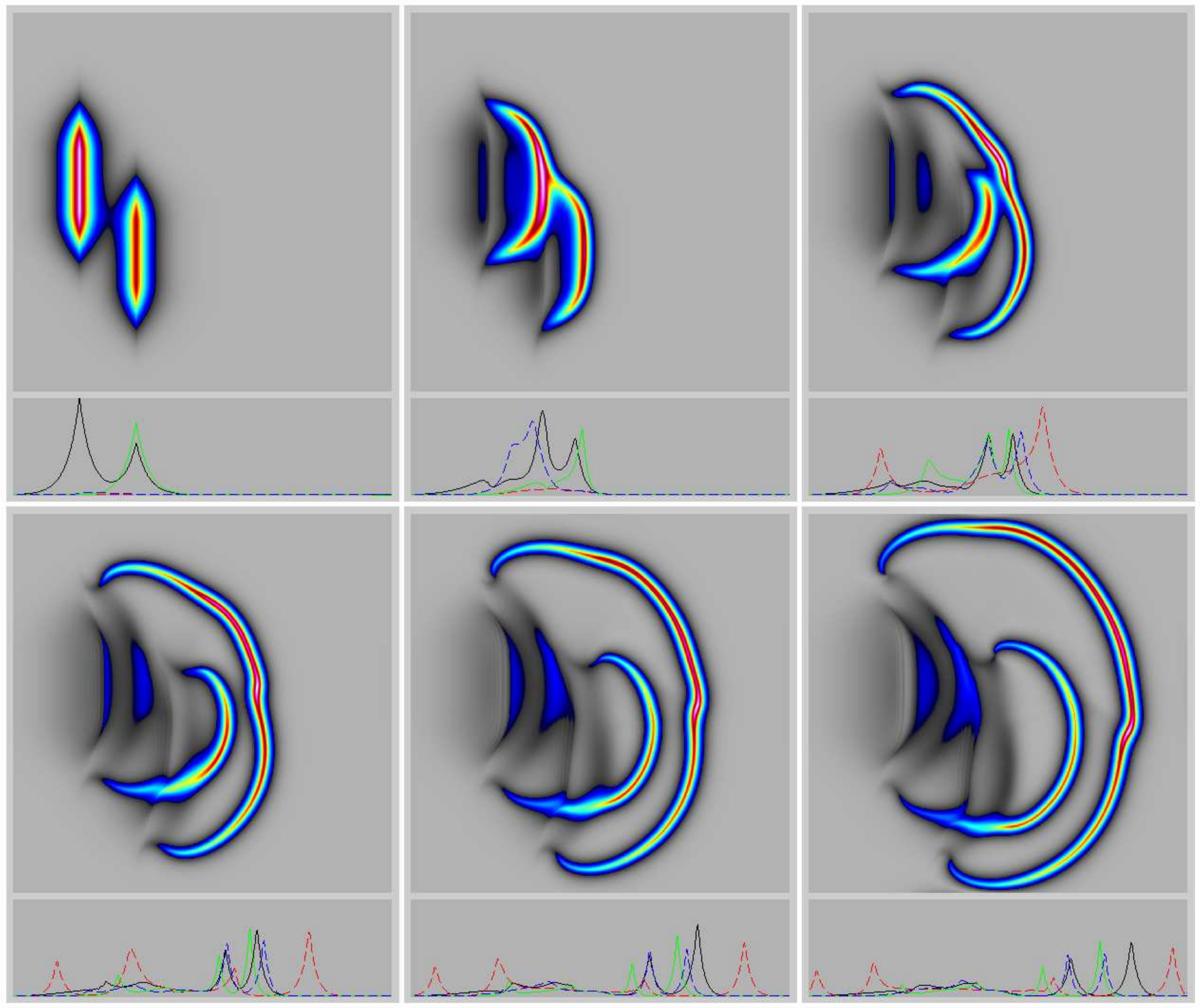}{2d}{parallel}{\sigma/2}{}
\figfour{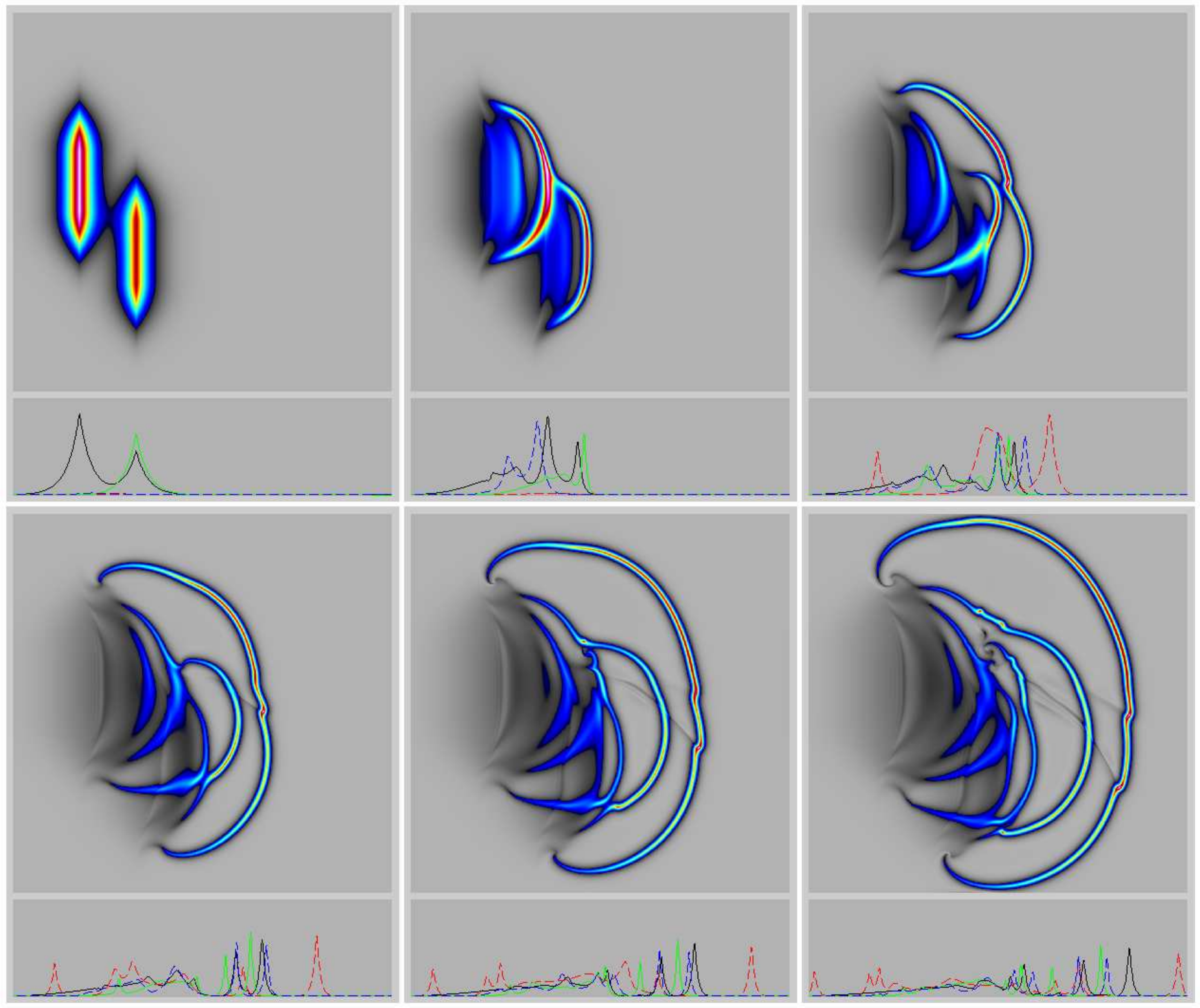}{2d}{parallel}{\sigma/4}{}
\figfour{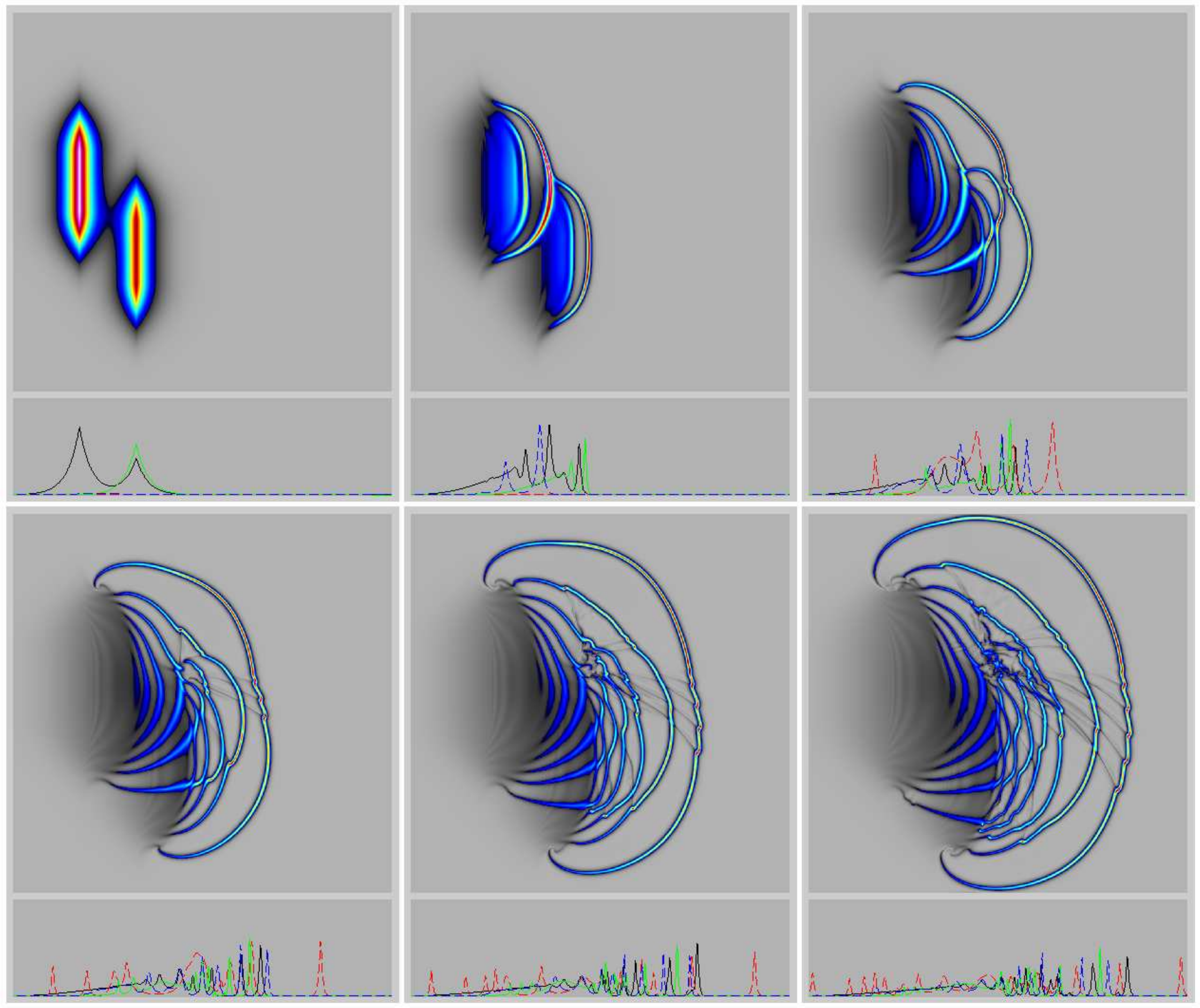}{2d}{parallel}{\sigma/8}{}

\figfourreverse{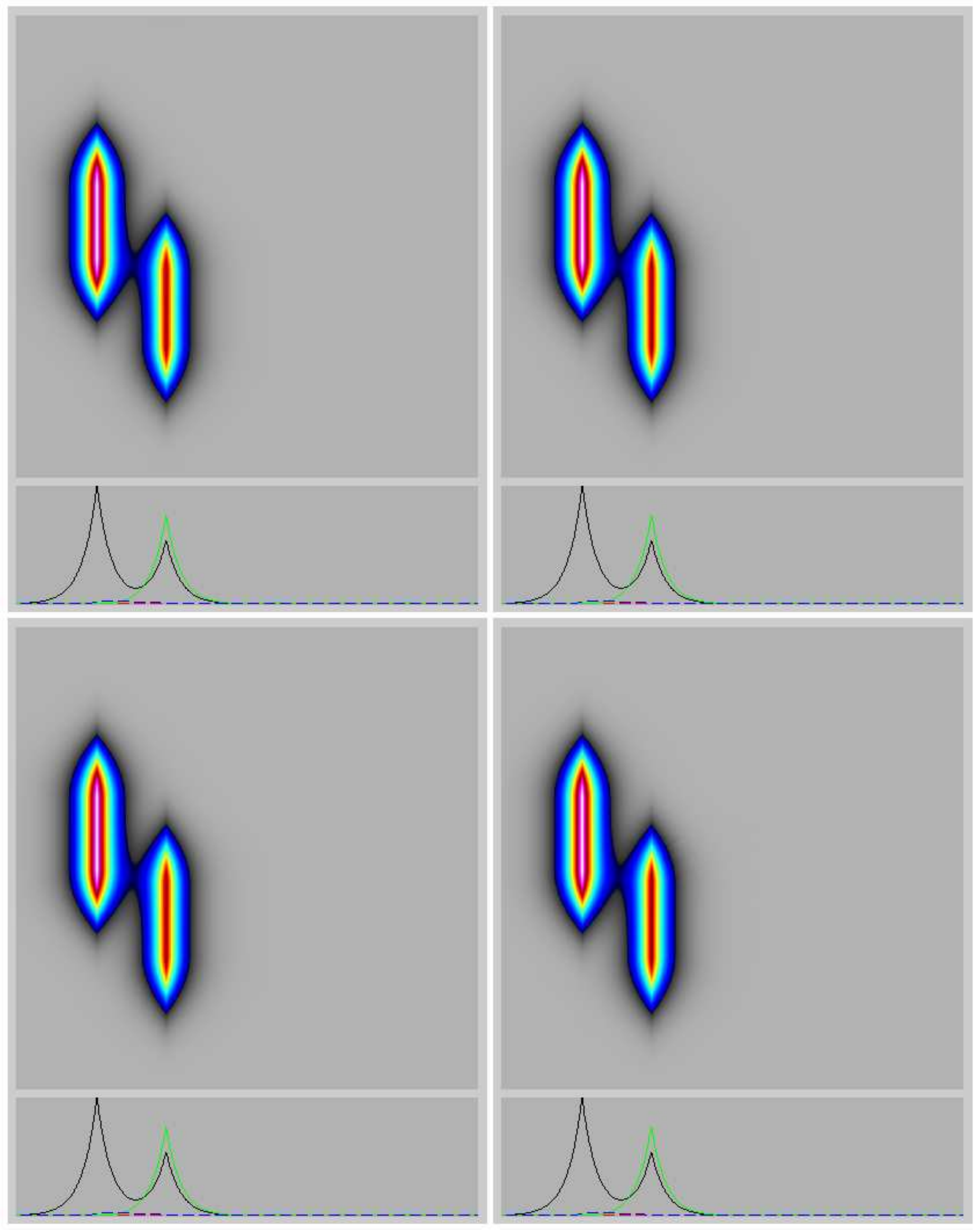}{2d}{parallel}{}

\subsection{Skew}\label{2d-skew}

Skew flows in Figures \ref{2d-skew-sigma-1}-\ref{2d-skew-sigma-8} begin with two peakon
segments of the same width, but oriented so that the one behind, which has
twice the amplitude (speed) of the one in front, overtakes the one ahead
by moving along the negative diagonal. Again, one sees integrity of the
$\alpha=\sigma$ case and reconnection with hot spots and memory wisps
trailing behind kinks in the main peakon segments after the overtaking
peakon segment transfers its momentum to the one ahead. This locally 1d
soliton elastic-collision rule seems to explain the momentum
transfer. Once again the lower panels show that the solution tends to
peakon profiles in each direction.

\paragraph{Memory wisps: another open problem.}
Figures \ref{2d-skew-sigma-1}-\ref{2d-skew-sigma-8} raises issues (such as the
production of memory wisps) for the 2d peakon segment interactions which
could lead to new research well beyond the scope of the present work. As
before, in the cases of $\alpha<\sigma$, for skew flows it seems that each
memory wisp is attached to a hot spot along a major peakon segment. The
memory wisps, by the way, are very low in amplitude, but they seem to also
possess the peakon exponential transverse profile. Hence, the stretching
motion along the wisp must considerably {\it faster\/}
than across it. This may be
because the hot spots keep contributing to the wisps trailing behind them.
Thus, the wisps may be seen as trailing residuals from the hot spots. The
memory wisp feature of the reconnection remains to be explained in more
detail, both numerically and analytically.

For small $\alpha<\sigma$, the central regions of the skew flows in Figures
\ref{2d-skew-sigma-2}-\ref{2d-skew-sigma-8} become very intricate (partially
mixed). Hence, one could expect that its reversibility is compromised. The
case $\alpha=\sigma/8$ is mostly reconstituted after the time reversal, but
some small-amplitude, high-frequency errors do remain, as observed in Figure
\ref{2d-skew-back}. For other values of alpha
($\alpha=\sigma\,,\sigma/2\,,\sigma/4$) the initial conditions reconstitute
quite well.

\figfour{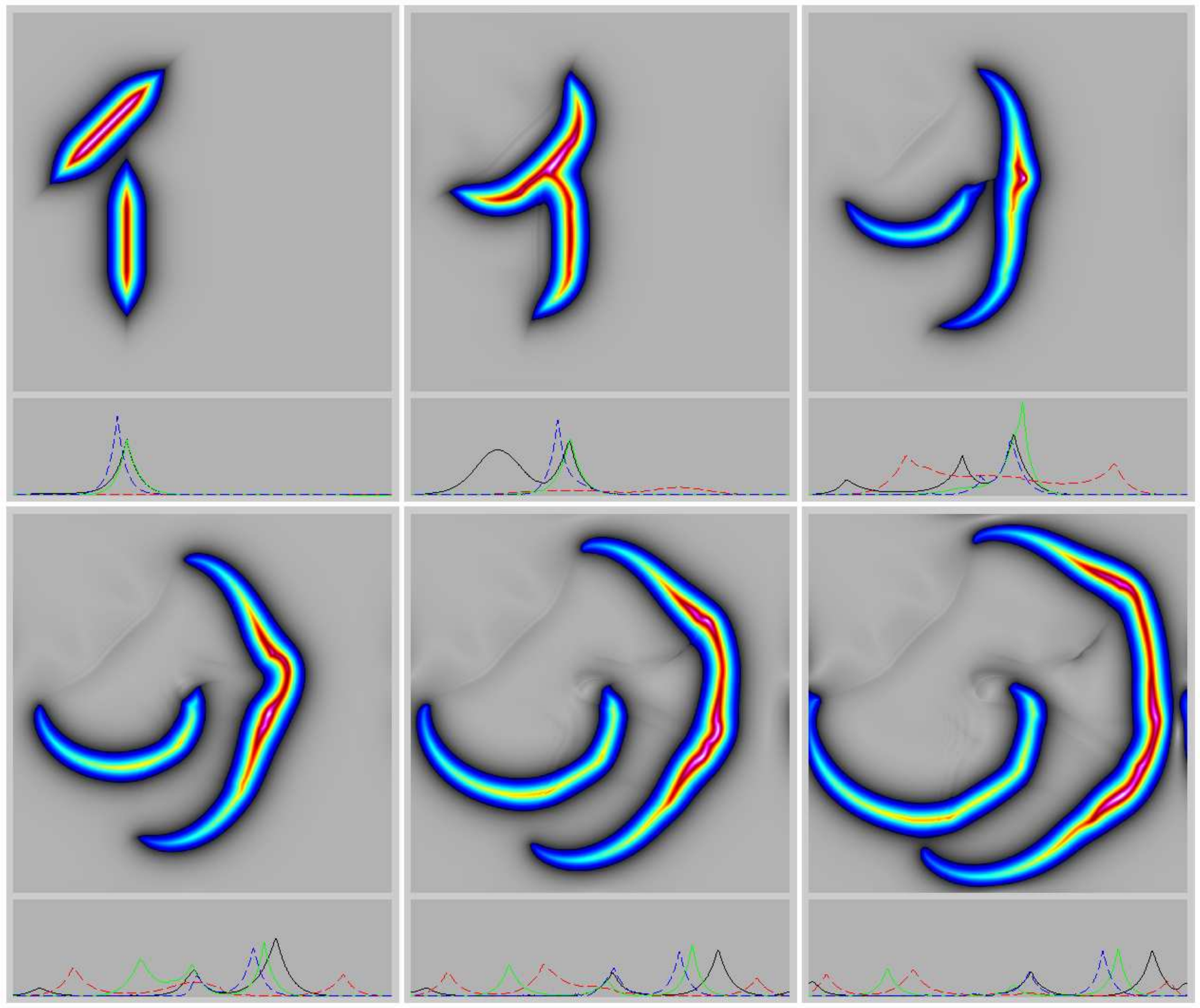}{2d}{skew}{\sigma  }{}
\figfour{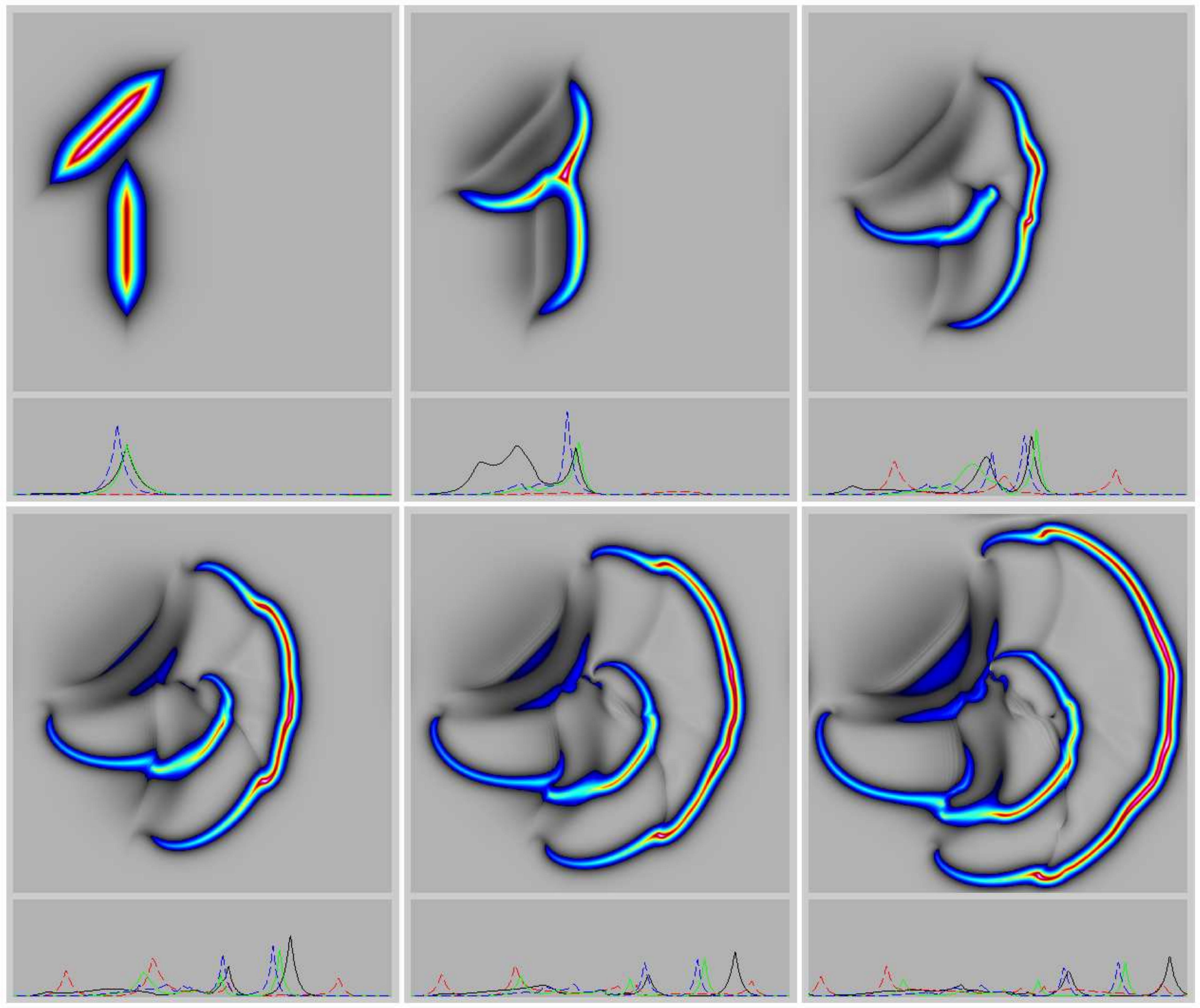}{2d}{skew}{\sigma/2}{}
\figfour{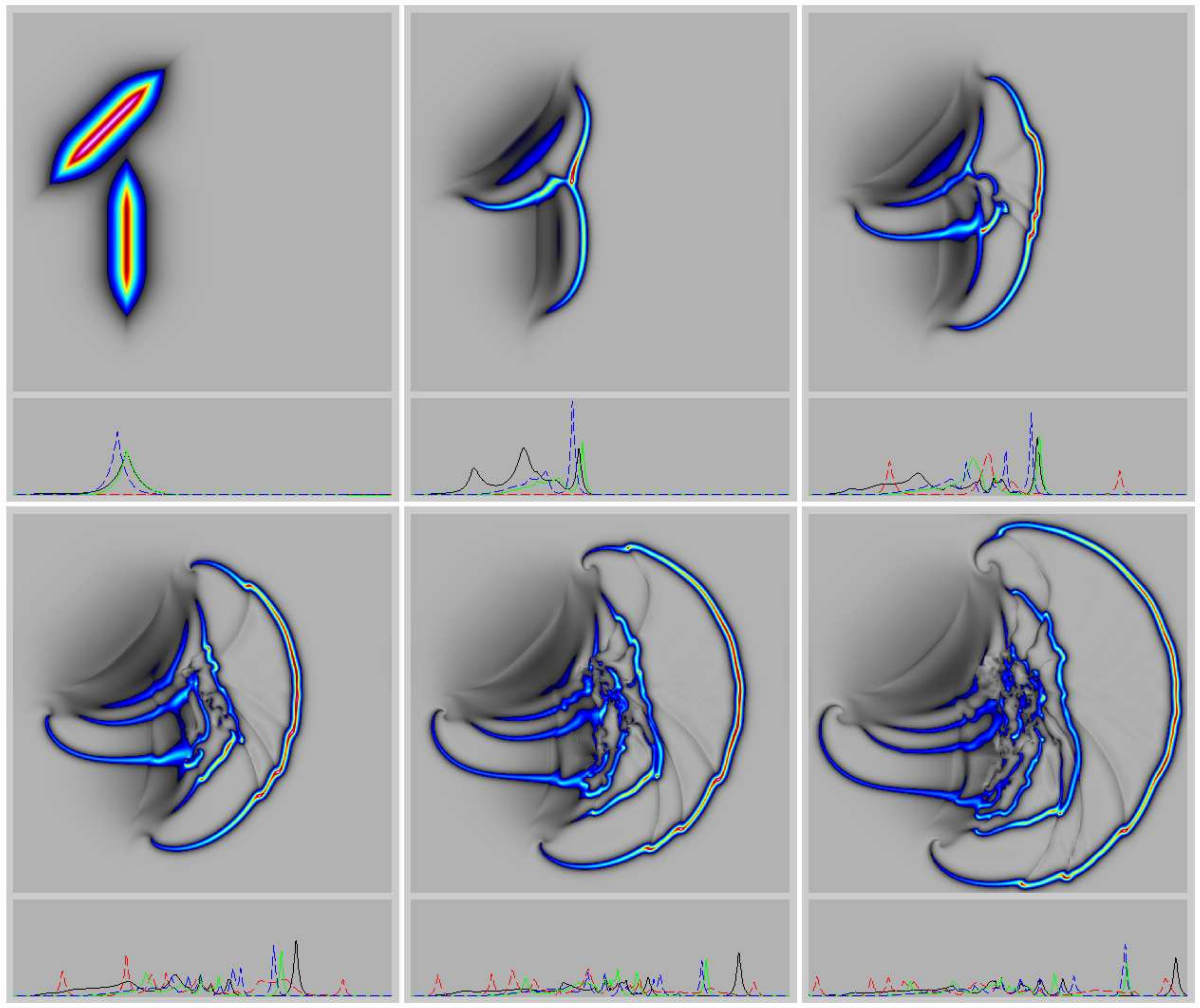}{2d}{skew}{\sigma/4}{}
\figfour{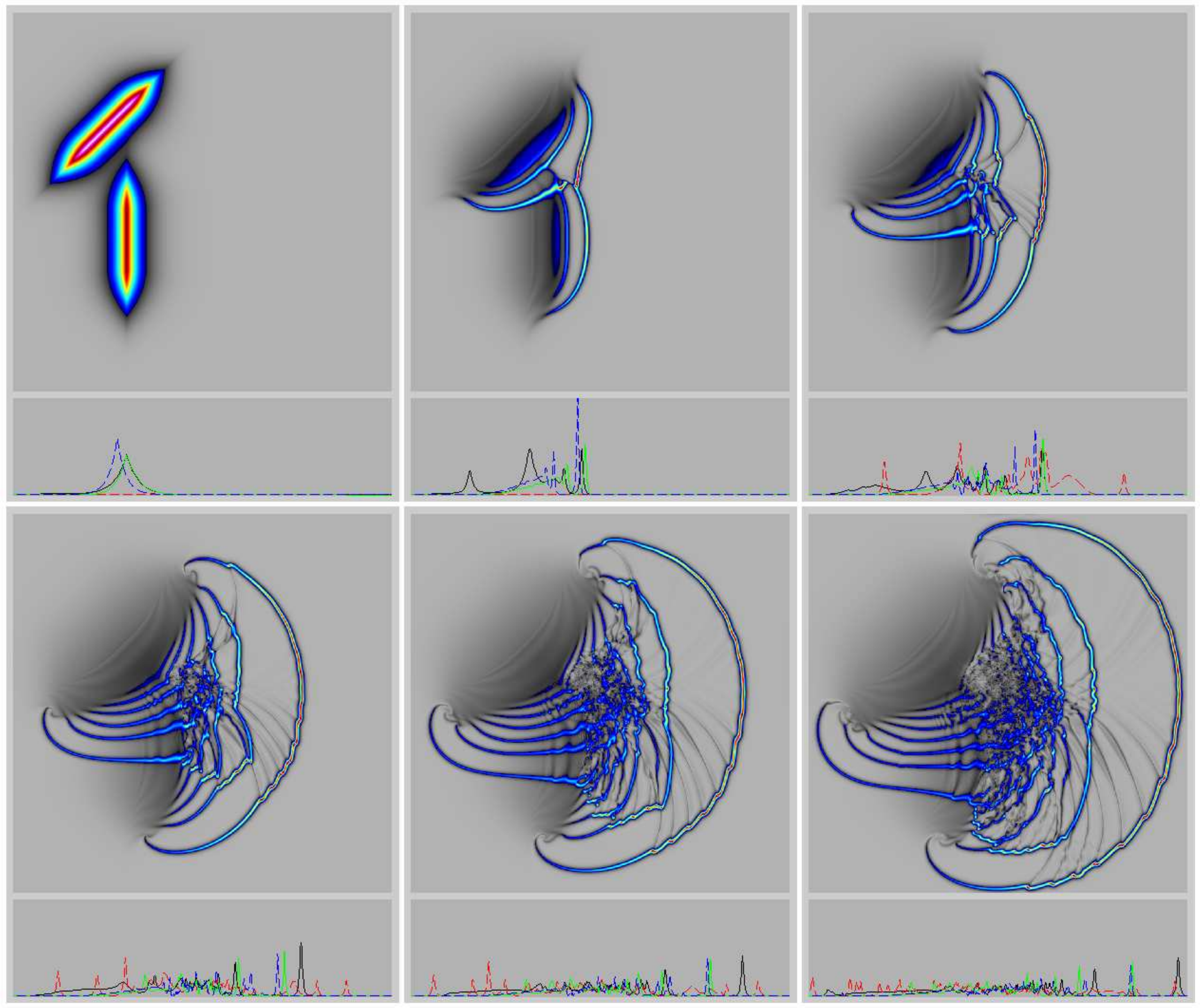}{2d}{skew}{\sigma/8}{}

\figfourreverse{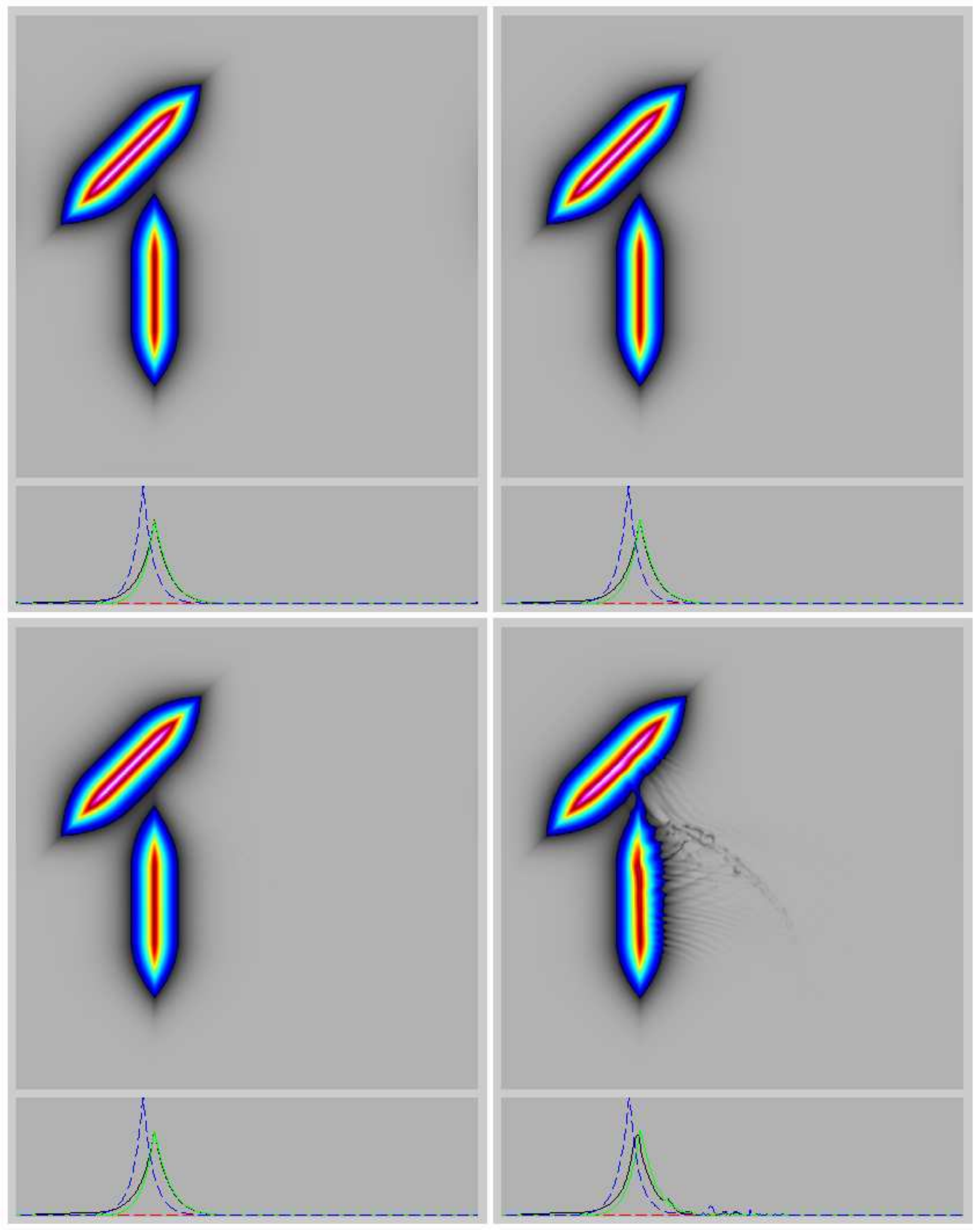}{2d}{skew}{}

\subsection{Wedge}\label{2d-wedge}

The wedge flows in Figures \ref{2d-wedge-sigma-1}-\ref{2d-wedge-sigma-8} are variants of
skew flows in Figures \ref{2d-skew-sigma-1}-\ref{2d-skew-sigma-8}, showing two
plates
colliding along opposite diagonals in the plane,
with reflection symmetry about the
horizontal axis. The wedge flows are convergent, and therefore they have
some head-on features that emerge on the left-hand side of the collisions.
They also show considerable acceleration along the midline, in forming
jets moving along the horizontal axis in both directions. This jet
formation is due to convergence of momentum which continues to build up
after the initial collision. The multiple wedge collisions occurring for
values of $\alpha<\sigma$ show successive strong accelerations due to
convergence. They also show enhanced stretching of the main peakon
segments. These collisions also produce complex patterns of
small-amplitude peakon wisps, trailing from hotspots at kinks along the
main peakon curves.

Again, the lower panels in Figures \ref{2d-skew-sigma-1}-\ref{2d-skew-sigma-8} show
primarily peakon profiles and the emergence of peakons from ramp-and-cliff
formations with inflection points of negative slope, in agreement with the
steepening lemma for CH in 1d.

The head-on features of the wedge collisions cause noticeable errors in
reversibility, of order 10 to 20 percent for $\alpha=\sigma/8$, as observed
in Figure \ref{2d-wedge-back}. However, these errors decrease for larger alpha.
As we will see in other plots, errors in reversibility tend to arise for
smaller values of $\alpha<\sigma$ whenever head-on collisions occur.

\figfour{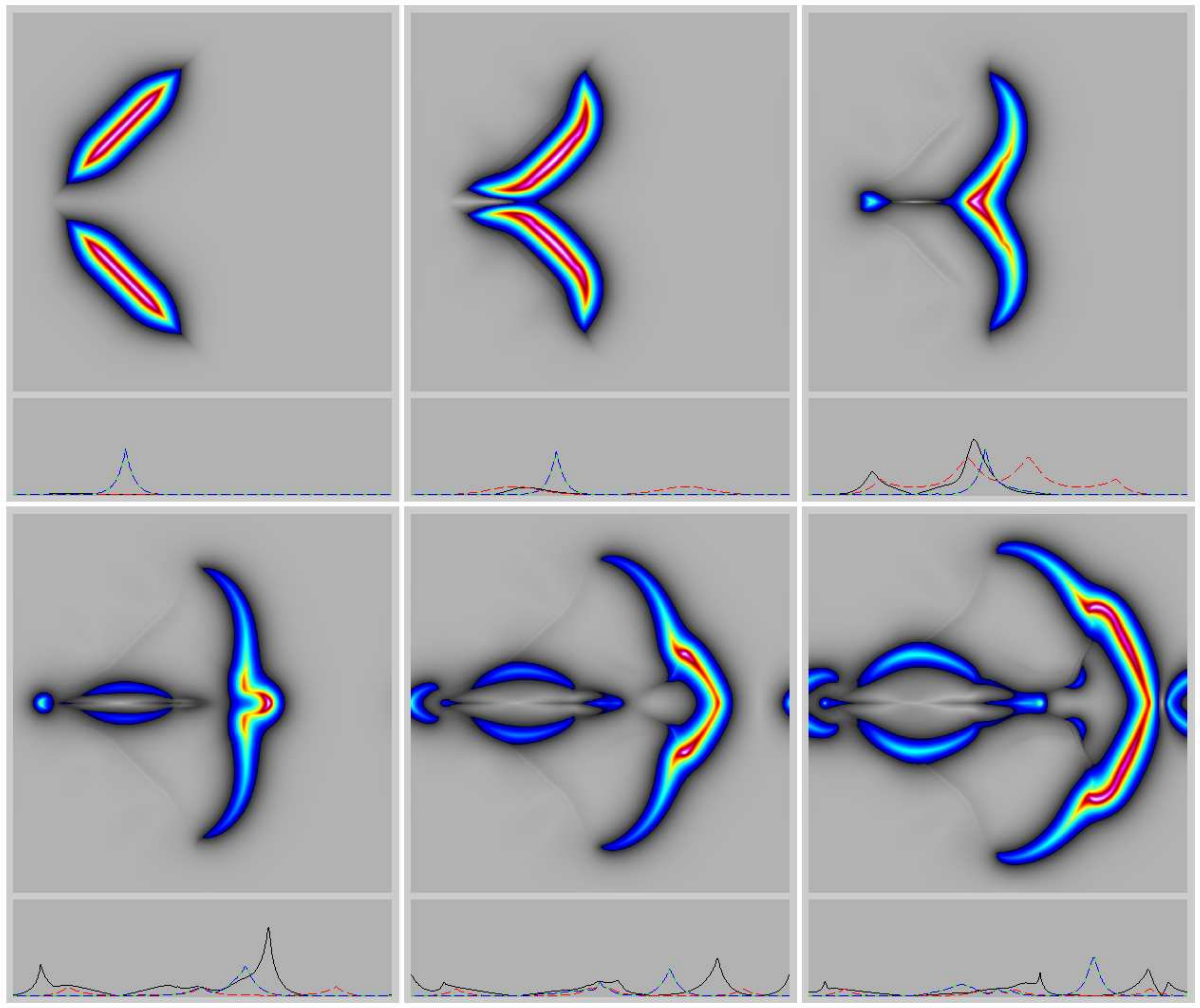}{2d}{wedge}{\sigma  }{}
\figfour{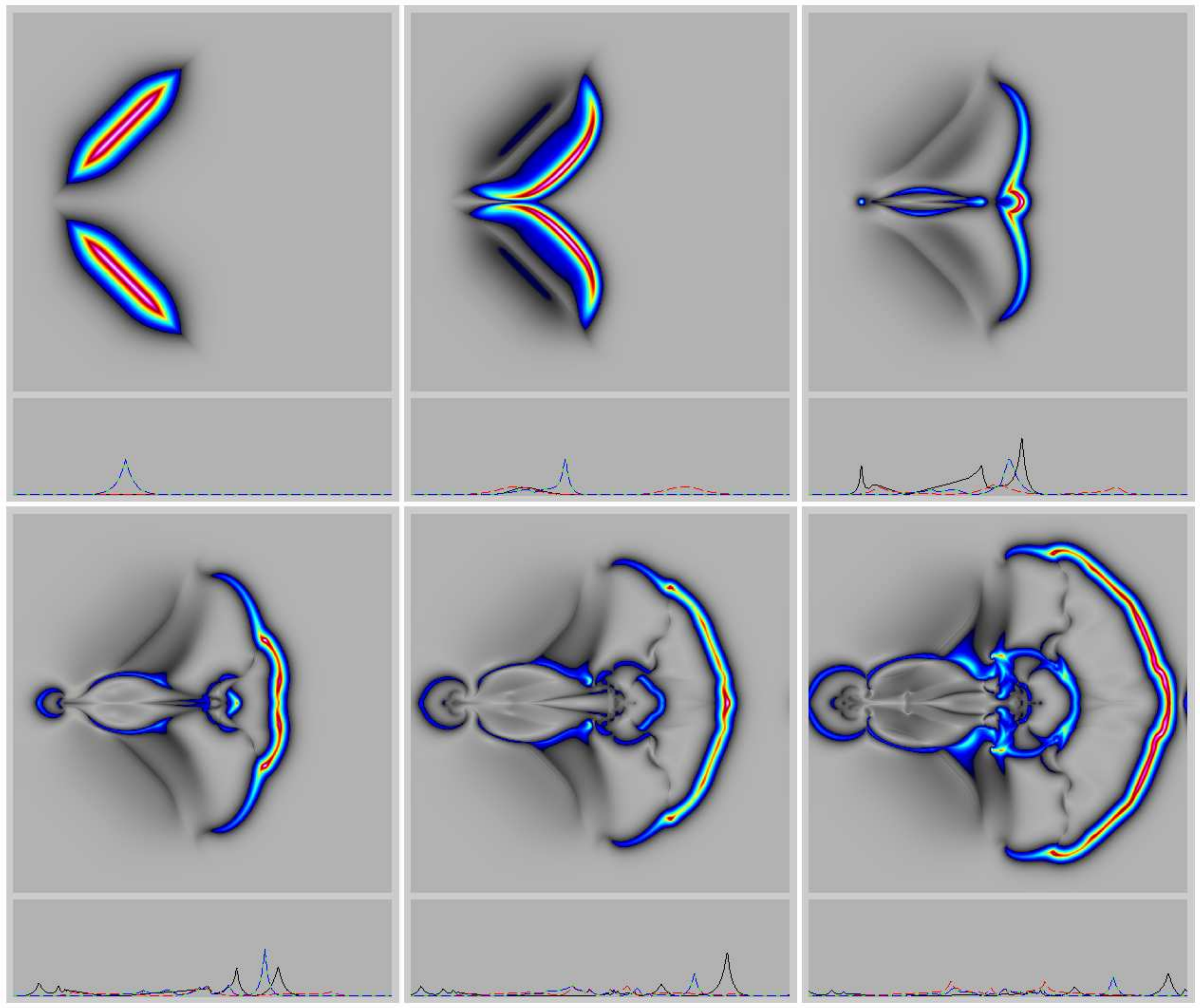}{2d}{wedge}{\sigma/2}{}
\figfour{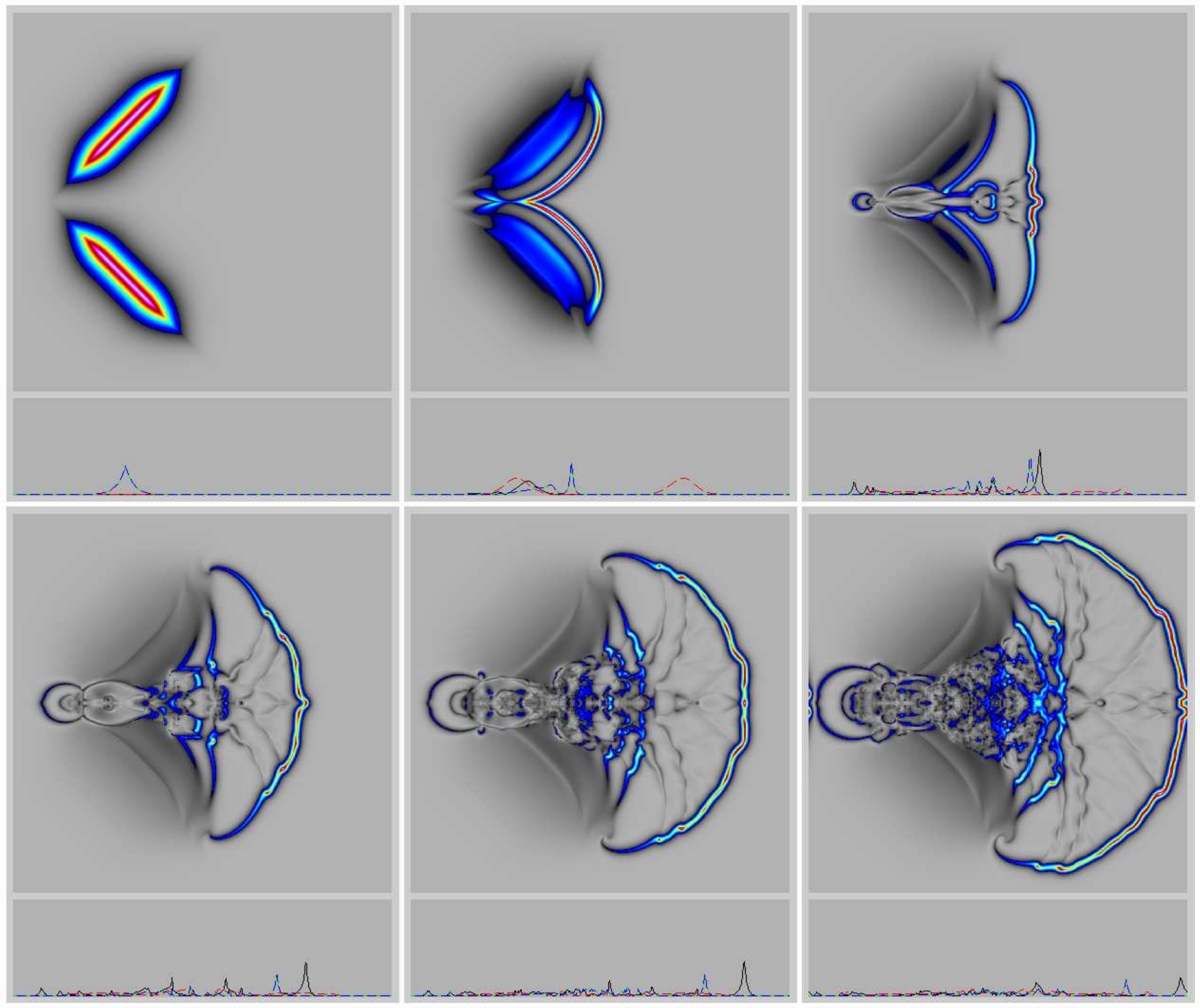}{2d}{wedge}{\sigma/4}{}
\figfour{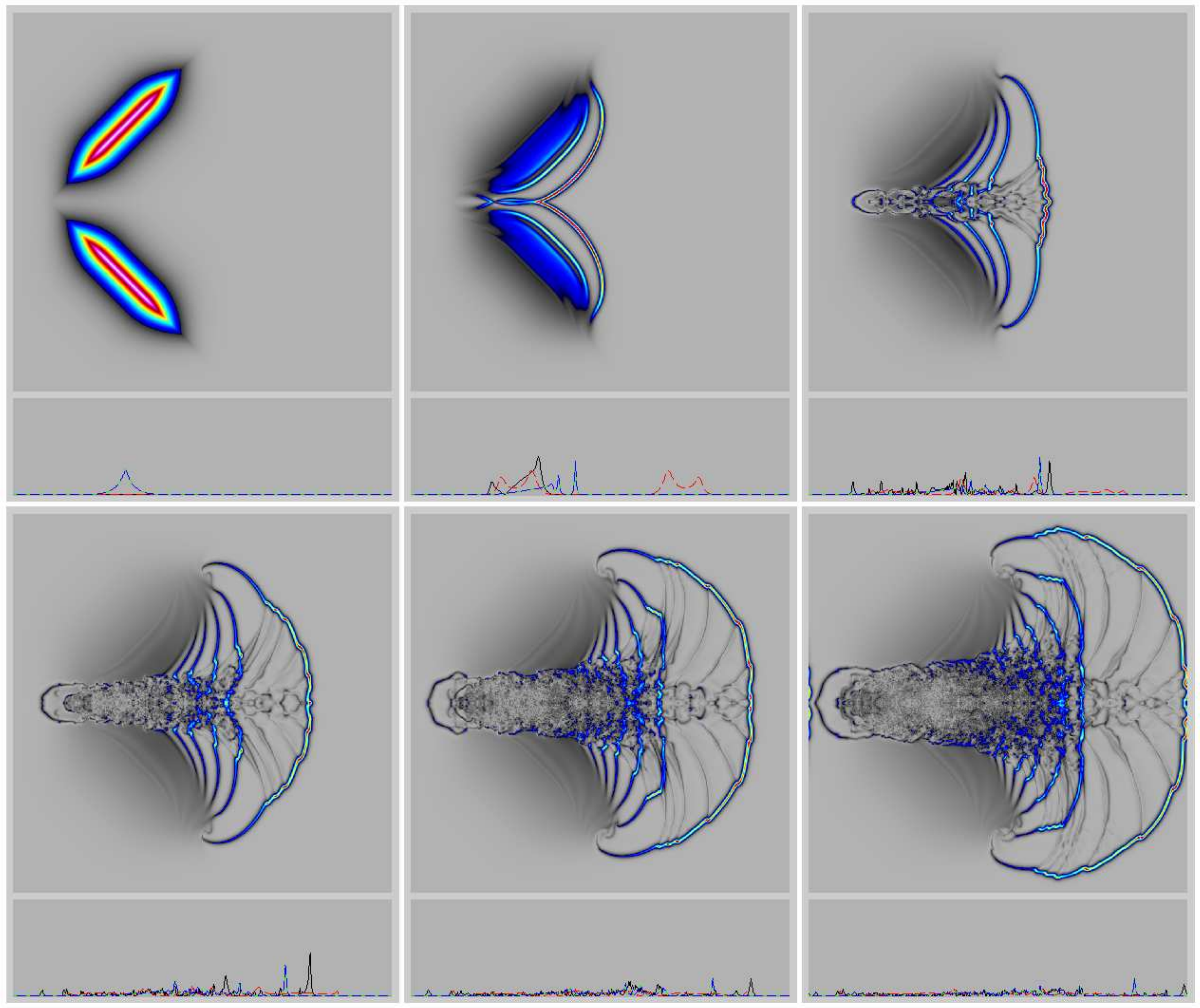}{2d}{wedge}{\sigma/8}{}

\figfourreverse{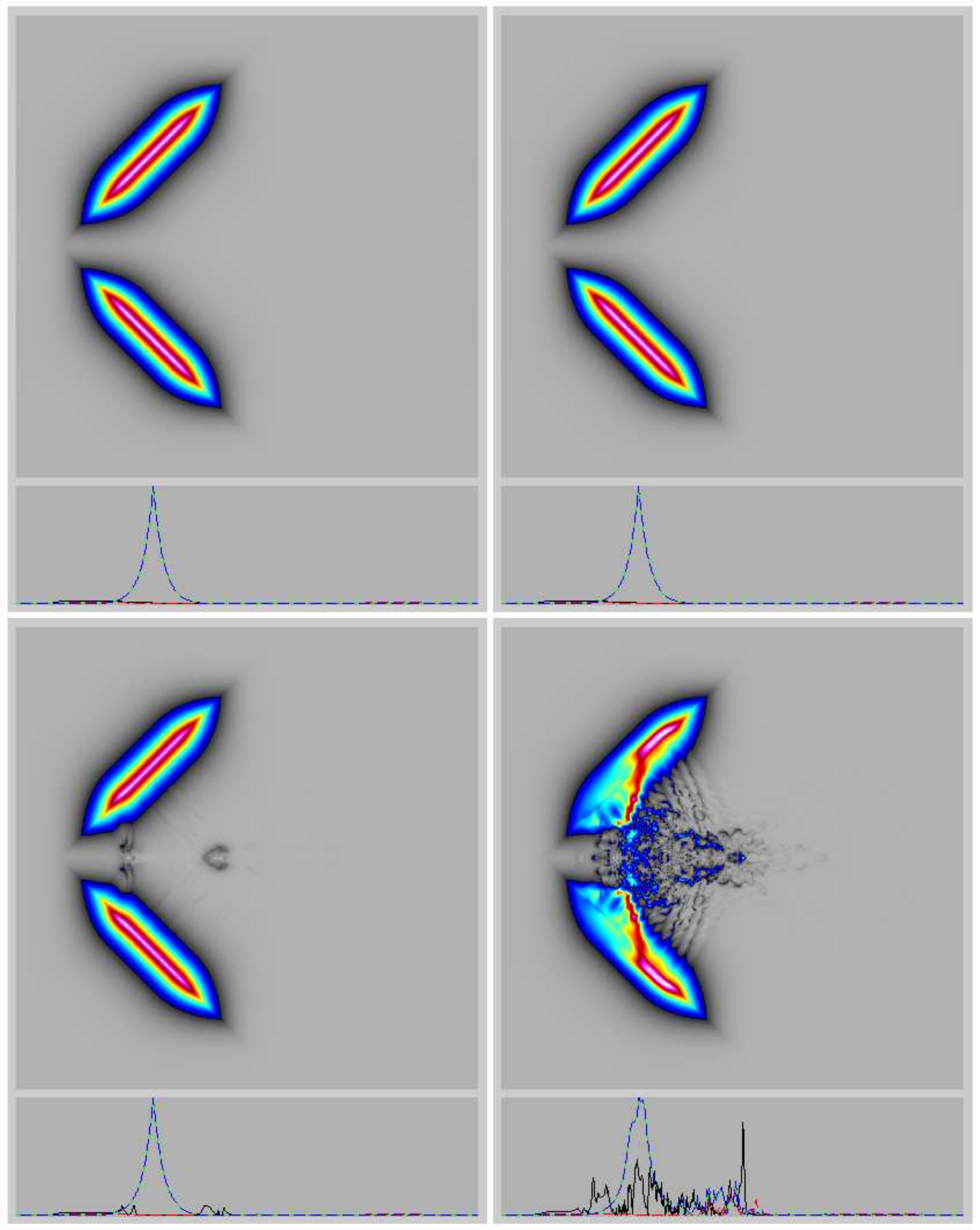}{2d}{wedge}{}

\subsection{Head-on}\label{2d-headon}

The head-on collisions of two offset peakon segments in
Figures \ref{2d-headon-sigma-1}-\ref{2d-headon-sigma-8} show great complexity. Some of this
complexity is due to the process of annihilation and recreation known to
occur in the purely 1d antisymmetric head-on collisions of a peakon with
its reflection, the antipeakon. At the moment a head-on collision
occurs, one sees the blue profiles in the lower panels of Figures
\ref{2d-headon-sigma-1}-\ref{2d-headon-sigma-8} becoming very steep, exactly as seen in 1d
peakon-antipeakon collisions. Thus, the locally 1d collision rules carry
over to the 2d case for head-on collisions.

The case $\alpha=\sigma$ in Figure \ref{2d-headon-sigma-1} shows the
annihilation and recreation expected from the 1d rules. Thereafter, a new
feature emerges as the recreated segments disconnect, then eventually
reconnect again with segment elements that did not participate in the
head-on part of the collision. The first disconnection occurs in the
presence of rapid rotation or circulation of the peakon segments, which
causes extreme stretching before the disconnection. The later reconnection
of these segments occurs less violently.

Once again, peakon profiles are ubiquitous in the lower panels of Figures
\ref{2d-headon-sigma-1}-\ref{2d-headon-sigma-8}.

Time reversal of the head-on collisions shows, for $\alpha=\sigma$, an
essentially complete restoration of the initial condition; see Figure
\ref{2d-headon-back}. However, the cases for smaller alpha
($\alpha=\sigma/4\,,\sigma/8$) show breakup and head-on collisions occurring
simultaneously. These simultaneous processes produce complex mixed states which
tend to reverse less accurately, as one might expect, because of the plethora
of peaked excitations at small scales.

\figfour{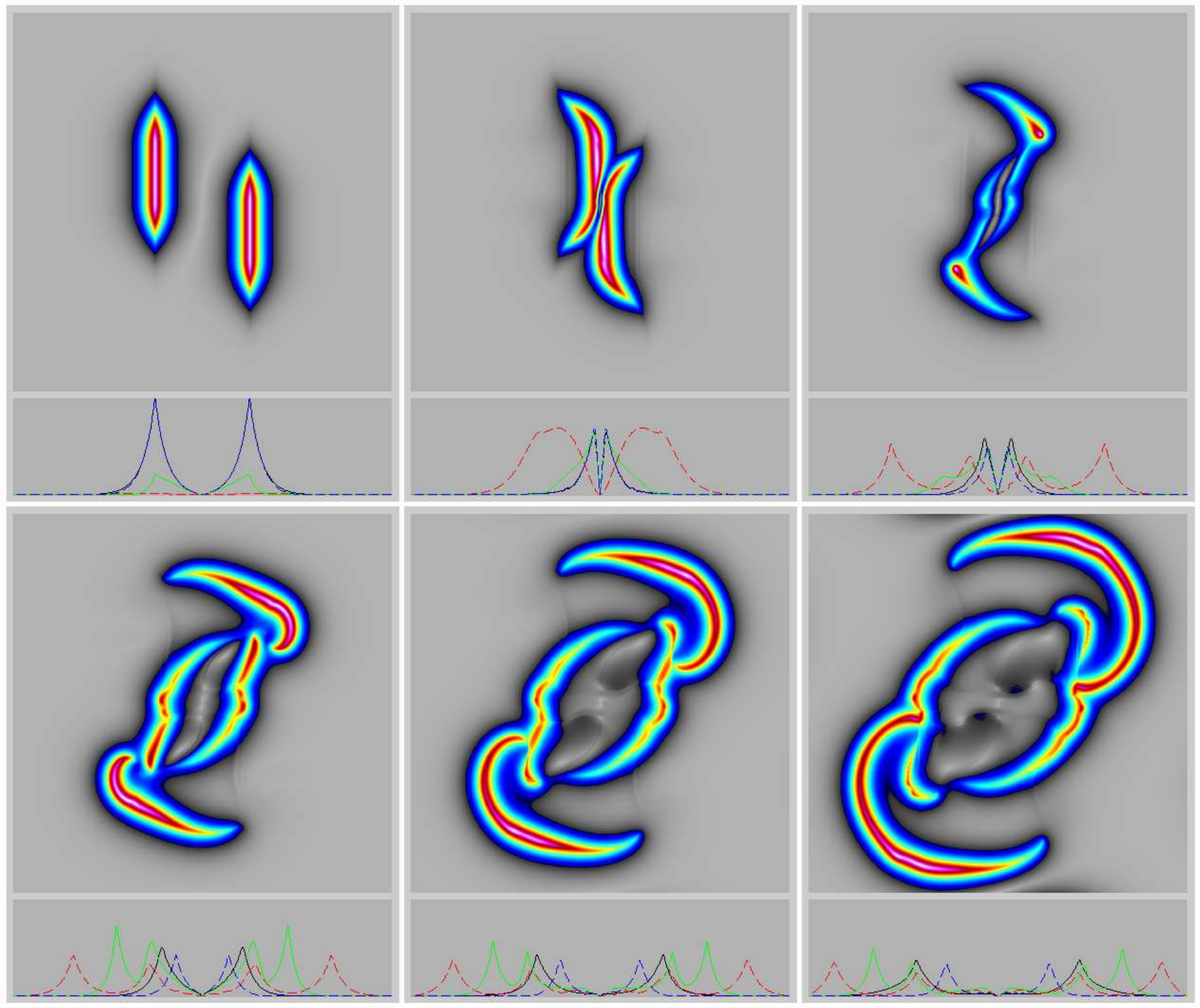}{2d}{head-on}{\sigma  }{}
\figfour{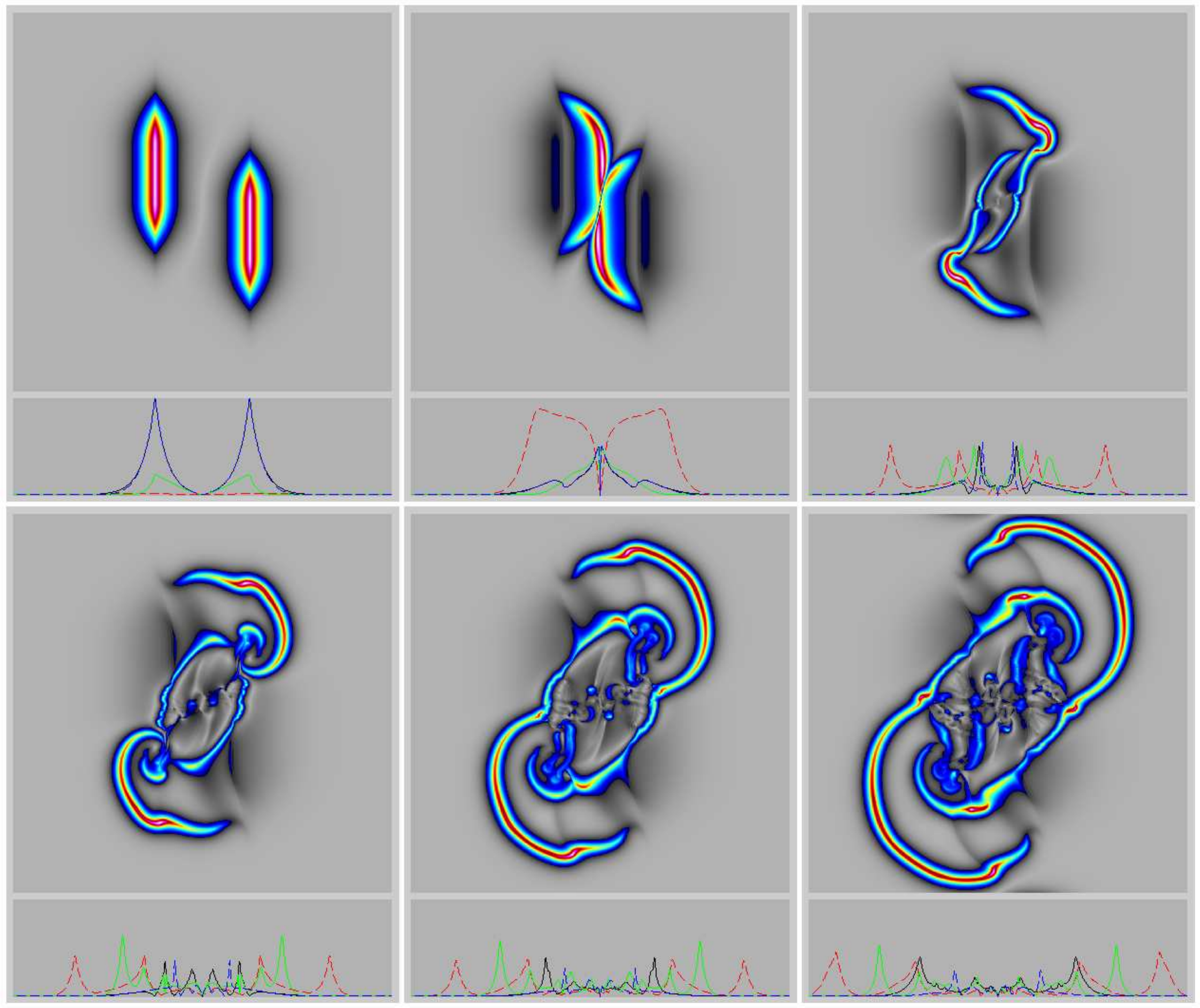}{2d}{head-on}{\sigma/2}{}
\figfour{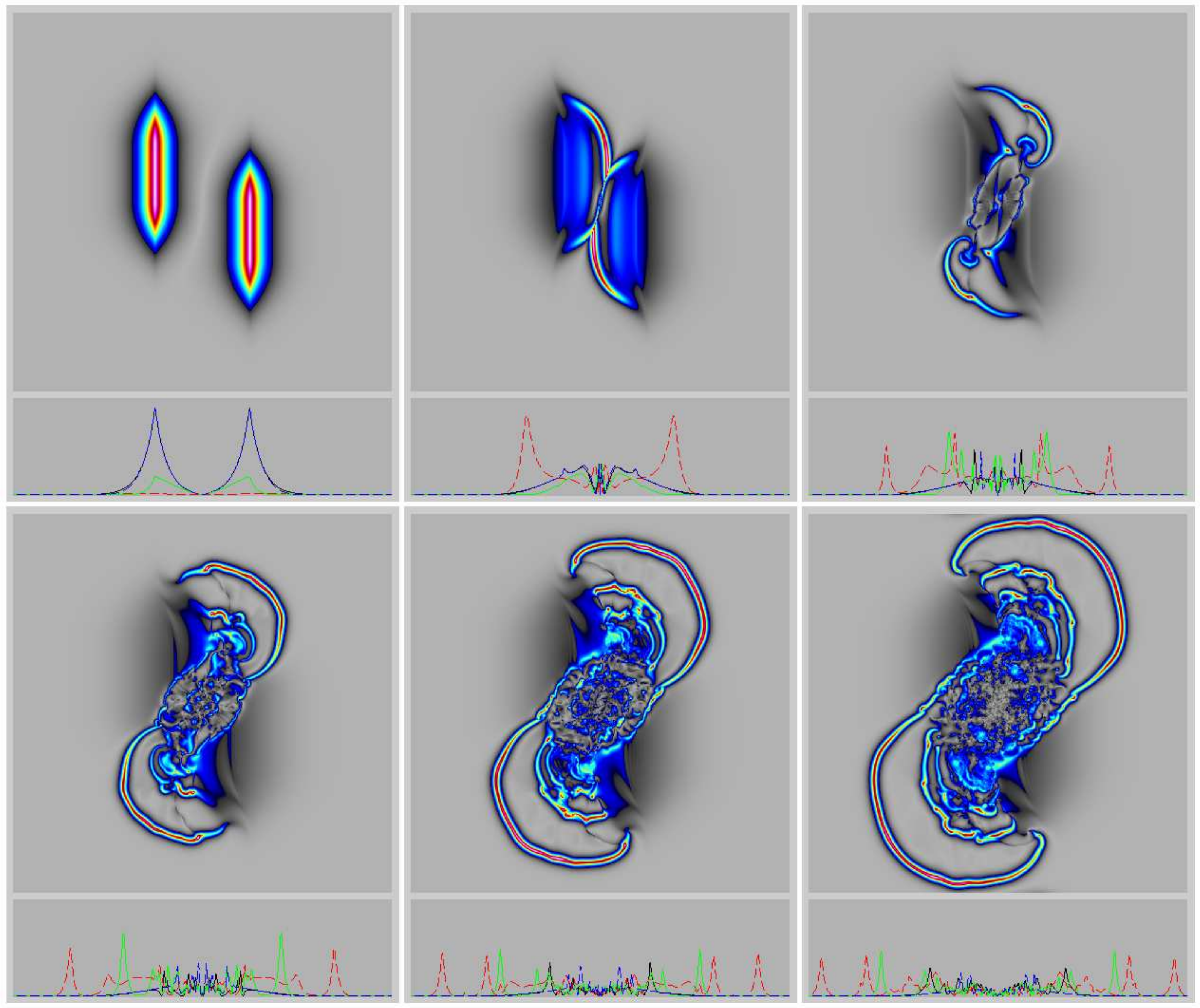}{2d}{head-on}{\sigma/4}{}
\figfour{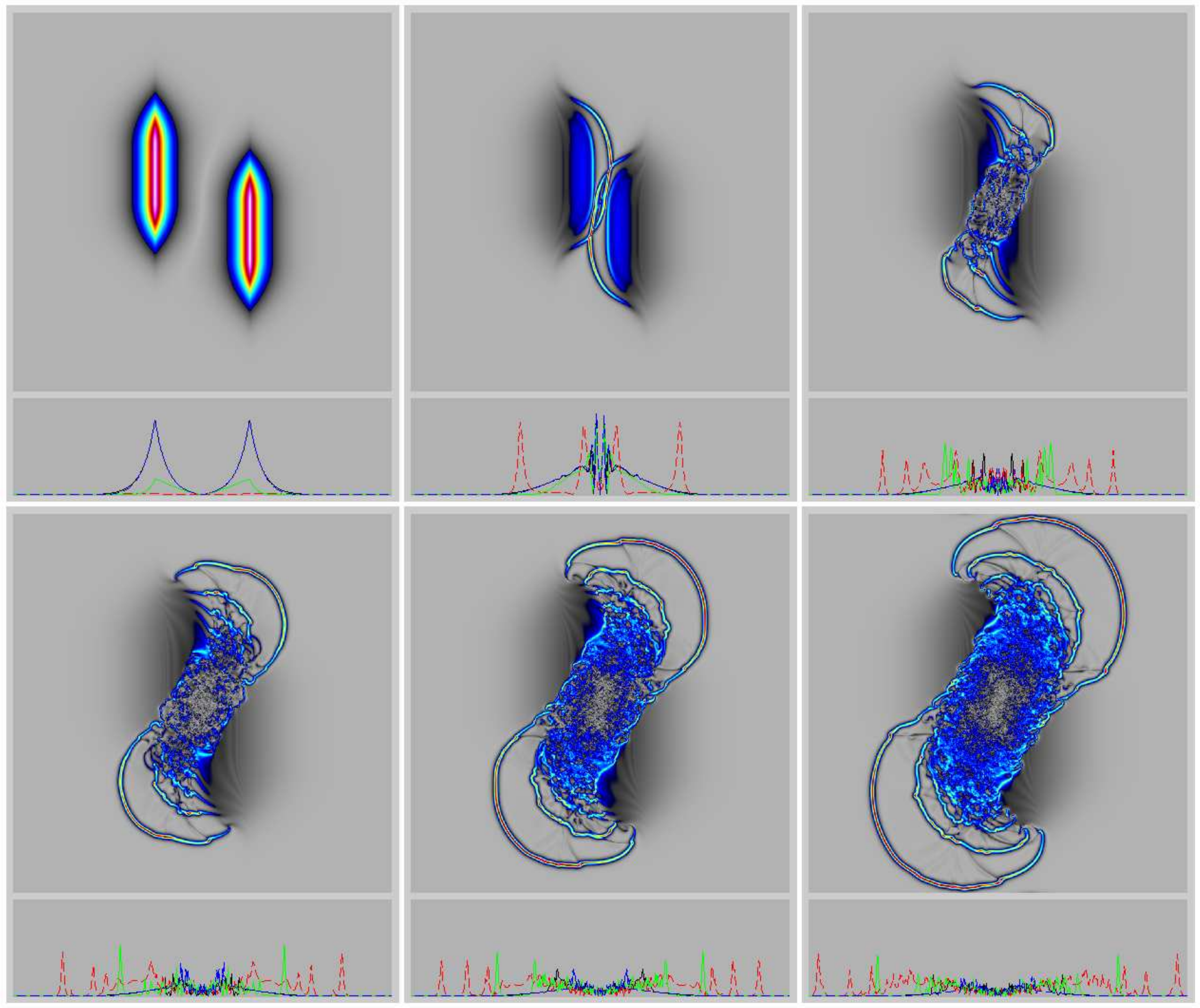}{2d}{head-on}{\sigma/8}{}

\figfourreverse{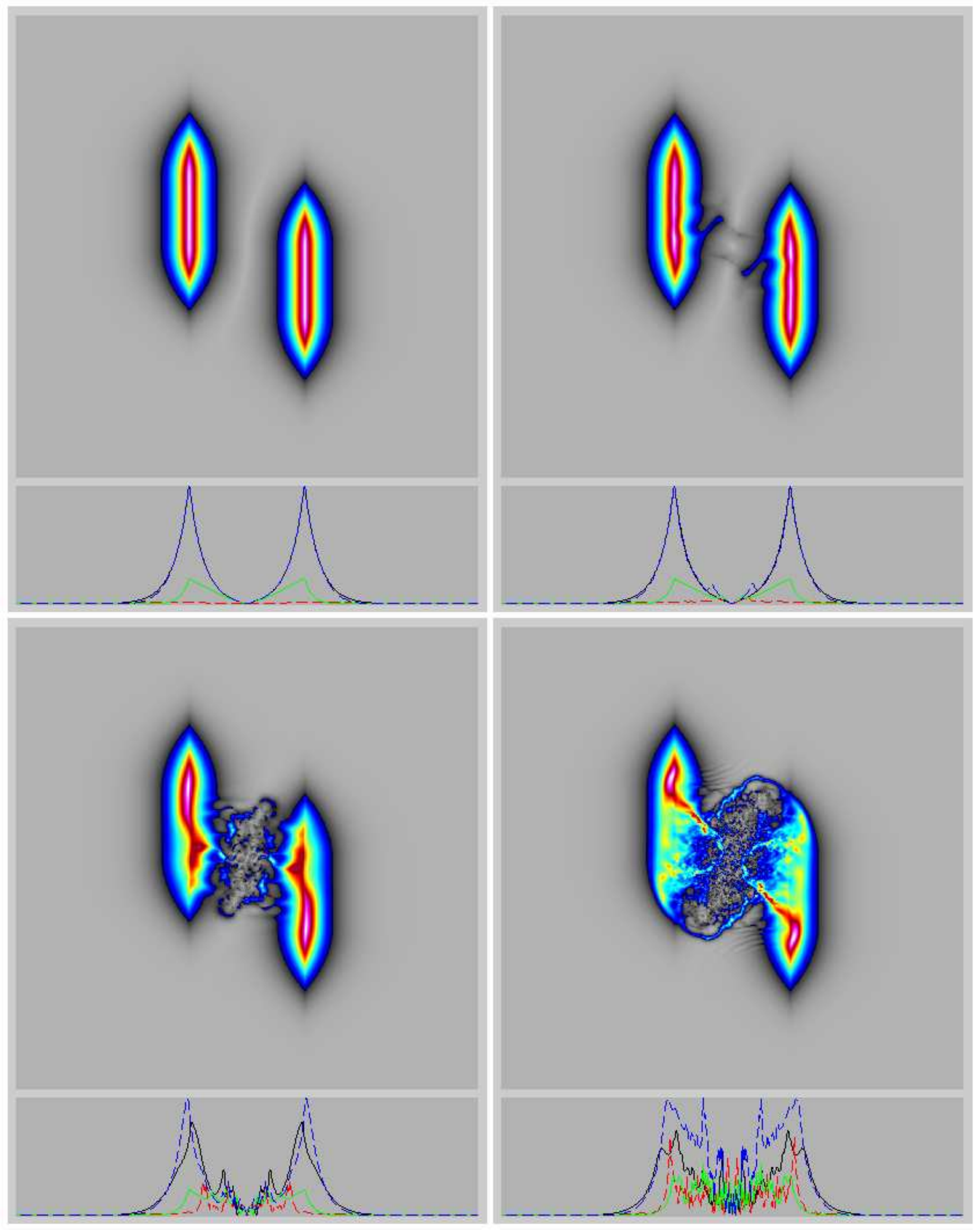}{2d}{head-on}{}

\subsection{Star}\label{2d-star}

Star flows in Figures \ref{2d-star-sigma-1}-\ref{2d-star-sigma-8} form a
variant of wedge flows with fivefold symmetry, instead of simple reflection
antisymmetry. The mutual rotation of the overtaking-collision evolution in
each case preserves the fivefold symmetry well, and is seen in Figure
\ref{2d-star-back} to be largely reversible. Figures
\ref{2d-star-sigma-1}-\ref{2d-star-sigma-8} each show many reconnections
(mergers), until eventually one peakon filament ring surrounds all
the others. Again, peakon segments are the ubiquitous feature of the
solution. If the evolution were allowed to proceed further, reconnections
would tend to produce additional concentric rings of peakon filaments.

\figfour{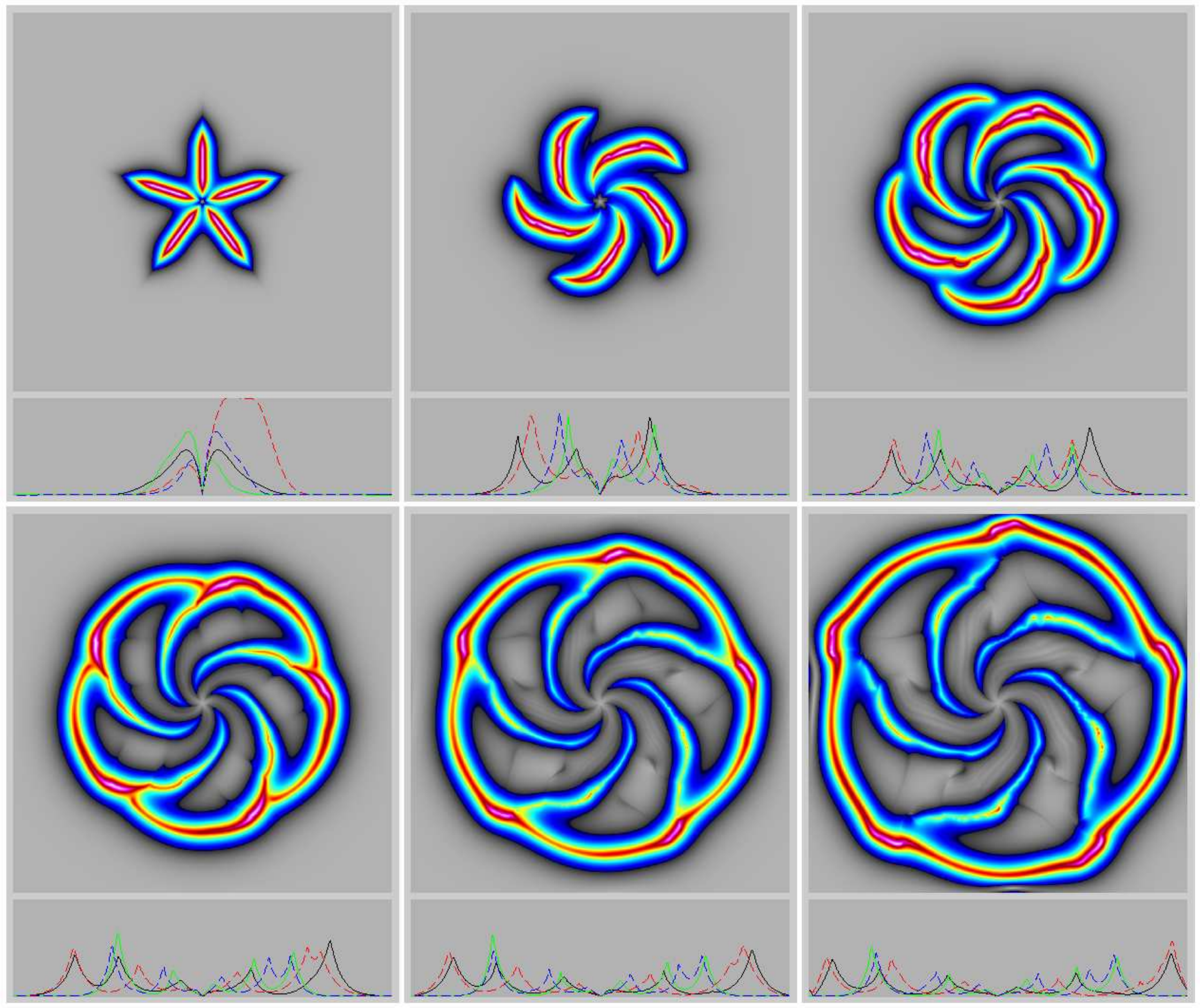}{2d}{star}{\sigma  }{}
\figfour{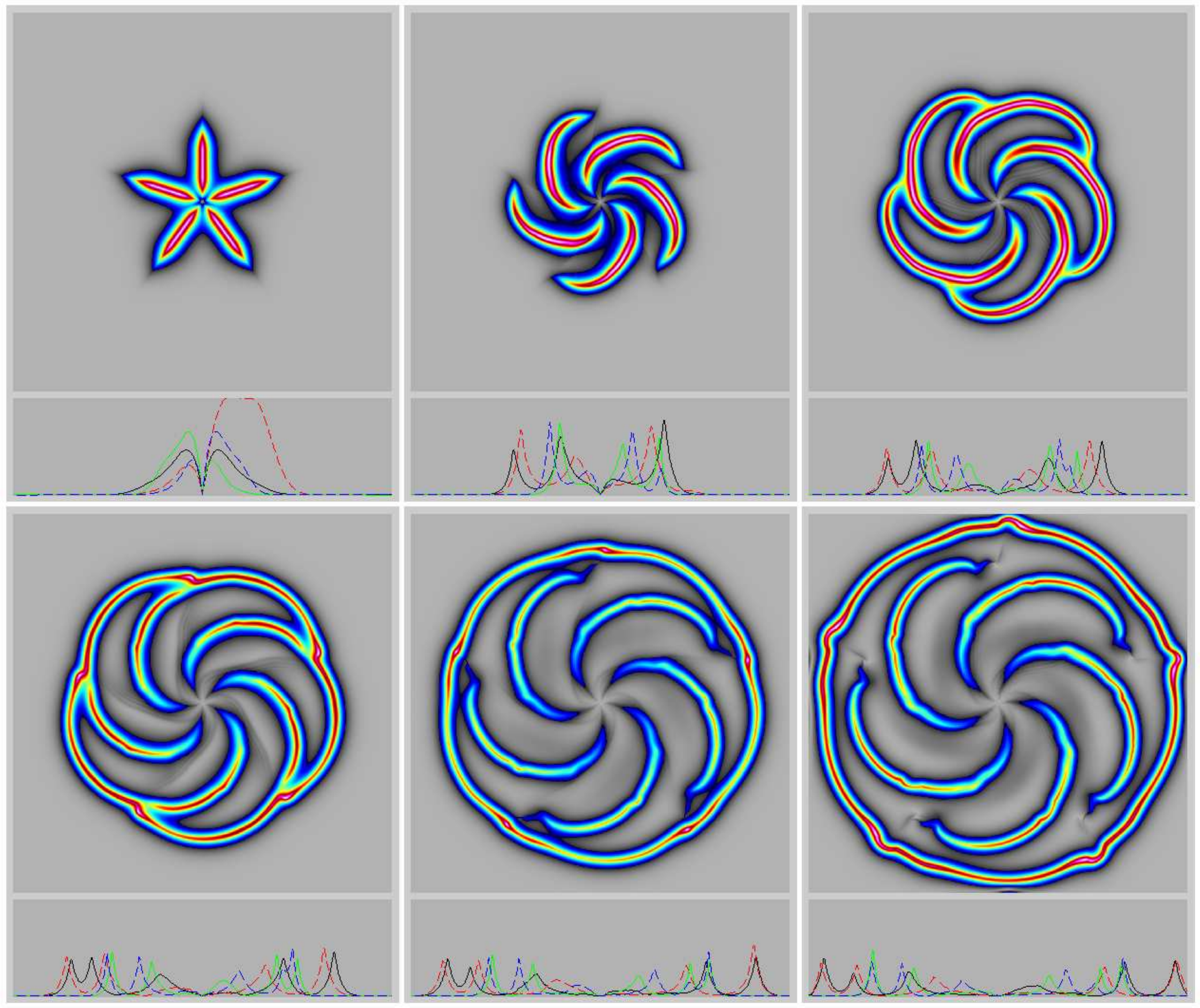}{2d}{star}{\sigma/2}{}
\figfour{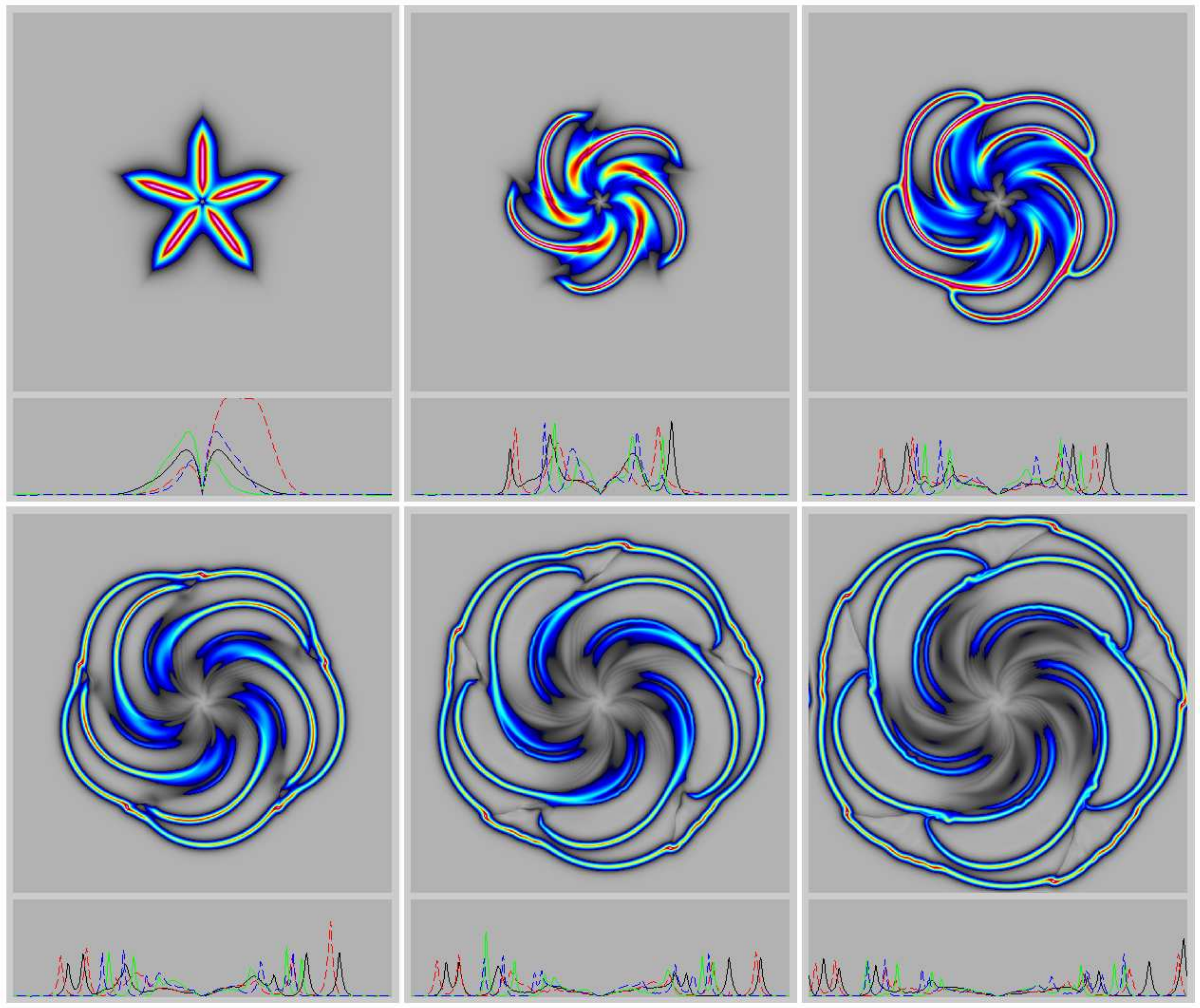}{2d}{star}{\sigma/4}{}
\figfour{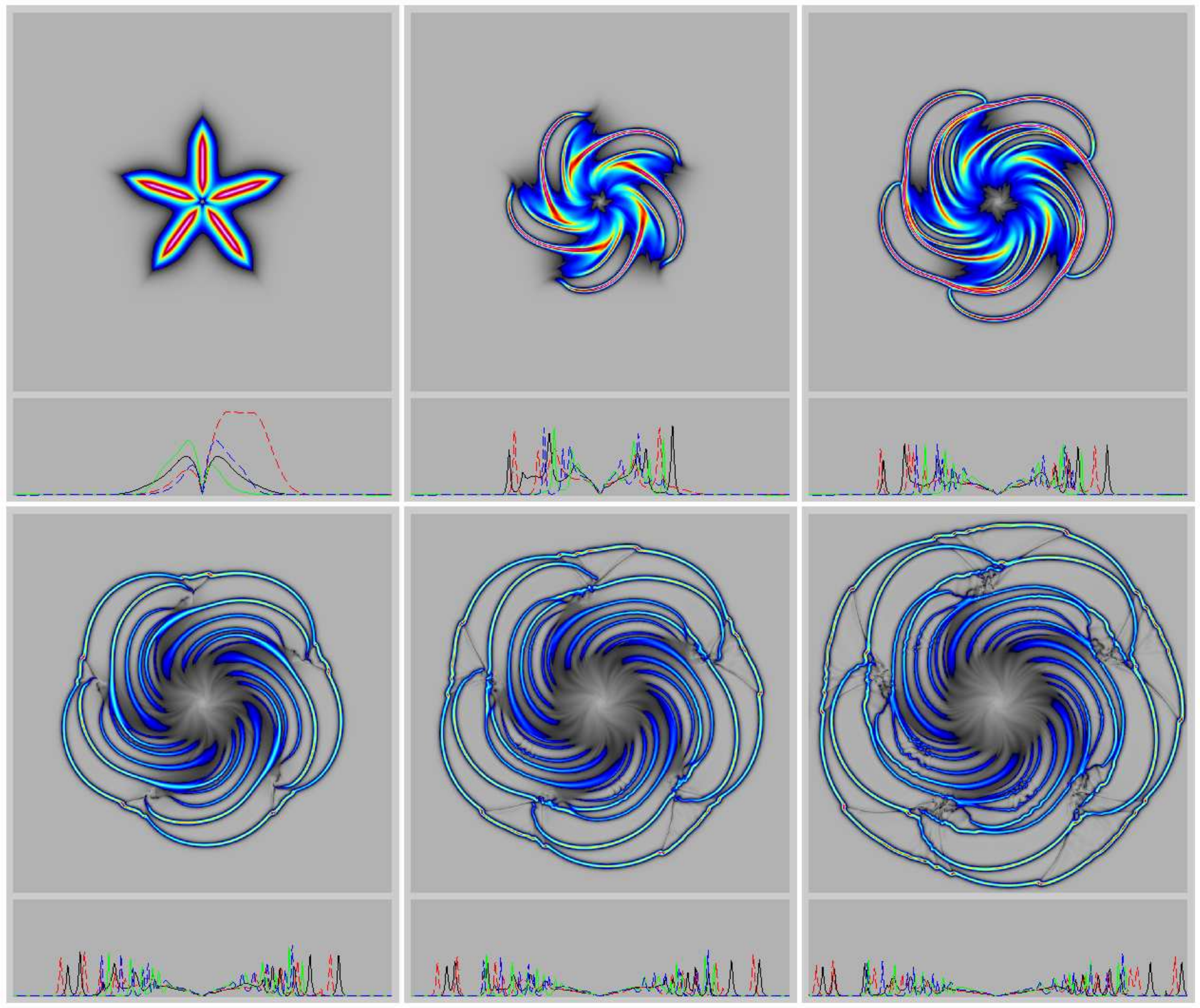}{2d}{star}{\sigma/8}{}

\figfourreverse{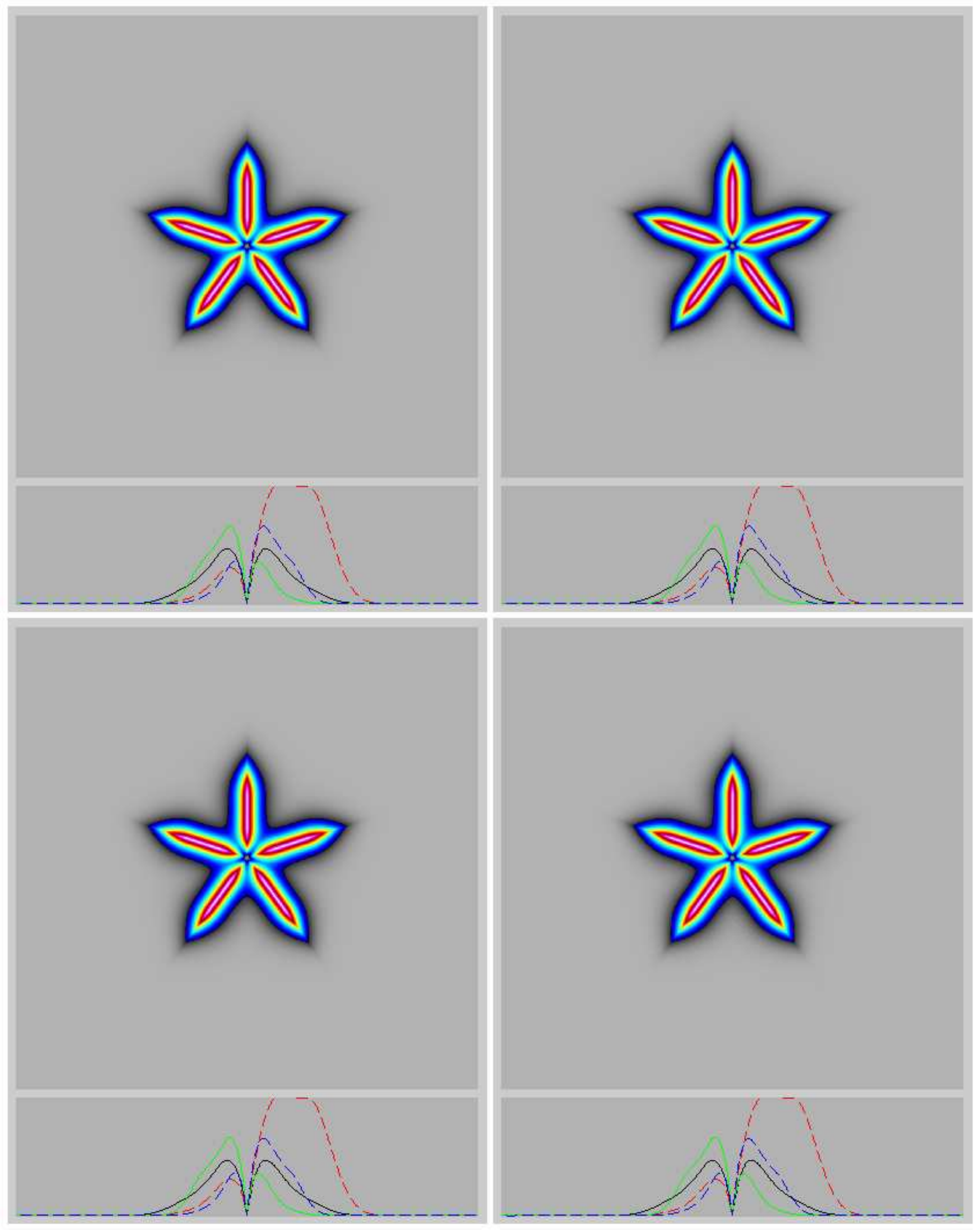}{2d}{star}{}

\subsection{Rotate}\label{2d-rotate}

Figures \ref{2d-rotate-sigma-1}-\ref{2d-rotate-sigma-8} show how angular
(azimuthal) motion couples to radial motion in the plane. Each evolution starts
with a circularly symmetric velocity distribution in a Gaussian (not peakon)
ring, or annulus, of width alpha, which is initially rotating rigidly at
constant angular velocity. The angular motion couples to the radial motion
by producing a sequence of outward and inward propagating circular peakons.
These circular peakons also rotate clockwise (in the same sense as the
initial condition) because of conservation of angular momentum.
See Holm et al. \cite{HoMuSt2004,HoPuSt2004} for more details about circular
peakons.

For $\alpha<\sigma$, the first inward propagating circular peakon in
Figures \ref{2d-rotate-sigma-2}-\ref{2d-rotate-sigma-8}  collapses to the
center, then reflects outward and overwhelms the formation of any
subsequent inward moving circular peakons. For $\alpha=\sigma$, no inward
moving peakons form during the time of the simulation.

Upon reflection from the center, the circular peakon is influenced by the
finite Cartesian mesh, especially for smaller alpha (narrower peakons).
The reflection interaction with the mesh distorts the outward propagating
solution, as seen in the animations, with greater distortion for smaller
alpha. This is a severe test of the numerics near the center of symmetry.
Nonetheless, time reversal from the final velocity profile reconstitutes
the initial condition to within less than one percent, as seen in Figure
\ref{2d-rotate-back}.
This illustrates that accurate time reversal to the initial condition does not
guarantee accuracy of the solution during the forward evolution. However, it
does show that even in a somewhat distorted solution, dissipation plays little
role in the numerical simulation.

\figfour{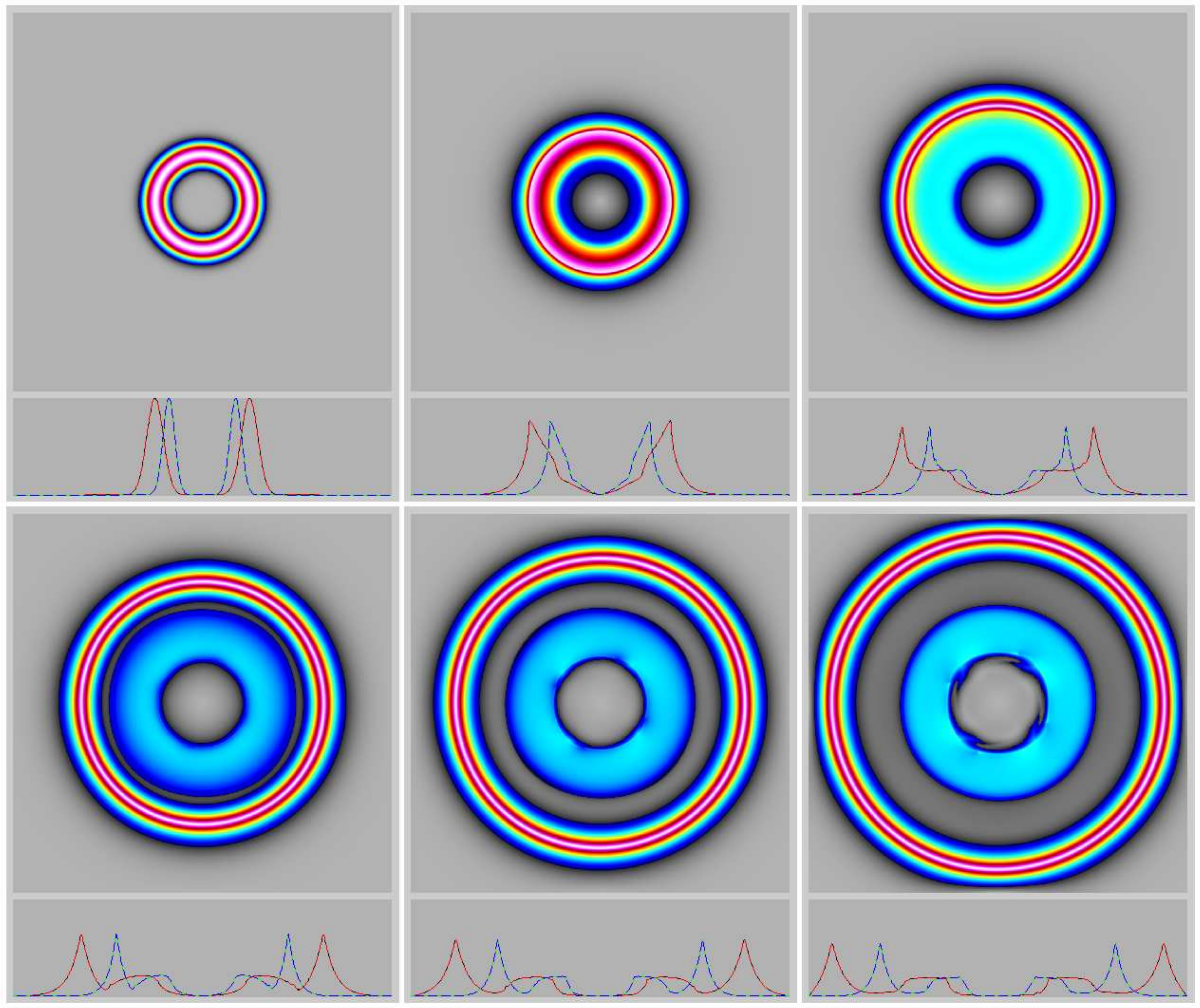}{2d}{rotate}{\sigma  }{}
\figfour{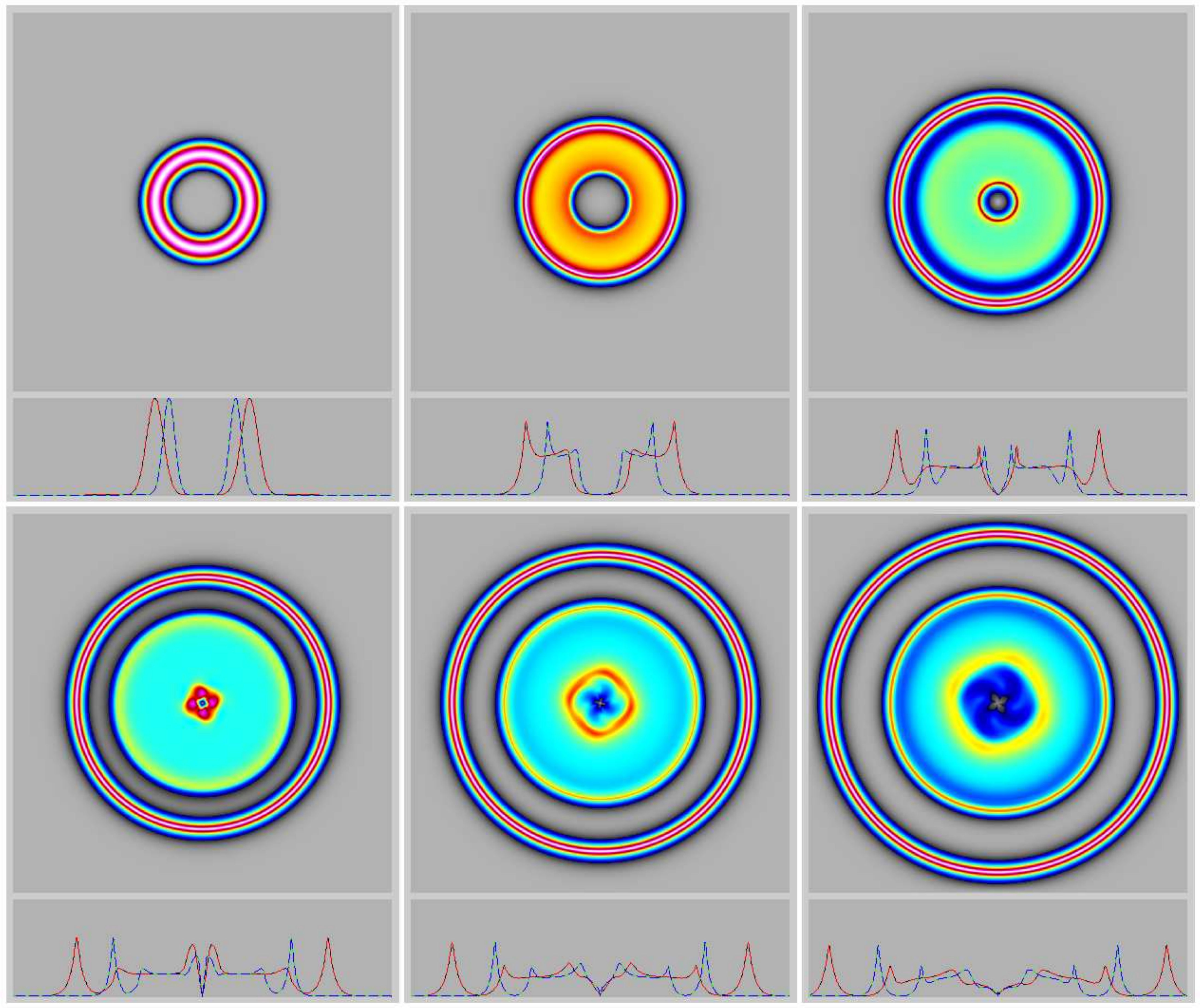}{2d}{rotate}{\sigma/2}{}
\figfour{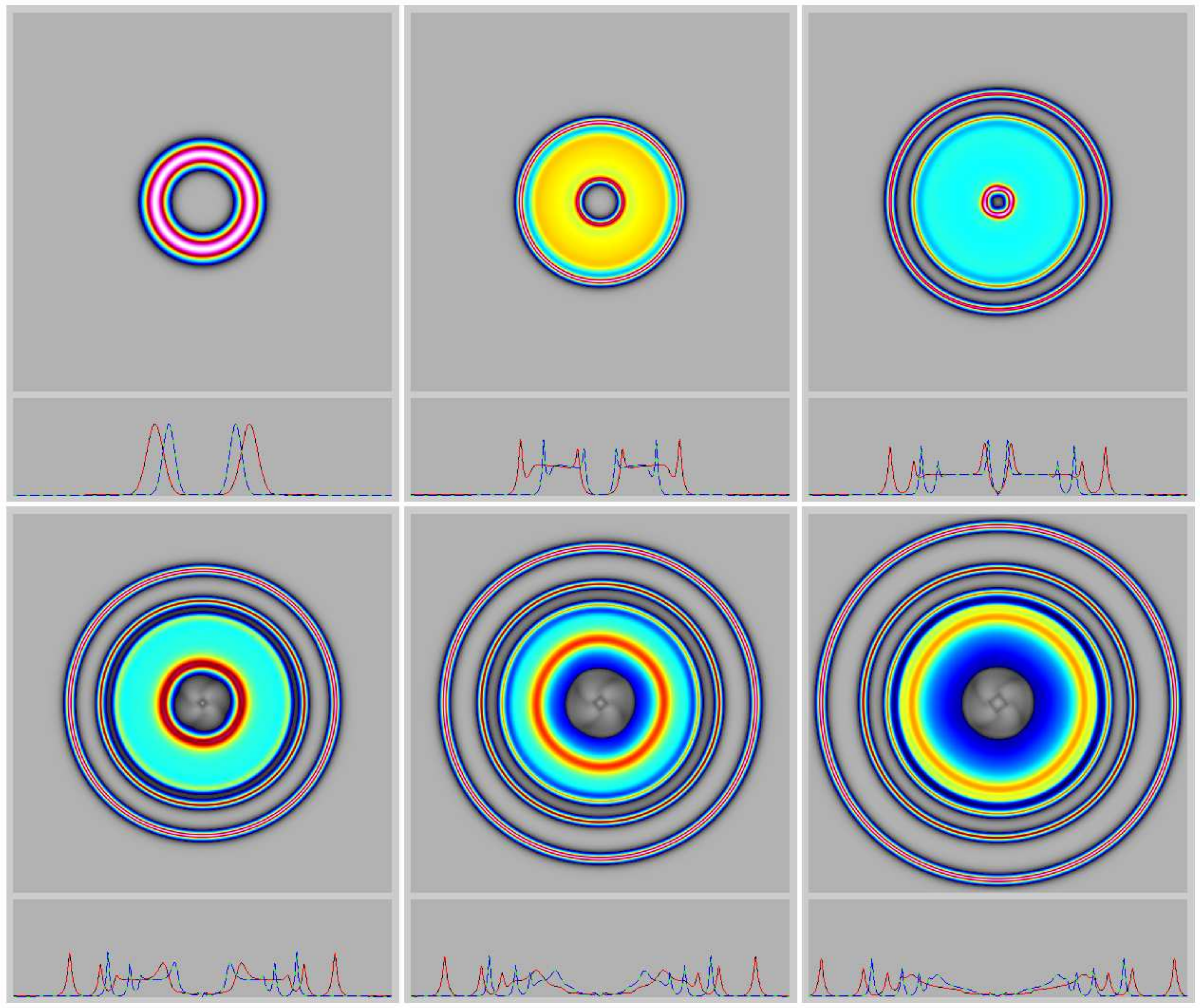}{2d}{rotate}{\sigma/4}{}
\figfour{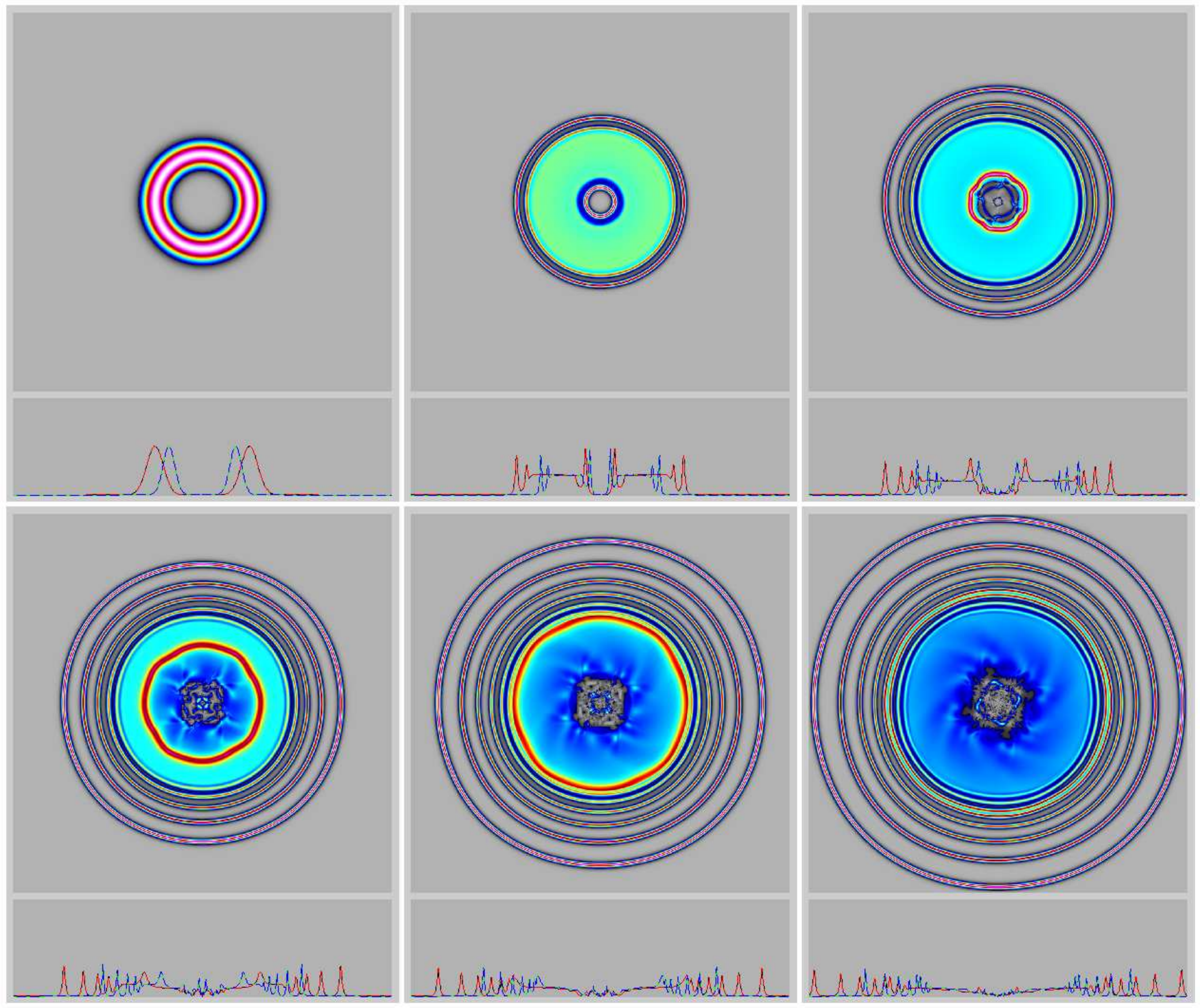}{2d}{rotate}{\sigma/8}{}

\figfourreverse{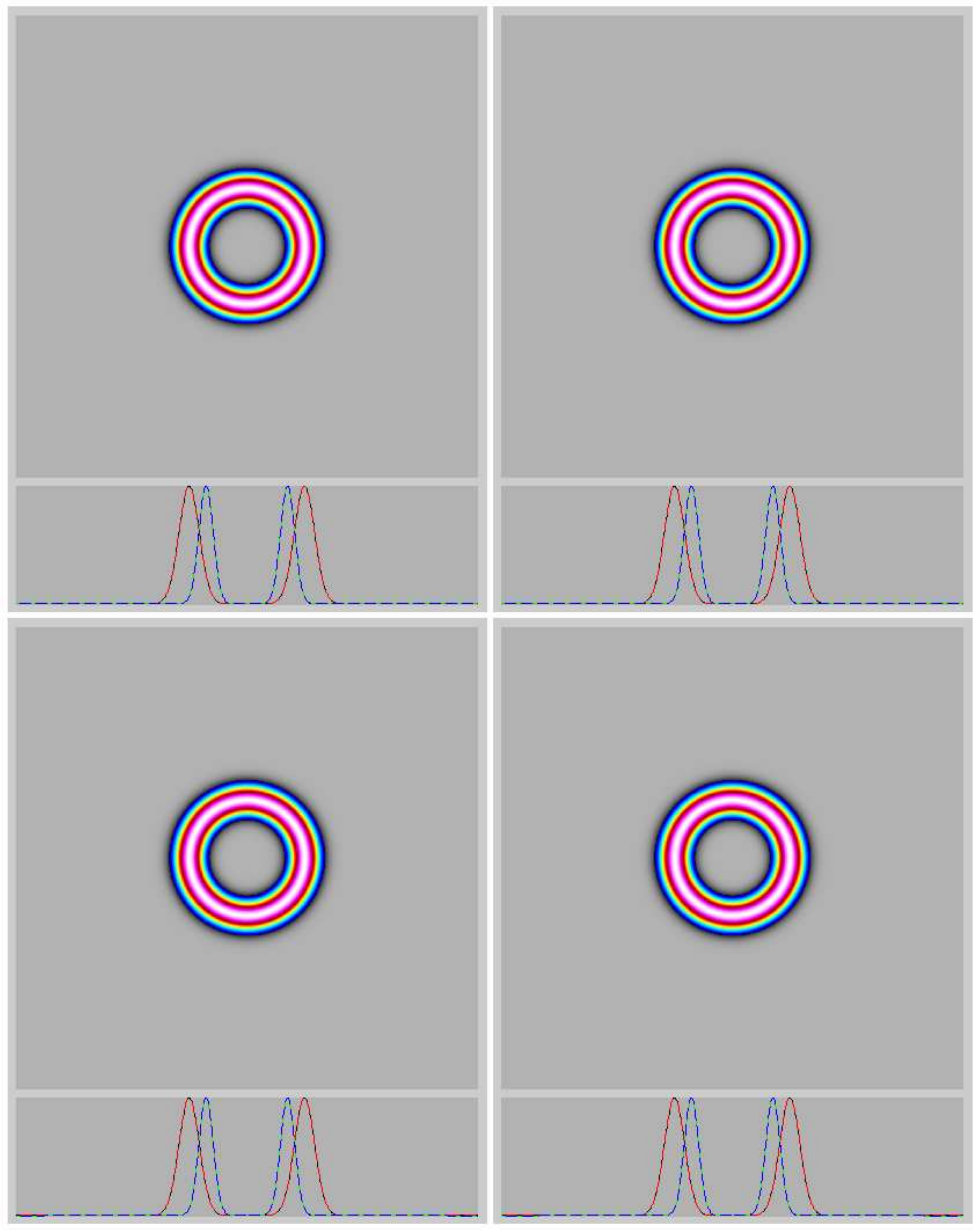}{2d}{rotate}{}

\subsection{Right}\label{2d-right}

In the simulations shown in Figures \ref{2d-right-sigma-1}-\ref{2d-right-sigma-8}, the
velocity distribution is initially a Gaussian ring in magnitude, uniformly
pointed rightward along the $x$ axis. The right outer side of the ring
produces diverging peakon contact curves, which slow as they propagate
outward. The left inner side of the ring, however, produces converging
peakon contact curves, which accelerate as they converge, undergo a strong
interaction along the axis, then break again into contact curves still
moving rightward, approaching the previous divergent peakon curves and
colliding with them from behind. These overtaking collisions impart
momentum but they do not produce reconnections.

After the collisions, a complex flow remains near the axis, in which one
also sees hot spots at kinks in the contact curves, with trailing memory
wisps behind them.

The lower panels show peakon profiles with high wavenumber oscillations
(possibly noise) in the complex flow region remaining behind near the axis.
Except for the smallest case of $\alpha=\sigma/8$, all of the time-reversed
runs reassemble into the Gaussian ring without significant distortion, as
diagnosed by the black profile in the lower panel of the time reversed
runs in Figure \ref{2d-right-back}.

\figfour{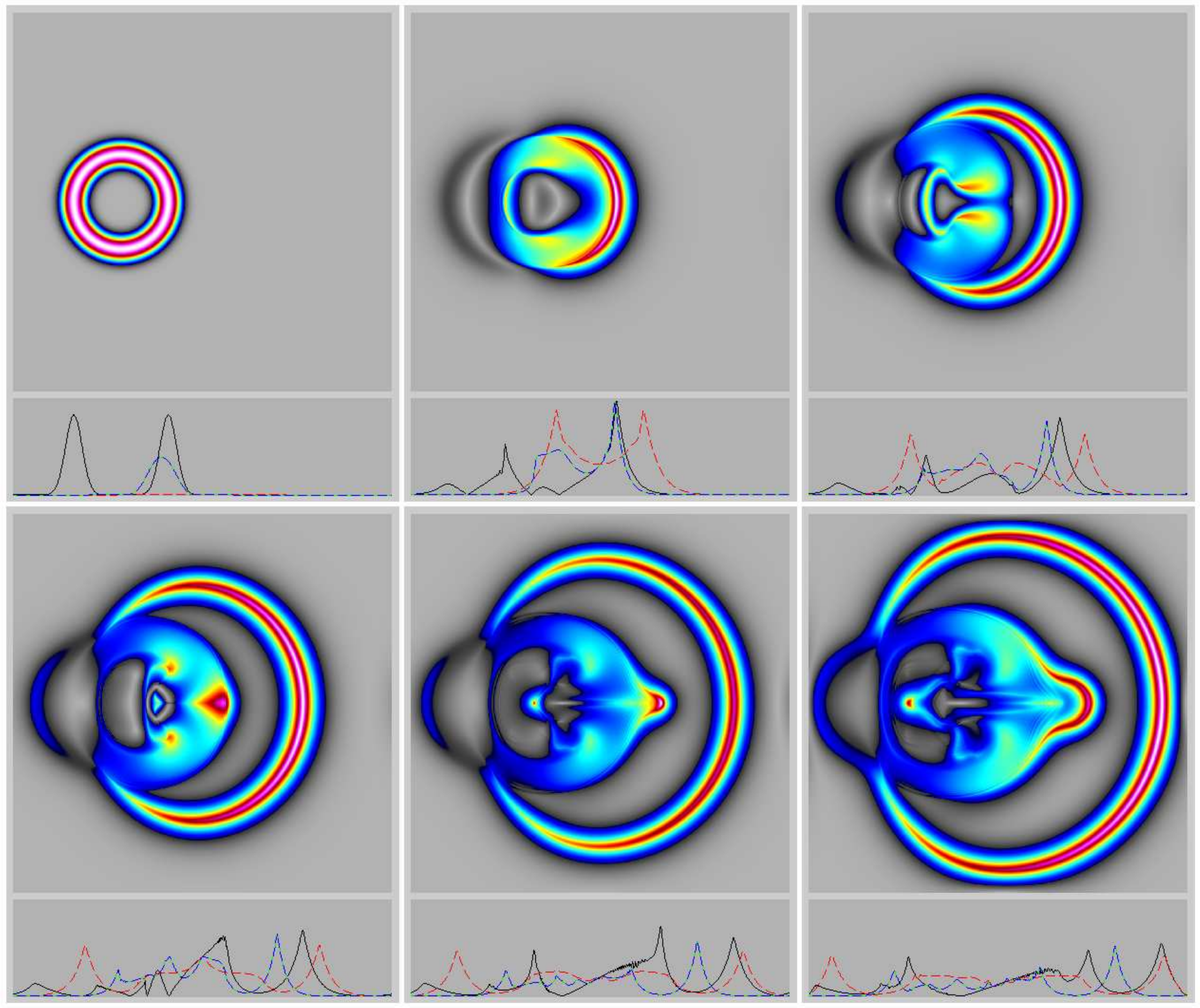}{2d}{right}{\sigma  }{}
\figfour{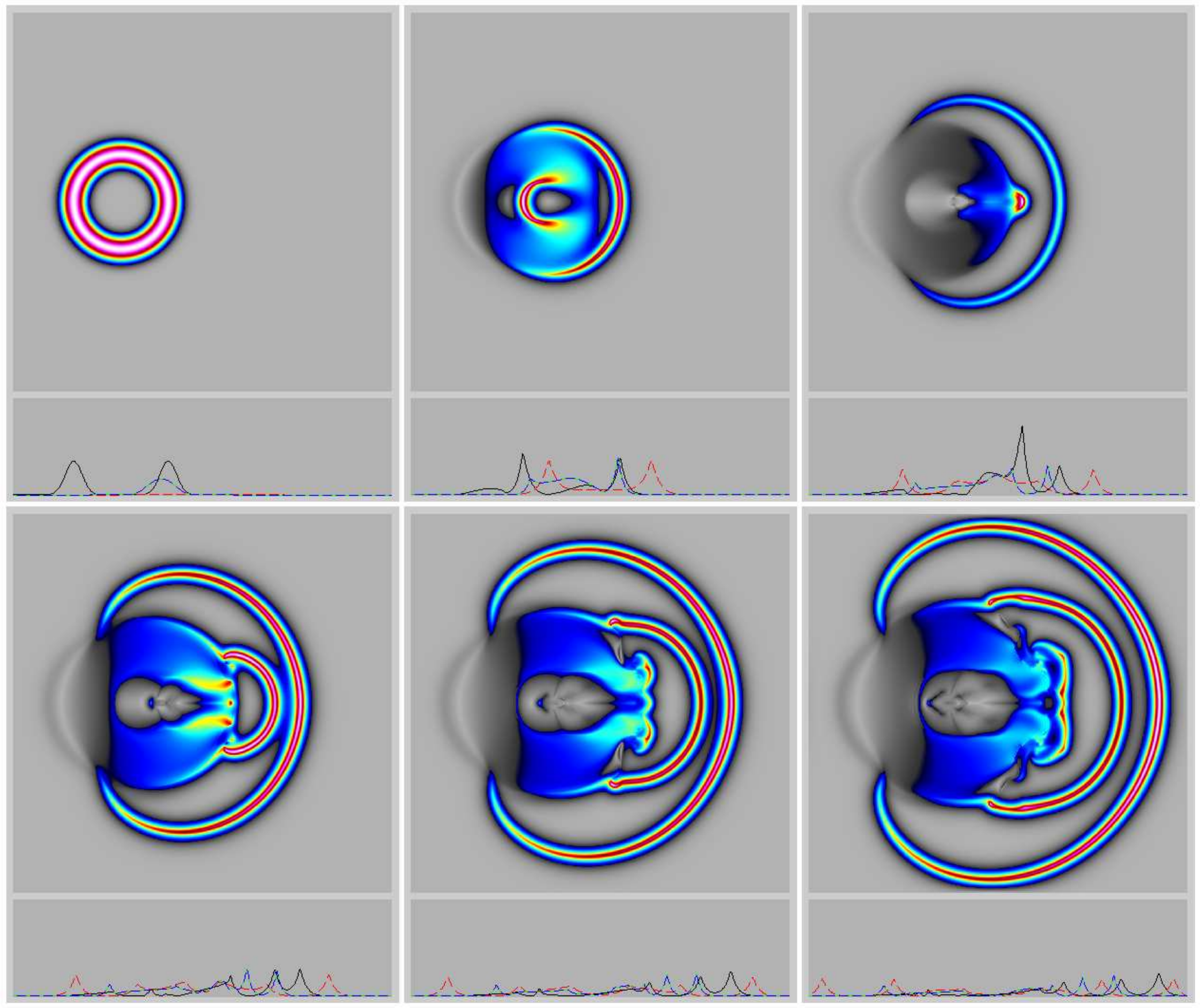}{2d}{right}{\sigma/2}{}
\figfour{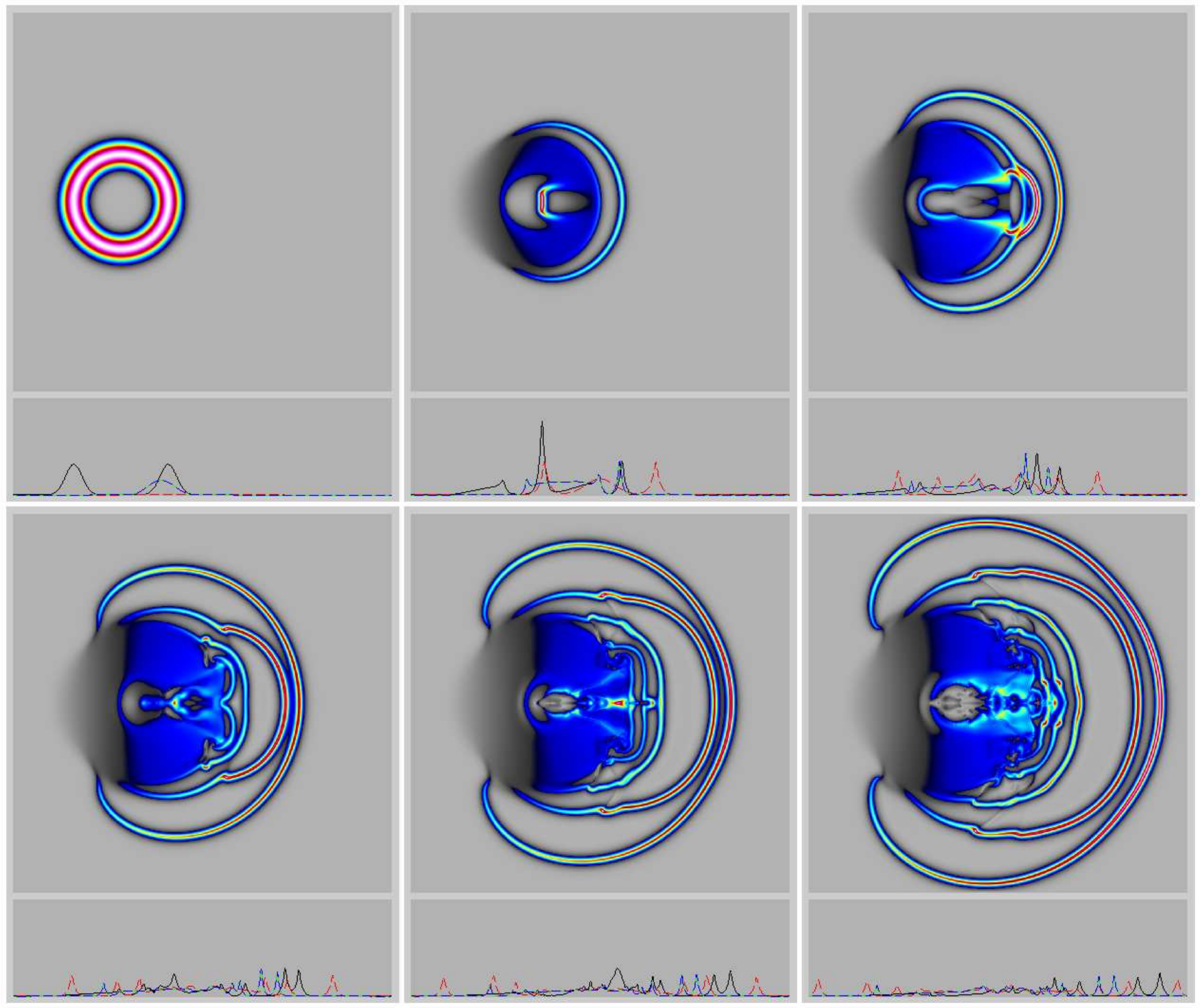}{2d}{right}{\sigma/4}{}
\figfour{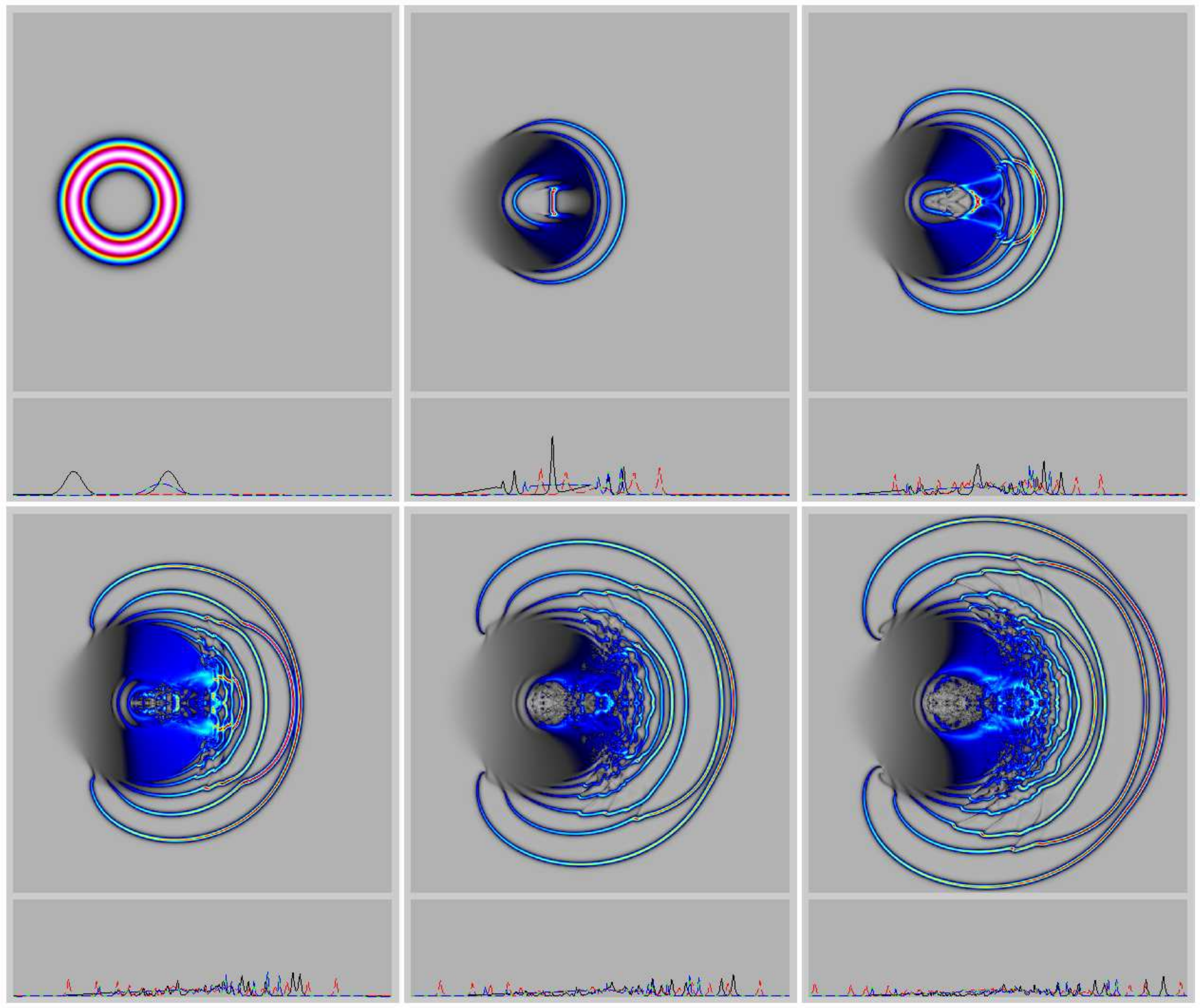}{2d}{right}{\sigma/8}{}

\figfourreverse{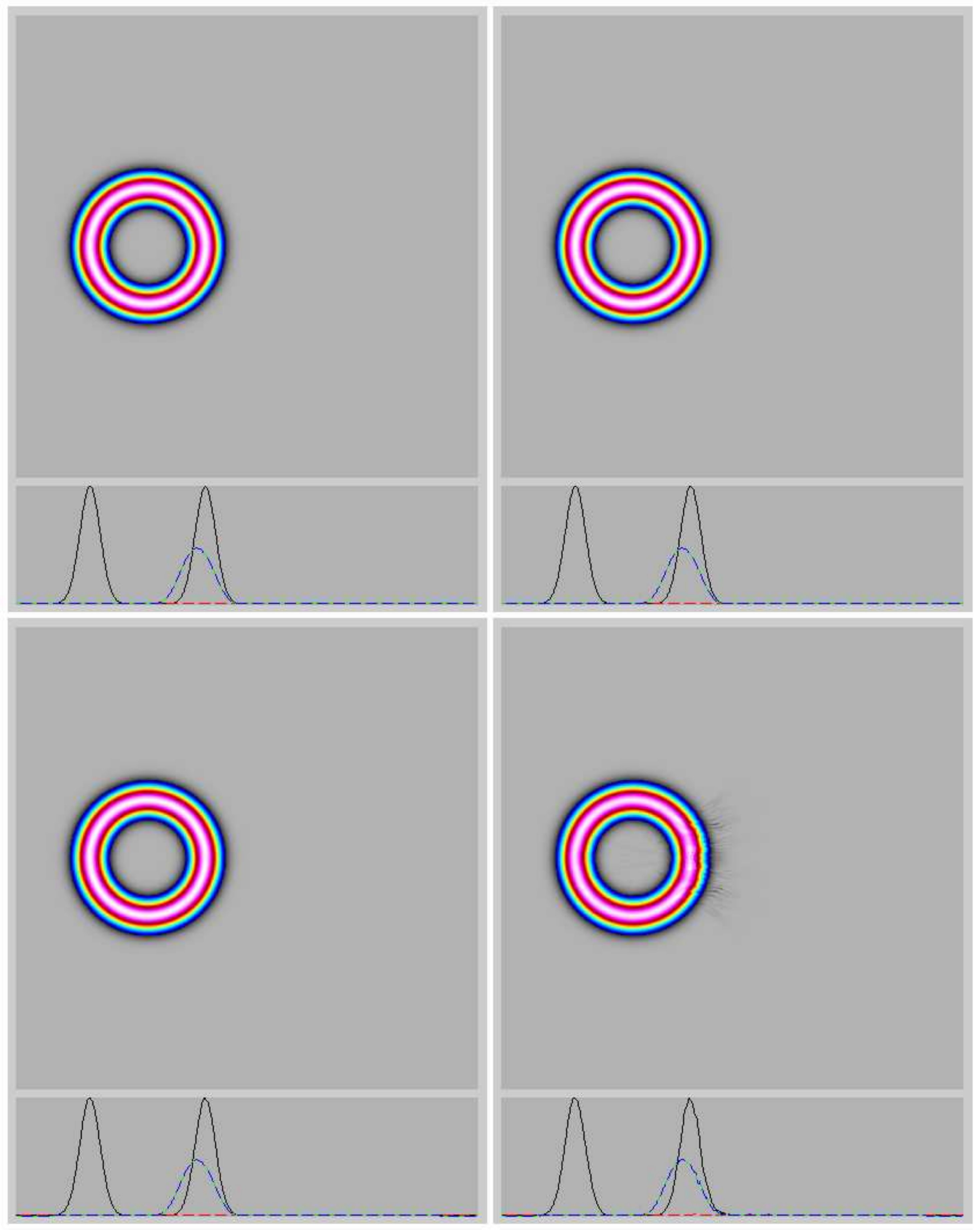}{2d}{right}{}

\subsection{Inout}\label{2d-inout}

In Figures \ref{2d-inout-sigma-1}-\ref{2d-inout-sigma-8} we start with an initial
Gaussian ring of width alpha in speed, with an angular distribution of
$-(\sin\theta,\cos\theta)\exp(-(r-r_0)^2/\sigma^2)$ for the direction of
the velocity. Consequently, the motion is inward along the positive
diagonal and outward along the negative diagonal. The outward motion
breaks into a sequence of curved peakon segments of width alpha, as usual.
The inward motion also produces peakon segments, which however undergo
head-on collisions, so that they annihilate, recreate and re-emerge moving
along the positive diagonal.

The blue profiles in the lower panels of Figures
\ref{2d-inout-sigma-1}-\ref{2d-inout-sigma-8} show the breakup of the outward motion into peakon profiles. The
green profiles on the lower panel show the head-on collisions of the
peakon profiles of the inward motion. The inward moving head-on collisions
leave a residue of complex flow.

Time reversal in this case shows severe distortion of the initial condition for
the smaller values $\alpha=\sigma/8$ and $\alpha=\sigma/4$, but not for the
larger values $\alpha=\sigma/2$ and $\alpha=\sigma$.
See Figure \ref{2d-inout-back}.

\figfour{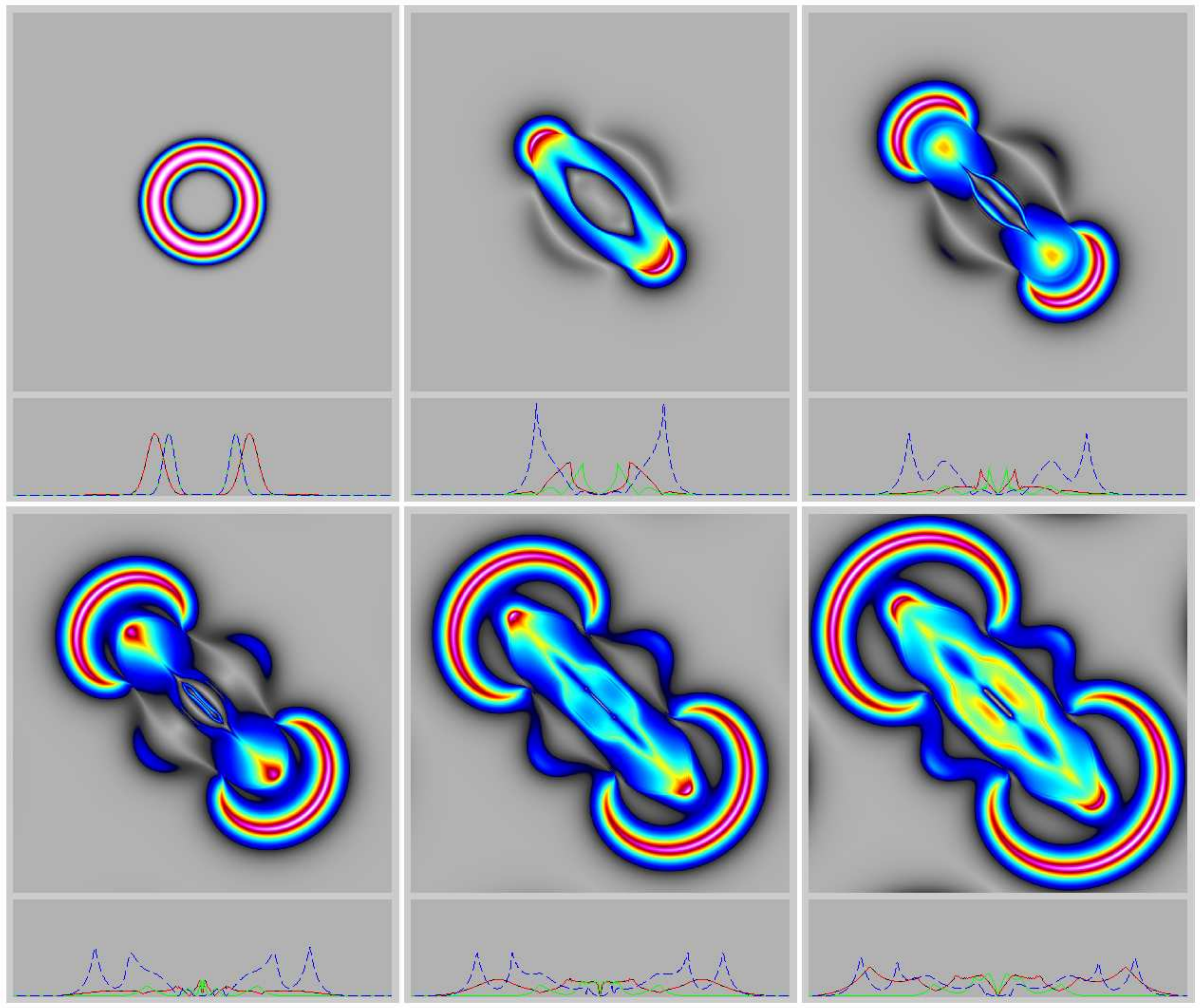}{2d}{inout}{\sigma  }{}
\figfour{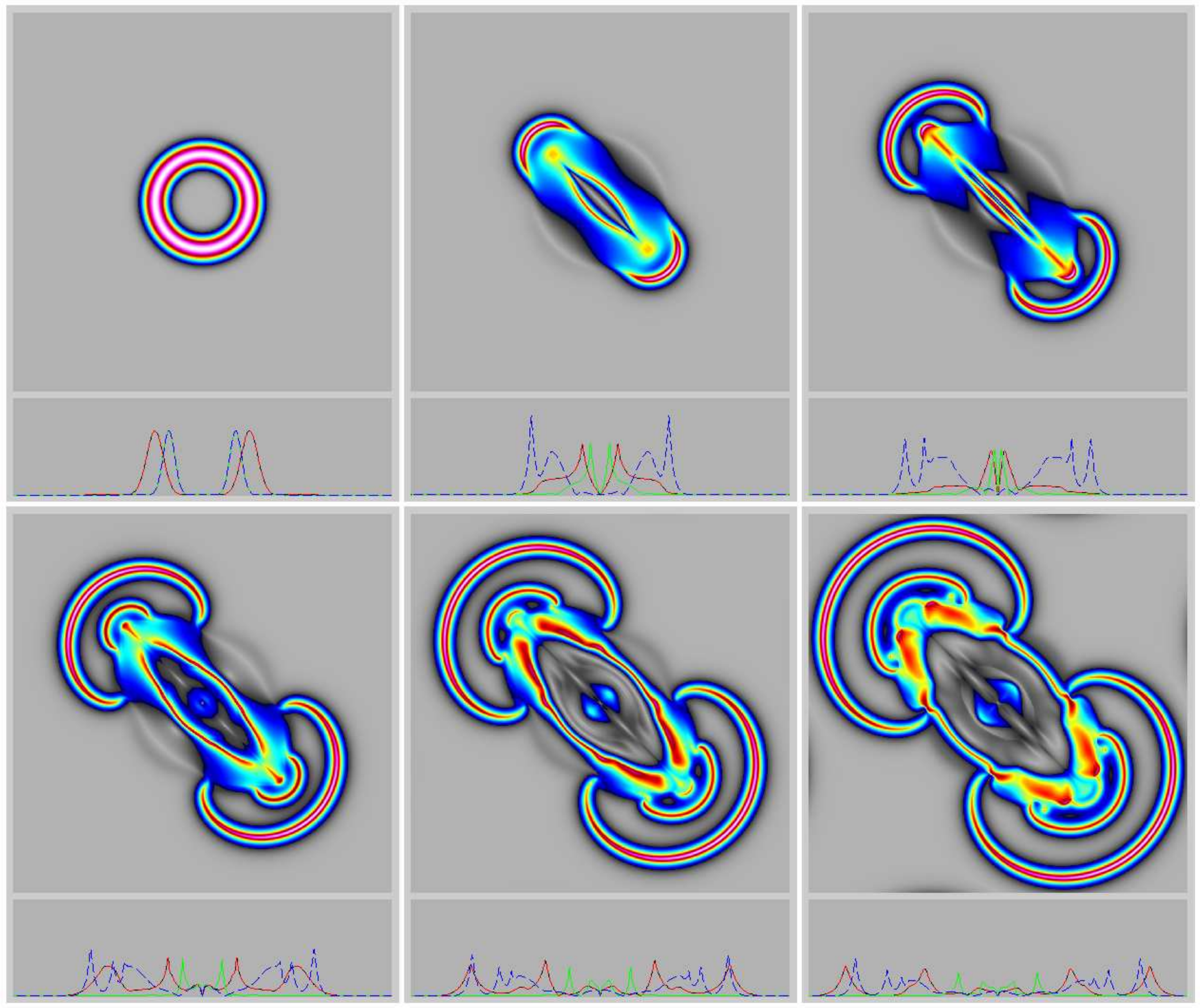}{2d}{inout}{\sigma/2}{}
\figfour{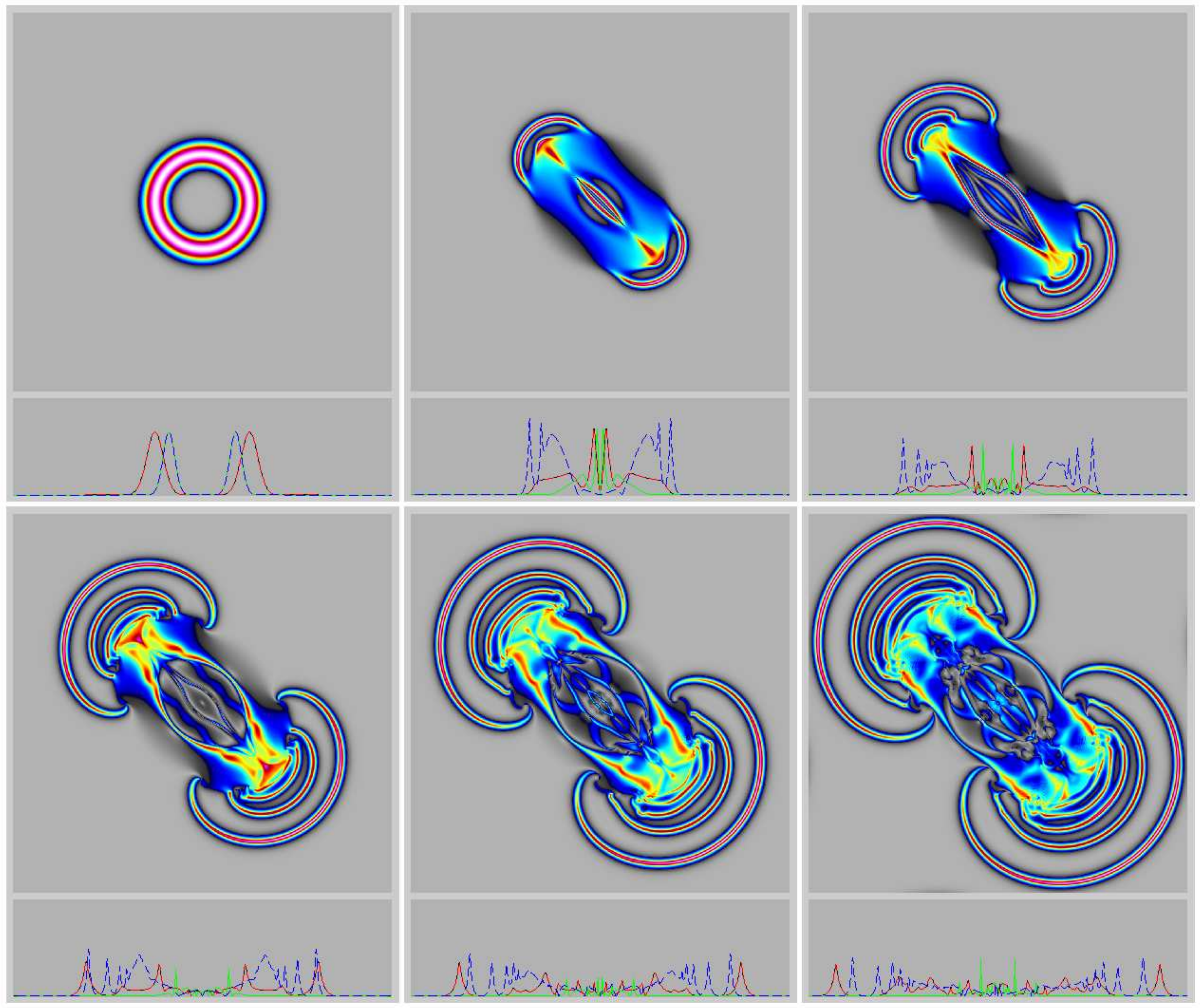}{2d}{inout}{\sigma/4}{}
\figfour{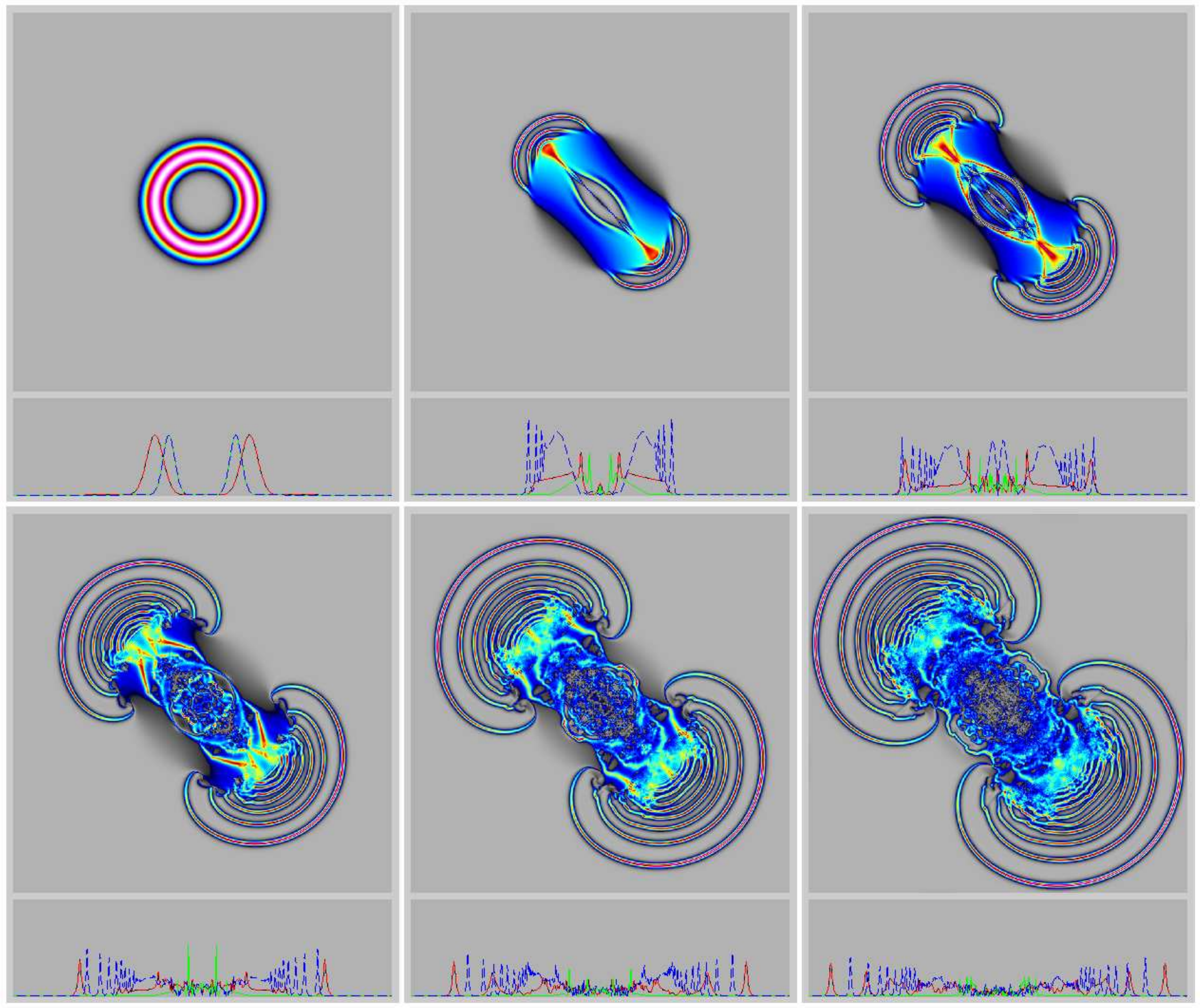}{2d}{inout}{\sigma/8}{}

\figfourreverse{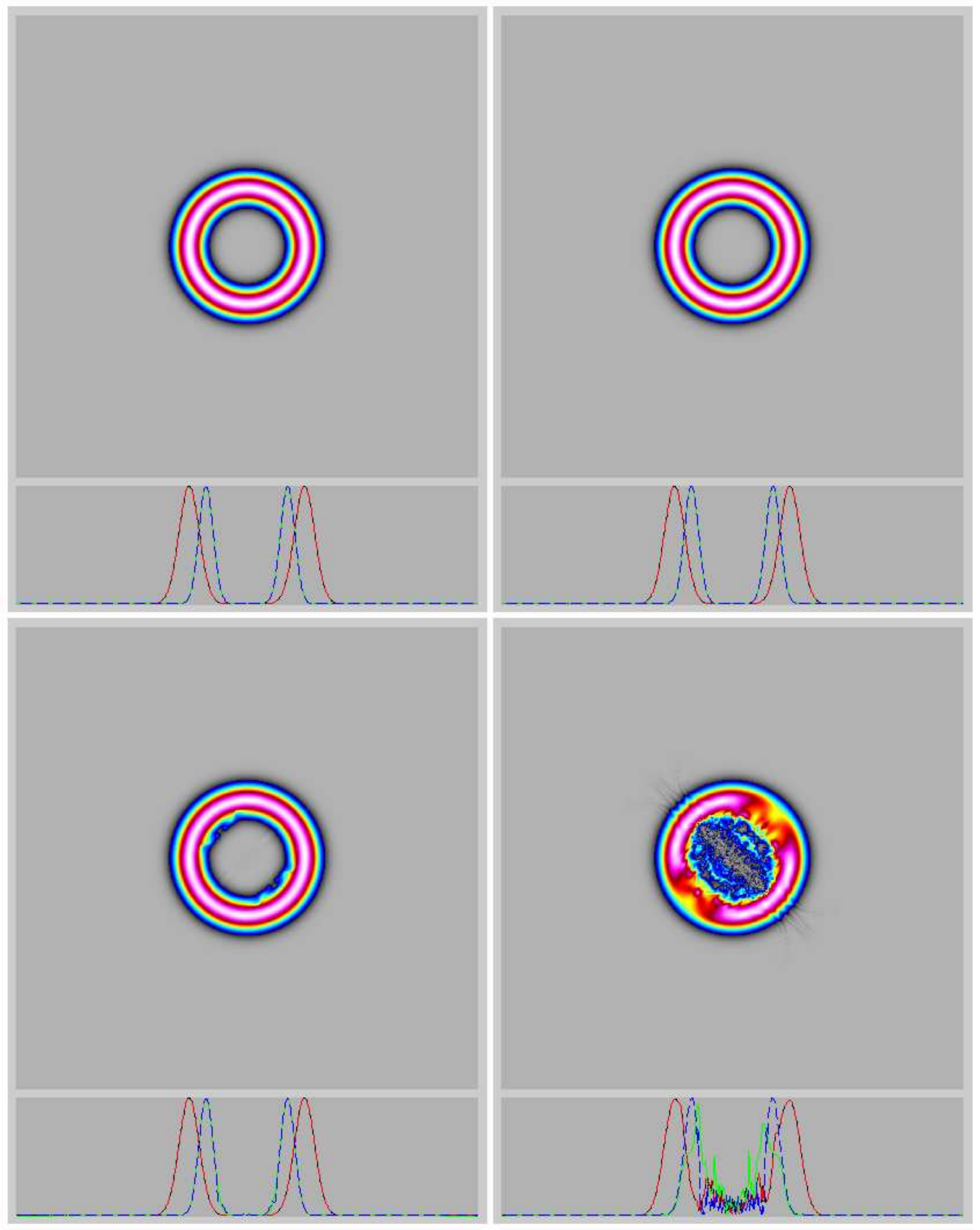}{2d}{inout}{}

\subsection{Time reversal}\label{2d-time-rev}

Table \ref{norm-2d} summarizes how well the initial velocity profiles
reconstitute upon reversing the EPDiff evolution from the final states.
Numbers in each table entry are the $L^1$ norm, $L^2$ norm, and max norm of
the difference between the initial velocity profile and its reconstituted
(time reversed) value.
The time reversed values are typically accurate to within one percent or
less except in cases where head-on collisions occur. See text for further
discussion.

\begin{table}
\caption{
$L^1$, $L^2$, and max norm of the difference between the initial velocity
profile and its reconstituted (time reversed) value for each 2d simulation.
\label{norm-2d}}
\begin{tabular}{|l|l|l|l|l|}
\hline
Simulation & $\alpha=\sigma$ & $\alpha=\sigma/2$ & $\alpha=\sigma/4$ & $\alpha=\sigma/8$\\
\hline
           & {\tt 3.62e-05} & {\tt 7.03e-05} & {\tt 0.000213} & {\tt 0.000577} \\
plate      & {\tt 0.000142} & {\tt 0.000199} & {\tt 0.000624} & {\tt 0.00207 } \\
           & {\tt 0.00296 } & {\tt 0.00279 } & {\tt 0.00371 } & {\tt 0.0147  } \\\hline
           & {\tt 7.86e-05} & {\tt 8.69e-05} & {\tt 0.00024 } & {\tt 0.000844} \\
parallel   & {\tt 0.000204} & {\tt 0.000217} & {\tt 0.000616} & {\tt 0.00236 } \\
           & {\tt 0.00301 } & {\tt 0.00283 } & {\tt 0.00358 } & {\tt 0.0161  } \\\hline
           & {\tt 0.00134 } & {\tt 0.000498} & {\tt 0.000719} & {\tt 0.00711 } \\
skew       & {\tt 0.0029  } & {\tt 0.00111 } & {\tt 0.00139 } & {\tt 0.0177  } \\
           & {\tt 0.0199  } & {\tt 0.0102  } & {\tt 0.00727 } & {\tt 0.148   } \\\hline
           & {\tt 0.000532} & {\tt 0.00061 } & {\tt 0.00451 } & {\tt 0.0567  } \\
wedge      & {\tt 0.000705} & {\tt 0.00108 } & {\tt 0.0123  } & {\tt 0.111   } \\
           & {\tt 0.00395 } & {\tt 0.00539 } & {\tt 0.148   } & {\tt 0.671   } \\\hline
           & {\tt 0.000179} & {\tt 0.00636 } & {\tt 0.0274  } & {\tt 0.0538  } \\
head-on    & {\tt 0.000332} & {\tt 0.0165  } & {\tt 0.0622  } & {\tt 0.111   } \\
           & {\tt 0.00335 } & {\tt 0.172   } & {\tt 0.43    } & {\tt 0.665   } \\\hline
           & {\tt 0.000207} & {\tt 0.00023 } & {\tt 0.000604} & {\tt 0.00152 } \\
star       & {\tt 0.000419} & {\tt 0.000442} & {\tt 0.000851} & {\tt 0.00237 } \\
           & {\tt 0.00531 } & {\tt 0.00476 } & {\tt 0.00515 } & {\tt 0.00883 } \\\hline
           & {\tt 7.21e-05} & {\tt 2.54e-05} & {\tt 4.81e-05} & {\tt 0.000293} \\
rotate     & {\tt 8.17e-05} & {\tt 3.95e-05} & {\tt 8.23e-05} & {\tt 0.00059 } \\
           & {\tt 0.000301} & {\tt 0.000139} & {\tt 0.000339} & {\tt 0.0027  } \\\hline
           & {\tt 0.00021 } & {\tt 0.000292} & {\tt 0.000583} & {\tt 0.0029  } \\
right      & {\tt 0.000372} & {\tt 0.000615} & {\tt 0.00137 } & {\tt 0.00696 } \\
           & {\tt 0.00545 } & {\tt 0.00422 } & {\tt 0.0106  } & {\tt 0.0892  } \\\hline
           & {\tt 0.000219} & {\tt 0.000245} & {\tt 0.0024  } & {\tt 0.0348  } \\
inout      & {\tt 0.00048 } & {\tt 0.000431} & {\tt 0.0103  } & {\tt 0.087   } \\
           & {\tt 0.00621 } & {\tt 0.00183 } & {\tt 0.137   } & {\tt 0.594   } \\\hline
\end{tabular}
\end{table}

\section{Numerical results for EPDiff in 3d}
\label{num-3d-sec}

\paragraph{Singular solutions of EPDiff in 3d.}
In this section, we extend our numerical solutions of EPDiff to 3d and discover
that the codimension-one singular solution behavior of EPDiff persists. Namely,
the singular solution behavior of EPDiff extends from curves in 2d to surfaces
in 3d. To our knowledge, this is the first example of numerical simulations of
the nonlinear interactions of contact discontinuities for fluid velocity in 3d.
Of course, the singular surface solutions of EPDiff in 3d have no interpretation
in the theory of internal waves. However, codimension-one singular solutions of
EPDiff in 3d may have applications elsewhere, for example in the physics of
liquid crystals in the inertial regime, as discussed for 1d evolution by Hunter
and Saxton \cite{HuSa1991}. The 3d evolution of contact surfaces may also be
regarded as ``growth'' from the viewpoint of imaging science, or evolution of
shapes in 3d. The present work focuses on the role of convergence and momentum
exchange in the evolution of contact surfaces.

Upcoming sections describe 3d simulations of evolution under EPDiff for each
of twelve initial velocity distributions. For each simulation we have figures
for $\alpha=\sigma$ and $\alpha=\sigma/2$. Due to memory and computational
limitations, our 3d simulations were run at one-fourth the resolution of our
2d simulations ($256^3$ instead of $1024^2$), and this lower resolution
precluded us from performing accurate computations at the smaller values
$\alpha=\sigma/4$ and $\alpha=\sigma/8$ that were simulated in 2d.

Each figure contains six frames, showing the initial magnitude (speed) of
velocity, $\modu$, in the upper left frame, followed by plots of $\modu$ at
future times, reading across and then down. For each
simulation, the domain is $[-1,1]\times[-1,1]\times[-1,1]$, and $x$ is toward
the right, $y$ toward the back, and $z$ toward the top. Individual plots
contain colored, partially transparent level surfaces of $\modu$ at
35\% (purple),
25\% (red),
15\% (green), and
 5\% (blue)
of its maximum value over all frames in the figure. The number of level
surfaces, and the colors, amount of transparency, and fractions of $\modu$
represented by each level surface, where chosen after considerable
experimentation so that they reveal maximal information with minimal clutter.

\begin{figure}
\begin{center}
   \leavevmode{\hbox{\epsfig{
       figure=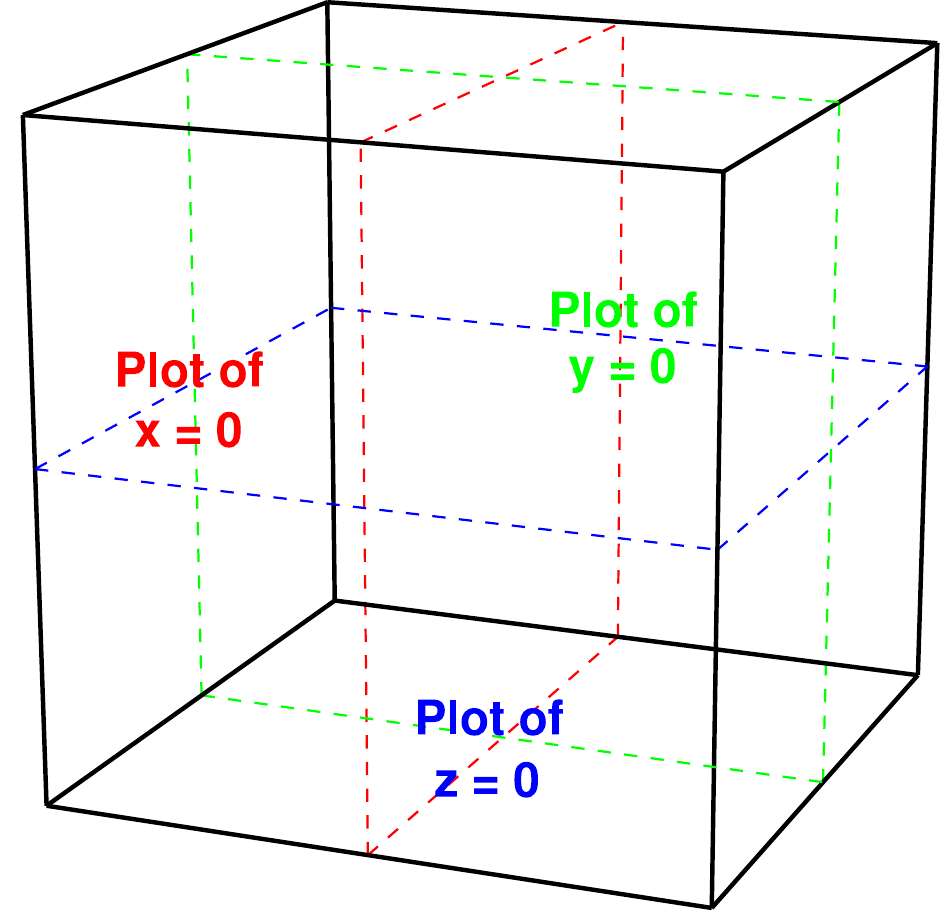, scale=0.75
   }}}
   \caption{\label{figure-scheme3d}
       Locations of 2d slices of $\modu$ shown in the left, back, and bottom
       panels in the upcoming 3d figures.
   }
\end{center}
\end{figure}

The left, back, and bottom panels of each frame show color contour plots of 2d
slices through the three dimensional data at $x=0$, $y=0$, and $z=0$,
respectively, as sketched in Figure \ref{figure-scheme3d}. The same color scheme
that was used for the 2d simulations, as described in Section \ref{num-2d-sec}
and illustrated in Figure \ref{color-bar}, is used here for the 2d slices of the
3d simulations. Note that the colors used for the 2d slices do not correspond to
the colors used for the 2d level surfaces. For example, red in the 2d slices
does not represent the same value of $\modu$ as the red level surface.

\subsection{Plate}\label{3d-plate}

The analog of a segment in 2d is a disc, or plate, in 3d. Figures
\ref{3d-plate-sigma-1} and \ref{3d-plate-sigma-2} show the evolutions at various values of alpha
in 3d from a confined initial velocity distribution in the shape of a
plate in speed and moving rightward (in the positive $x$ direction).
The initial plate distribution is chosen so it
falls off exponentially in the normal $x$ direction
and it falls off at the edges (or rim) of the plate as a Gaussian, as we did
in 2d at the endpoints of the segments.
Because the distribution moves with flow, it expands rightward and
outward.

The bottom panels in Figures \ref{3d-plate-sigma-1} and \ref{3d-plate-sigma-2} show the
propagation in a horizontal section (at the midplane) $x=0$ of the cube of
size 2. This is an invariant plane, by symmetry, for this initial
value problem. The invariant midplane section
essentially reproduces the 2d evolution of the peakon segments; note the
similarity in patterns at late time on the bottom panel in 3d and the
corresponding interactions in 2d.

The bottom panel in Figure \ref{3d-plate-sigma-2} with $\alpha=\sigma/2$
shows the plate expanding and decomposing into several peakon contact
surfaces. The left panel shows a vertical section at $x=0$. The initial
circular disc propagates rightward through this vertical section and
maintains its circular symmetry as it expands. The back panel shows a
vertical section at $y=0$. By symmetry, the results on the vertical
section shown on the back panel in this case are the same as those on the
bottom panel ($z=0$ horizontal section).

The disc shaped velocity distribution balloons out rightward and expands
cylindrically into one peakon contact surface in Figure
\ref{3d-plate-sigma-1} for $\alpha=\sigma$, and into several peakon contact surfaces of
width $\alpha=\sigma/2$ in Figure \ref{3d-plate-sigma-2}.
Again, this is consistent with the similar evolution restricted
to lower spatial dimensions.

\figfourthree{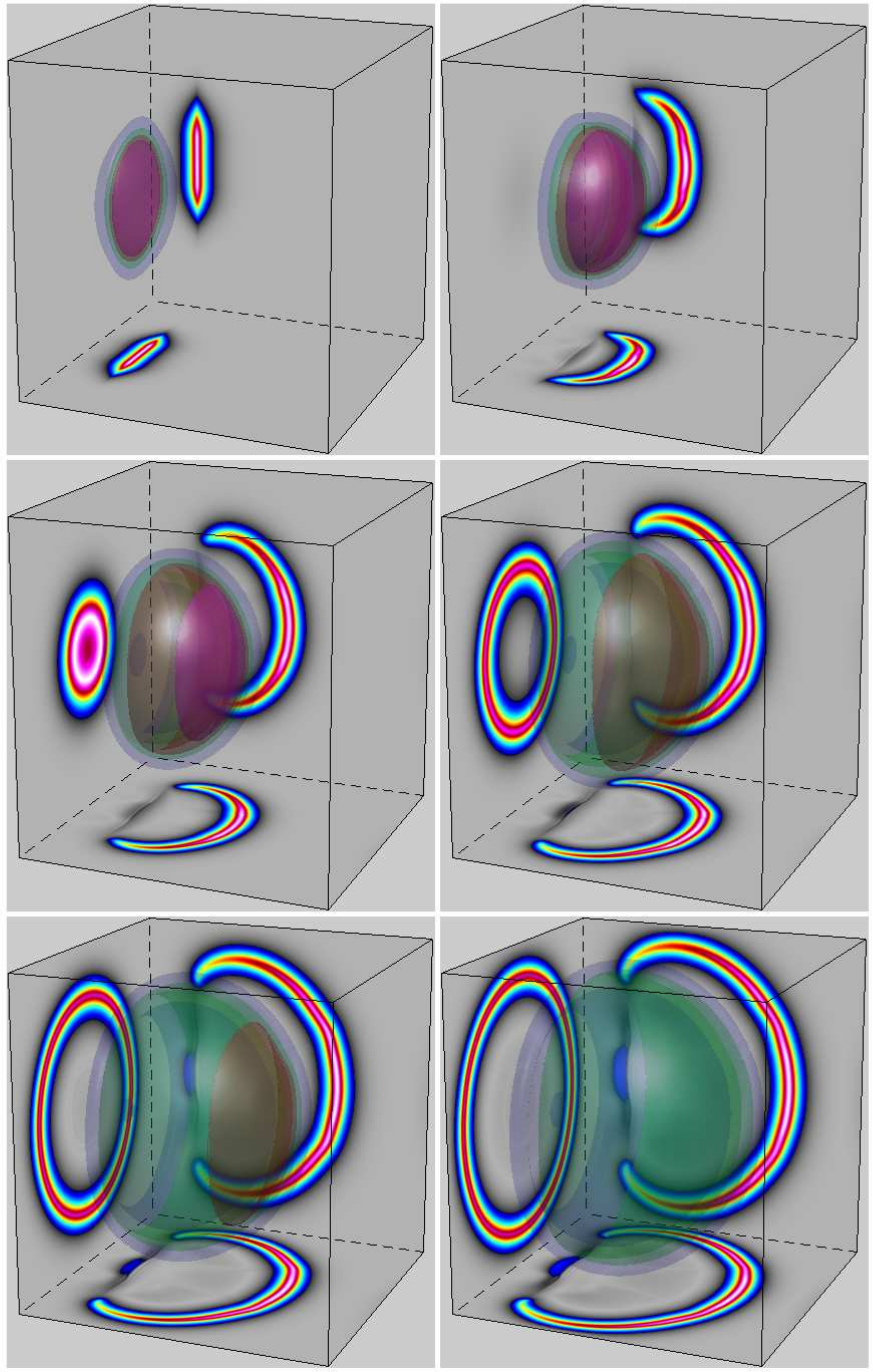}{3d}{plate}{\sigma  }{}
\figfourthree{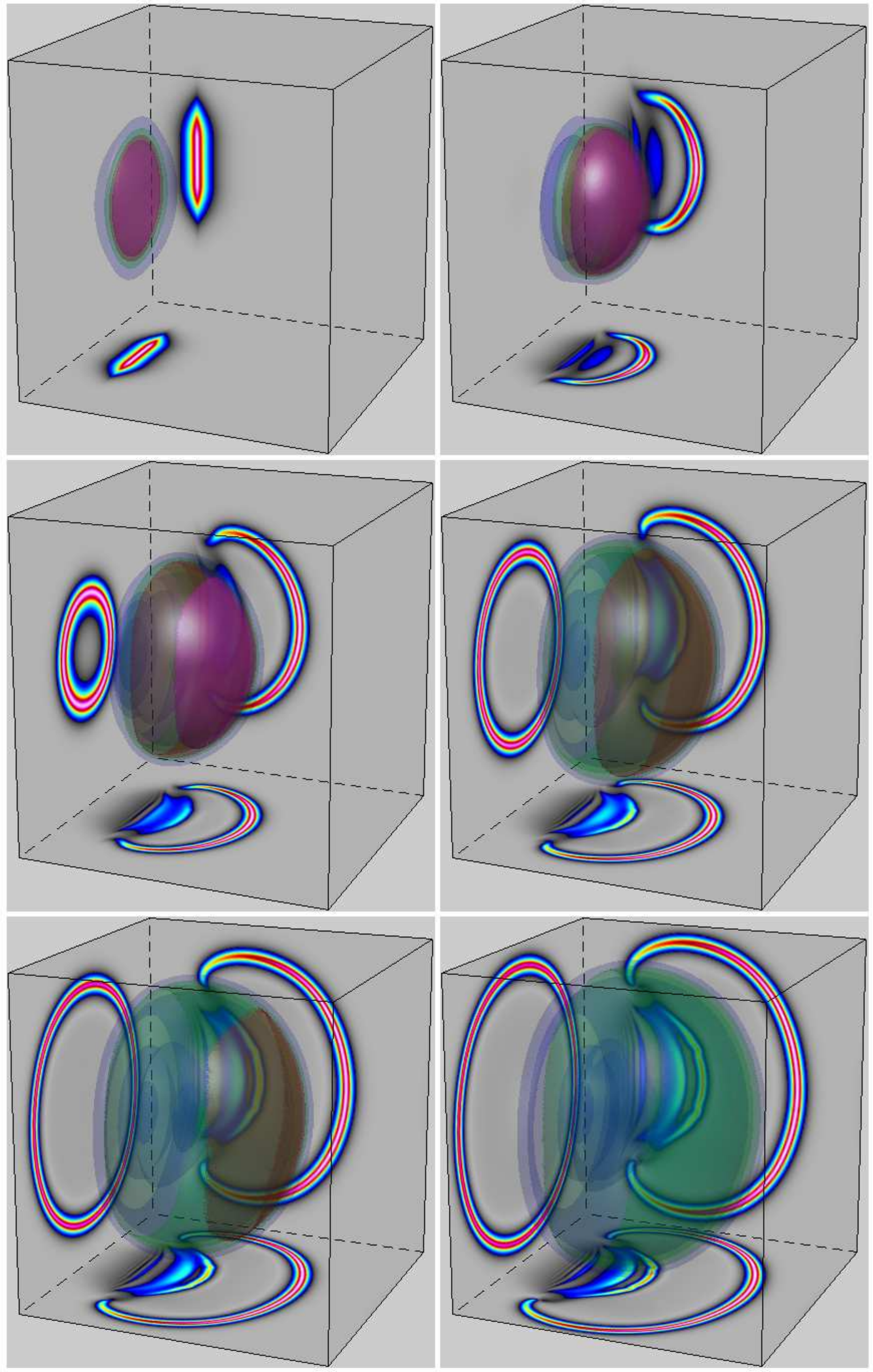}{3d}{plate}{\sigma/2}{}

\subsection{Parallel}\label{3d-parallel}

Figures \ref{3d-parallel-sigma-1} and \ref{3d-parallel-sigma-2} each show two plate-like velocity
distributions of the same diameter, initialized moving rightward so
that the one behind, with twice the speed, will overtake the one ahead.
The discs are initially offset at 45 degrees in the $y$-$z$ plane; so this
evolution has no reflection symmetry about any of the midplanes.

Both plates balloon outward and decompose into peakon contact surfaces,
and the overtaking collision shows two reconnections of the surfaces. One
reconnection occurs at first contact and the other evolves late in the
simulation, as seen on both the bottom and back panels. The left panel
shows a transverse vertical section of the first reconnection. As the two
discs propagate past the plane $x=0$, their peakon contact surfaces
develop into a perimeter that surrounds the interior interactions.

\figfourthree{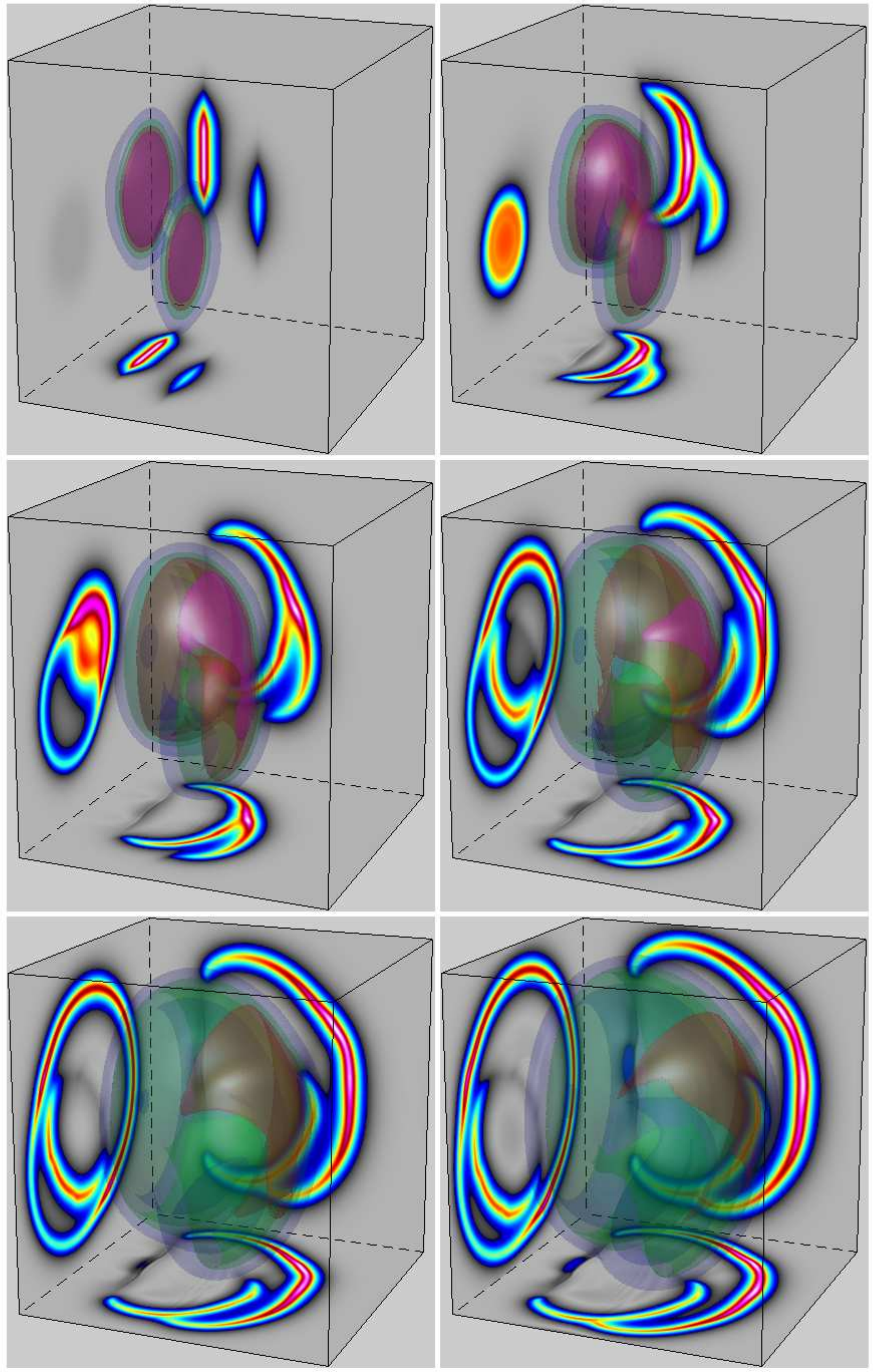}{3d}{parallel}{\sigma  }{}
\figfourthree{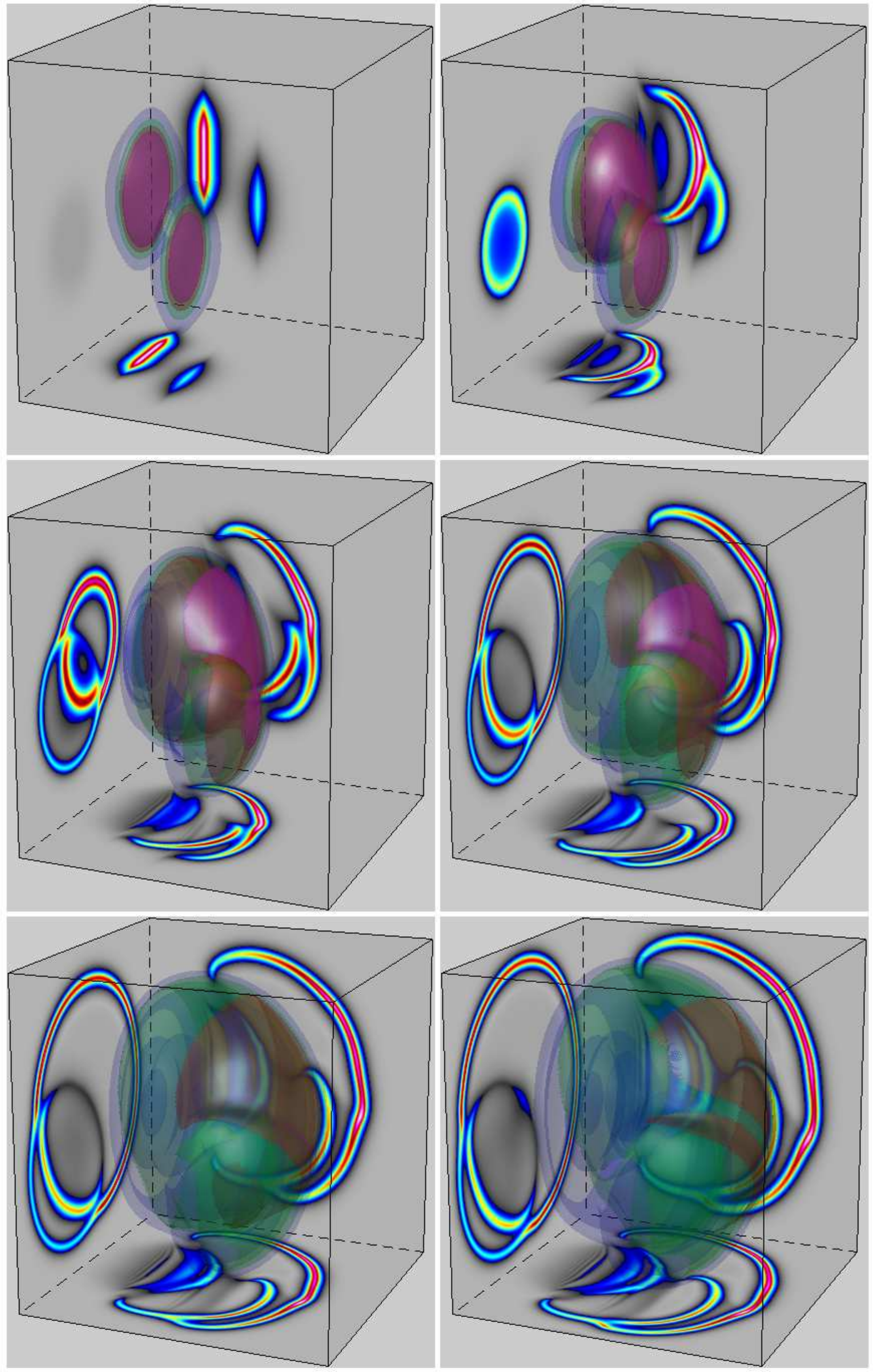}{3d}{parallel}{\sigma/2}{}

\subsection{Skew}\label{3d-skew}

The skew overtaking collision in Figures \ref{3d-skew-sigma-1} and \ref{3d-skew-sigma-2} show both
the first and second reconnections as the two initial plate distributions
collide and interact. Again, the sections shown on the bottom and back
panels are very reminiscent of the corresponding skew collisions of peakon
segments in 2d. The memory wisps that appear so clearly in the 2d skew
overtaking collisions are less distinct here, likely because of the
coarser resolution ($256^3$, versus $1024^3$ in 2d).

\figfourthree{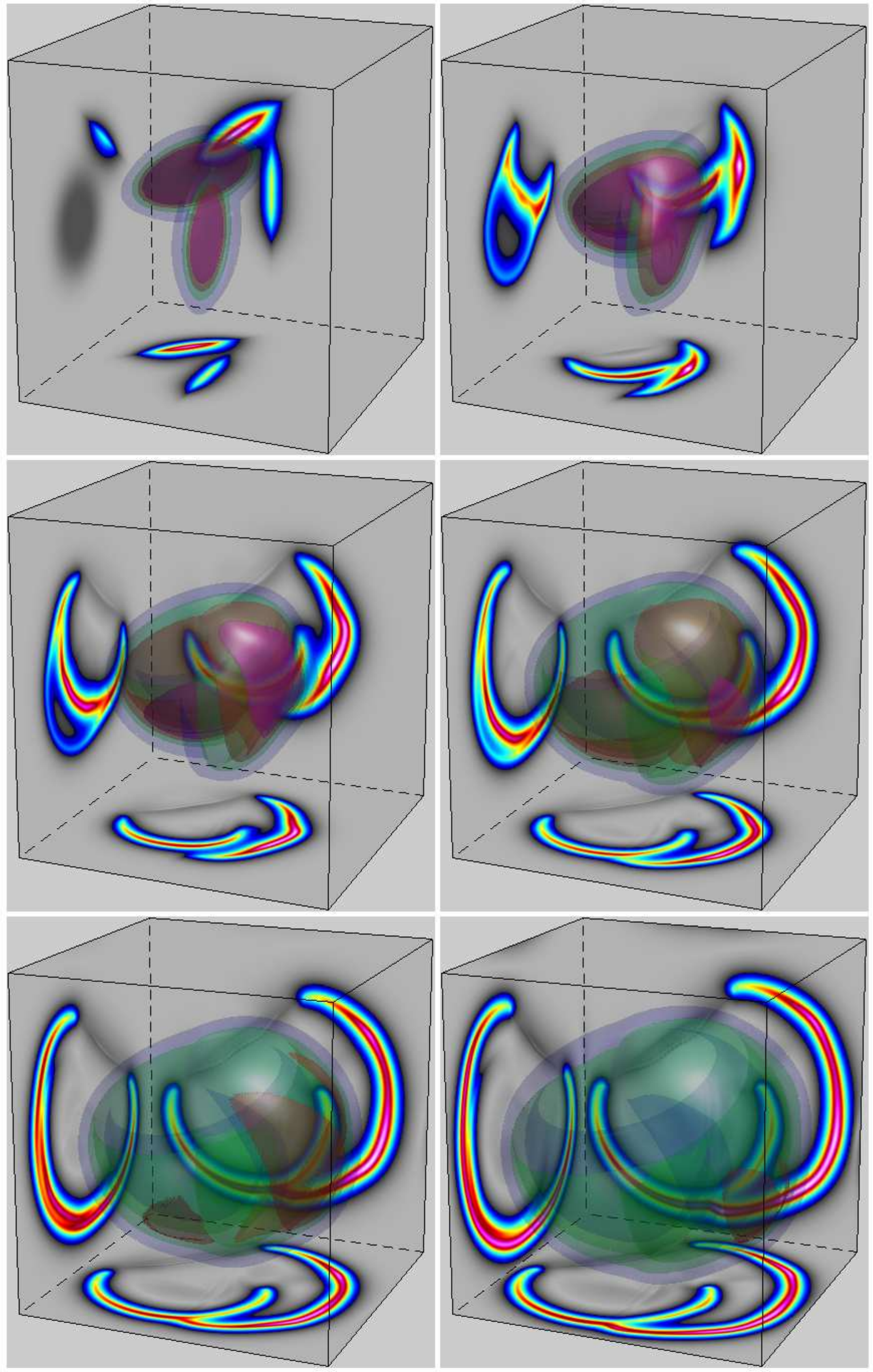}{3d}{skew}{\sigma  }{}
\figfourthree{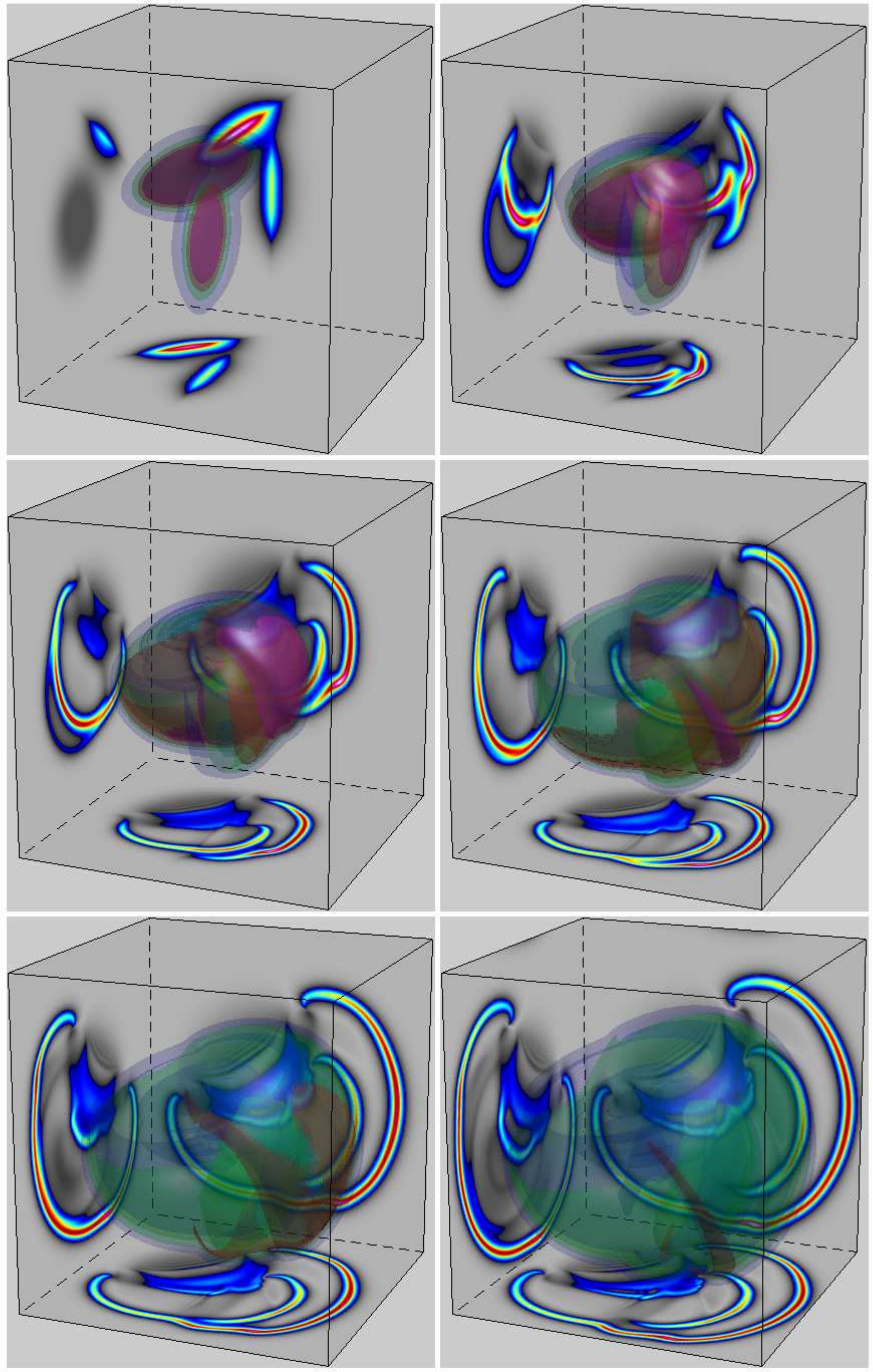}{3d}{skew}{\sigma/2}{}

\subsection{Wedge}\label{3d-wedge}

In Figures \ref{3d-wedge-sigma-1} and \ref{3d-wedge-sigma-2}, two plate distributions of velocity
approach the $z=0$ horizontal midplane from above and below, each at 45
degrees. Their ``wedge'' collision occurs at the horizontal midplane and
has reflection symmetry about the $xz$ vertical midplane, at $y=0$.

The evolution shown on the back panels in these figures is similar to the
2d wedge collision of peakon segments. The bottom panel shows the creation
of peakon contact surfaces which are
reminiscent of the peakon contact segments in
the 2d ``right'' figure. Namely, peakon contact surfaces emerge from the
collision and reconnect to encircle the collision region. Each successive
peakon contact surface (whose intersection with the horizontal midplane is
shown as a curve in the bottom panel) reconnects at the rear and produces a
wavefront that encircles the collision region. The wavefront motion is
less intense in the rear because it is moving slower at the rear than
in the front. We see one major collision at the midplane followed by an
emission of rightward moving peakon contact surfaces and their later
reconnections in the rear. At later times, the hot spots show pronounced
trailing memory wisps.

\figfourthree{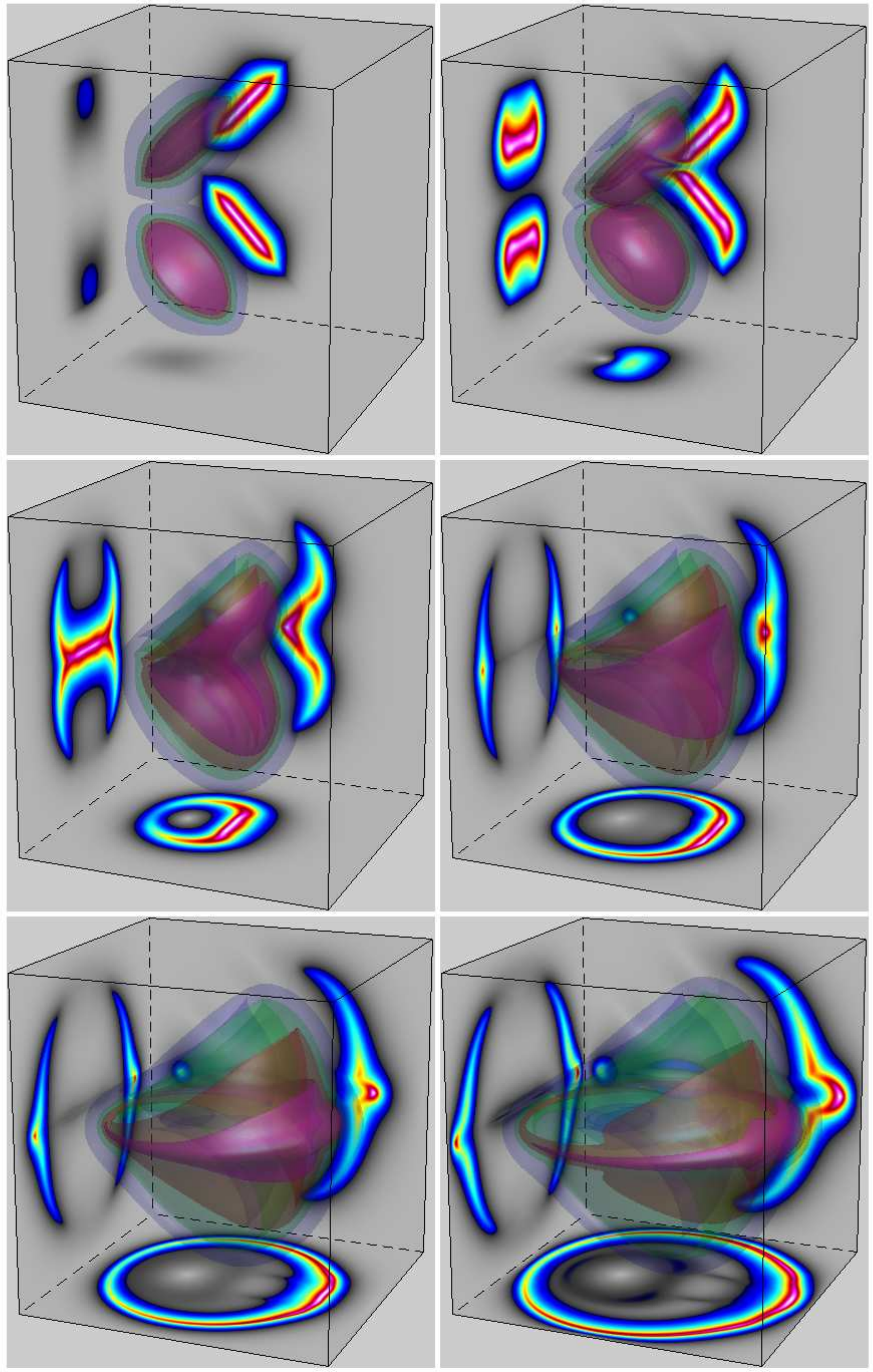}{3d}{wedge}{\sigma  }{}
\figfourthree{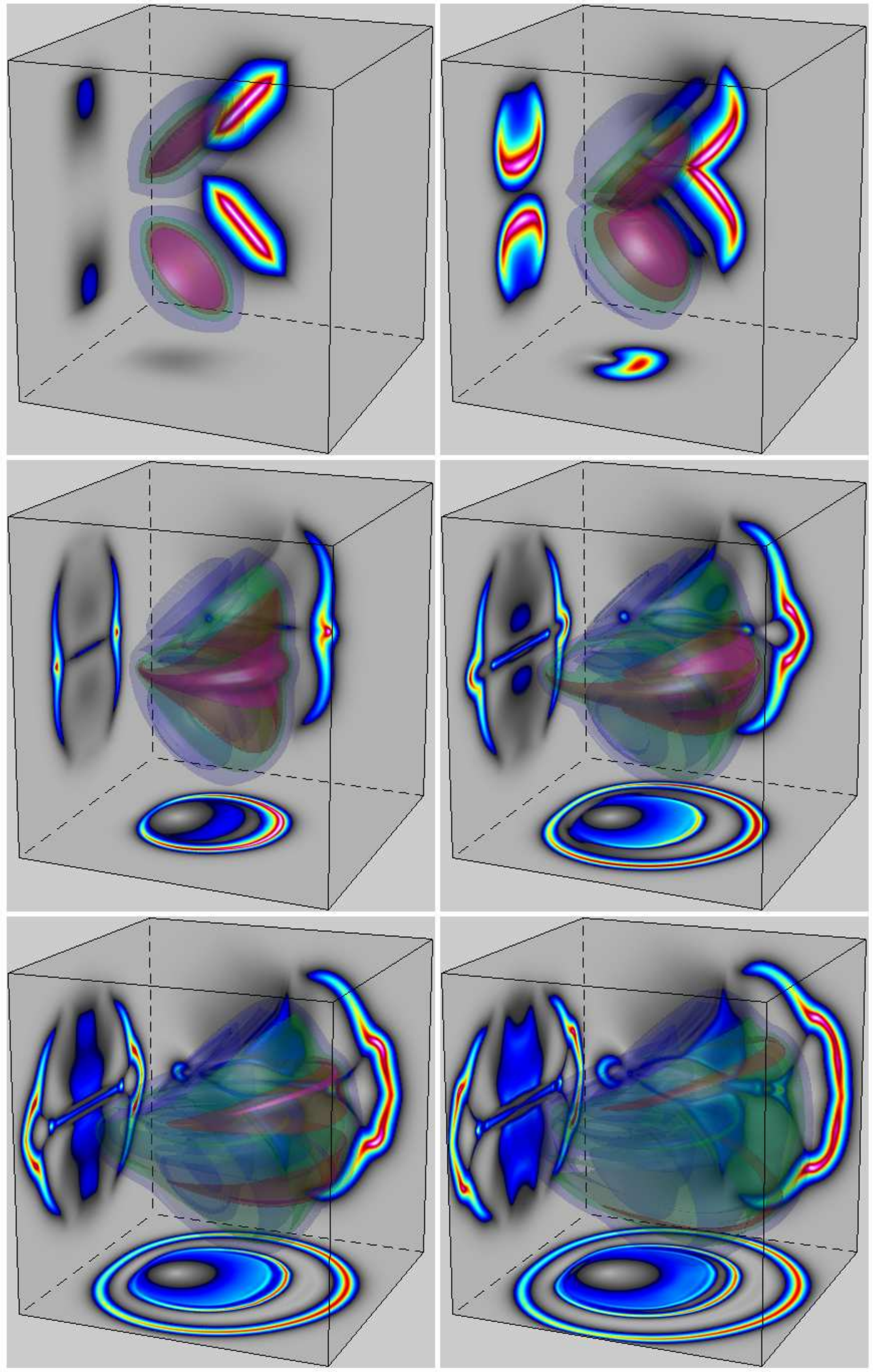}{3d}{wedge}{\sigma/2}{}

\subsection{Head-on}\label{3d-headon}

Figures \ref{3d-headon-sigma-1} and \ref{3d-headon-sigma-2} show the head-on collision of two
identical plate-like distributions offset by the same distance at 45
degrees when projected into the $yz$ plane. Hence, the evolution in the
sections shown in the back and bottom panels are identical, and all three
panels have reflection and rotation symmetries.

The back and bottom panels of Figures \ref{3d-headon-sigma-1} and \ref{3d-headon-sigma-2} have
symmetry under reflection about one diagonal and rotation by $\pi$ about
the other. Likewise, the left panels are reflection symmetric about both
diagonals. These symmetries help diagnose the complex evolution occurring
as the two peakon contact segments balloon outward toward each other and
collide head-on. As expected from earlier animations, the segments
annihilate and re-emerge, leaving behind a complex residual flow where the
head-on collision occurred.

The left panels of Figures \ref{3d-headon-sigma-1} and
\ref{3d-headon-sigma-2} show that the reconnection after the collision
occurs results in a peakon contact segment surrounding the central region.

\figfourthree{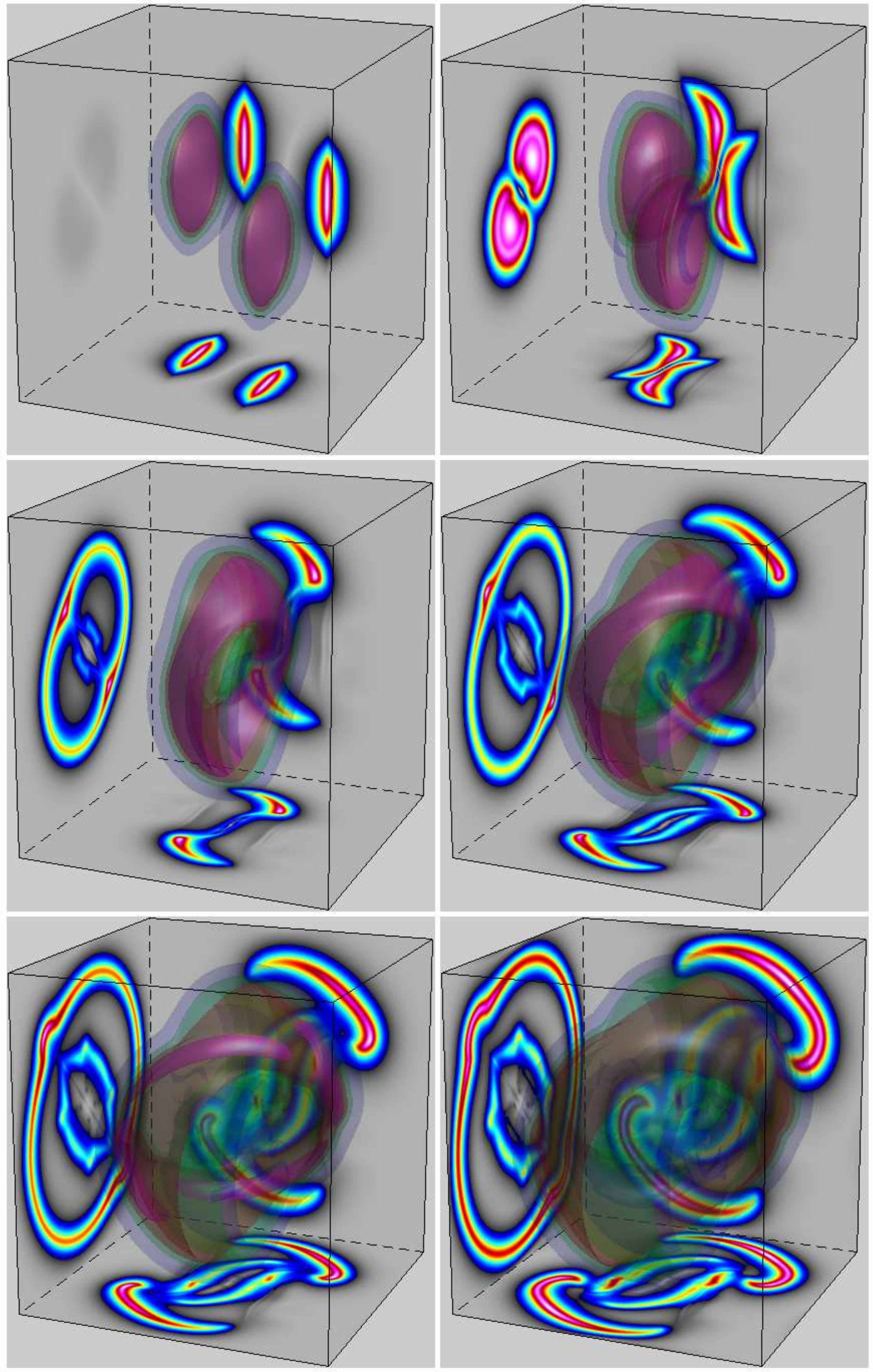}{3d}{head-on}{\sigma  }{}
\figfourthree{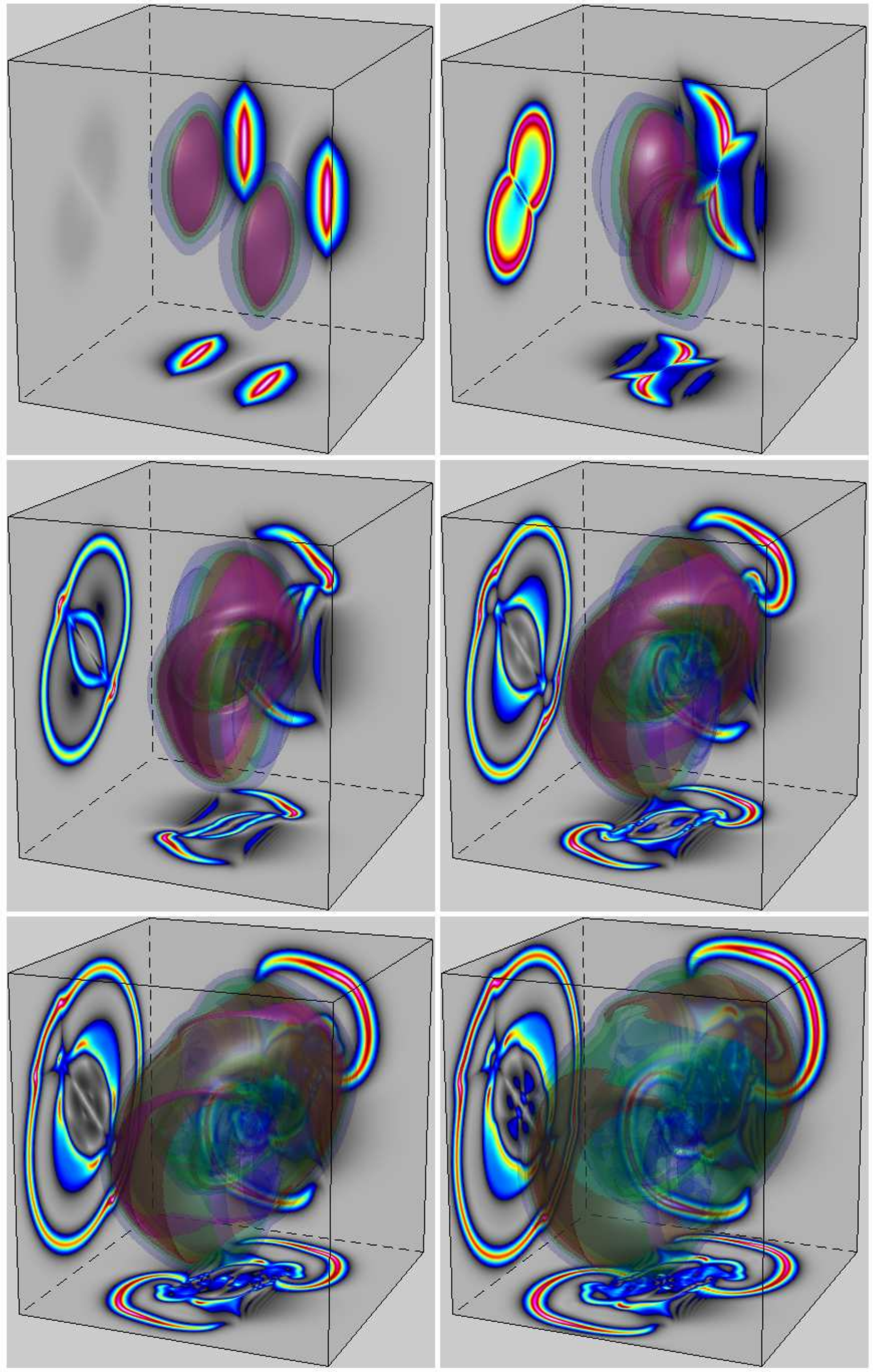}{3d}{head-on}{\sigma/2}{}

\subsection{Rotate}\label{3d-rotate}

For the simulations shown in
Figures \ref{3d-rotate-sigma-1} and \ref{3d-rotate-sigma-2}, an
initially spherical Gaussian shell is rigidly rotating about its vertical
axis, like a planet. This angular motion couples to radial motion with
cylindrical symmetry about the axis of rotation.
The evolution remains cylindrically symmetric about
this axis and shows convergence to it, as seen in the bottom panels of
Figures \ref{3d-rotate-sigma-1} and \ref{3d-rotate-sigma-2}. The left and
back panels show identical emergence of peakon contact segments. This
evolution shows the coupling between the angular and radial motion in the
cylindrically symmetric case. The outward velocities expand essentially
like an oblate sphere, and may become less oblate with time. The midplane
$z=0$ shown on the bottom panel is an invariant section, because of
up-down symmetry, and it shows the cylindrical convergence and expansion
driven by the initial angular rotation.

\figfourthree{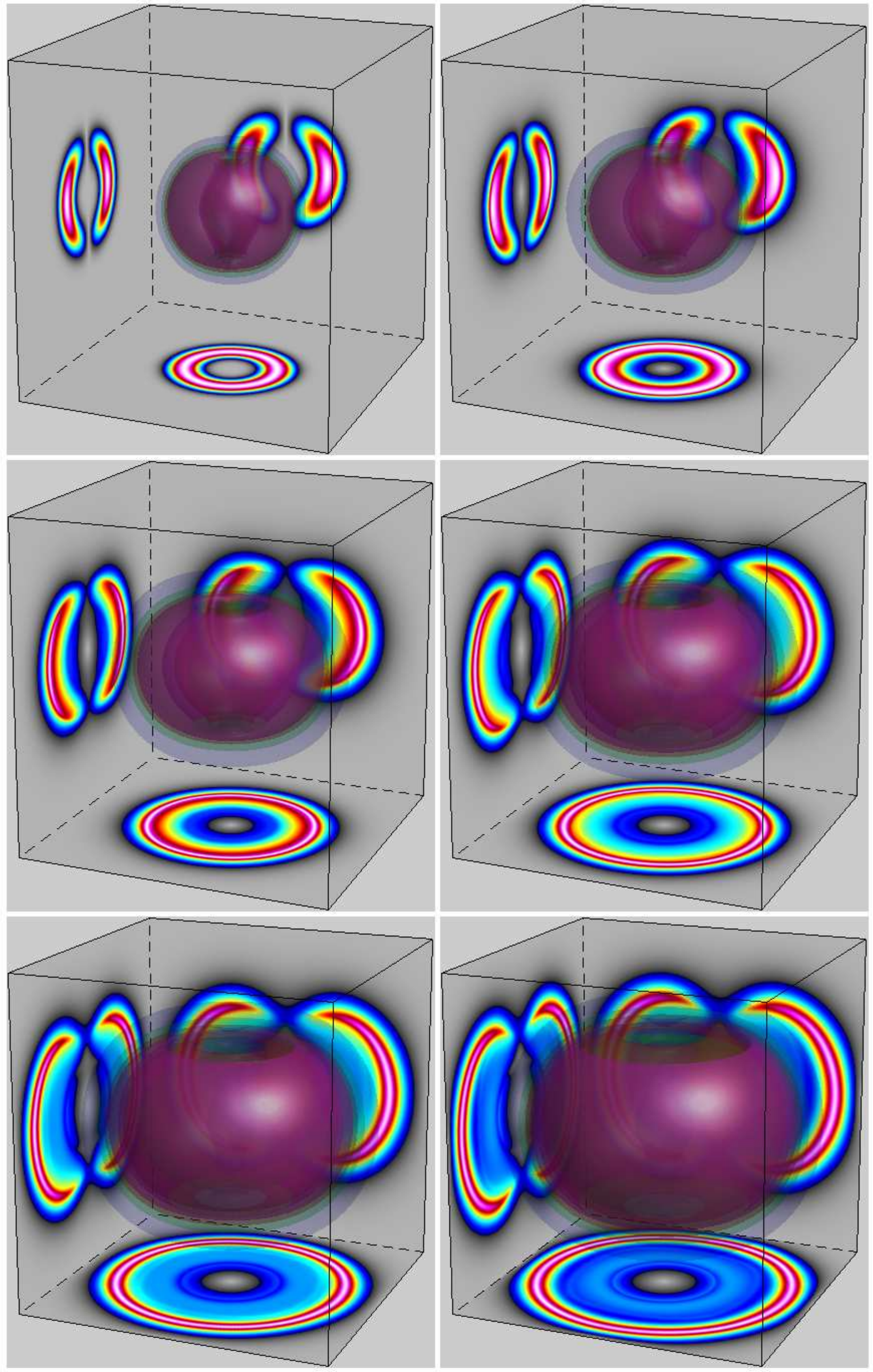}{3d}{rotate}{\sigma  }{}
\figfourthree{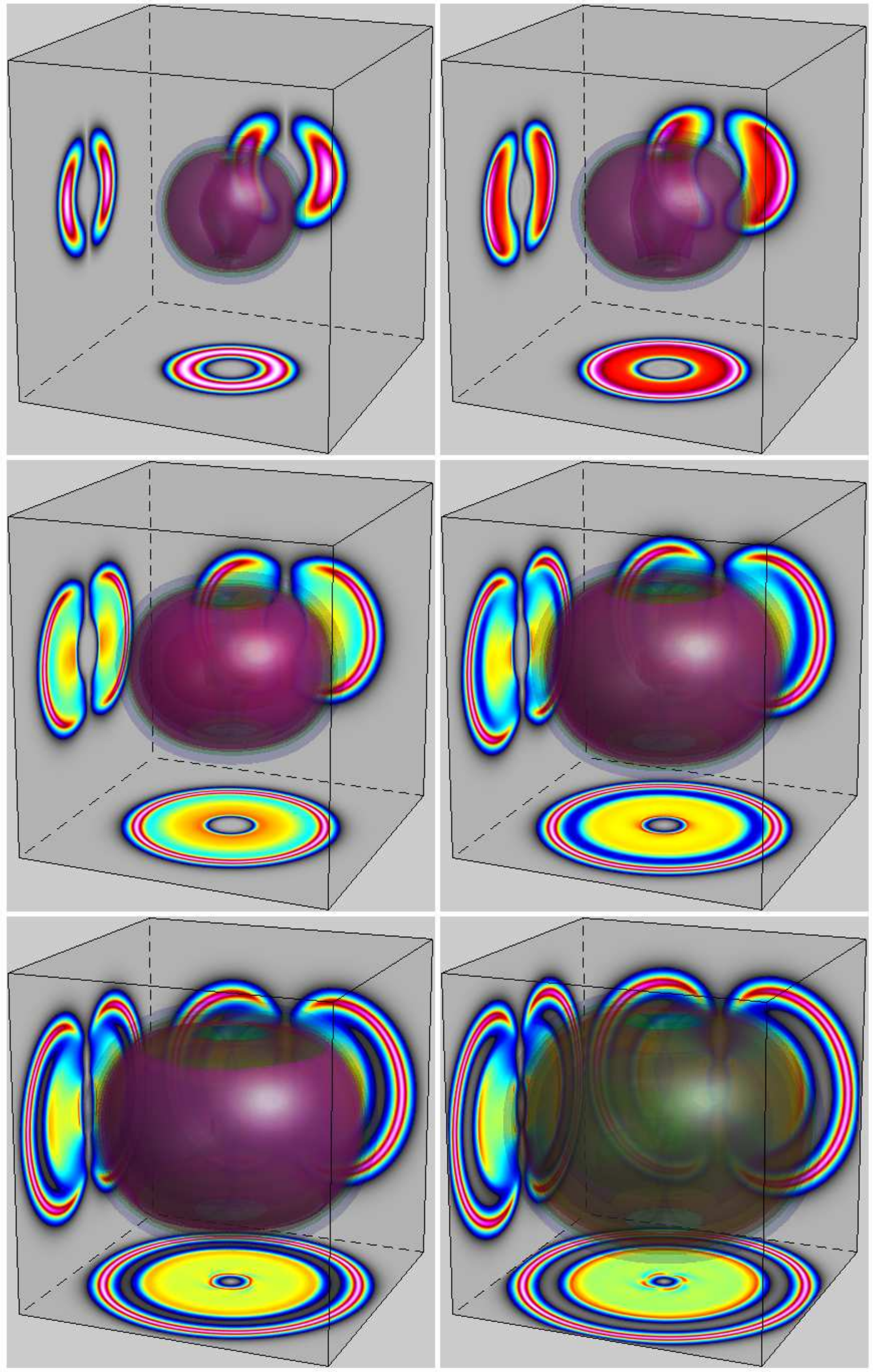}{3d}{rotate}{\sigma/2}{}

\subsection{Right}\label{3d-right}

In Figures \ref{3d-right-sigma-1} and \ref{3d-right-sigma-2}, a Gaussian shell in speed is initially
moving rightward. The peakon contact surfaces emerging on its outer right
side are diverging, while those emerging on its inner left side are
converging. The acceleration due to this convergence leads to an
overtaking collision that imparts rightward momentum to the diverging
peakon contact segments. By cylindrical symmetry about the $x$ axis, the
bottom and back panels show the same motion. The left panels show how this
cylindrically symmetric motion moves through the vertical midplane at
$x=0$. The midplane $z=0$ is invariant and the motion in this plane mimics
the motion observed in the corresponding 2d problem.

This shared behavior
is one of the main conclusions of this paper: 3d numerical results have
planar slices which show the corresponding 2d momentum transfer
behavior, and 2d numerical results have linear slices which show the
corresponding 1d momentum transfer behavior. This reduction principle
allows the complex interactions of contact wave surfaces to be analyzed
as elastic collisions showing local momentum transfer. This is a feature
of the collective behavior of singular solutions, because such solutions
possess no internal degrees of freedom.

\figfourthree{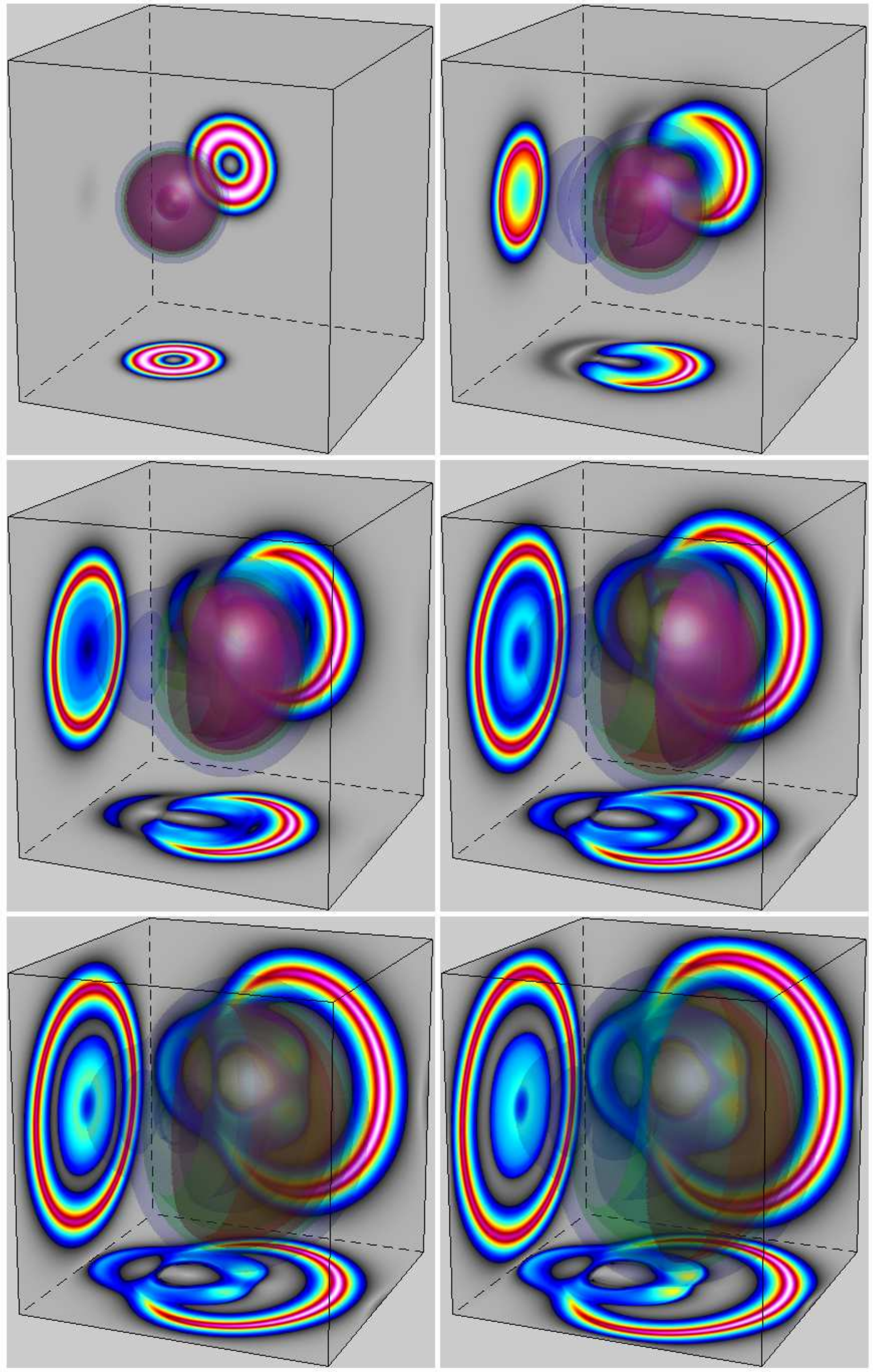}{3d}{right}{\sigma  }{}
\figfourthree{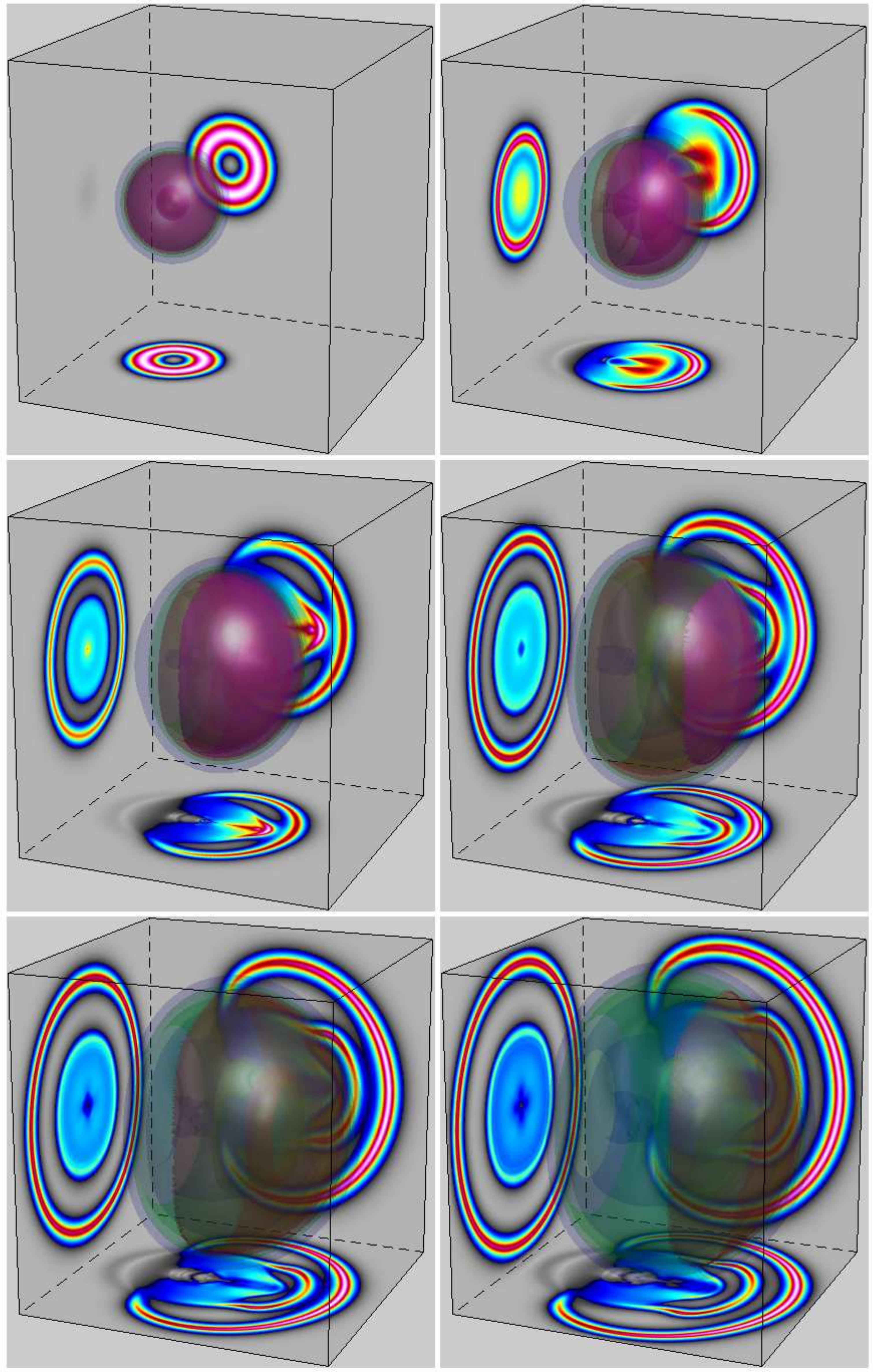}{3d}{right}{\sigma/2}{}

\subsection{Inout}\label{3d-inout}

In Figures \ref{3d-inout-sigma-1} and \ref{3d-inout-sigma-2}, a 3d Gaussian shell of speed is
initially given the 2d ``inout'' velocity distribution,
$-(\sin\theta,\cos\theta)\exp(-(r-r_0)^2/\sigma^2)$, on each horizontal
plane, weighted by the cosine of the polar angle, as in the rigid rotation
simulation. Consequently, the bottom panel shows the same ``in-out'' 2d
motion as before, now at one-fourth the resolution and only for
$\alpha=\sigma/2$ and $\alpha=\sigma$. The left and back panels show
identical motion, consisting of head-on collisions in the center and a
wedge-like collision at the top and bottom. The bottom panel for
$\alpha=\sigma/2$ seems to agree better with the corresponding 2d case,
than for $\alpha=\sigma$, which has less of a collision that in the 2d
case.

\figfourthree{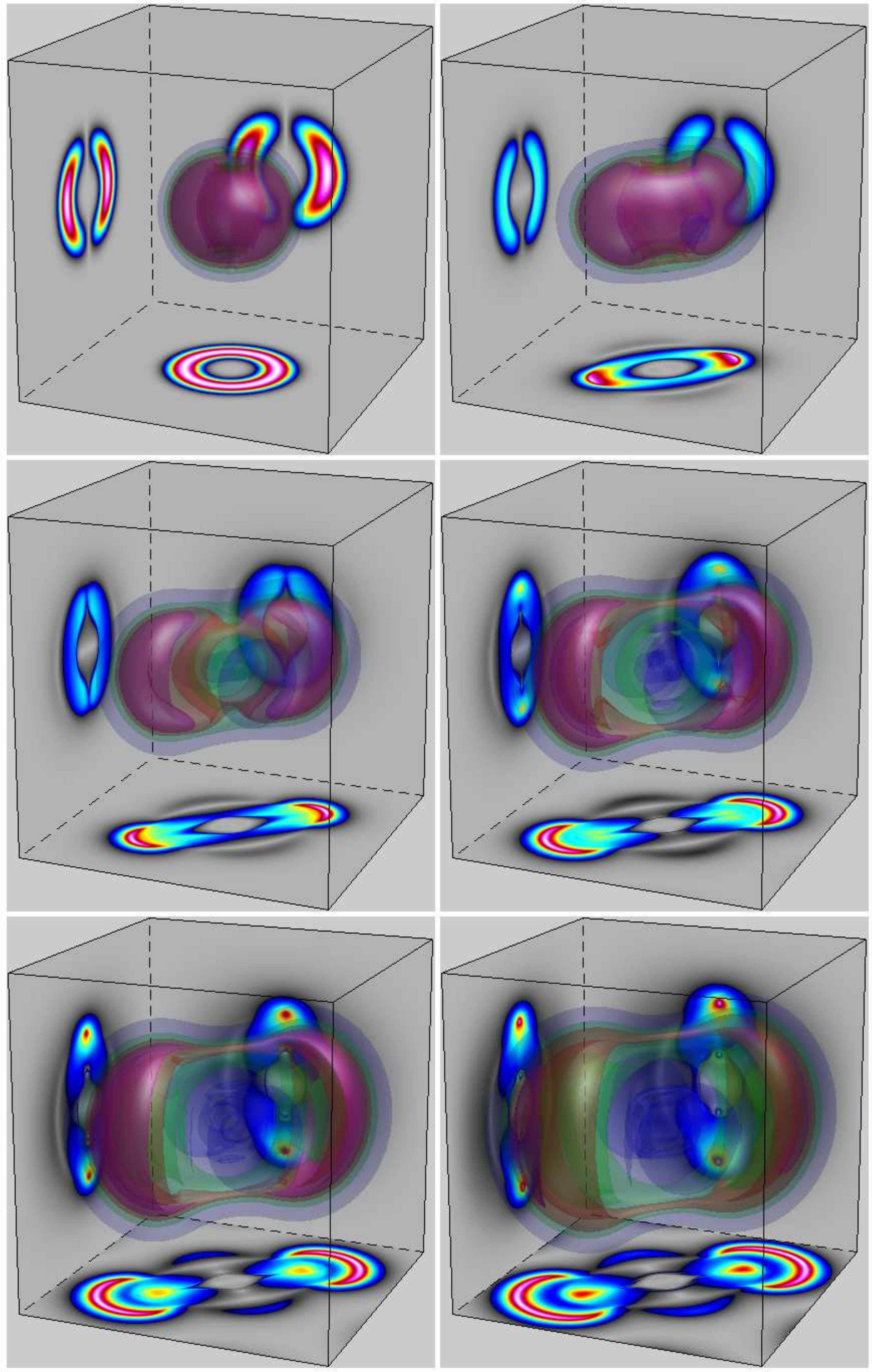}{3d}{inout}{\sigma  }{}
\figfourthree{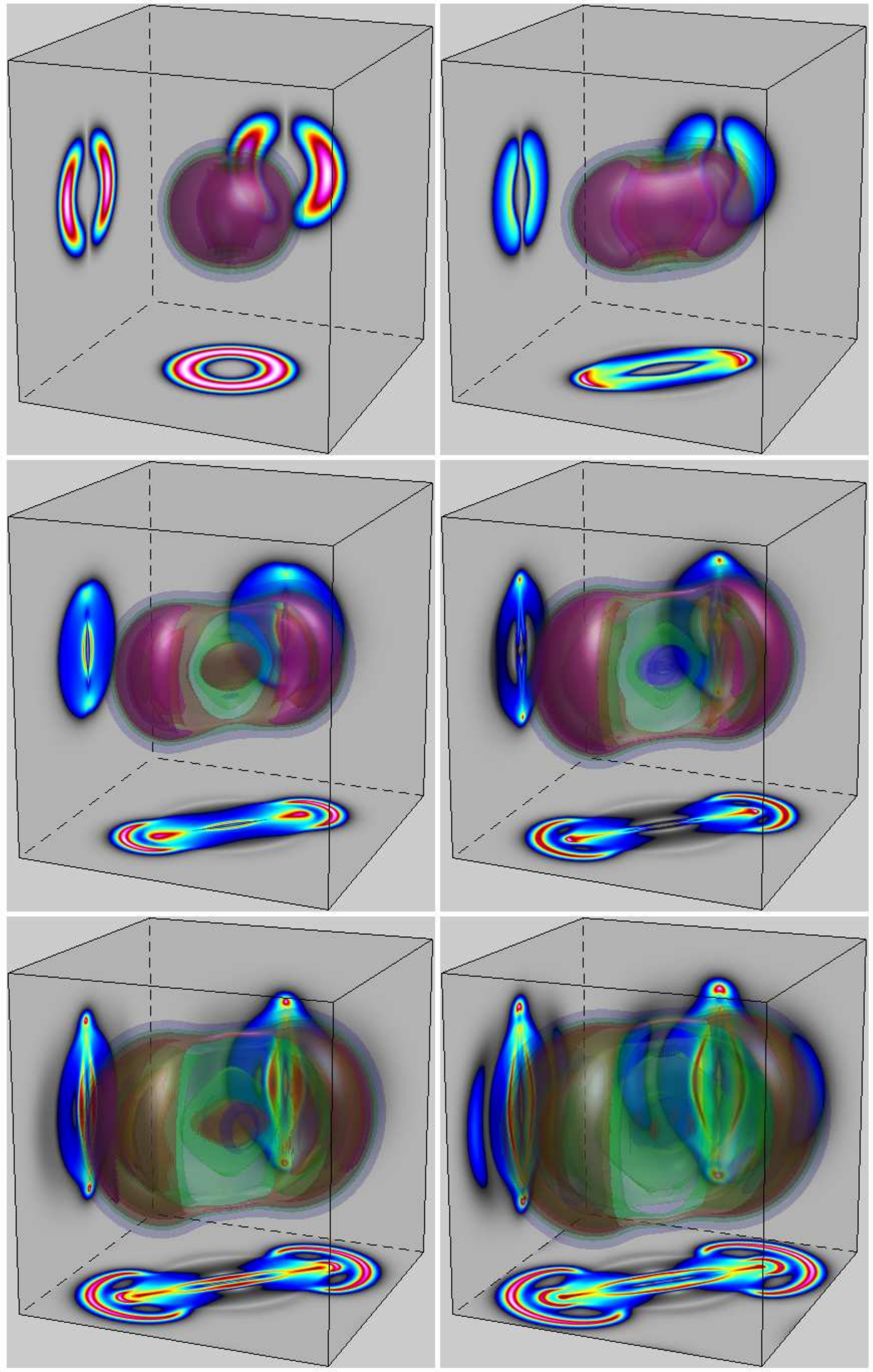}{3d}{inout}{\sigma/2}{}

\subsection{Wheel}\label{3d-wheel}

In Figures \ref{3d-wheel-sigma-1} and \ref{3d-wheel-sigma-2}, the speed is initially distributed as
a Gaussian within a toroidal annulus rotating rigidly about its axis of
rotational symmetry, oriented along the $x$ axis.
This ``wheel'' initial distribution has reflection symmetry about its axis
of rotational symmetry, oriented along the
midplane $x=0$. The motion on this invariant symmetry plane $x=0$ shows
circularly symmetric collapse and radial expansion, seen in the left
panel. (Some distortion is seen upon reflection from the $x$ axis of
cylindrical symmetry, again due to mesh effects.)

The identical back and bottom panels show a head-on collision occurring
on the horizontal midplane, followed by re-emergence in 3d for
$\alpha=\sigma/2$ and the development of a pair of strong hot spots
(actually a funnel shape in 3d) followed by emergence of a perimeter of
peakon contact segments surrounding the interior region.

\figfourthree{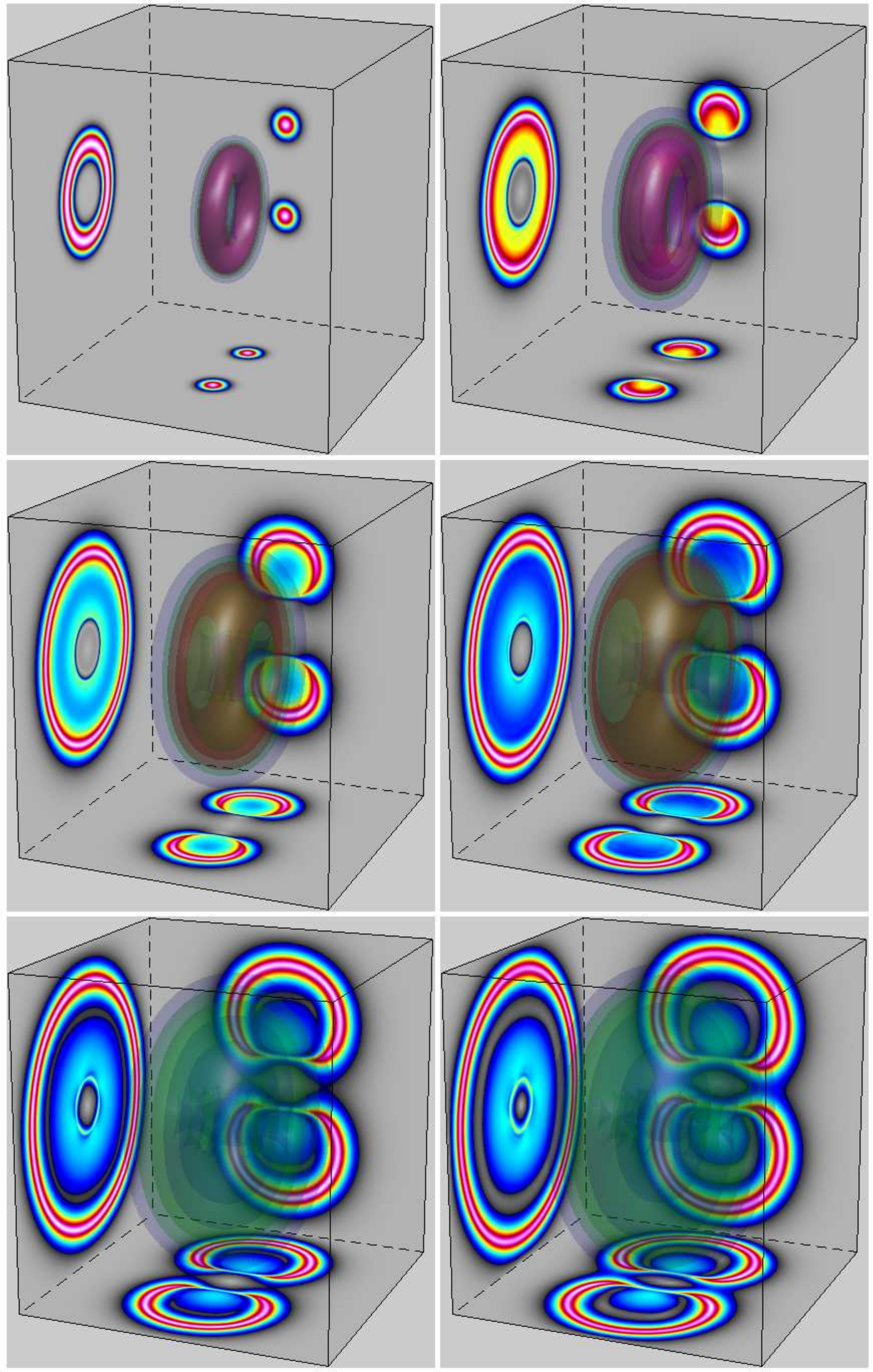}{3d}{wheel}{\sigma  }{}
\figfourthree{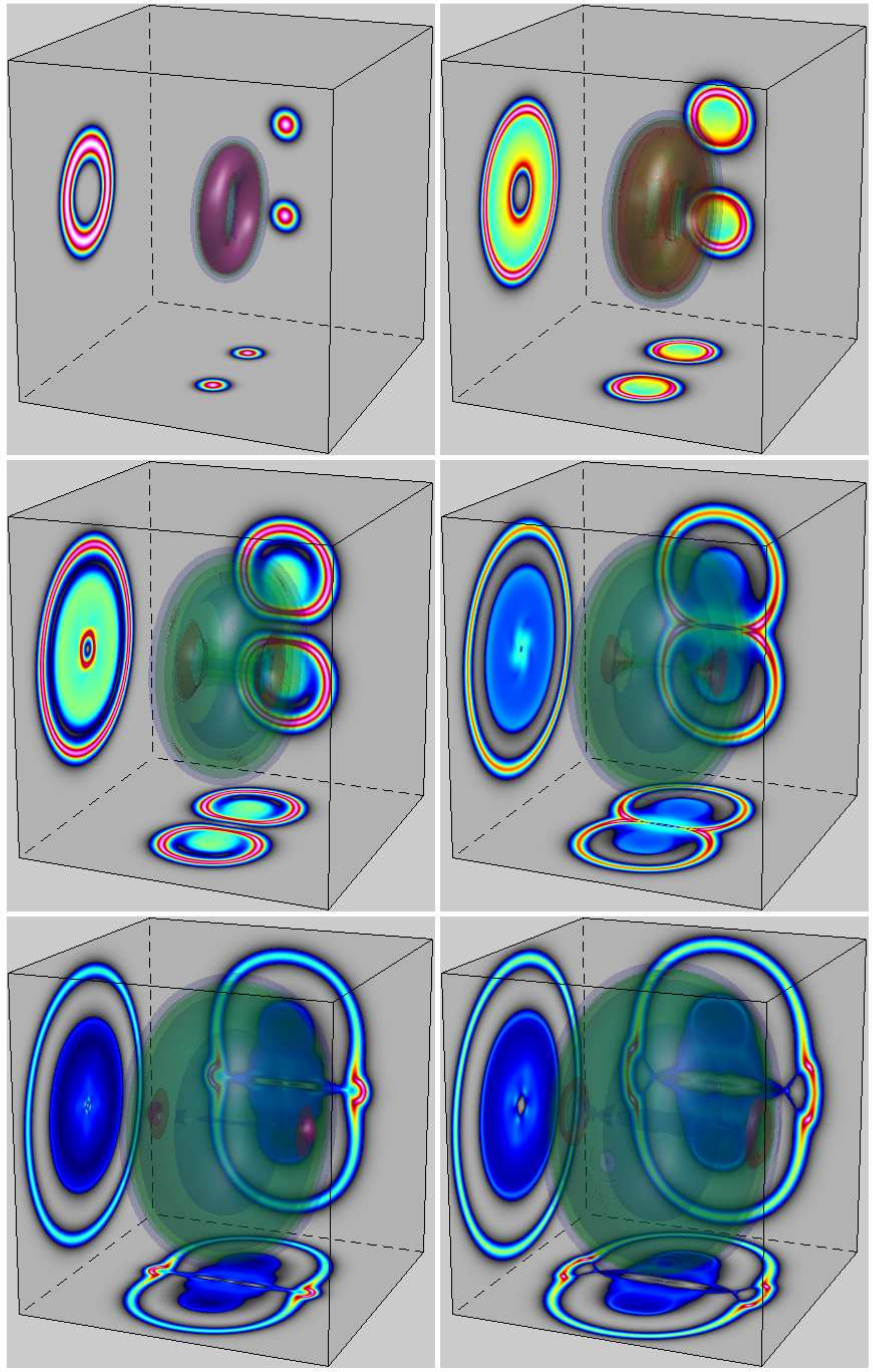}{3d}{wheel}{\sigma/2}{}

\subsection{Wheels}\label{3d-wheels}

Figures \ref{3d-wheels-sigma-1} and \ref{3d-wheels-sigma-2} show simulations of the interactions of
two coaxial tori (wheels) of velocity, both rotating about the $x$ axis,
but offset from each other in the $x$ direction. The cylindrically
symmetric evolution from this initial state shows axial convergence,
radial expansion and reconnection of peakon contact surfaces, via a series
of head-on collisions. The exchange of momentum in these
interactions is especially dramatic. Eventually, an outward-expanding
peakon contact surface emerges and surrounds the entire interaction region.

\figfourthree{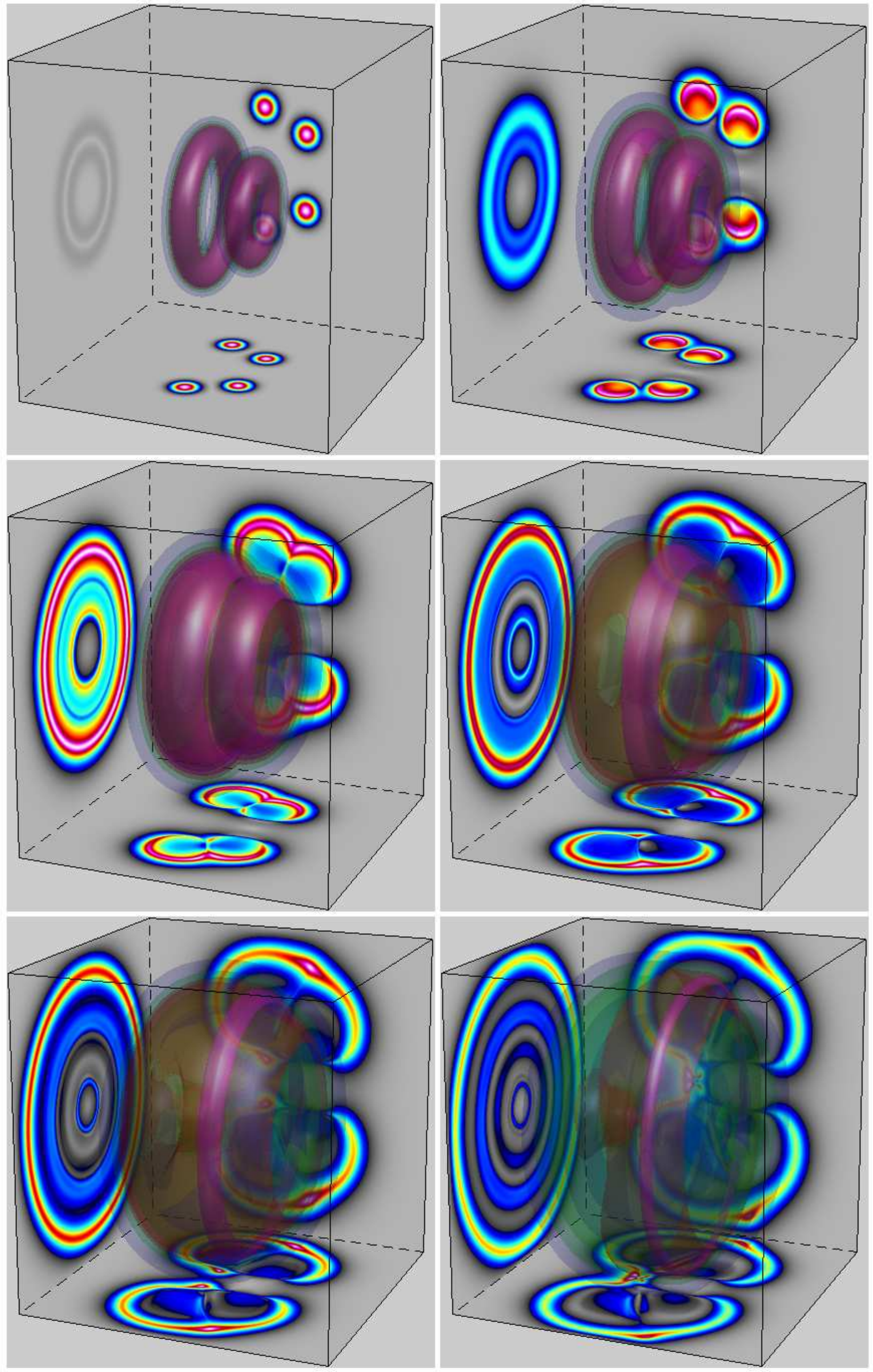}{3d}{wheels}{\sigma  }{}
\figfourthree{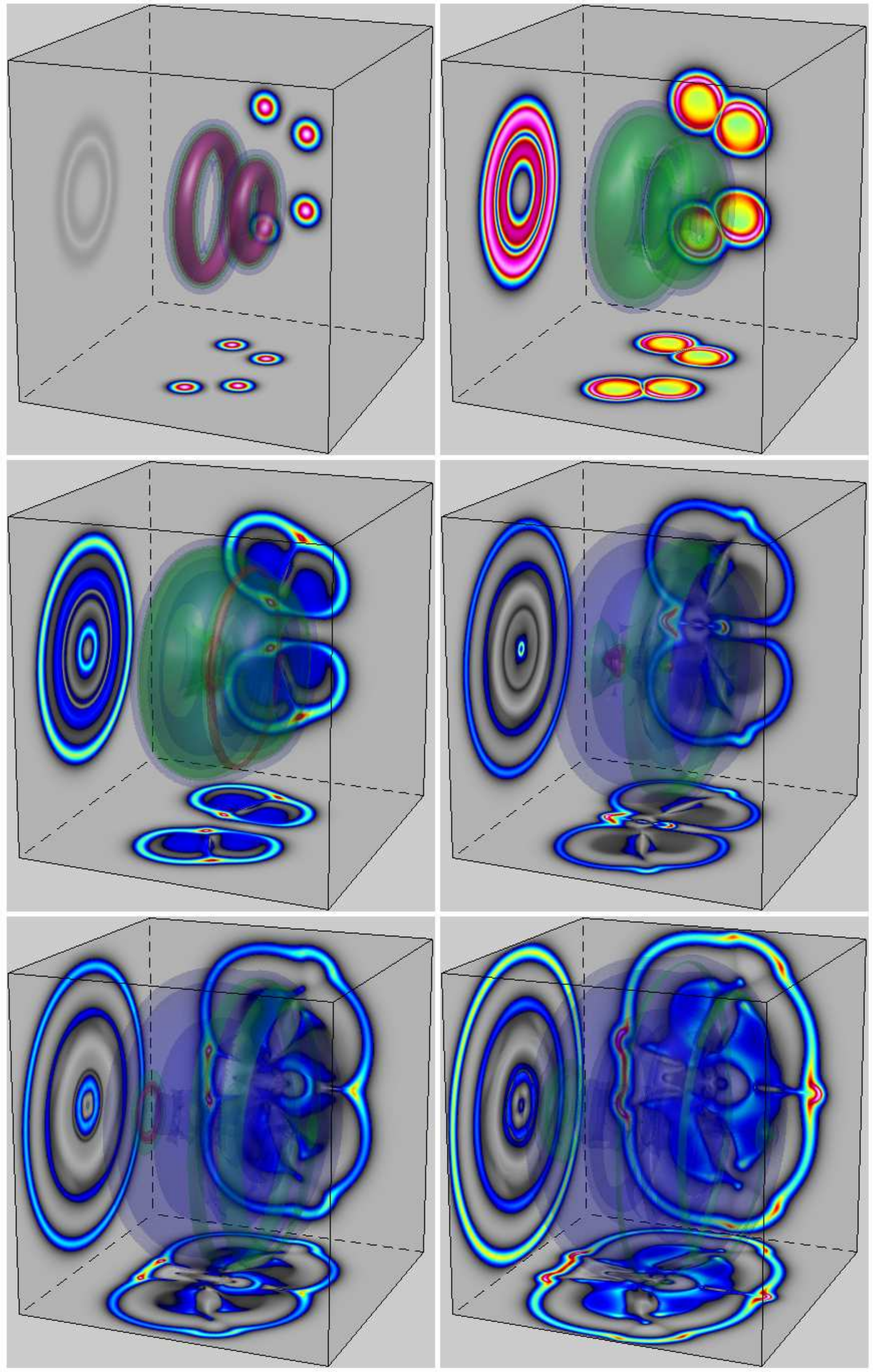}{3d}{wheels}{\sigma/2}{}

\subsection{Torus}\label{3d-torus}

In Figures \ref{3d-torus-sigma-1} and \ref{3d-torus-sigma-2}, a torus-shaped Gaussian velocity
distribution (wheel) is initially moving uniformly rightward in the
direction of its symmetry axis. Cylindrically symmetric rightward moving
fronts form, then expand, collide, and reconnect at the axis of symmetry.
This is a ``cylindrically symmetric wedge'' collision that funnels into
the axis, then forms jets that accelerate in both forward and backward
directions along the $x$ axis.

The three dimensional images of peakon contact surfaces are particularly
vivid in the animations to which Figures \ref{3d-torus-sigma-1}
and \ref{3d-torus-sigma-2} belong. Because of the cylindrical symmetry, the
back and bottom panels show the same motion in different perspectives. The
left panel also shows the collapse to the axis of symmetry. Because of the
curvature of the peakon contact surfaces, this cylindrical collision
imparts axial momentum both forward and backward, which is especially clear
in the case $\alpha=\sigma$/2.

\figfourthree{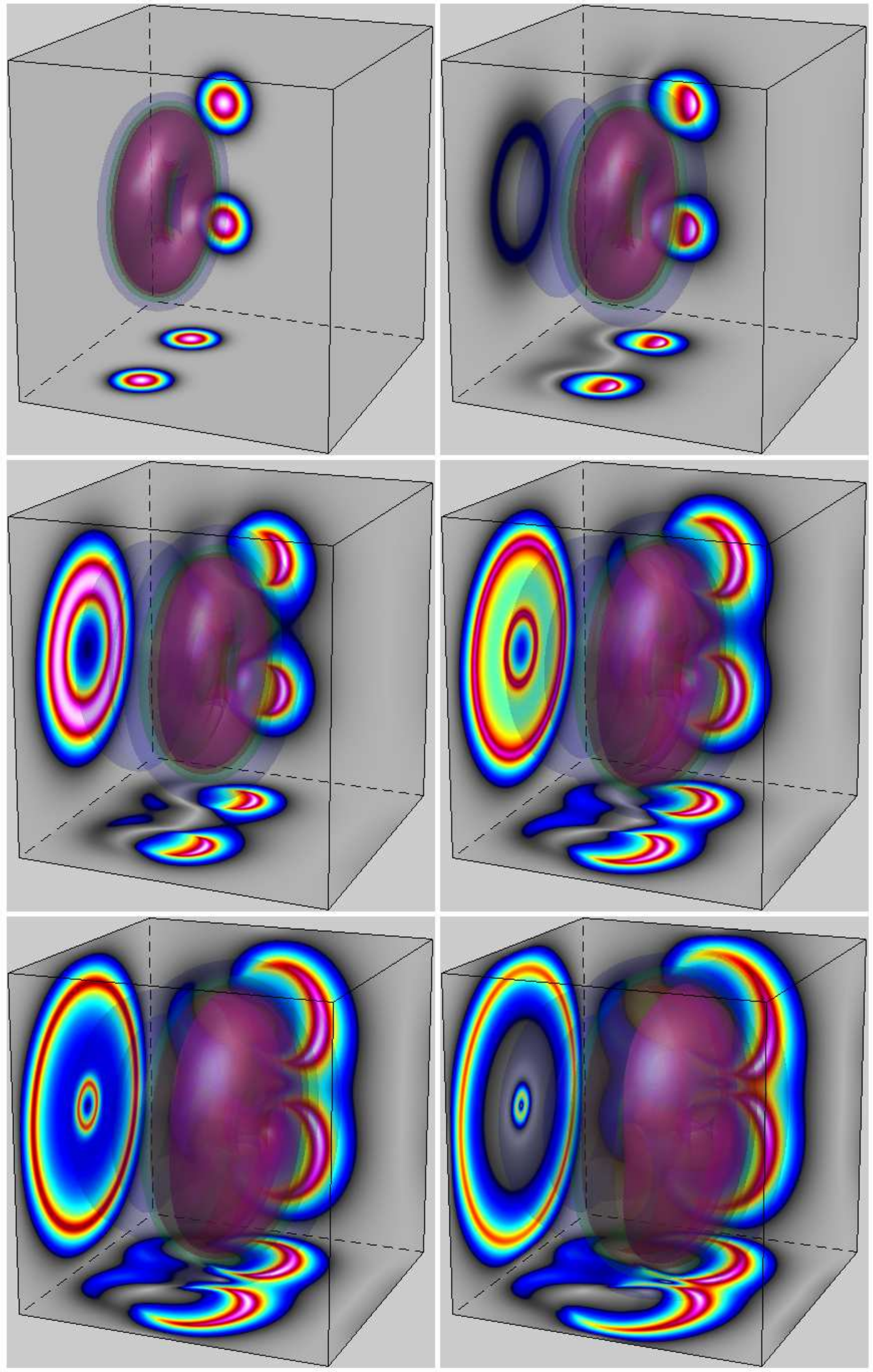}{3d}{torus}{\sigma  }{}
\figfourthree{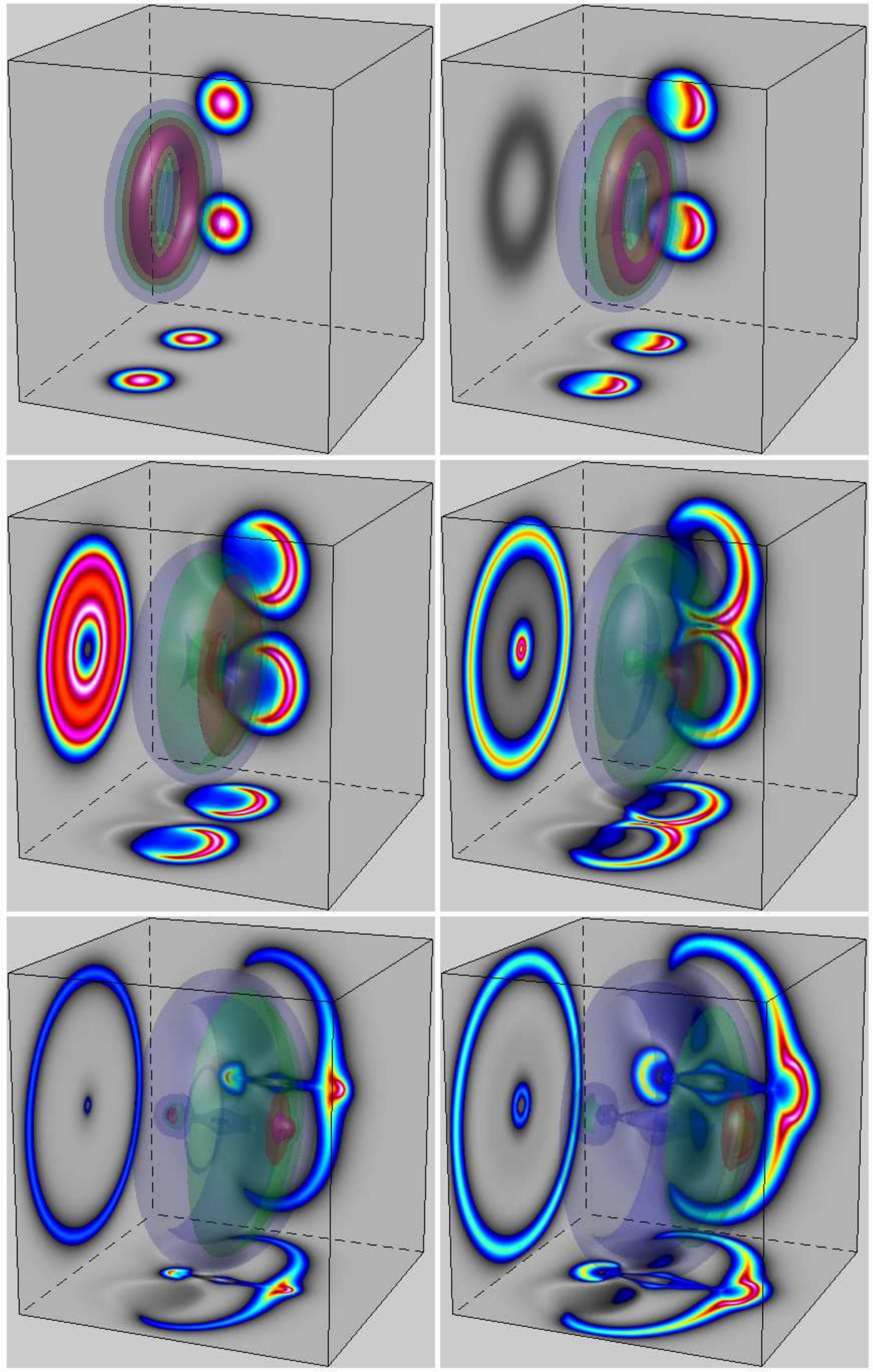}{3d}{torus}{\sigma/2}{}

\subsection{Tori}\label{3d-tori}

In Figures \ref{3d-tori-sigma-1} and \ref{3d-tori-sigma-2}, two linked tori of speed are started along
their axes of rotational symmetry in orthogonal directions, one rightward and
one upward. They undergo a series of wedge-like collisions leading to many
reconnections. They also undergo overlapping collisions that impart momentum,
but do not reconnect the peakon contact segments. Eventually, a single outward
moving peakon contact segment will surround the interaction region.

\figfourthree{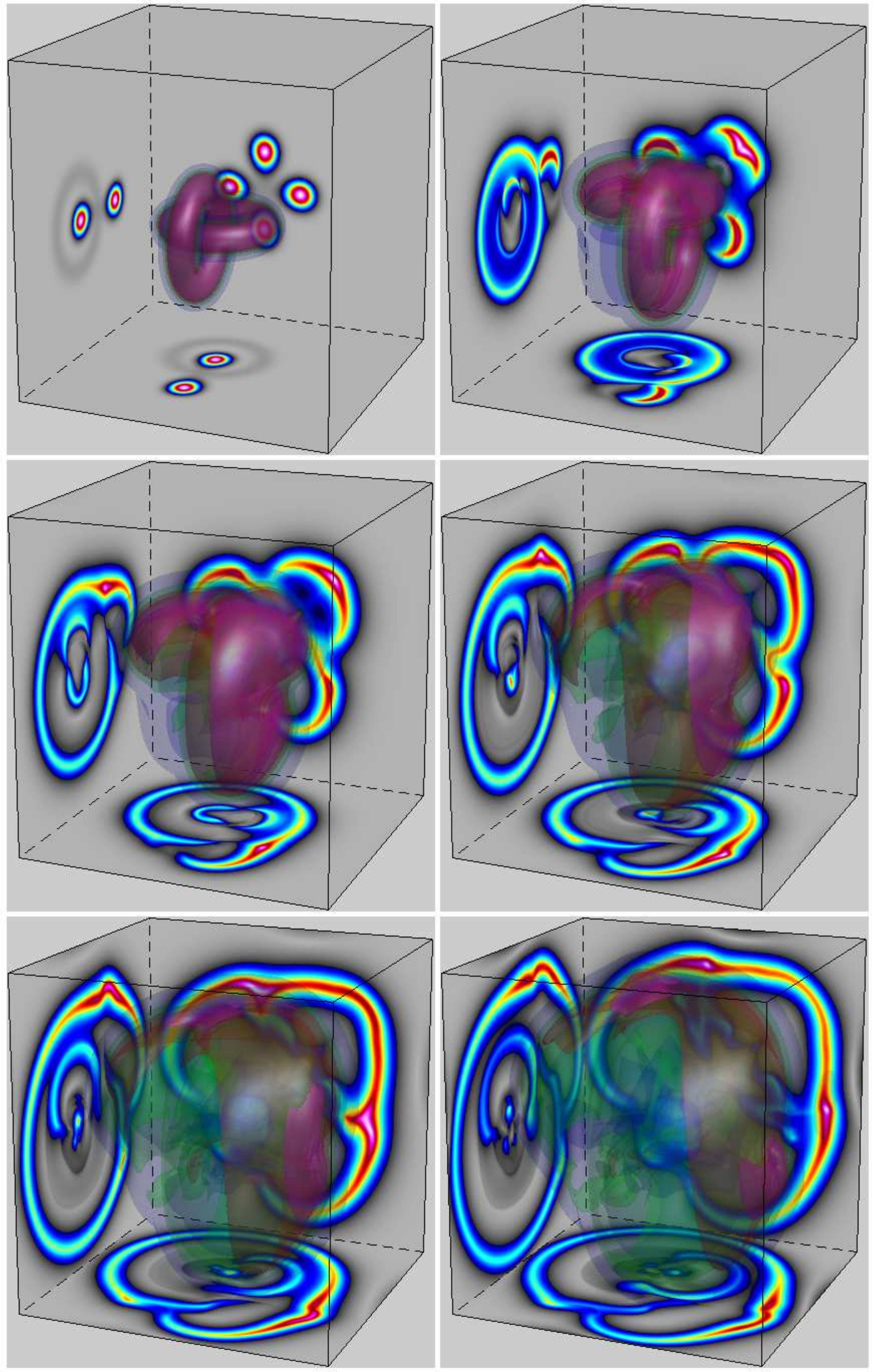}{3d}{tori}{\sigma  }{}
\figfourthree{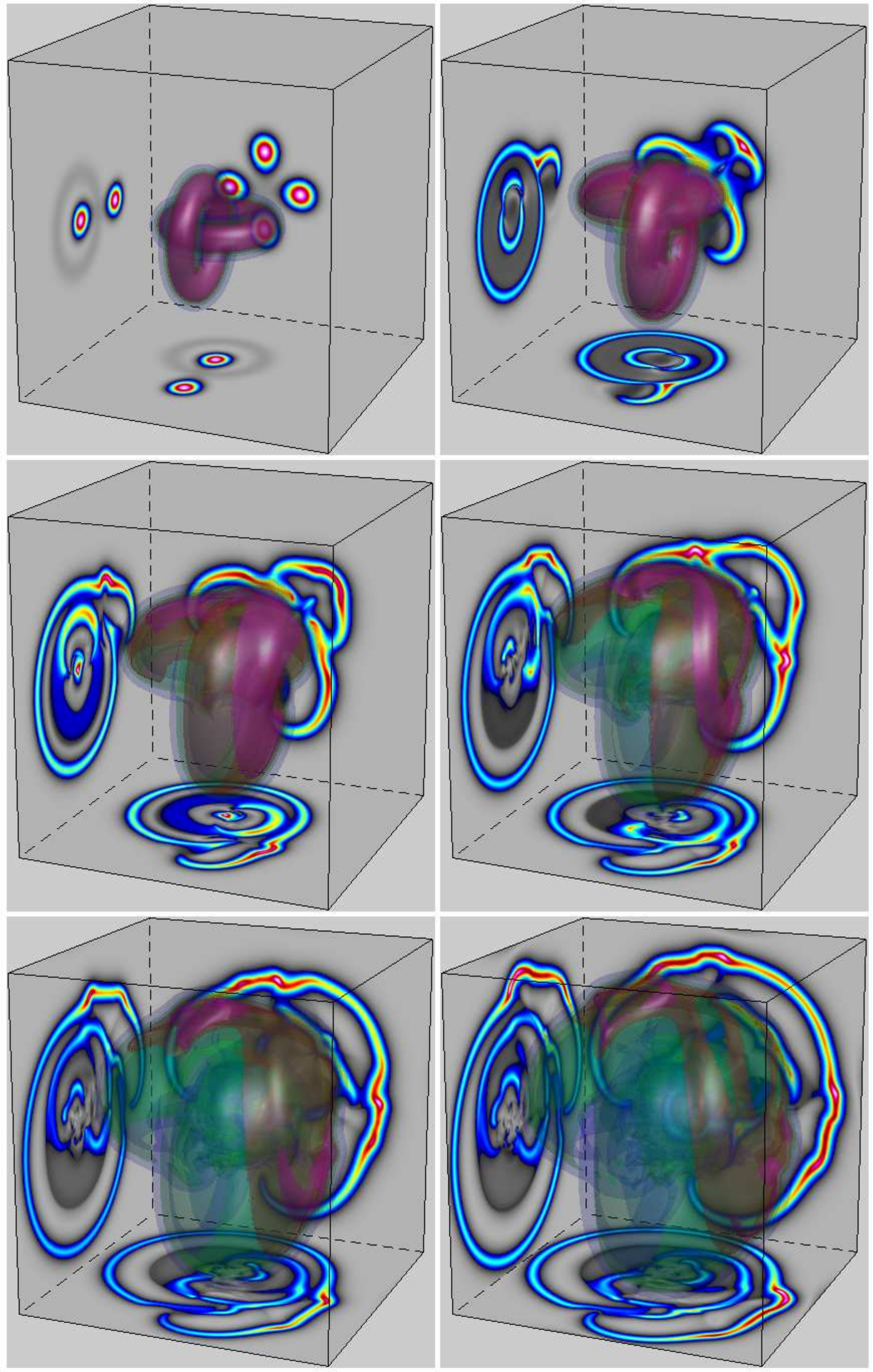}{3d}{tori}{\sigma/2}{}

\subsection{Time reversal}\label{3d-time-rev}

Table \ref{norm-2d} summarizes how well the initial velocity profiles
reconstitute upon reversing the EPDiff evolution from the final states.
Numbers in each table entry are the $L^1$ norm, $L^2$ norm, and max norm of
the difference between the initial velocity profile and its reconstituted
(time reversed) value.
We do not show contour plots of the time reversed runs in 3d because they
are all visually indistinguishable at the larger values of $\alpha$ chosen for
our 3d simulations.
The time reversed values are typically accurate to within one percent or
less.

\begin{table}
\caption{
$L^1$, $L^2$, and max norm of the difference between the initial velocity
profile and its reconstituted (time reversed) value for each 3d simulation.
\label{norm-3d}}
\begin{tabular}{|l|l|l|}
\hline
Simulation & $\alpha=\sigma$ & $\alpha=\sigma/2$\\
\hline
           & {\tt 0.000105} & {\tt 9.26e-05} \\
plate      & {\tt 0.000595} & {\tt 0.000553} \\
           & {\tt 0.0109  } & {\tt 0.01    } \\\hline
           & {\tt 0.000165} & {\tt 0.000146} \\
parallel   & {\tt 0.000686} & {\tt 0.00064 } \\
           & {\tt 0.0109  } & {\tt 0.01    } \\\hline
           & {\tt 9.81e-05} & {\tt 8.68e-05} \\
skew       & {\tt 0.000435} & {\tt 0.000411} \\
           & {\tt 0.00793 } & {\tt 0.00753 } \\\hline
           & {\tt 0.00029 } & {\tt 0.000135} \\
wedge      & {\tt 0.000513} & {\tt 0.000398} \\
           & {\tt 0.00545 } & {\tt 0.00522 } \\\hline
           & {\tt 0.000193} & {\tt 0.000178} \\
head-on    & {\tt 0.000772} & {\tt 0.000738} \\
           & {\tt 0.00885 } & {\tt 0.00829 } \\\hline
           & {\tt 9.36e-05} & {\tt 6.13e-05} \\
rotate     & {\tt 0.000151} & {\tt 0.000102} \\
           & {\tt 0.000625} & {\tt 0.000433} \\\hline
           & {\tt 6.08e-05} & {\tt 4.01e-05} \\
right      & {\tt 0.000128} & {\tt 8.88e-05} \\
           & {\tt 0.00131 } & {\tt 0.00103 } \\\hline
           & {\tt 9.4e-05 } & {\tt 6.19e-05} \\
inout      & {\tt 0.000151} & {\tt 0.000103} \\
           & {\tt 0.00063 } & {\tt 0.000436} \\\hline
           & {\tt 5.87e-05} & {\tt 3.98e-05} \\
wheel      & {\tt 0.000176} & {\tt 0.000124} \\
           & {\tt 0.0019  } & {\tt 0.00138 } \\\hline
           & {\tt 0.000123} & {\tt 8.51e-05} \\
wheels     & {\tt 0.000259} & {\tt 0.000184} \\
           & {\tt 0.00193 } & {\tt 0.0014 }  \\\hline
           & {\tt 7.92e-05} & {\tt 5.45e-05} \\
torus      & {\tt 0.000109} & {\tt 7.87e-05} \\
           & {\tt 0.000523} & {\tt 0.000391} \\\hline
           & {\tt 0.00011 } & {\tt 7.48e-05} \\
tori       & {\tt 0.000248} & {\tt 0.000173} \\
           & {\tt 0.0019  } & {\tt 0.00137 } \\\hline
\end{tabular}
\end{table}

\section{Conclusions, future directions, and outstanding problems}
\label{future-direx-sec}

By a sequence of approximations applied to the Euler-Poincar\'e
variational principle for the multilayer columnar motion (MLCM) of an
incompressible fluid, we derived a hierarchy of EP equations. Several new
MLCM equations belong to this hierarchy, as well as the standard Boussinesq
equations for shallow water waves, which are recovered upon specialization
to weak nonlinearity.

The EPDiff equation (\ref{H1-EP-eqn-Intro}) was derived in the limiting
case of a single layer undergoing strongly nonlinear motion in the absence
of linear dispersion. In 1d, the EPDiff equation restricts to the
dispersionless case of the CH equation for nonlinear shallow water waves.
The dispersionless CH equation possesses singular soliton solutions whose
velocity possesses a sharp peak (jump in slope) moving at a speed equal to
its height. The corresponding momentum for a train of such peakons is a set
of delta functions at the locations of the peaks in velocity. Thus, this
momentum density is distributed as points on the line which evolve under
the action (by $\hbox{Ad}^*$) of the diffeomorphisms (smooth invertible
maps).

\paragraph{A geometrical version of the soliton paradigm.}
The local description of the $\hbox{Ad}^*$ action of the smooth invertible
maps is the EPDiff $\hbox{ad}^*$ equation (\ref{H1-EP-eqn-Intro}), which
holds in {\it any} number of dimensions. Hence, EPDiff  allows comparison
of the behavior of its singular solutions in 1d, 2d and 3d. Numerically,
the singular solutions of EPDiff of codimension one (points on the line,
curves on the plane, surfaces in a volume) are found to be stable.
Moreover, they are found to dominate the intial value problem, essentially
by following the soliton paradigm. Namely, the ``singular solution
content'' of a given initially continuous distribution of velocity emerges
under the evolution of EPDiff and retains its integrity under collision
interactions. In 1d, these singular solutions are true solitons (the
peakons) for the dispersionless CH shallow water equation. In this case,
the inverse scattering transform for CH determines its soliton content for
an arbitrary initial condition and also gives the soliton collision
rules. In 2d and 3d, the theory for determining how these singular
solutions will emerge and how they will survive their collisions has not
yet been developed. This is the major open problem for EPDiff. One clue for
approaching it is to keep in mind that the singular solutions in
(\ref{EP-sing-mom}) form an {\it invariant manifold} for the EPDiff
equations.

\paragraph{Our numerical findings.}
We studied the initial value problem for EPDiff in various scenarios in 2d
and 3d. We found that the key solution behavior of the IVP for EPDiff is
the breakup of an initially smooth confined velocity distribution into
singular solutions supported on codimension-one delta function densities
moving with the velocity of the flow. We extended the solution ansatz
for peakon momentum density supported on points on the line to the
case of singular momentum density supported on smoothly embedded sets,
e.g.~curves in the plane, or surfaces in three dimensional space. We showed
that this singular solution ansatz for EPDiff reduces to Hamilton's
canonical equations for the vector parameters defining these surfaces. The
underlying geometrical reason why this reduction occurs was explained in
Holm \& Marsden \cite{HoMa2004} by recognizing that the singular solution
ansatz (\ref{H1-EP-eqn-Riemann}) is a momentum map for the (right) action of
diffeomorphisms on distributions defined as smoothly embedded subspaces of
a manifold. Thus, the singular solutions evolve by $\hbox{Ad}^*$ action on
embedded subspaces along a curve $g(t)$ in the diffeomorphisms which is a
geodesic path. As we explained, such a curve is a geodesic if and only if
the corresponding momentum satisfies the EPDiff equation
(\ref{H1-EP-eqn-Intro}).

Remarkably, our numerical results showed that only the codimension-one
singular solutions emerge in the IVP. Being defined on delta functions,
these solutions have no internal degrees of freedom. Consequently, their
local interactions may be characterized as elastic collisions of contact
surfaces in which momentum is exchanged. Across these contact surfaces, the
slope of the velocity has a jump which moves with the flow. The collision
rules for these interactions in 2d and 3d may be built up from the soliton
collision rules in 1d. That is, a linear section transverse to a 2d
solution shows 1d elastic momentum exchange behavior, and a planar section
transverse to a 3d solution shows the corresponding 2d behavior. This
reduction to lower dimensional behavior holds especially well on
reflection-invariant sections. For example, the midplane $z=0$ is invariant
in Figure
\ref{3d-right-sigma-2} and the motion in this plane projected on the
bottom panel of Figure \ref{3d-right-sigma-2} mimics the motion observed in
the corresponding 2d problem in Figure \ref{2d-right-sigma-2}.

Thus, 3d numerical results have planar slices which show the corresponding
2d momentum transfer behavior. Likewise, 2d numerical results have linear
slices which show the corresponding 1d momentum transfer behavior. Such a
reduction principle allows the complex interactions of contact wave
surfaces to be analyzed as elastic collisions showing local momentum
transfer. This principle for collective behavior based on simple momentum
exchange in collision interactions arises for the singular solutions,
because such solutions possess no internal degrees of freedom.

Two new numerical features emerge in 2d and 3d. These are the reconnections
of the wavefronts, which occur due to momentum exchange in these nonlinear
collisions, and the remarkable ``memory wisps'' that arise to guarantee
reversibility of those collisions. (These memory wisps are a bit
reminiscent of the emission of neutrinos, which preserve detailed balance
in beta decay.) The memory wisp feature of the reconnections remains to be
explained in more detail, both numerically and analytically.

\paragraph{EPDiff applications: solitons, turbulence and
medical images.} From the viewpoint of nonlinear PDE analysis, the EPDiff
equation, being nonlinear and nonlocal, escapes classification. Its
nonlocality requires solving an elliptic problem for determining its
velocity from its momentum at each time step. It's worth repeating that the
EPDiff nonlinearity is geometric, because this is the key to understanding
its motion. Namely, it is reversible geodesic motion in the diffeomorphisms
acting on smoothly embedded subspaces. Physically, the singular solutions
are contact surfaces (jumps in the velocity derivative that move with the
flow). The corresponding EPDiff equation on the volume-preserving
diffeomorphisms is the Lagrangian Averaged Euler (LAE-$\alpha$) equation
derived first in \cite{HoMaRa1998}, which was first derived as a geometrical
extension of the CH equation. Upon adding Navier-Stokes viscosity, this
became the LANS-$\alpha$ model for incompressible turbulence in
\cite{Chen-etal1998}. For a review of the properties of LANS-$\alpha$
solutions and their relation to Navier-Stokes analysis, see
\cite{FoHoTi2001} and
\cite{MaSh2001}.

By a remarkable coincidence, L.~Younes derived the EPDiff equation
(\ref{H1-EP-eqn-Intro}) as the evolution for the template matching approach
in medical imaging in \cite{Yo1998}. See also D. Mumford's discussion of
the same problem in \cite{Mu1998}. Thus, the singular solutions we discuss
here, and their momentum exchange paradigm, should be expected to develop
increasing interest in medical imaging science. Medical imaging must solve
an optimization problem rather than the IVP. However, our solution of the
IVP may help guide intuition in medical imaging, especially the idea of
encoding information in an image by its momentum, rather than just by the
positions of its outlines and their associated intensities. Thus, EPDiff
represents a crossroads of endeavor in mathematics where methods of fluid
dynamics and imaging science may transfer technologies. See
\cite{HoRaTrYo2004} for more discussions of this new paradigm in image
processing.

\paragraph{Many open questions remain.}
Many open problems and other future applications remain for the EPDiff
equation. For example, its analysis requires additional developments of PDE
methods. In particular, while its smooth solutions satisfy a local
existence theorem analogous to the Ebin-Marsden theorem for the Euler fluid
equations, its IVP inevitably develops singular solutions. The implications
of this observation are mentioned in \cite{HoMa2004} as perhaps indicating
an incompleteness of the geodesic flows on the diffeomorphisms, which opens
many future opportunities for analysis of the emergence of measure values
solutions from smooth initial conditions in nonlinear nonlocal PDEs.

In addition, many open questions remain for the practical problem of
internal wavefronts emerging in shallow water dynamics. For example, there
remains the issues of boundary and topography interactions, including
diffraction and refraction. Moreover, for the applications in turbulence
modeling there remains a variety of open questions about singular vortex
interactions, which are responsible for the famous vortex stretching that
drives the cascade of energy and vorticity in turbulence. A host of other
problems also remains for the applications of EPDiff in medical imaging,
particularly for the statistical treatment of the information encoded in
the linear space of their momenta. Finally, the development of numerical
approaches that are fully capable of tracking the singular solutions of
EPDiff, perhaps by using geometrical methods which incorporate discrete
exterior calculus and variational multisymplectic integration methods. See
\cite{DeHiLeMa2004} and \cite{LeMaOrWe2003} for descriptions of these
promising methods, which seem to lie on the horizon for the next
applications of singular solutions.

\subsection*{Acknowledgements}

We are grateful to many friends and colleagues for their enormous help and
encouragement during the course of this work. These include R.~Lowrie and
B.~A.~Wingate, whose encouragement inspired the beginning of this endeavor
in the context of Lagrangian averaged models of turbulence.
We are also grateful to R.~Camassa, A.~Hirani, M.~Leok, S. T. Li,
J.~E.~Marsden, M.~I.~Miller, T.~Ratnanather, and L.~Younes for helpful
discussions and challenging observations. Finally, we thank the Los Alamos
Turbulence Working Group (TWG) and the members of the TWG external advisory
committee for continuing discussions, challenges, and encouragement.

This work was supported by the United States Department of Energy under
contracts W-7405-ENG-36 and the Applied Mathematical Sciences
Program KC-07-01-01. We are also grateful for partial funding from
Applied Mathematical Sciences Division, Office of Advanced Scientific
Computing Research, DOE Office of Science.



\begin{thebibliography}{}
\bibitem{footnote0}
This process may be extended to higher moments, at the cost of introducing
additional variables, which we decline.
\bibitem{footnote2}
Contributions arising when integrating by parts vanish, upon requiring the usual
condition that the fluid velocities and their variations are tangential to the
fixed horizontal boundaries.
\end{thebibliography}
\end{document}